\documentclass{article}

\usepackage[letterpaper,textheight=9in,textwidth=6.5in,top=1in,
            headheight=14pt,headsep=25pt,footskip=30pt]{geometry}

\usepackage{fancyhdr}
\newcommand{\shorttitle}{\textit{Neural Estimation of Summary Statistics for SBI}}
\fancyheadoffset{0pt}
\makeatletter

\renewcommand{\normalsize}{%
  \@setfontsize\normalsize\@xpt\@xipt
  \abovedisplayskip      7\p@ \@plus 2\p@ \@minus 5\p@
  \abovedisplayshortskip \z@ \@plus 3\p@
  \belowdisplayskip      \abovedisplayskip
  \belowdisplayshortskip 4\p@ \@plus 3\p@ \@minus 3\p@
}
\normalsize
\renewcommand{\small}{%
  \@setfontsize\small\@ixpt\@xpt
  \abovedisplayskip      6\p@ \@plus 1.5\p@ \@minus 4\p@
  \abovedisplayshortskip \z@  \@plus 2\p@
  \belowdisplayskip      \abovedisplayskip
  \belowdisplayshortskip 3\p@ \@plus 2\p@   \@minus 2\p@
}
\renewcommand{\footnotesize}{\@setfontsize\footnotesize\@ixpt\@xpt}
\renewcommand{\scriptsize}{\@setfontsize\scriptsize\@viipt\@viiipt}
\renewcommand{\tiny}{\@setfontsize\tiny\@vipt\@viipt}
\renewcommand{\large}{\@setfontsize\large\@xiipt{14}}
\renewcommand{\Large}{\@setfontsize\Large\@xivpt{16}}
\renewcommand{\LARGE}{\@setfontsize\LARGE\@xviipt{20}}
\renewcommand{\huge}{\@setfontsize\huge\@xxpt{23}}
\renewcommand{\Huge}{\@setfontsize\Huge\@xxvpt{28}}

\renewcommand{\section}{%
  \@startsection{section}{1}{\z@}%
                {-2.0ex \@plus -0.5ex \@minus -0.2ex}%
                { 1.5ex \@plus  0.3ex \@minus  0.2ex}%
                {\large\bf\raggedright}%
}
\renewcommand{\subsection}{%
  \@startsection{subsection}{2}{\z@}%
                {-1.8ex \@plus -0.5ex \@minus -0.2ex}%
                { 0.8ex \@plus  0.2ex}%
                {\normalsize\bf\raggedright}%
}
\renewcommand{\subsubsection}{%
  \@startsection{subsubsection}{3}{\z@}%
                {-1.5ex \@plus -0.5ex \@minus -0.2ex}%
                { 0.5ex \@plus  0.2ex}%
                {\normalsize\bf\raggedright}%
}
\renewcommand{\paragraph}{%
  \@startsection{paragraph}{4}{\z@}%
                {1.5ex \@plus 0.5ex \@minus 0.2ex}%
                {-1em}%
                {\normalsize\bf}%
}
\renewcommand{\subparagraph}{%
  \@startsection{subparagraph}{5}{\z@}%
                {1.5ex \@plus 0.5ex \@minus 0.2ex}%
                {-1em}%
                {\normalsize\bf}%
}

\renewcommand{\textfraction     }{0.1}
\renewcommand{\floatpagefraction}{0.7}

\newlength{\@abovecaptionskip}
\newlength{\@belowcaptionskip}
\renewenvironment{table}
  {\setlength{\abovecaptionskip}{\@belowcaptionskip}%
   \setlength{\belowcaptionskip}{\@abovecaptionskip}%
   \@float{table}}
  {\end@float}

\renewcommand{\footnoterule}{\kern-3\p@ \hrule width 12pc \kern 2.6\p@}

\def\@listi  {\leftmargin\leftmargini}
\def\@listii {\leftmargin\leftmarginii
              \labelwidth\leftmarginii
              \advance\labelwidth-\labelsep
              \topsep  2\p@ \@plus 1\p@    \@minus 0.5\p@
              \parsep  1\p@ \@plus 0.5\p@ \@minus 0.5\p@
              \itemsep \parsep}
\def\@listiii{\leftmargin\leftmarginiii
              \labelwidth\leftmarginiii
              \advance\labelwidth-\labelsep
              \topsep    1\p@ \@plus 0.5\p@ \@minus 0.5\p@
              \parsep    \z@
              \partopsep 0.5\p@ \@plus 0\p@ \@minus 0.5\p@
              \itemsep \topsep}
\def\@listiv {\leftmargin\leftmarginiv
              \labelwidth\leftmarginiv
              \advance\labelwidth-\labelsep}
\def\@listv  {\leftmargin\leftmarginv
              \labelwidth\leftmarginv
              \advance\labelwidth-\labelsep}

\def\keywordname{{\bfseries \emph{Keywords}}}%
\def\keywords#1{\par\addvspace\medskipamount{\rightskip=0pt plus1cm
\def\and{\ifhmode\unskip\nobreak\fi\ $\cdot$
}\noindent\keywordname\enspace\ignorespaces#1\par}}

\renewcommand{\maketitle}{%
  \par
  \begingroup
    \renewcommand{\thefootnote}{\fnsymbol{footnote}}
    \renewcommand{\@makefnmark}{\hbox to \z@{$^{\@thefnmark}$\hss}}
    \long\def\@makefntext##1{%
      \parindent 1em\noindent
      \hbox to 1.8em{\hss $\m@th ^{\@thefnmark}$}##1
    }
    \thispagestyle{empty}
    \@maketitle
    \@thanks
  \endgroup
  \let\maketitle\relax
  \let\thanks\relax
}

\renewcommand{\@maketitle}{%
  \vbox{%
    \hsize\textwidth
    \linewidth\hsize
    \vskip 0.1in
    \centering
    {\LARGE\bfseries \@title\par}
    \vskip 0.42in \@minus 0.1in
    \def\And{%
      \end{tabular}\hfil\linebreak[0]\hfil%
      \begin{tabular}[t]{c}\bf\rule{\z@}{24\p@}\ignorespaces%
    }
    \def\AND{%
      \end{tabular}\hfil\linebreak[4]\hfil%
      \begin{tabular}[t]{c}\bf\rule{\z@}{24\p@}\ignorespaces%
    }
    \begin{tabular}[t]{c}\bf\rule{\z@}{24\p@}\@author\end{tabular}%
    \vskip 0.25in
  }%
}

\renewenvironment{abstract}
{
  \centerline
  {\large \bfseries \scshape Abstract}
  \begin{quote}
}
{
  \end{quote}
}

\makeatother

\usepackage[utf8]{inputenc}
\usepackage[T1]{fontenc}
\usepackage{textcomp}

\usepackage{amsmath,amssymb,amsfonts,amsthm}
\usepackage{mathtools}
\usepackage{mathrsfs}
\usepackage{extarrows}
\usepackage{nicefrac}

\usepackage{booktabs}
\usepackage{array}
\usepackage{multirow}
\usepackage{longtable}
\usepackage{calc}
\usepackage{siunitx}
\usepackage{graphicx}
\usepackage{float}
\usepackage{caption}
\usepackage{subcaption}

\usepackage{etoolbox}
\usepackage[dvipsnames,svgnames,x11names]{xcolor}
\usepackage[ruled,vlined]{algorithm2e}

\usepackage{natbib}
\usepackage{url}
\usepackage{microtype}
\usepackage{hyperref}
\usepackage{doi}

\hypersetup{
  pdftitle={Neural Estimation of Summary Statistics for Simulation-Based Inference with Non-asymptotic Guarantees},
  pdfauthor={Rui Li, Rong Tang},
  pdfkeywords={M-estimation; Generalization error; Conditional distribution estimation; Curse of dimensionality; Amortized inference},
  colorlinks=true,
  linkcolor={blue},
  filecolor={Maroon},
  citecolor={Blue},
  urlcolor={Blue},
}

\newtheorem{theorem}{Theorem}
\newtheorem{assumption}{Assumption}
\newtheorem{remark}{Remark}
 \newtheorem{property}{Property}
\newtheorem{corollary}{Corollary}
\newtheorem{lemma}{Lemma}

\newtheorem*{definition*}{Definition}

\def\m{\mathcal}
\def\mb{\mathbb}

\def\ms{\mathscr}
 \def\dd{{\rm d}}

\def\wt{\widetilde}
\def\wh{\widehat}
\def\ov{\overline}

\newcommand{\mnorm}[1]{{\vert\kern-0.25ex\vert\kern-0.25ex\vert #1
    \vert\kern-0.25ex\vert\kern-0.25ex\vert}}
\newcommand{\bmnorm}[1]{{\big\vert\kern-0.25ex\big\vert\kern-0.25ex\big\vert #1
    \big\vert\kern-0.25ex\big\vert\kern-0.25ex\big\vert}}
\newcommand{\Bmnorm}[1]{{\Big\vert\kern-0.25ex\Big\vert\kern-0.25ex\Big\vert #1
    \Big\vert\kern-0.25ex\Big\vert\kern-0.25ex\Big\vert}}
\newcommand{\bbmnorm}[1]{{\bigg\vert\kern-0.25ex\bigg\vert\kern-0.25ex\bigg\vert #1
    \bigg\vert\kern-0.25ex\bigg\vert\kern-0.25ex\bigg\vert}}
\newcommand{\BBmnorm}[1]{{\Bigg\vert\kern-0.25ex\Bigg\vert\kern-0.25ex\Bigg\vert #1
    \Bigg\vert\kern-0.25ex\Bigg\vert\kern-0.25ex\Bigg\vert}}

\newcommand{\N}{\mathbb{N}}
\newcommand{\cX}{\mathcal{X}}
\newcommand{\R}{\mathbb{R}}
\newcommand{\E}{\mathbb{E}}

\newcommand{\norm}[1]{\|#1\|}
\newcommand{\abs}[1]{|#1|}

\newcommand{\calN}{\mathcal{N}}

\newcommand{\Xn}{{X}^{(n)}}
 \newcommand{\ttheta}{{\theta}}

\newcommand{\bE}{\mathbb{E}}

\newcommand{\cG}{\mathcal{G}}
\newcommand{\hW}{\widehat{W}}
\newcommand{\hS}{\widehat{S}}
\newcommand{\muws}{\mu_{W}(\cdot| S(\Xn))}

\newcommand{\hmuws}{\mu_{\widehat{W}}(\cdot|\widehat{S}(\Xn))}

\newcommand{\smuws}{\mu_{W^*}(\cdot|S^*(\Xn))}

\newcommand{\post}{p(\theta|\Xn)}
\newcommand{\pos}{\mu^*_{\theta|\Xn}(\cdot|\Xn)}
\newcommand{\mmd}{{\rm MMD}}

\newcommand{\EX}{\mathbb{E}_{\Xn\sim \m P^{(n)}}}

\newcommand{\EXandth}{\mathbb{E}_{\theta\sim \Pi_{\theta},\Xn\sim \m P^{(n)}_{\theta} }}
\newcommand{\ellws}{\ell(S, W, \theta, \Xn)}

\newcommand{\ellhat}{\ell(\hS, \hW, \theta, \Xn)}
\newcommand{\ellhati}{\ell(\hS, \hW, \ttheta_i, \Xn_i)}
\newcommand{\ellstar}{\ell( S^*, W^*, \theta, \Xn)}
\newcommand{\ellstari}{\ell( S^*, W^*, \ttheta_i, \Xn_i)}
\newcommand{\sgws}{g^*_{W,S}}
\newcommand{\sgwsh}{g^*_{\widehat{W},\widehat{S}}}

\newcommand{\epx}{\varepsilon_X}
\newcommand{\epth}{\varepsilon_{\theta}}

\newcommand{\cS}{\ms{S}}
\newcommand{\cW}{\ms{W}}

\makeatletter
\def\codex@tocname{toc}
\newcommand{\appendixstart}{%
  \let\codex@addcontentsline\addcontentsline
  \def\addcontentsline##1##2##3{\codex@addtoc{##1}{##2}{##3}}%
}
\newcommand{\codex@addtoc}[3]{%
  \def\codex@tmp{#1}%
  \ifx\codex@tmp\codex@tocname
    \codex@addcontentsline{apx}{#2}{#3}%
  \else
    \codex@addcontentsline{#1}{#2}{#3}%
  \fi
}
\newcommand{\appendixcontents}{%
  \section*{\appendixname
    \@mkboth{\MakeUppercase\appendixname}{\MakeUppercase\appendixname}}%
  \@starttoc{apx}%
}
\makeatother

\title{Neural Estimation of Summary Statistics \\
       for Simulation-Based Inference \\
       with Non-asymptotic Guarantees}

\author{Rui Li \\
        Department of Mathematics \\
        Hong Kong University of Science and Technology \\
        \texttt{rlico@connect.ust.hk}
        \And
        Rong Tang{\setcounter{footnote}{1}\thanks{Corresponding author.}} \\
        Department of Mathematics \\
        Hong Kong University of Science and Technology \\
        \texttt{martang@ust.hk}}

\begin{document}
\maketitle
\setcounter{footnote}{0}

\begin{abstract}
Simulation-based inference (SBI) enables Bayesian inference when a simulator is available but the likelihood is intractable. Modern SBI methods train neural networks on simulated data to obtain an amortized approximation of the posterior, making inference for new observations fast. A central challenge in SBI is to extract informative, low-dimensional summary statistics; without them, performance can degrade rapidly as the observation dimension grows.
We therefore highlight summary learning as a critical component of SBI and propose a general framework that jointly learns a neural-network–parameterized summary and an amortized posterior estimator.  We further introduce a local refinement strategy that adapts the amortized estimator to a specific observation, thereby reducing the amortization gap.  Both amortized training and local refinement accommodate implicit posterior estimators without tractable densities. On the theoretical side, we analyze the finite-sample behavior of the joint learning procedure through the lens of nonparametric conditional distribution estimation, and show on representative examples how low-dimensional summaries can mitigate the classical curse of dimensionality. Empirically, our results support this motivation and demonstrate favorable performance relative to existing methods.
\end{abstract}




\section{Introduction}

Simulation-Based Inference (SBI) refers to a class of  Bayesian inference methods for models specified by stochastic simulators with intractable likelihoods. In this setting, observed data $\Xn=(X_1,X_2,\cdots,X_n) \in \mathbb{R}^{n \times d_X}$ are generated by a forward simulator $\m P_{\theta}^{(n)}$ parameterized by $\theta$. While we can readily draw samples by running the simulator,   the likelihood function $\m L(\theta;  \Xn)$  cannot be evaluated in closed form \citep{beaumont2010approximate,beaumont2002approximate,lintusaari2017fundamentals}. Given a prior $\Pi_{\theta}$ with density $\pi(\theta)$ and an observation $\Xn_*$, Bayesian inference aims to characterize the posterior distribution $\mu^*_{\theta|\Xn}(\cdot|\Xn_*) \propto \pi(\theta)\m L(\theta;  \Xn_*)$. However, the intractability of the likelihood precludes the use of standard likelihood-based  methods such as Markov Chain Monte Carlo \citep{hastings1970monte,roberts1998optimal}. Instead,  SBI leverages simulated data to train neural networks that map observations to approximate posterior distributions, thereby avoiding explicit likelihood evaluations.

The standard SBI workflow first generates a training set $\{(\theta_j,  X_j^{(n)})\}_{j=1}^N$,  with the parameter $\theta_j\sim\Pi_{\theta}$, and the simulated data $X_j^{(n)}$ generated  from the simulator $\m P_{\theta_j}^{(n)}$. These pairs are then used to train a  network to directly estimate the posterior \citep{papamakarios2016fast}, or alternatively, to estimate the likelihood \citep{papamakarios2019sequential} or likelihood ratio \citep{hermans2020likelihood} for use with standard sampling methods.  In the posterior-estimation paradigm, SBI targets a global inference mapping $\Xn\to \wh \mu(\cdot|\Xn)$ that approximates the true posterior $\mu^*_{\theta|\Xn}(\cdot|\Xn)$.  From a statistical perspective, this corresponds to a conditional distribution estimation of the response $\theta$ given the covariate $\Xn$, based on paired samples $\{(\theta_j,  X_j^{(n)})\}_{j=1}^N$. 
 However, classical minimax theory and associated optimal estimators for nonparametric conditional distribution estimation~\citep{10.1214/23-AOS2270,10.1214/22-EJS2037,tang2025minimax}  typically suffer from the curse of dimensionality,  where the convergence rates deteriorate rapidly as the dimension of the conditioning variable grows. In the amortized SBI setting, this challenge is exacerbated.  The ``input'' to the inference model is the whole dataset $\Xn=\{X_1,X_2,\cdots, X_n\}$, whose dimension is ${\rm dim}(\Xn_*)=nd_X$ (with $d_X$ being the per-observation dimension).
As the sample size $n$ grows, the dimension of the conditioning variable grows linearly, leading to slow convergence from a nonparametric perspective.  
 To mitigate this issue, it is necessary to exploit additional structure of the problem, rather than treating amortized SBI as a fully ``nonparametric'' conditional distribution estimation  task. The prevailing solution in SBI is to introduce a low-dimensional summary statistic $S(\cdot)$~\citep{radev2020bayesflow,jiang2017learning,creel2017neural,barnes2011considerate,chen2021neural,aakesson2021convolutional,hjelm2018learning,pmlr-v202-chen23h} that compresses the high-dimensional data vector $\Xn$ into a low-dimensional summary $S(\Xn)$, and then approximate the target posterior by the summary-conditioned posterior $\mu^*_{\theta|S(\Xn)}(\cdot|S(\Xn))$.

The choice of the summary $S(\cdot)$ is critical, as it balances low dimensionality against informativeness about $\theta$. 
  Since exact sufficiency is generally unattainable, a central topic is the construction of  ``informative'' statistics, as a replacement for the (unattainable) sufficient ones.   Given the training set, we seek to learn a  low-dimensional summary mapping $\wh S(\cdot)$ such that,  on average over $\Xn$,  the summary-conditioned posterior $\mu^*_{\theta|\wh S(\Xn)}(\cdot|\wh S(\Xn))$ is as close as possible to the full-data posterior  $\mu^*_{\theta|\Xn}(\cdot|\Xn)$. In other words, $\wh S(\cdot)$ is trained to preserve as much information about $\theta$ as possible while compressing $\Xn$.

\subsection{Main Contribution}
In this paper, we propose a unified  discrepancy-minimization framework for learning summary statistics. The key idea is to learn a summary map (``encoder'') $S(\cdot)$ by minimizing  a tractable proxy for the expected discrepancy between the full-data posterior (i.e., the distribution of $\theta$ given $\Xn$) and the summary-conditioned posterior (i.e., the distribution of $\theta$ given $S(\Xn)$). The discrepancy can be chosen from several options, including KL divergence, maximum mean discrepancy (MMD) and sliced MMD. To make these objectives estimable from simulations, we introduce an auxiliary neural ``decoder'', commonly called an inference network in the SBI literature, $W(\cdot)$, which is trained jointly with $S(\cdot)$. The learned pair $(\wh S(\cdot),\wh W(\cdot))$, fitted on the training set  $\{(\theta_j,  X_j^{(n)})\}_{j=1}^N$, yields an amortized posterior estimator $\Xn\to\mu_{\wh W}(\cdot| \wh S(\Xn))$  that generalizes across datasets.   A main methodological contribution is to enable implicit inference networks within this framework. With the KL objective and a conditional normalizing-flow inference network, our formulation recovers BayesFlow-style training \citep{radev2020bayesflow}. In contrast, the MMD and sliced-MMD objectives can be estimated directly from samples generated by the  inference network. The inference network therefore need not have a tractable density and may instead be defined by a flexible implicit generator.  Beyond amortized training, we further introduce two complementary local refinement strategies to reduce the amortization gap for a given
observation, which remain compatible with an implicit inference network. The first keeps the learned summary fixed and uses samples from the amortized posterior to initialize multiple ABC--MCMC chains. This places the chains near posterior-relevant regions and can improve the exploration of multiple modes captured by the amortized approximation. The second refinement instead updates the ``amortized'' summary map itself by sequential training, making it  more predictive of  $\theta$ for simulated datasets ``similar to'' the given observation.

  On the theoretical side, we study the finite-sample behavior of the joint summary-learning and amortized posterior-estimation procedures, from the perspective of nonparametric conditional distribution estimation.  We  establish an oracle inequality that provides  a general tool to verify the statistical accuracy  of the amortized estimator and apply the result to two settings: (1) i.i.d. data from exponential families, where sufficient statistics exist, and (ii) i.i.d. data from smooth densities with $n\to \infty$,  where we use the  Bernstein-von Mises theorem  to construct ``approximately sufficient'' statistics.  In both cases, we derive non-asymptotic guarantees with explicit dependence on the training size $N$ and the within-dataset sample size $n$. Crucially, we show that the amortized posterior estimator avoids the classical curse of dimensionality in $n$ (or the dimension of the conditioning variable $\Xn$)  for conditional distribution estimation: to achieve a target accuracy, e.g., $\frac{1}{n}$, measured by expected MMD, the required training size  $N$ grows only polynomially  with $n$.

\subsection{Related Works}\label{sec: Related Works}
Recent advances in SBI have benefited from the  representational capacity of neural networks, which can learn complex mappings from data to inferential targets. Here we focus on Neural Posterior Estimation (NPE), a paradigm that directly approximates the posterior distribution. For a comprehensive overview of other SBI methods, we refer readers to \cite{Deistler2025SimulationBasedIA,zammit2025neural}. The development of NPE methods has closely followed advances in conditional density estimation, from early mixture density networks \citep{papamakarios2016fast,NIPS2017_addfa9b7,greenberg2019automatic} to more expressive architectures like normalizing flows \citep{papamakarios2019sequential}, generative adversarial networks \citep{wang2023adversarialbayesiansimulation}, and score-based generative models \citep{geffner2023compositional,linhart2024diffusion,jiang2025langevin,arruda2026diffusionmodelssimulationbasedinference,sharrock2022sequential}.  A key difficulty arises when the observations are high-dimensional, motivating the common practice of compressing them into low-dimensional summary statistics before inference.

  Many statistical methods have been proposed to  learn informative summary statistics automatically or semi-automatically~\citep{radev2020bayesflow,jiang2017learning,barnes2011considerate,chen2021neural,aakesson2021convolutional,hjelm2018learning,pmlr-v202-chen23h,belghazi2018mutual,pacchiardi2022score,huang2023learning}. To name a few of the most relevant to ours, ~\cite{jiang2017learning} propose using neural approximations of the posterior mean as summaries.  While elegant and simple, this approach can be inadequate in settings where  sufficiently informative  summary statistics are higher-dimensional than the parameter. ~\cite{radev2020bayesflow,radev2023bayesflow} introduce BayesFlow, which comprises a summary network that maps high-dimensional data to a low-dimensional representation and an inference network that maps this representation to an approximate posterior.  Our perspective differs from that of~\cite{radev2020bayesflow} in several important aspects. In particular,  BayesFlow is designed for amortized inference. It learns a  global posterior estimator that works across datasets, with the summary network primarily serving to structure this global  estimator.  By contrast, we view the central object as the learning of a global summary mapping. This mapping is learned with the explicit goal of preserving posterior-relevant information and can subsequently be reused for local refinement of the amortized posterior for a specific observed dataset.  Moreover, BayesFlow can be viewed as a special case of our framework for  learning summaries when the discrepancy between the summary-conditioned posterior and the true posterior is measured by the KL divergence. Our framework naturally accommodates other discrepancies, such as MMD and sliced MMD.  Other approaches learn summaries by maximizing  parameter--summary mutual information through tractable surrogates, including the Donsker–Varadhan representation \citep{belghazi2018mutual} and Jensen–Shannon-based estimators \citep{chen2021neural, hjelm2018learning}. It is worth noting that maximizing the mutual information is equivalent to minimizing the expected KL divergence between the summary-conditioned posterior and the full-data posterior,  linking these approaches to the same ``KL-divergence between posteriors'' paradigm as BayesFlow.

Despite the empirical success of neural methods for SBI,
the available  rigorous theory is still relatively limited in clarifying why these approaches can capture complex data structure, and in characterizing when, and in what sense, they can mitigate the curse of dimensionality in amortized inference.
    Our work contributes to this direction by providing non-asymptotic guarantees for the joint summary and amortized posterior learning procedure.

Besides neural approaches, there is also a rich literature on SBI that does not rely on training neural networks, such as approximate Bayesian computation (ABC) methods with sequential Monte Carlo~\citep{toni2009approximate,del2012adaptive}, discrepancy-based  ABC methods~\citep{wang2022approximate,bernton2019approximate,park2016k2}, and Bayesian synthetic likelihood~\citep{frazier2023bayesian,price2018bayesian,wood2010statistical}. These methods remain   effective in many applications. Neural methods, in contrast, are attractive when amortization across datasets is beneficial, and  they can leverage flexible function approximation to automatically learn rich, data-driven summaries that scale with data complexity.

\subsection{Notation}

Let $\theta \in \Theta \subseteq \mathbb{R}^d$ be the parameter of interest, with prior $\Pi_{\theta}$ and density $\pi(\theta)$. Given $\theta$, the dataset $\Xn = (X_1,\dots,X_n) \in \mathcal{X}^{n} \subseteq \mathbb{R}^{d_X n}$ is generated from the simulator  $\m P_{\theta}^{(n)}$. We treat $\theta$ as the response and $\Xn$ as covariates; their joint distribution is thus $\Pi_{\theta}\m P_{\theta}^{(n)}$, and the marginal, or the prior-predictive, distribution of $\Xn$ is $\m P^{(n)} = \mathbb{E}_{\theta\sim\Pi_{\theta}}[\m P_{\theta}^{(n)}]$. The conditional distribution of $X$ given $Y=y$ is written as $\mu^*_{X|Y}(\cdot|y)$. For a measure $\nu$ on $\mathbb{R}^d$ and a measurable map $G:\mathbb{R}^d \to \mathbb{R}^{d'}$, the pushforward  measure $G_\#\nu$ is defined by $G(z) \sim G_\#\nu$ when $z \sim \nu$. The $n$-fold product of a measure $\mu$ is denoted by $\mu^{\otimes n}$. For a function $f: \mathcal{X} \times \Theta \to \mathbb{R}$, $\nabla_\theta f(x,\theta)$ and $\mathrm{Hess}_\theta f(x,\theta)$ denote its gradient and Hessian with respect to $\theta$. For a function class $\m F$ with a pseudometric $d_N$, we write  $\mathcal{N}(\mathcal{F}, d_N, u)$ for the $u$-covering number of $\mathcal{F}$ with respect to $d_N$, defined as the minimum number of balls of radius $u$ under the pseudometric $d_N$ required to cover $\mathcal{F}$. Generic positive constants are written as $C, c$ and may change from line to line. We also use the following standard notation:  $[a] = \{1,\dots,a\}$; $0_d$ is the $d$-dimensional zero vector; $\mathbb{N}_0^d$ denotes the set of $d$-tuples of non-negative integers; for $j \in \mathbb{N}_0^d$,  we write $|j| = \sum_i j_i$ and denote $\partial^{|j|}f/\partial\theta^j$ the mixed partial derivatives of $f$; $B(c,r)$ denotes the Euclidean ball centered at $c$ with radius $r$.

\subsection{Organization}
The rest of the paper is organized as follows. 
Section~\ref{sec: summary statistics learning and posterior inference} introduces our main framework for jointly learning summary statistics and  amortized posterior estimators, along with local refinement procedures. Section~\ref{sec: Theoretical investigation} provides theoretical guarantees through an oracle inequality, with applications to exponential families and large-sample regimes.  Section~\ref{sec: Numerical Experiments} reports numerical experiments on benchmark problems, and Section~\ref{sec: Discussion} concludes with a discussion. Proofs and implementation details are deferred to the Supplementary Material.
 
\section{Summary Learning and Posterior Inference }
\label{sec: summary statistics learning and posterior inference}

 Let $\ms S$ be a class of summary maps from $\Xn \in \mathbb{R}^{n \times d_X}$ to a low-dimensional vector in $\mathbb{R}^p$.
  Our goal is to learn a statistic $S(\cdot)\in \ms S$ so that the summary-conditioned posterior $\mu^*_{\theta|S(\Xn)}(\cdot|S(\Xn))$ is close to the full posterior $\mu^*_{\theta|\Xn}(\cdot|\Xn)$ on average over $\Xn$.  Formally,  we aim to minimize $\mb{E}_{X^{(n)}\sim \m P^{(n)}}\big[\m D\big(\mu^*_{\theta|\Xn}(\cdot|\Xn),\mu^*_{\theta|S(\Xn)}(\cdot|S(\Xn))\big)\big]$ at the population level, where recall $\m P^{(n)}$ is the prior-predictive distribution   and $\m D$ is a suitable discrepancy between distributions. In this sense, the summary statistic $S$ acts as a compression map from a high-dimensional dataset to a low-dimensional summary.  A widely used approach to data compression is the autoencoder~\citep{hinton2006reducing}, which learns an encoder to compress data, and a decoder to reconstruct the data from  the compressed representation. Our goal here is related but different. Rather than reconstructing the raw data, we seek a compressed representation that preserves the information about the parameter $\theta$ contained in the data. To this end, we introduce an inference-network class $\ms W$, where for any $s\in \mb R^p$ and $W\in \ms W$, one can uniquely identify a distribution on $\mb R^d$, denoted as $\mu_{W}(\cdot|s)$, which, when $s=S(X^{(n)})$, is intended to approximate the posterior $\mu^*_{\theta|\Xn}(\cdot|\Xn)$.  For example, $W(\cdot)$ may specify a conditional density $ \theta \rightarrow W(\theta, s)$,  so that $\mu_{W}(\cdot|s)$ has a density function $W(\theta,s)$. Alternatively, $W$ could define a conditional generative map, where for a fixed base distribution $\nu$, the pushforward measure $W(\cdot,s)_{\#}\nu$ yields $\mu_{W}(\cdot|s)$.

Intuitively, $S(\cdot)$ acts as the ``encoder'' to compress the data into a low-dimensional summary, and $W(\cdot)$ acts as the ``decoder'' to reconstruct the information about the parameter contained in the data from the compressed summary. The quality of this reconstruction is judged not by data-level reconstruction error as in an autoencoder, but by how well $\mu_{W}(\cdot|S(X^{(n)}))$ matches the full posterior in a distributional sense. The learning procedure then proceeds using the training set $\{(\theta_j,X_j^{(n)})\}_{j=1}^N$ i.i.d. drawn from $\Pi_{\theta}\m P^{(n)}_{\theta}$. Given a loss $\ell(S,W,\theta,X^{(n)})$ that scores the pair $(S,W)$ with a single sample  $(\theta,X^{(n)})$, we solve an M-estimation  problem: 
\begin{equation}\label{M:est}
    (\wh S,\wh W)={\arg\min}_{S\in \ms{S}, W\in \ms W} \frac{1}{N}\sum_{j=1}^N \ell(S,W,\theta_j, X_{j}^{(n)}).
\end{equation}
The training objective $\frac{1}{N}\sum_{j=1}^N \ell(S,W, \theta_j, X_{j}^{(n)})$ can be viewed as a ``reconstruction loss'' incurred on the training set.  We aim to  design the loss function $\ell$ such that the population risk $\m R(S,W)=\mb{E}_{\theta\sim \Pi_{\theta}, X^{(n)}\sim \m P_{\theta}^{(n)}}[\ell(S,W,\theta,X^{(n)})]$  suitably controls  the  expected discrepancy of $\mb{E}_{X^{(n)}\sim \m P^{(n)}}[\m D(\mu^*_{\theta|\Xn}(\cdot|\Xn),\mu_{W}(\cdot|S(X^{(n)})))]$. To formalize this, we require the following properties. In Section~\ref{sec:choice}, we discuss concrete choices of $\m D$ and $\ell$ that ensure these properties.
 
\begin{property}\label{pro1}{\textbf{(Learning the summary-conditioned posterior).}}
For any $S\in \ms S$,  minimizing $\m R(S,W)$ over $W\in \ms W$ is equivalent to minimizing the expected discrepancy between the posterior given $S(X^{(n)})$ and $\mu_{W}(\cdot|S(X^{(n)}))$, that is, $\forall\, S\in \ms S$,  $$   \underset{W\in \ms W}{\arg\min} \m R(S,W)= \underset{W\in \ms W}{\arg\min} \mb{E}_{X^{(n)}\sim \m P^{(n)}}\big[\m D\big(\mu^*_{\theta|S(\Xn)}(\cdot|S(\Xn)),\mu_{W}(\cdot|S(X^{(n)}))\big)\big].$$
 
\end{property}

 \begin{property}\label{pro2}{\textbf{(Learning summary statistics).}}
If $\underset{W\in \ms W}{\min} \mb{E}_{X^{(n)}\sim \m P^{(n)}}\big[\m D(\mu^*_{\theta|S(\Xn)}(\cdot|S(\Xn)),\\\mu_{W}(\cdot|S(X^{(n)})))\big]=0$ holds for any $S\in \ms S$,  then minimizing $\underset{W\in \ms W}{\min} \m R(S,W)$ is equivalent to minimizing the expected discrepancy between  the posteriors given $X^{(n)}$ and $S(X^{(n)})$, that is, 
$$\underset {S\in \ms S}{\arg\min}\underset{W\in \ms W}{\min} \m R(S,W)= \underset{S\in \ms S}{\arg\min}\, \mb{E}_{X^{(n)}\sim \m P^{(n)}}\big[\m D\big(\mu^*_{\theta|\Xn}(\cdot|\Xn),\mu^*_{\theta|S(\Xn)}(\cdot|S(\Xn))\big)\big].$$
 
 \end{property}

 \begin{property}\label{pro3}{\textbf{(Summary-conditioned posterior leads to a better estimate).}}
For any $S\in \ms S$ and $W\in \ms W$, it holds that   $\mb{E}_{X^{(n)}\sim \m P^{(n)}}\big[\m D\big(\mu^*_{\theta|\Xn}(\cdot|\Xn),\mu^*_{\theta|S(\Xn)}(\cdot|S(\Xn))\big)\big]  \leq \mb{E}_{X^{(n)}\sim \m P^{(n)}}\big[\m D\big(\mu^*_{\theta|\Xn}(\cdot|\Xn),\mu_{W}(\cdot|S(X^{(n)}))\big)\big].$
 
\end{property}
 
 \noindent  
Property~\ref{pro1} states that for a fixed summary $S$,  the optimal $W$ learns the summary-conditioned posterior. Property~\ref{pro2} ensures that if the inference-network (or decoder) class $\ms W$ is sufficiently expressive to approximate the summary-conditioned posterior exactly, the joint optimization problem reduces to finding the summary that minimizes the discrepancy between the full and the summary-conditioned posterior.  Here, the decoder $W(\cdot)$ acts as an inference network to recover the posterior given the information bottleneck $S(X^{(n)})$, while the summary's role is to decide what information passes through the bottleneck. Our target is to choose $S$ that preserves as much posterior-relevant information as possible, as reflected by the optimal decoder's fit.  Property~\ref{pro3} states that the amortized posterior estimator cannot outperform the summary-conditioned posterior on average. This directly motivates our local refinement strategy where we refine $\mu_{\wh W}(\cdot|\wh S(\Xn_*))$ towards $\mu^*_{\theta|\wh S(\Xn)}(\cdot|\wh S(\Xn_*))$ to  close the residual gap caused by finite capacity or imperfect global training of $\wh W$.

We next discuss choices of $\ell$ and $\m D$ that guarantee these properties, then present the local refinement algorithm and computational details for selecting  $\ms S$ and $\ms W$.

\subsection{Choices of Discrepancy and Loss Function}\label{sec:choice}

A key step is to select a suitable discrepancy measure $\m D$ for which we can construct a reconstruction loss function $\ell$ whose expectation satisfies Properties~\ref{pro1}-\ref{pro3}.  We consider three choices: KL divergence, maximum mean discrepancy (MMD), and sliced MMD.

\subsubsection{KL Divergence} 

A principled choice for $\m D$ is the Kullback-Leibler (KL) divergence: ${\rm KL}(\mu||\nu)=\int \log\frac{\dd\mu}{\dd \nu}\,\dd \mu$. Minimizing the KL divergence between the true posterior and the inference network is equivalent to minimizing the negative log-likelihood (NLL), the standard  training objective in  SBI \citep{papamakarios2016fast,papamakarios2019sequential,radev2020bayesflow}. The following theorem shows that an NLL-type loss yields the desired properties.

\begin{theorem}\label{th1}
Let $\mathcal{S}$ be a class of summary maps
$S:\mathbb{R}^{n\times d_X}\to\mathbb{R}^p$, and let $\mathcal{W}$ be a
class of conditional densities on $\mathbb{R}^d$, with
$\mu_W(\cdot| s)$ having density $W(\cdot,s)$.
Assume the posterior has a Lebesgue density $p(\theta| X^{(n)})$
for $P^{(n)}$-almost every $X^{(n)}$. For each $S\in\mathcal{S}$, let
$p_S(\theta| S(X^{(n)}))$ denote the summary-conditioned posterior
density and assume $\mathbb{E}_{\theta\sim\Pi_\theta,\,X^{(n)}\sim P_\theta^{(n)}}
[
|\log p(\theta| X^{(n)})|
+|\log p_S(\theta| S(X^{(n)}))|]<\infty$. Then Properties~1--3 hold with $D(\mu,\nu)=\operatorname{KL}(\mu\|\nu)$
and $\ell(S,W,\theta,X^{(n)})=-\log W(\theta,S(X^{(n)}))$.
\end{theorem}

 Theorem~\ref{th1} implies that, with a sufficiently expressive $\ms W$,  at the population level, learning summaries with the NLL loss is equivalent to minimizing the expected KL divergence between the full posterior and the summary-conditioned posterior.  However, a key limitation of using KL divergence  is that $\ms W$ must provide tractable, normalized conditional density functions. For example,~\cite{radev2020bayesflow} model the amortized posterior with a conditional normalizing flow,  representable as a conditional generative model $G(\cdot, S(X^{(n)}))_{\#}\nu$,
 where for any $s \in \mathbb{R}^p$, the map $\theta \mapsto G(\theta, s)$ is invertible. This invertibility yields a closed-form expression for the conditional density function but can also restrict the expressiveness of the conditional generative model. 
  This motivates us to explore alternative discrepancy measures $\mathcal{D}$ that do not require explicit densities and thereby allow more flexible $\ms W$.

  \subsubsection{Maximum Mean Discrepancy}  

To avoid explicit conditional densities, we need a density-free discrepancy measure. A popular choice is the maximum mean discrepancy (MMD) \citep{JMLR:v13:gretton12a}, defined as  ${\rm MMD}^2_{\kappa}(\mu,\nu)=\mb{E}_{X,X'\sim \mu}[\kappa(X,X')]+\mb{E}_{Y,Y'\sim \nu}[\kappa(Y,Y')]-2\cdot\mb{E}_{X\sim \mu,Y\sim\nu}[\kappa(X,Y)],$ where $\kappa$ is a positive-definite kernel inducing a reproducing kernel Hilbert space (RKHS). For universal kernels such as the Gaussian kernel $\kappa(x,y)=\exp(-\|x-y\|^2/(2h^2))/(2\pi h^2)^{d/2}$, 
  $\text{MMD}=0$ if and only if $\mu=\nu$. The following theorem  then integrates MMD into our framework.
  \begin{theorem}\label{th2}
Let $\kappa:\mb R^d\times \mb R^d\to \mb R$ be a bounded positive-definite kernel, $\ms S$ be a class of summary maps $S:\mb R^{n\times d_X}\to \mb R^p$ and $\ms W$ be a class so that each  $W\in \ms W$ and $s\in \mb R^p$ define a distribution  $\mu_{W}(\cdot|s)$ on $\mb R^d$.  Then Properties~\ref{pro1}--\ref{pro3} hold with discrepancy $\m D(\mu,\nu)={\rm MMD}^2_{\kappa}(\mu,\nu)$ and  loss    $\ell(S,W,\theta, \Xn)=\mb{E}_{\theta',\theta^{''}\sim \mu_{W}(\cdot|S(\Xn))}[\kappa(\theta',\theta^{''})]-2\cdot \mb{E}_{\theta'\sim  \mu_{W}(\cdot|S(\Xn))}[\kappa(\theta, \theta')].$
 
  \end{theorem}
Unlike Theorem~\ref{th1}, Theorem~\ref{th2} does not require $\ms W$ to specify tractable  conditional densities, which greatly expands the range of admissible models.  For instance, $\mu_{W}(\cdot|S(\Xn))$ can be defined as a conditional generative model by pushing forward an isotropic Gaussian through a neural generator  $W(\cdot, S(\Xn))$ without any invertibility constraints. Compared with normalizing flows, such architectures are simpler, more parameter-efficient, and easier to train while maintaining strong expressiveness. However, using MMD also introduces computational challenges. The loss involves expectations over $\mu_{W}(\cdot|S(\Xn))$, which should be approximated via Monte Carlo sampling. The cost of accurate estimation can scale poorly with the parameter dimension $d$.  For example, with a Gaussian kernel of bandwidth $h$,  obtaining a reliable Monte Carlo approximation of the kernel expectations may require $O(h^{-d})$ samples from  $\mu_{W}(\cdot|S(X^{(n)}))$, reflecting a curse of dimensionality  effect in $d$.

\subsubsection{Sliced Maximum Mean Discrepancy}

To alleviate the computational cost of MMD, we employ sliced variants~\citep{kolouri2019generalized} that compare distributions via random one-dimensional projections. Specifically, we project samples onto random directions $\lambda\sim {\rm Unif}(\mb S_1^{d-1})$, where  ${\rm Unif}(\mb S_1^{d-1})$ is the uniform distribution on the $(d-1)$-sphere,  then evaluate a univariate MMD along each projection, and finally average over directions. Formally, the sliced MMD (SMMD) is defined as ${\rm SMMD}^2_{\kappa}(\mu,\nu)=\mb{E}_{\lambda\sim {\rm Unif}(\mb S_1^{d-1})}[{\rm MMD}^2_{\kappa}(\lambda^T_{\#}\mu,\lambda^T_{\#}\nu)]$, where $\lambda^T_{\#}\mu$ is the push-forward distribution of $\theta \rightarrow \lambda^T\theta$ for $\theta\sim \mu$. If the kernel is universal, then the Cram\'{e}r-Wold theorem  implies that ${\rm SMMD}^2_{\kappa}(\mu,\nu)=0$ if and only if $\mu=\nu$. The next theorem  then incorporates SMMD into our framework.

   \begin{theorem}\label{th3}
       Let $\kappa:\mb R\times \mb R\to \mb R$ be a bounded positive-definite kernel. Under the assumptions of Theorem~\ref{th2}, Properties~\ref{pro1}--\ref{pro3} hold for   $\m D(\cdot,\cdot)={\rm SMMD}^2_{\kappa}(\cdot,\cdot)$ and   loss   
$        \ell(S,W,\theta, \Xn) = \mb{E}_{\lambda\sim {\rm Unif}(\mb S_1^{d-1})}\Big[
            \mb{E}_{\theta',\theta''\sim \mu_{W}(\cdot|S(\Xn))}\!\left[\kappa(\lambda^T\theta',\lambda^T\theta'')\right]  
  - 2\cdot\mb{E}_{\theta'\sim \mu_{W}(\cdot|S(\Xn))}\!\left[\kappa(\lambda^T\theta, \lambda^T\theta')\right]\Big].$

   \end{theorem}

   Unlike full $d$-dimensional MMD, SMMD replaces high-dimensional kernel evaluations with cheap univariate computations, yielding much better scaling in high dimensions. The stochastic gradient of the loss function $   \ell(S,W,\theta, X^{(n)})$  for the training parameters can be obtained via Monte Carlo by sampling $\lambda\sim\rm Unif(\mb S_1^{d-1})$ and $\theta\sim\mu_{W}(\cdot|S(X^{(n)}))$.   When $\kappa$ is a $1$-dimensional Gaussian kernel with bandwidth $h$, the number of $\theta$-samples needed for an informative gradient  grows only linearly with $h^{-1}$. Averaging over random directions    provides complementary views of the distributions and helps stabilize the stochastic gradient.

\subsection{Local Refinement for Specific Observations}

The estimator $\Xn\mapsto\mu_{\wh W}(\cdot|\wh S(\Xn))$ defines an amortized inference procedure, but being trained for average performance over the prior, it may lack high precision for a specific observation $\Xn_*$. To mitigate this amortization gap, we consider two forms of local refinement.  One type of refinement keeps the summary map fixed and improves the posterior estimate toward the summary-conditioned posterior.   The other type of refinement updates the amortized summary mapping itself. We draw a new training set that more closely resembles the observed dataset and retrain the model, so that the resulting summary map is more predictive of parameters for samples ``similar to'' the observation.  

\subsubsection{ Refinement of the Amortized Posterior Estimator.}\label{sec:refine1}
 Property~\ref{pro3} indicates that the fitted amortized estimator   cannot outperform  the summary-conditioned posterior on average. Learning a globally effective summary  can also be easier than learning a globally accurate posterior estimator.  For example, in an exponential family with sufficient statistic $T(X^{(n)})=\sum_{i=1}^n t(X_i)$, the same summary $T(\cdot)$ is optimal for datasets from all parameters. Thus, even datasets simulated at $\theta$ far from the true $\theta^*$ that generated $X^{(n)}_*$ can inform summary learning by helping identify data features predictive of $\theta$ across the parameter space. In contrast, learning a global posterior estimator is intrinsically harder, as it must map each dataset (or summary) to a distribution whose shape can vary widely across the data/summary space, shifting from diffuse to sharp, symmetric to skewed, or unimodal to multimodal.  This motivates us to directly target $\mu^*_{\theta|\wh S(\Xn)}(\cdot|\wh S(\Xn_*))$ by leveraging the  local observation $X^{(n)}_*$ together with the coarse amortized estimate $\Xn\to\mu_{\wh W}(\cdot|\wh S(\Xn))$. We perform this local refinement using ABC-MCMC~\citep{marjoram2003markov}.

\begin{figure}[h]
    \centering
    \includegraphics[width=0.8\textwidth]{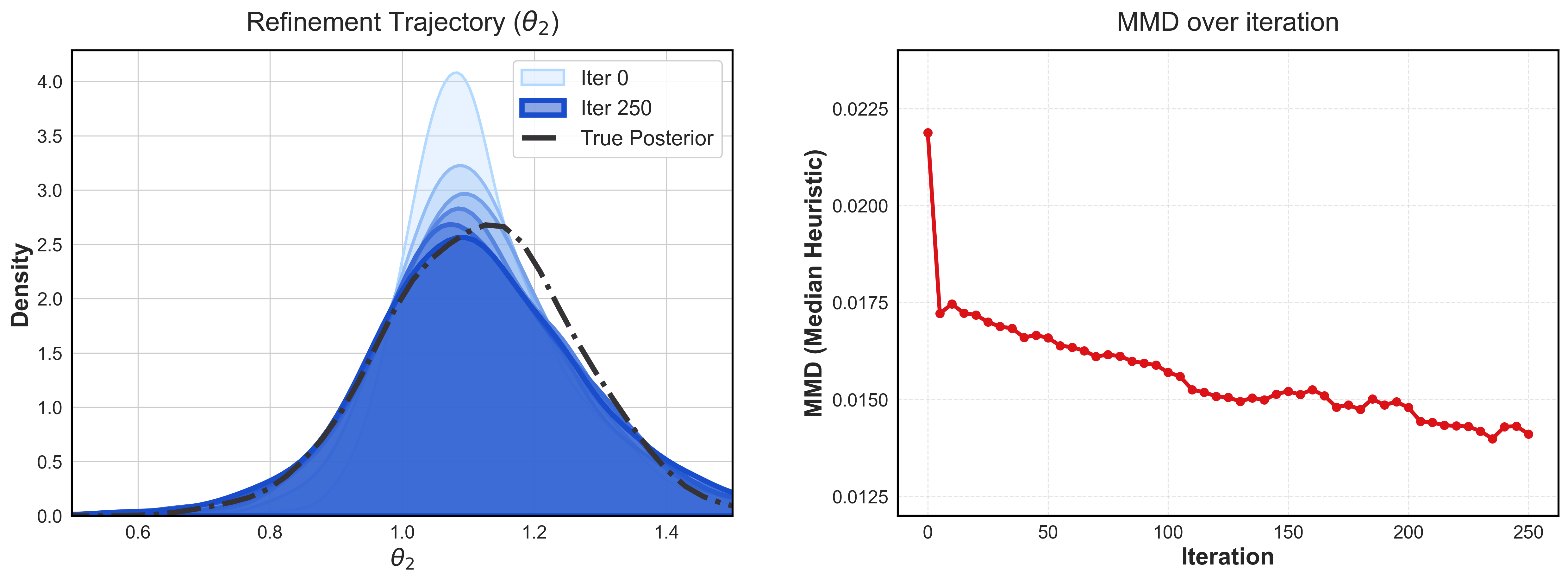}
    \caption{Refinement results for the toy example (Section~\ref{subsec: toy example}). The left panel displays the marginal density of $\theta_2$: the black curve is the true marginal posterior, while the blue gradient curves represent the marginal densities of particles at different refinement rounds. We perform $250$ rounds of local MCMC refinement, plotting densities every $50$ rounds—lighter blue shades correspond to earlier rounds, and darker shades to later rounds. Round 0 specifically refers to the marginal of the amortized posterior estimator $\mu_{\wh W}(\cdot|\wh S(\Xn_*))$. The right panel shows the trends of  MMD with median heuristic bandwidth \citep{lueckmann2021benchmarking} between the target posterior and the empirical distribution of particles from each MCMC step. The MMD decreases rapidly and almost monotonically over the first $100$ iterations and then stabilizes, indicating that each MCMC step makes small, targeted adjustments to the particles, progressively moving their empirical distribution closer to the target posterior.}
    \label{fig:mmd}
         
\end{figure}

Given a summary statistic $\wh S(\cdot)$ and a small enough threshold $\epsilon>0$, ABC-MCMC targets the ABC posterior density:
\(
p_{\epsilon}(\theta \mid \widehat S(X^{(n)}_*))
\propto \pi(\theta)\cdot \mathbb{E}_{X^{(n)} \sim \mathcal{P}_{\theta}^{(n)}}\big[
k (\frac{\|\widehat S(X^{(n)}) - \widehat S(X^{(n)}_*)\|}{\epsilon})
 \big],
\)
where $k(\cdot)$ is either compactly supported (e.g., $\boldsymbol{1}\{0\leq x<1\}$) or   light-tailed (e.g., $e^{-x^2}$). This defines  a kernel-based approximation to $\mu^*_{\theta|\wh S(\Xn)}(\cdot|\wh S(\Xn_*))$. The expectation term  acts as a kernel smoother 
in the conditioning variable $\wh S(X^{(n)})$, assigning higher weight to parameter values $\theta$ that more frequently generate summaries near the observed summary $\widehat S(X^{(n)}_*)$. Exact conditioning on  $\wh S(X^{(n)})=
\wh S(X^{(n)}_*)$ is then replaced by kernel-weighted averaging over summaries in a neighborhood of  $\wh S(X^{(n)}_*) $. The coarse estimator   $\mu_{\wh W}(\cdot|\wh S(X^{(n)}_*))$  enables a data-driven way to choose the threshold $\epsilon$. Since $\epsilon$ should be scaled to the variability of the summaries,  we first draw a set of parameter samples from  $\mu_{\wh W}(\cdot|\wh S(X^{(n)}_*))$, simulate a dataset for each sample, compute the Euclidean distances between their summaries  and the observed  summary $\widehat S(X^{(n)}_*)$, and then set $\epsilon$ to a chosen quantile (e.g., the $0.01$-quantile) of these distances  aggregated across simulations. Furthermore, to further utilize the shape information of  $\mu_{\wh W}(\cdot|\wh S(X^{(n)}_*))$, we initialize ABC-MCMC with a set of particles drawn from $\mu_{\wh W}(\cdot|\wh S(X^{(n)}_*))$ and run multiple chains.   We then use a  proposal distribution $f(\cdot|\cdot)$ that proposes local moves, e.g., $f(\cdot|\theta)=N(\theta, \sigma^2I_d)$. At each step, each chain proposes a local move via  $f(\cdot|\theta)$ from its current particle $\theta$ and accepts or rejects the proposal according to the Metropolis–Hastings acceptance ratio, which combines the prior density ratio, the proposal density ratio, and the change in the simulator’s ability to generate simulated data whose summaries are close to the observed summary. In this way, the chain concentrates mass in regions of parameter space that better reproduce the observed summary, thereby refining the initial amortized approximation for the  dataset at hand. The procedure is summarized in Algorithm~\ref{algo:refinement} (without \emph{Refine+}) and Figure~\ref{fig:mmd} illustrates the process. 
 
\begin{algorithm}[h]
\small
\caption{Local Refinement (\emph{Refine+} includes summary retraining)}\label{algo:refinement}
\KwIn{Data $\Xn_*$, estimator $(\widehat S,\widehat W)$, simulator $\theta\mapsto\mathcal{P}_{\theta}^{(n)}$, prior density $\pi(\cdot)$, training size $N$; Params: $N_{\text{train}}$,  $N_{\text{ref}}$, $N_{\text{chain}}$, $\tau$, $L$, $k(\cdot)$, $f(\cdot|\cdot)$.}
\KwOut{Refined particles $\{\vartheta_i\}_{i=1}^{N_{\text{chain}}}$}
Let  $\wh \mu(\dd\theta)=\mu_{\widehat W}(\dd\theta|\widehat S(\Xn_*))$\;
\If{Refine+}{
    Draw new training set $\{(\theta'_j,\Xn_j{}')\}_{j=1}^{N_{\rm train}} \sim \wh \mu(\dd\theta) \mathcal{P}_{\theta}^{(n)}$;
fit a density estimator $q(\cdot)$ from $\{\theta'_j\}_{j=1}^{N_{\rm train}}$; merge with the original training set and retrain to obtain $(\widehat S_{\text{new}},\widehat W_{\text{new}})$; update
    $\widehat S = \widehat S_{\text{new}}$ and $\wh \mu(\dd\theta)\propto \mu_{\widehat W_{\text{new}}}(\dd\theta|\widehat S_{\text{new}}(\Xn_*)) \cdot \frac{\pi(\theta)}{N_{\rm train}q(\theta) +N\pi(\theta)}$;
} 

\BlankLine
\emph{Adaptive Threshold:} Draw $\{(\wt\theta_j,\wt X^{(n)}_j)\}_{j=1}^{N_{\text{ref}}} \sim \wh\mu(\dd\theta)\mathcal{P}_{\theta}^{(n)}$; set $\epsilon = \tau$-th quantile of $\{\|\widehat S(\wt X^{(n)}_j)-\widehat S(\Xn_*)\|\}_{j=1}^{N_{\text{ref}}}$;

\BlankLine
\emph{Initialize (Per Chain, Parallel):}
\For{$i=1$ \textbf{to} $N_{\text{chain}}$}{
    $\vartheta_i\sim \wh\mu$; $\ov{X}^{(n)}_i \sim \mathcal{P}_{\vartheta_i}^{(n)}$;
    $R_i = k\left(\frac{\|\widehat S(\ov{X}^{(n)}_i)-\widehat S(\Xn_*)\|}{\epsilon}\right)$;  Repeating until $\pi(\vartheta_i)R_i>0$\;
}

\BlankLine
\emph{Parallel MCMC:}
\For{$i=1$ \textbf{to} $N_{\text{chain}}$ (parallel) and $l=1$ \textbf{to} $L$ {(Iter per chain)}}{
    $\vartheta_{i\Delta} \sim f(\cdot|\vartheta_i)$; $\Xn_{i\Delta} \sim \mathcal{P}_{\vartheta_{i\Delta}}^{(n)}$;
    ${R}_{i\Delta} = k\Big(\frac{\|\widehat S(\Xn_{i\Delta})-\widehat S(\Xn_*)\|}{\epsilon}\Big)$\; 

    Update $(\vartheta_i,R_i) =(\vartheta_{i\Delta}, {R}_{i\Delta})$ with probability
    $ \min\left(1,\frac{\pi(\vartheta_{i\Delta})f(\vartheta_i|\vartheta_{i\Delta})}{\pi(\vartheta_i)f(\vartheta_{i\Delta}|\vartheta_i)}\frac{{R}_{i\Delta}}{R_i}\right)$\;
  
}

\Return $\{\vartheta_i\}_{i=1}^{N_{\text{chain}}}$
\end{algorithm}

 \subsubsection{Refinement of the Amortized Summary}\label{sec:refine2}

The refinement in Section~\ref{sec:refine1} implicitly assumes that the learned summary is predictive for $\theta$ across datasets $\Xn\sim \m P^{(n)}$. Under a diffuse prior, such a globally informative summary may not exist, and  the summary-conditioned posterior $\mu^*_{\theta|\wh S(\Xn)}(\cdot| \widehat S(\Xn_*))$ itself can be far from the true posterior.
We therefore perform one round of sequential training, a standard technique in SBI for handling vague priors \citep{papamakarios2016fast}. Specifically, we generate an additional training set of size $N_{\rm train}$ by drawing parameters from the amortized posterior
\(Q_\theta(\dd\theta)=\mu_{\wh W}(\dd\theta|\wh S(\Xn_*))\)
and simulating the corresponding datasets. We then merge these new pairs with the original training set to form a combined dataset of size $N+N_{\rm train}$, and retrain the model to obtain updated networks $\wh W_{\rm new}$ and $\wh S_{\rm new}$ that better capture the local posterior structure around $\Xn_*$.  To  correct for sampling $\theta$ from $\frac{N_{\rm train}}{N+N_{\rm train}} Q_{\theta} +\frac{N}{N+N_{\rm train}} \Pi_{\theta}$ rather than $\Pi_{\theta}$,  we use importance weights $ \pi(\theta)/(N_{\rm train} q(\theta) + N \pi(\theta))$ when sampling from the retrained inference network, where $q(\cdot)$ denotes the density of $ Q_\theta $ (estimated via KDE when not available in closed form). We  then use the refined summary $\wh S_{\rm new}$ with the ABC-MCMC refinement to produce the final posterior estimate. Here, the reweighted sequential estimator is used only as a coarse, locally adapted approximation that provides a warm start for the refinement, and sequential training is
primarily a way to refine the summary around the observed dataset rather than the final inference stage. Therefore, approximate importance weights constructed from a KDE estimate
are sufficient for our purpose.
Provided that the chains converge, the influence of the
initialization vanishes, and the chains target the ABC posterior induced by the refined summary. Thus, an analytically tractable inference-network density is not required during either amortized training or local refinement. The full procedure is summarized in Algorithm~\ref{algo:refinement} (with \emph{Refine+}).

\subsection{Computation}\label{sec:comp}

 We summarize key practical choices for implementing the framework below.

 \emph{Summary and Inference Networks.} The architecture of $W$ depends on the chosen discrepancy $\m D$. For instance, if $\m D$ is the KL divergence, $W$ must output a tractable conditional density (e.g.,  a normalizing flow), enabling direct evaluation of the negative log-likelihood. If $\m D$ is a divergence that can be estimated from samples (e.g., MMD), $W$ can be a conditional generative map, such as a simple MLP. The summary network should be chosen to exploit the structure of the data. For i.i.d. observations, we can use permutation-invariant architectures such as DeepSets \citep{zaheer2017deep} to respect exchangeability. For sequential or time-series data, we can use convolutional or recurrent architectures (CNNs/LSTMs) \citep{lecun1998gradient,hochreiter1997long} to capture temporal dependencies.

 \emph{Discrepancy.}  KL-based training has a simple objective, but training invertible networks is computationally expensive, and the logarithmic NLL term can destabilize optimization, especially for highly concentrated posteriors.  MMD-based training enables a simpler decoder network $W$, yet its gradient estimates can be high-variance, especially for small bandwidths $h$. This creates a tradeoff: smaller bandwidths increase MMD's ability to distinguish nearby distributions, yet they also make gradient estimates noisier. We mitigate these issues with two practical solutions:  (i) using a mixture of bandwidths to improve robustness across scales; (ii) employing sliced MMD, which projects samples onto random  directions and computes MMD in lower dimensions, providing more stable gradient estimates.

\emph{Output dimension and regularization.} The output dimension $p$ of $S$ is a key hyperparameter. We adopt a simple heuristic: choose $p$ moderately large and use regularization to control the effective dimension.  Instead of directly penalizing $S$'s parameters, which may force it to discard information too aggressively, we impose an $\ell_1$ penalty on the parameters of $W$. This encourages the decoder network to rely on only a few directions in the encoded space  thereby implicitly pushing the learned summaries to concentrate on a low-dimensional subspace. For example, in the toy example of Section \ref{subsec: toy example}, the minimal sufficient statistic is five-dimensional, but we set $p=10$. However, the first five principal components of the  learned summaries  explain over $95\%$ of the variance (see Appendix B.3.3 for details). Thus, it is safe in practice to set $p$ to a moderately large value. Consistent with~\cite{chen2021neural}, we use $p=2d$ in all experiments, where $d$ is the parameter dimension. 
 
 \vspace{-0.5em}
\section{Theoretical Investigation}
\label{sec: Theoretical investigation}

In this section, we study the learned summary statistic and the resulting amortized posterior estimator through the lens of nonparametric conditional distribution estimation, specializing to the MMD criterion with a Gaussian kernel:
 $\kappa_h(x,y)=\frac{1}{(2\pi h^2)^{\frac{d}{2}}}\exp(-\frac{\|x-y\|^2}{2h^2})$, and write $\mmd^2_h(\cdot,\cdot)=\mmd^2_{\kappa_h}(\cdot,\cdot)$. The Gaussian-kernel MMD  can be expressed as the squared $L_2$ distance between the kernel-smoothed densities: 
\begin{equation}\label{expressMMD}
    \begin{aligned}
              & \mmd_h^2(\mu, \nu) =       \mathbb{E}_{\theta,\theta' \sim \mu}[\kappa_{h}(\theta,\theta')]
        -2\cdot\mathbb{E}_{\theta \sim \mu, \theta'
        \sim \nu}[\kappa_{h}(\theta,\theta')]
      +
        \mathbb{E}_{\theta,\theta' \sim \nu}[\kappa_{h}(\theta,\theta')]\\
        &= \int \Big( 
                \mathbb{E}_{\theta \sim \mu}[\kappa_{h/\sqrt{2}}(\theta,y)] 
                - 
                \mathbb{E}_{\theta \sim \nu}[\kappa_{h/\sqrt{2}}(\theta,y)]\Big)^2 \dd y.
    \end{aligned}
\end{equation}
As $h\to 0$, $\mmd^2_h(\mu, \nu)$ converges to the squared $L_2$ distance between the densities of $\mu$ and $\nu$. We assess the posterior approximation quality of $\mu_{\wh W}(\cdot|\wh S(\Xn))$ using the expected MMD over the covariate marginal $\m P^{(n)}$, namely $\mb{E}_{\Xn \sim \m P^{(n)}}[\mmd^2_h(\mu^*_{\theta|\Xn}(\cdot|\Xn),\mu_{\wh W}(\cdot|\wh S(\Xn)))]$. This frames our task as conditional distribution estimation: given $N$ i.i.d.\ pairs $\{(\theta_j,\Xn_j)\}_{j=1}^N$ drawn from $\Pi_{\theta}\m P_{\theta}^{(n)}$, we estimate the conditional distribution $\mu^*_{\theta|\Xn}(\cdot|\Xn)$ of $\theta\in \mb R^d$ given the high-dimensional covariate $\Xn\in \mb R^{n\times d_X}$.
 Classical smoothness-based theory typically fixes the conditioning dimension and assumes that the target density belongs to a fixed H\"older  or controlled-complexity class \citep{10.1214/22-EJS2037,10.1214/23-AOS2270}. In our setting, however, both
the conditioning variable and the target distribution vary with \(n\). As \(n\) increases, the ambient dimension
of \(\Xn\) grows as \(nd_X\) and the posterior may become increasingly
concentrated. Consequently,  the
sequence of posteriors does not belong to a common smoothness class with regularity bounds uniform in \(n\). Moreover,  if \(\Xn\) were treated as an unrestricted covariate in
\(\mathbb{R}^{nd_X}\), a naive application of classical fixed-dimensional
minimax rates would suggest that the training size \(N\) grows exponentially with \(nd_X\). Some recent analyses
reduce this ambient-dimensional dependence by assuming that the data or
conditioning variables lie on low-dimensional manifolds
\citep{loaiza2024deep,tang2025minimax,chae2023likelihood}. Our setting
exploits a different source of low-dimensional structure. Rather than
assuming that \(\Xn\) itself lies on a low-dimensional manifold, we use
structure in the generating model that may allow the posterior-relevant
information in \(\Xn\) to be represented by an informative low-dimensional
summary.

Given a summary $S\in \ms S$ and a posterior model $\mu_W(\cdot|S(\Xn))$ with $W \in \ms W$, we use the MMD-based loss from Theorem~\ref{th2}: $\ell(S,W,\theta,\Xn) 
= 
\mathbb{E}_{\theta'
,\theta^{''} \sim \mu_{W}(\cdot|S(\Xn)) }[\kappa_{h}(\theta',\theta^{''})]
        -2\cdot\mathbb{E}_{\theta' \sim \mu_{W}(\cdot|S(\Xn)) }[\kappa_{h}(\theta,\theta')].$
 The summary statistic and posterior estimator are then obtained via
the following M-estimation: $    (\wh S,\wh W)={\arg\min}_{S\in \ms S,W\in \ms W} \frac{1}{N}\sum_{j=1}^N \ell(S,W,\theta_j,\Xn_j),$ which defines the amortized posterior estimator $\Xn\mapsto\mu_{\wh W}(\cdot|\wh S(\Xn))$. The population risk $\m R(S,W) = \mb{E}_{\theta\sim \Pi_{\theta},\Xn\sim \m P_{\theta}^{(n)}}[\ell(S,W,\theta,\Xn)]$ satisfies $\m R(S,W) = \E_{\Xn\sim \m P^{(n)}} [\mmd^2_h(\mu^*_{\theta|\Xn}(\cdot|\Xn), \mu_{W}(\cdot| S(\Xn)))] - C_p$ for a constant $C_p$ independent of $S$ and $W$, so controlling $\m R(S,W)$ also controls the expected MMD between the true and learned posteriors and, by Theorem~\ref{th2}, that between the true and the summary-conditioned posterior. In what follows, we derive an oracle inequality that quantifies the approximation quality of our amortized estimator, then apply it to exponential-family models and large-sample regimes. Throughout, we treat the parameter dimension $d$ and the per-observation dimension $d_X$ as fixed constants, while allowing both the within-dataset sample size $n$ and training size $N$ to grow to infinity.

\subsection{Oracle Inequality}\label{sec:general}

In this subsection, we analyze the finite-sample performance of the summary statistic and posterior estimator via an oracle inequality~\citep{rigollet2023highdimensionalstatistics}. We begin by imposing an assumption that controls the complexity of the approximation families $\ms S$ and $\ms W$.

\begin{assumption}\label{Assumpgeneral}
Let $\ms S$ be a class of summary maps $S: \m X^{n}\to \mb R^p$, and $\ms W$ be a class of inference networks such that each
$W\in \ms W$ and $s\in \mb R^p$  induces a distribution $\mu_{W}(\cdot|s)$ on $\mb R^d$.  There exist $\varepsilon\geq 0$ and $T_1, T_2 \geq 1$ such that
\begin{enumerate}
    \item $ \underset{S\in \ms S, W\in \ms W}{\min} \E_{\Xn\sim \m P^{(n)}}[\mmd^2_h(\mu^*_{\theta|\Xn}(\cdot|\Xn), \mu_{W}(\cdot|S(\Xn)))] \leq \varepsilon $.
  
 \item  For any sample $\{(\theta_j,\Xn_j)\}_{j=1}^N$, define the empirical pseudometric $d_N$ on $\ms S\times \ms W$ by $d_N((S,W),(S',W'))= \sqrt{\frac{1}{N} \sum_{j=1}^N \big(\ell(S,W, \theta_j,\Xn_j) - \ell(S',W',\theta_j,\Xn_j)\big)^2}$. Then for all $u$ satisfying $\varepsilon \leq u \leq \sup_{(S,W),(S',W')\in \cS \times \cW } d_N((S,W),(S',W'))$, the covering number satisfies $\calN( \cS\times \cW, d_N, u) \leq \left( \frac{T_1}{u} \right)^{T_2}.$ 
\end{enumerate}
\end{assumption}
Here, $\varepsilon$ is an upper bound on the minimal expected MMD attainable over the model class $\ms S \times \ms W$, and it typically decreases as the models become richer. The terms $T_1$ and $T_2$ quantify the metric entropy of $\ms S \times \ms W$, which grows with the expressiveness of the function classes. Together, $\varepsilon, T_1, T_2$ determine the non-asymptotic estimation accuracy of the estimator $\Xn\mapsto\mu_{\wh{W}}(\cdot|\wh S(\Xn))$.

\begin{theorem}[Oracle inequality]\label{thm:Generalization Error} 
Under Assumption~\ref{Assumpgeneral}, there exist constants $C_1$ and $C_2$ depending only on $d$ such that with probability at least $1-\frac{1}{N}$,  
 \begin{equation*}
    \begin{aligned}
        &\E_{\Xn\sim \m P^{(n)}} [\mmd^2_h(\mu^*_{\theta|\Xn}(\cdot|\Xn), \mu^*_{\theta|\wh S(\Xn)}(\cdot|\wh S(\Xn)))]\\
  & \leq
  \E_{\Xn\sim \m P^{(n)}} [\mmd^2_h(\mu^*_{\theta|\Xn}(\cdot|\Xn), \mu_{\wh{W}}(\cdot|\wh S(\Xn)))] \leq 
  C_1\varepsilon   +C_2 \frac{T_2\log(T_1N(2+h^d))}{Nh^d}.
    \end{aligned}
\end{equation*}

\end{theorem}
\begin{remark} 
The probability $1-\frac{1}{N}$  is taken over the random training sample  $\{(\theta_j,\Xn_j)\}_{j=1}^N \overset{i.i.d}{\sim} \Pi_{\theta}\m P_{\theta}^{(n)}$.  The bound decomposes into an approximation error $\varepsilon$ and a finite-sample estimation error $\frac{T_2 \log(T_1 N (2+h^d))}{Nh^d}$.
The approximation error has two sources. Compression error arises when the summary \(S\) discards posterior-relevant information, while inference-network approximation error arises when \(W\) cannot   represent the summary-conditioned posterior accurately. 
If the summary class $\ms S$ contains no approximately sufficient low-dimensional summary, the compression error may remain positive, preventing exact recovery of the full posterior. However, when $\ms W$ is sufficiently expressive, the learned summary can approach the minimum compression error attainable within $\ms S$ as $N$ grows. If such a best-in-class summary is identifying, the corresponding posterior may still concentrate around the true parameter without exactly reproducing the full-data posterior, although information lost through compression may reduce statistical efficiency. Increasing the summary dimension or enriching the summary class may reduce the compression error, while the resulting summary-conditioned posterior can be more complex and  therefore require a richer inference network class $\ms W$. This increases the complexity parameters \(T_1\) and \(T_2\), requiring a larger $N$ to control the estimation error.
\end{remark}
\begin{remark} 
In regular identifiable models, the posterior typically concentrates around the true parameter as \(n\) increases. This shrinking scale creates an additional challenge for MMD. To remain sensitive to the local posterior shape, the bandwidth \(h\) should be comparable to the posterior spread and may therefore need to decrease with \(n\). If \(h\) is much larger than the posterior spread, it may oversmooth the posterior and  wash out differences at its local scale. If \(h\) is much smaller, it provides little additional resolution while increasing the finite-training-sample error through the factor \(h^{-d}\) in Theorem~\ref{thm:Generalization Error}. Thus, when the posterior variance is of order \(n^{-1}\), the natural posterior-scale choice is \(h\asymp n^{-1/2}\). A simple Gaussian example makes this trade-off explicit. Suppose the true and approximate posteriors are \(p_1=N(\mu(X^{(n)}),n^{-1}I_d)\) and \(p_2=N(\widehat{\mu}(X^{(n)}),n^{-1}I_d)\). When the difference between their centers is small relative to \((h^2+2/n)^{1/2}\), a direct calculation gives
\(\|\widehat{\mu}(X^{(n)})-\mu(X^{(n)})\|^2
\asymp (h^2+2/n)^{d/2+1}\operatorname{MMD}_h^2(p_1,p_2)\).
A larger \(h\) therefore weakens the ability of MMD to control errors in the posterior center. Combining this relation with the leading estimation term \((Nh^d)^{-1}\) gives the local center-error proxy
\((h^2+2/n)^{d/2+1}/(Nh^d)\), up to logarithmic and complexity factors. Its bandwidth-dependent component is minimized at \(h^2=d/n\), giving \(h\asymp n^{-1/2}\) for fixed \(d\).
\end{remark}

\begin{remark}
Theorem~\ref{thm:Generalization Error} bounds the MMD risk averaged over $\Xn \sim \m P^{(n)}$, which may induce an amortization gap at a given $\Xn_*$. However, if the training set is instead generated from  $Q_{\theta}\m P_{\theta}^{(n)}$ with a proposal $Q_{\theta}$ having density $q(\cdot)$ (as is typical in sequential training), then Theorem~\ref{thm:Generalization Error} controls $\E_{X^{(n)}\sim \mb{E}_{Q_{\theta}}[\m P_{\theta}^{(n)}]} [\mmd^2_h({\mu^*{}^{\frac{q}{\pi}}_{\theta|\Xn}(\cdot|\Xn)}, \mu^*{}^{\frac{q}{\pi}}_{\theta|\wh S(\Xn)}(\cdot|\wh S(\Xn)))]$, where we write $\mu^\rho$ for the reweighted measure $\mu^\rho(\dd\theta)=\frac{\rho(\theta)\mu(\dd\theta)}{\int \rho(\theta)\mu(\dd\theta)}$ and the conditional distributions are still defined under the original joint law $\Pi_{\theta}\m P_{\theta}^{(n)}$. If $Q_{\theta}$ places more mass near the $\theta$ generating $\Xn_*$, the proposal-predictive data distribution concentrates on datasets similar to $\Xn_*$, making the bound more informative locally. Moreover, this may reduce the approximation error, since  $\mb{E}_{Q_{\theta}}[\m P_{\theta}^{(n)}]$ is expected to be sharper than $\m P^{(n)}=\mb{E}_{\Pi_{\theta}}[\m P_{\theta}^{(n)}]$, making it easier to approximate a globally predictive summary over this narrower family of datasets.
 
\end{remark}

 \begin{remark}
 Theorem~\ref{thm:Generalization Error} focuses on the first, amortized stage of our method. It controls, on average over \(X^{(n)}\), how closely the learned amortized posterior and the summary-conditioned posterior approximate the full posterior. The latter is the best posterior that can be reached while keeping the learned summary fixed. In the second stage, local ABC--MCMC refinement is applied to a particular observation \(X_*^{(n)}\). It moves the amortized approximation toward the ABC posterior conditional on the learned summary and can thereby reduce the inference-network approximation error. However, it introduces additional error due to the positive ABC tolerance and finite Monte Carlo sampling. Our oracle inequality and existing ABC theory therefore address complementary parts of the procedure. The former controls summary learning and amortized posterior approximation, while the latter characterizes the effects of the ABC tolerance and Monte Carlo sampling~\citep{frazier2018asymptotic,li2018asymptotic}.
 \end{remark}

We now apply Theorem~\ref{thm:Generalization Error} to two canonical i.i.d. settings. The i.i.d. setting has been  explored in the SBI literature~\citep{papamakarios2019sequential,linhart2024diffusion} for its convenient exchangeable structure.  Even in this regime,  general non-asymptotic theory for amortized Bayesian inference remains limited.  Moreover, methodologically, our framework can also be applied in non-i.i.d. settings. We then consider: (1) $\m P_\theta$ belongs to an exponential family; and (2) a   large-sample regime where the log-density is three times differentiable and  the sample size $n$ goes to infinity. In both cases we specify $\ms S$, $\ms W$ that mitigate the usual curse of dimensionality, so that the required training size $N$ scales only polynomially with $n$ rather than exponentially as in classical nonparametric conditional density estimation.

 \subsection{Example 1: Exponential Family Models}
Exponential families include many common models and admit tractable sufficient statistics. Their densities take the form $p(x|\theta)=h(x)\,\exp\!\big\{\eta(\theta)^{\top} t(x) - A( \theta)\big\}$, where $x\in \mb R^{d_X}$ and $t(x)=(t_1(x),\cdots, t_k(x))$.  We assume $\Xn\sim \m P_{\theta}^{(n)}=\m P_{\theta}^{\otimes n} $ are i.i.d. generated from an exponential family model $\m P_{\theta}$. Following standard density estimation literature~\citep{donoho1996density,singh2018nonparametric}, we take $\Theta=[0,1]^d$ (bounded parameter space) and impose mild regularity conditions on the prior $\Pi_{\theta}$ and the data-generating model; see Assumption 1 in Appendix A for details.  Briefly, we assume that $\Pi_{\theta}$ has a positive density whose logarithm is Lipschitz on $\Theta$; and that $\m P_{\theta}$ is an exponential family model with bounded-dimensional $t(\cdot)$, compact data space, and uniformly bounded, Lipschitz-continuous $\eta(\cdot)$, $A(\cdot)$ and $t(\cdot)$.  Importantly, we assume only the existence of functions $\eta(\cdot)$, $t(\cdot)$, $A(\cdot)$, $h(\cdot)$,  their precise forms, as well as the minimal sufficient-statistic dimension, are unknown. Thus, Assumption 1 is essentially nonparametric, as it imposes only regularities on these unknown functions without specifying any parametric form.  Then we have the following  Corollary.

\begin{corollary}[Exponential Family Model]\label{thm:Exponential family result}  
Let $\{(\theta_j,\Xn_j)\}_{j=1}^N$ be i.i.d. samples from $\Pi_{\theta}\m P_{\theta}^{(n)}$.  For any $\varepsilon>0$ and $0<h<1$, there exists a model class $\cS\times \cW$ (depending only on $n$, $\varepsilon$, $h$, $d$, $d_X$ and constants in Assumption 1 of Appendix A), such that for any $\Pi_{\theta}\m P_{\theta}^{(n)}$ satisfying Assumption 1,  if:
\vspace{-0.5em}
\begin{align*}
    N\geq C \log^5 \big(1+\frac{n}{\min(\varepsilon,1) h}\big)  \Big(1+h\sqrt{\log\big(1+\frac{1}{\varepsilon}\big)}\Big)^{\frac{d\ov d}{2}+d}\varepsilon^{-\frac{\ov d}{2}-1}h^{-d(\ov d+1)} \max (n^d,n^{d_X+1},h^{-2d}),
\end{align*}
\vspace{-0.5em}
with $\ov d=\max(d,d_X)$ and $C$ depending only on $d,d_X$ and constants in the assumption,  then with probability at least $1-\frac{1}{N}$,   $\E_{\Xn\sim \m P^{(n)}} [\mmd^2_h \left(\mu^*_{\theta|\Xn}(\cdot|\Xn), \mu_{\wh{W}}(\cdot|\wh S(\Xn))\right)] \leq \varepsilon$.
\end{corollary}

\begin{remark}
   Here $\ms S$ is a ReLU neural network  class whose outputs have dimension $p$, which serves as the upper bound on the sufficient-statistic dimension in Assumption 1. The class $\ms W$ consists of  ReLU neural networks with input dimension $p+d$ and scalar output.
   For fixed grid points $\{\theta_j^*\}_{j=1}^J \subset \Theta$, 
   $\mu_{W}(\cdot|S(\Xn))$ is the discrete distribution over $\{\theta_j^*\}_{j=1}^J$ with softmax scores $W(S(\Xn),\theta_j^*)$.
Network  sizes (depth and  trainable-parameters number) and the grid resolution (depending on $n$, $h$ and $\varepsilon$) are selected to achieve an approximation error of $\m O(\varepsilon)$ for any  $\Pi_{\theta}\m P_{\theta}^{(n)}$ satisfying Assumption 1.  Further details are given in Appendix C.4. Note that this  discrete approximation and ReLU network construction are mainly used for  technical convenience in theoretical analysis.  The assumption that the data space $\m X$ is compact is also made for technical simplicity, and can be replaced by a light-tail condition on $\m P_{\theta}$ (e.g. exponential decay),  at the cost of extra logarithmic factors in $\varepsilon^{-1}$ in the bound.
\end{remark}

Corollary~\ref{thm:Exponential family result} shows that when the true likelihood lies in an exponential family,  and standard regularity conditions hold,  any prescribed target error in MMD can be attained with a training size $N$ that depends only polynomially on  $\varepsilon^{-1}$, $n$ and $h^{-1}$.   In particular, if we take $h=\frac{1}{\sqrt{n}}$, then, omitting logarithmic terms,  the required $N$ in Corollary~\ref{thm:Exponential family result} is $\m O(\varepsilon^{-\frac{\ov d}{2}-1}n^{\frac{d(\ov d+1)}{2}}\max (n^d,n^{d_X+1}))$.  Note that the exponent in this polynomial dependence on  $n$ is likely conservative and may be improved under stronger assumptions, for instance, higher-order smoothness of  $t(\cdot),\eta(\cdot),A(\cdot)$. The key reason for this polynomial dependence is that the exponential family admits an additive sufficient statistic of the form
 $ T(\Xn) \;=\; n^{-1}\sum_{i=1}^n t(X_i)$, with $t(X) =  (t_1(X),t_2(X),\ldots,t_k(X) )$, which mitigates the curse of dimensionality in the raw data vector. Moreover,  Corollary~\ref{thm:Exponential family result}  does not require $k=d$ and enables accurate approximation even for curved exponential families, thereby improving upon approaches that use the posterior mean as the summary statistic.

\subsection{Example 2: Large-sample Regime}\label{sec:large_sample}
Example 1 relied on exact low-dimensional sufficient statistics in exponential families. We now relax this restriction and consider general smooth parametric models in the large-sample regime ($n \to \infty$). Our analysis assumes the standard regularity conditions for a Bernstein--von Mises (BvM) theorem, detailed in Assumption 2 of Appendix A. These include: (i)   $\m P_{\theta}$ has a smooth enough log-density function; (ii) The prior has a smooth log-density and sufficiently light tails; (iii) The data distribution has exponential–type tails, so very large observations are rare; (iv) KL divergence identifies the parameter, and the Fisher information changes smoothly with $\theta$. Assumption~2 allows us to derive BvM  results, implying that for large $n$, 
the true posterior $\mu^*_{\theta|\Xn}(\cdot|\Xn)$ is approximately Gaussian $    N(\widehat{\theta}(\Xn),\, n^{-1} I(\theta)^{-1})$, where $\widehat\theta(\Xn)$ is the maximum likelihood estimator (MLE) and $I(\theta)$ is the Fisher information matrix. Therefore, we set the kernel bandwidth $h = \wt h/\sqrt{n}$ to match the posterior scale. With the reparameterization $z = \sqrt{n}\,(\theta - \widehat{\theta}(\Xn))$, the posterior 
$\mu^*_{z|\Xn}(\cdot|\Xn) $ of $z$
converges to $N(0_d, I(\theta)^{-1})$, which has constant-order variances. Under this rescaling, the MMD satisfies $   n^{-\frac{d}{2}}
\mmd_{h}^2\big(\mu^*_{\theta|\Xn}(\cdot|\Xn),\, \mu_{\wh W}(\cdot| S(\Xn))\big)
 =
\mmd_{\wt h}^2\!\big(\mu^*_{z|\Xn}(\cdot|\Xn),\,
[\sqrt{n}(\cdot-\widehat{\theta}(\Xn))]_{\#} \mu_{\wh W}(\cdot | S(\Xn))\big).$
This motivates evaluating the posterior estimator with the rescaled criterion of
$n^{-\frac{d}{2}}
\mmd^2_{h}\big(\mu^*_{\theta|\Xn}(\cdot|\Xn),\, \mu_{\wh W}(\cdot| S(\Xn))\big)$.
  Furthermore, the approximate Gaussianity  of the posterior enables the construction of effective low-dimensional summary statistics,  thereby mitigating the curse of dimensionality. The next corollary shows that, with appropriately chosen approximation families and a training sample size $N$ that scales polynomially in $n$ and $\wt h$, one can achieve an error of order $\m O(\frac{\log n}{n})$  in  terms of the rescaled expected MMD.

\begin{corollary}[Large-Sample Regime] \label{corollary: large sample setting}
Let $n\geq n_b$ for a sufficiently large constant $n_b$, and let  $\{(\theta_i,\Xn_i)\}_{i=1}^N$ be  i.i.d. samples from $\Pi_{\theta}\m P_{\theta}^{(n)}$.  Let the bandwidth satisfy $h = \wt h/\sqrt{n}$ with $c_6 n^{-c_7} \leq \wt h \leq c_8$ for constants $c_6,c_7,c_8>0$. There exists a model class $\cS \times \cW$  depending only on $n$, $d$, $d_X$ and constants in Assumption~2 of Appendix~A.2 such that,  for any $\Pi_{\theta}\m P_{\theta}^{(n)}$ satisfying Assumption~2, if  $    N \geq C\,  \frac{n^{2+d_X+d/2}}{\wt h^d}(\log n)^{\delta_1},$ then with probability at least $1-\frac{1}{N}$,  
\begin{equation*}
 n^{-\frac{d}{2}} \E_{\Xn\sim \m P^{(n)}} [\mmd^2_{h}(\mu^*_{\theta|\Xn}(\cdot|\Xn), \mu_{\wh{W}}(\cdot|\wh S(\Xn)))] \leq C_1\frac{\log n}{n},
\end{equation*}
where $\delta_1$, $C$, $C_1$ are constants independent of $(n,N)$. The explicit form of $\delta_1$ and the dependence of $C,C_1$ are given in the formal version of this Corollary in Appendix~A.2.
\end{corollary}

\begin{remark}
   Here $\ms S$  consists of ReLU  networks with $d$-dimensional output, and $\ms W$  consists of ReLU networks with input dimension $d$ and output dimension $d^2$.
    For each $(S,W)$, $\mu_{W}(\cdot|S(\Xn))$ is a Gaussian distribution with mean $S(\Xn)$ and covariance $n^{-1}W(S(\Xn)) \allowbreak W(S(\Xn))^T$.  Because the estimator is amortized, the proof must control approximation error across datasets rather than at a single observation. We first establish a Bernstein--von Mises bound on a high-probability set of datasets and carefully control its complement, obtaining a rescaled expected MMD error of order \(\mathcal O(\log n/n)\) between the true posterior and its data-dependent Gaussian approximation. We then choose the sizes of \(\mathcal S\) and \(\mathcal W\) so that the networks approximate the corresponding Gaussian mean and covariance, viewed as functions of \(\Xn\), at the same order. Finally, we calculate the resulting complexity terms \(T_1\) and \(T_2\) and apply Theorem~\ref{thm:Generalization Error}. Further details are provided in Appendix~C.5.
   
\end{remark}
\begin{remark}
 A ReLU network is not the only possible choice for the summary network. In practice, for i.i.d. datasets we use DeepSets~\citep{zaheer2017deep}, which respect the exchangeability of the summary over datapoints. Here we employ ReLU networks for technical convenience, as their approximation and generalization properties are well understood, allowing us to derive explicit bounds on the complexity terms $T_1$ and $T_2$ in the oracle inequality (Theorem~\ref{thm:Generalization Error}), and the ReLU class already suffices to establish the desired polynomial bounds.
    \end{remark}
\noindent The result in Corollary~\ref{corollary: large sample setting} differs from Corollary~\ref{thm:Exponential family result} as we require $n$ to be sufficiently large, and the attainable target tolerance is of order $\m O(\frac{\log n}{n})$ 
rather than an arbitrary $\varepsilon>0$. The reason is that, for general models,
low-dimensional sufficient statistics typically do not exist.  We therefore resort to the BvM theorem, stating that in large samples,  the posterior $\mu^*_{\theta|\Xn}(\cdot|\Xn)$ is close to a Gaussian, and  the posterior mean  serves as an ``approximately sufficient'' summary in the large-sample regime. This approximation introduces an irreducible error, which is the discrepancy between the true posterior and its Gaussian limit. Therefore, we fix the target tolerance at or above this approximation error, which is $\m  O(\frac{\log n}{n})$ in our analysis. Notably,  achieving this tolerance only requires $N$ to grow polynomially in $n$. In particular, if we take $h=\frac{1}{\sqrt{n}}$ (i.e., $\wt h=1$), then, omitting logarithmic terms, the required $N$ scales as $\m O(n^{2+d_X+d/2})$. This is smaller than the bound implied by Corollary~\ref{thm:Exponential family result}. For example, if $d=d_X=1$, choosing $h=\frac{1}{\sqrt{n}}$ and $\varepsilon=n^{\frac{d}{2}} \cdot \frac{\log n}{n}$ in Corollary~\ref{thm:Exponential family result} (note that this corollary considers the unrescaled MMD) yields a requirement of $N=\m O(n^{\frac{15}{4}})$ (omitting logarithmic terms), whereas Corollary~\ref{corollary: large sample setting} only requires $N=\m O(n^{\frac{7}{2}})$ to achieve the same order of error. The gap arises because, in Corollary~\ref{thm:Exponential family result}, we construct networks to  approximate the posterior directly, whereas in Corollary~\ref{corollary: large sample setting} we approximate a simpler Gaussian  limit. Approximating this Gaussian target admits simpler network architectures, which in turn reduces the excess risk for a fixed $N$, and consequently lowers the required $N$ to achieve a fixed error.

\vspace{-0.5em}

\section{Numerical Experiments}
\label{sec: Numerical Experiments}

We evaluate our framework against three baselines: BayesFlow \citep{radev2020bayesflow}, which can be regarded as a KL-based special case of our framework without refinement; DNN-ABC \citep{jiang2017learning}, which learns summaries via posterior mean estimation; and SNPE-C \citep{greenberg2019automatic}, a neural posterior estimation method based on sequential training. Experimental settings are in Appendix~\ref{appendix:Implementation Details} with replication code available at \url{https://github.com/anonymous-submission-202617/SummaryLearningForSBI}. Hereafter, \emph{Refine} and \emph{Refine+} denote the local refinement algorithms of Sections~\ref{sec:refine1} and~\ref{sec:refine2}.

\renewcommand{\floatpagefraction}{0.8}
\renewcommand{\textfraction}{0.1}
\subsection{Toy Example With Multimodal Posterior}\label{subsec: toy example}

\begin{figure}[ht]
  \centering
  \includegraphics[width=0.75\textwidth]{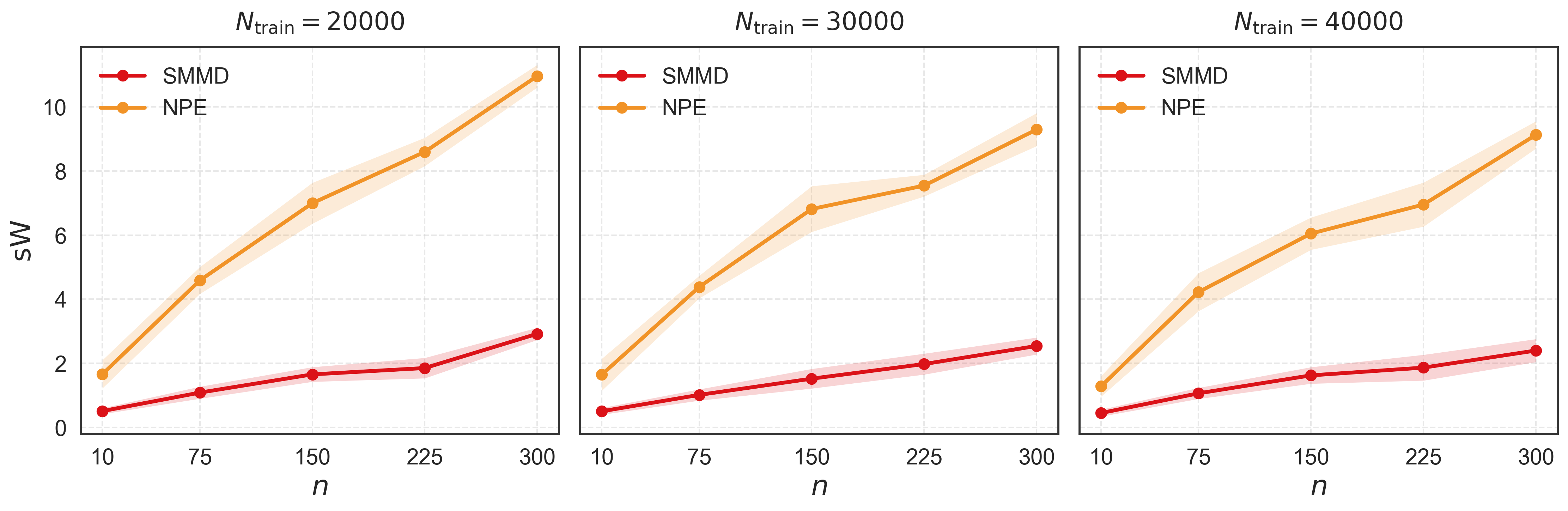}
\caption{Sliced-Wasserstein (sW) distance between the estimated and ground-truth posteriors, as a function of the observation count \(n\) under different training sizes \(N\). Since the posterior contracts at the root-\(n\) rate, we report the rescaled distance \(\sqrt{n}\times\mathrm{sW}\), which compares the posteriors on the root-\(n\) local scale. The lines and shaded regions show the mean and standard deviation over \(25\) test datasets.}
  \label{fig:toy_sample_size}
  \vspace{-1em}
\end{figure}

We first evaluate our method on the benchmark of \citet{papamakarios2019sequential}, which has tractable likelihood but a complex posterior with four symmetric modes and $5$ parameters. To increase difficulty, we map each $X_i\in\mathbb{R}^2$ onto the  sphere $\mathbb{S}^2$ by inverse stereographic projection,  making standard moment-based summaries insufficient. To isolate the role of summary networks, we train SNPE-C~\citep{greenberg2019automatic} directly on $X^{(n)}$ without a summary.  Since observations are i.i.d., we also include two observation-wise baselines: NLE~\citep{papamakarios2019sequential}, which learns and multiplies
single-observation likelihood factors, and the tall-data neural posterior score estimator of \citet{linhart2024diffusion}, which we refer to as TD-NPSE. It combines evaluations of a posterior-score network trained on single observations through a Gaussian approximation to the tall-posterior diffusion process.
Both methods are trained on parameter–single-observation pairs, and the same trained network can be reused for different $n$ values. We report MMD and sliced Wasserstein distance (sW) between each learned posterior and the ground-truth posterior, averaged over $10$ runs. As shown in Table~\ref{tab:toy_table}, SNPE-C without a summary performs poorly for these high-dimensional observations. Moreover, the summary-based methods achieve lower MMD and sW errors than observation-wise baselines  for every \(n\) considered. One possible explanation is that NLE and TD-NPSE estimate quantities from single observations and then combine them across the full dataset. Systematic errors in the learned likelihood factors or posterior scores may therefore accumulate as \(n\) increases. Moreover, the ambient-space NLE flow is structurally misspecified because the transformed data lie on a lower-dimensional manifold, which likely further contributes to NLE's poor performance. ABC--MCMC refinement provides an additional improvement to the summary-based estimators with only small extra computational cost. Following~\cite{linhart2024diffusion}, we also examine how the error changes with \(n\) under different training budgets, comparing SMMD with NPE, implemented as SNPE-C without sequential training. As Figure~\ref{fig:toy_sample_size} shows,  over the range considered, NPE without a summary deteriorates much more rapidly with increasing \(n\) than SMMD. Notably, even at the smallest budget, \(N=20{,}000\), SMMD achieves substantially lower sW values than NPE trained with larger budgets. This pattern is  consistent with our theoretical finding that low-dimensional summaries can mitigate the effect of increasing sample size.

\begin{table}[t]
\footnotesize
\centering
\caption{Results for the toy example.}
\label{tab:toy_table}
\setlength{\tabcolsep}{4pt}
\renewcommand{\arraystretch}{0.92}
\resizebox{\linewidth}{!}{%
\begin{tabular}{@{}l *{10}{S[table-format=2.2]}@{}}
\toprule
& \multicolumn{2}{c}{\textbf{SMMD}} & \multicolumn{2}{c}{\textbf{MMD}} & \multicolumn{2}{c}{\textbf{BayesFlow}} & \textbf{DNN-ABC} & \textbf{SNPE-C} & \textbf{NLE} & \textbf{TD-NPSE} \\
\cmidrule(lr){2-3} \cmidrule(lr){4-5} \cmidrule(lr){6-7} \cmidrule(l){8-8} \cmidrule(l){9-9} \cmidrule(l){10-10} \cmidrule(l){11-11}
& \textit{Amort.} & \textit{Refined} & \textit{Amort.} & \textit{Refined} & \textit{Amort.} & \textit{Refined} & & & & \\
\midrule
$\text{MMD}$ ($\times 10^{-2}$), $n=25$ & 2.76 & 1.98 & 2.51 & {\bfseries1.96} & 2.20 & 1.97 & 9.47 & 8.04 & 10.85 & 7.09 \\
MMD ($\times 10^{-2}$), $n=50$  & 2.47 & {\bfseries 1.81} & 2.49 & 1.93 & 2.44 & 1.86 & 9.71 & 10.27 & 18.05 & 11.02 \\
MMD ($\times 10^{-2}$), $n=100$ & 2.34 & {\bfseries 1.51} & 2.38 & 1.89 & 2.53 & 1.99 & 9.85 & 9.87 & 14.91 & 12.45 \\
\midrule
sW ($\times \sqrt{n})$, $n=25$ & 0.63 & {0.56} & 0.57 & {\bfseries 0.51} & 0.55 & 0.52 & 3.28 & 2.87 & 3.41 & 1.51 \\
sW ($\times \sqrt{n})$, $n=50$ & 0.70 & {\bfseries 0.65} & 0.72 & 0.67 & 0.90 & 0.72 & 4.41 & 3.89 & 4.77 & 2.01 \\
sW ($\times \sqrt{n})$, $n=100$ & 1.13 & {\bfseries 1.01} & 1.41 & 1.13 & 1.50 & 1.10 & 6.46 & 5.85 & 7.08 & 3.53 \\
\bottomrule
\end{tabular}}
\\[2pt]
\par\footnotesize\emph{Notes:}
SMMD, MMD and BayesFlow denote our approach with $\mathcal{D}$ given by sliced MMD, MMD, and KL divergence respectively.
``Amort.'' indicates amortized estimators before refinement; ``Refined'' indicates ABC-MCMC local refinement without sequential training.  Rows labeled ``$\text{MMD}$'' report the squared MMD with a median-heuristic bandwidth that accounts for the sign symmetry of $(\theta_3, \theta_4)$; this differs from the MMD training objective. Rows labeled ``sW'' report  sliced Wasserstein distances, scaled by \(\sqrt{n}\) to account for posterior contraction and place the discrepancy on the same scale for varying $n$.
 
\end{table}

\subsection{SIR Model With Small Training Budget}\label{subsec: sir}

\begin{figure}[!t]
    \centering
    \includegraphics[width=0.80\textwidth]{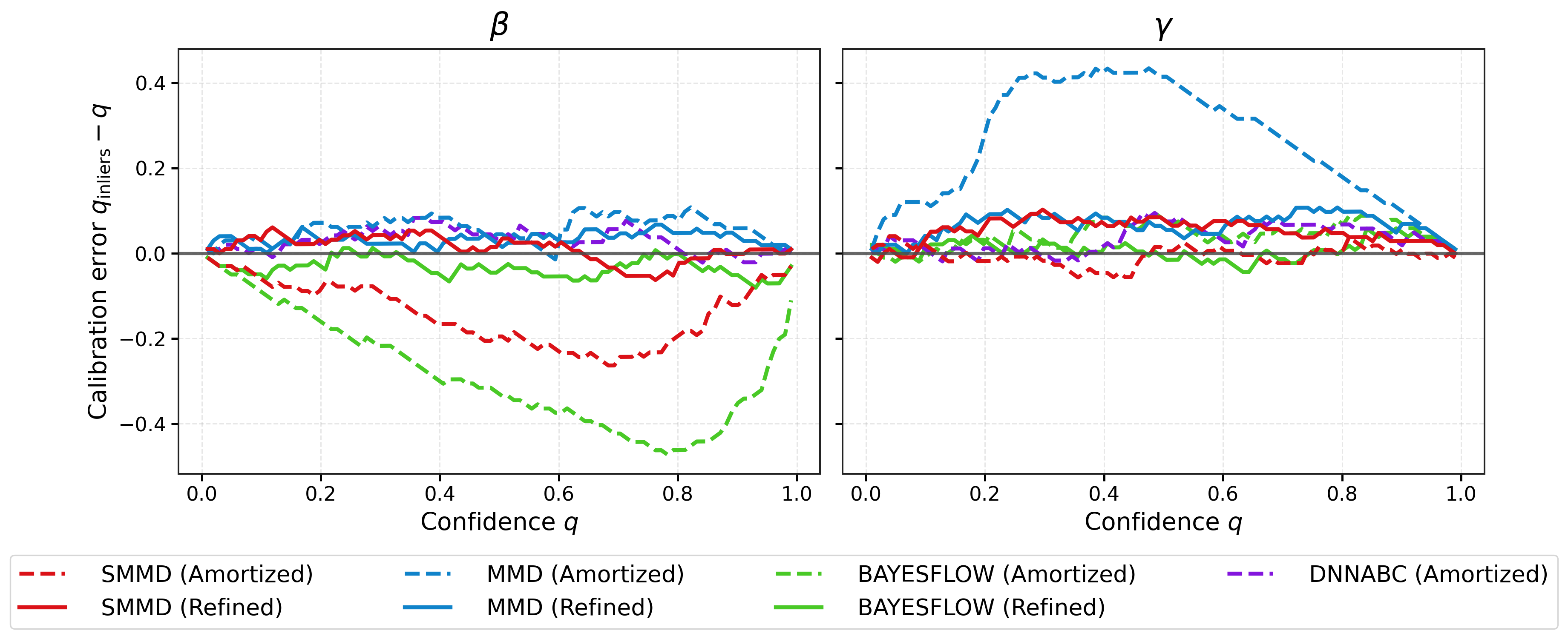}
    \caption{Calibration curves for the SIR benchmark. The $x$-axis shows the nominal credibility level, while the $y$-axis shows empirical coverage minus nominal coverage; zero indicates perfect calibration.}
    \label{fig:sir_calibration}
    \vspace{-0.5em}
\end{figure}

\begin{figure}[!t]
    \centering
    \includegraphics[width=0.75\textwidth]{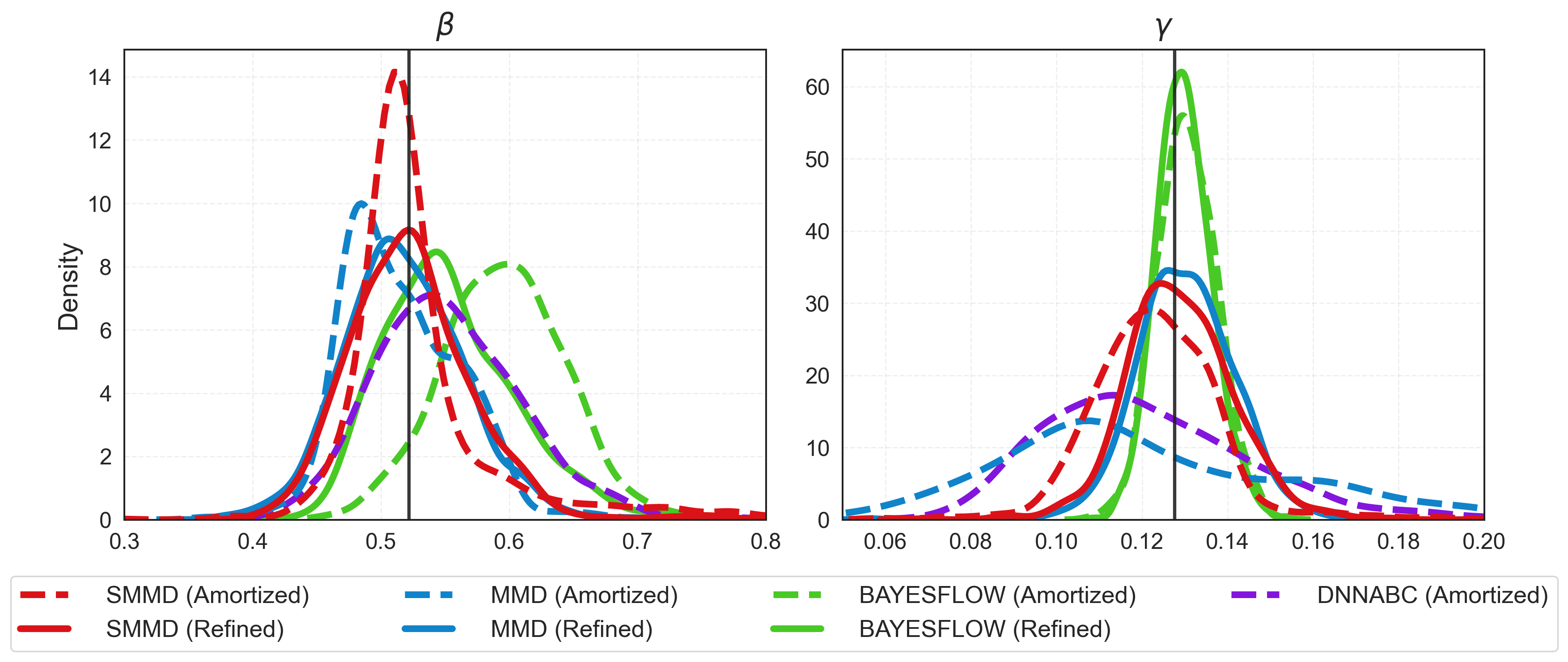}
    \caption{Illustration of the approximate posterior before and after refinement for SIR.}
    \label{fig:sir_refine}
 
\end{figure}

In the susceptible--infected--recovered (SIR) model, the transmission rate ($\beta$) and recovery rate ($\gamma$) are inferred from noisy observations of the infected population \citep{lueckmann2021benchmarking}. We increase the observation count from ($n=10$) to ($n=80$) and increase the observation noise to make the benchmark more challenging. We use a small training set of size $N=768$, so the training data sparsely cover the prior-predictive distribution, and a test  observation may have few similar training samples. This setting challenges global amortized inference. We assess prior-averaged calibration using the prior-predictive sampling protocol underlying simulation-based calibration (SBC) \citep{talts2018validating}. For each  replication, we draw a parameter from the prior, simulate a corresponding dataset, and perform posterior inference.  Table~\ref{tab:sir} reports quantitative results from \(500\) replications. Figure~\ref{fig:sir_calibration} reports the calibration curves and Figure~\ref{fig:sir_refine} shows marginal posteriors from a single replication, with and without ABC-MCMC refinement.   Overall, refinement reduces bias and generally brings empirical coverage closer to nominal levels, with the largest gains occurring when the amortized estimator exhibits substantial undercoverage or overcoverage. The refined posteriors are often conservative and exhibit some overcoverage, possibly because of the smoothing induced by the finite ABC tolerance. Although this indicates residual miscalibration, it can substantially alleviate severe undercoverage in the amortized approximation. These results suggest that, even when the training budget is insufficient to learn an accurate amortized posterior, the globally learned summary map can retain posterior-relevant information that supports effective observation-specific refinement.

\begin{table}[t]
\footnotesize
\centering
\caption{Results for the SIR model.}
\label{tab:sir}
\setlength{\tabcolsep}{8pt}
\renewcommand{\arraystretch}{0.95}
\begin{tabular}{@{}ll S[table-format=1.2] *{4}{S[table-format=1.2]}@{}}
\toprule
& & \multicolumn{1}{c}{\textbf{Bias}} & \multicolumn{2}{c}{\textbf{Coverage}} & \multicolumn{2}{c}{\textbf{Cal err}} \\
& & \multicolumn{1}{c}{($\times 10^{-2}$)} & \multicolumn{2}{c}{} & \multicolumn{2}{c}{($\times 10^{-1}$)} \\
\cmidrule(lr){4-5} \cmidrule(lr){6-7}
& & & {$\beta$} & {$\gamma$} & {$\beta$} & {$\gamma$} \\
\midrule
\multirow{2}{*}{\textbf{SMMD}} & Amort.  & 4.29 & 0.89 & 0.95 & 1.39 & 0.15 \\
                               & Refined & {\bfseries 3.42} & 0.96 & 0.98 & 0.25 & 0.60 \\
\cmidrule{2-7}
\multirow{2}{*}{\textbf{MMD}}  & Amort.  & 3.92 & 0.98 & 1.00 & 0.62 & 2.74 \\
                               & Refined & 3.53 & 0.97 & 1.00 & 0.33 & 0.74 \\
\cmidrule{2-7}
\multirow{2}{*}{\textbf{BayesFlow}} & Amort.  & 7.10 & 0.68 & 0.98 & 3.06 & 0.46 \\
                                    & Refined & 4.20 & 0.88 & 0.99 & 0.35 & 0.14 \\
\cmidrule{2-7}
\multicolumn{2}{@{}l}{\textbf{DNN-ABC}} & 5.08 & 0.95 & 0.98 & 0.32 & 0.30 \\
\bottomrule
\end{tabular}
\\[2pt]
\par\footnotesize\emph{Notes:}
Bias is the mean absolute deviation between posterior means and true parameters, averaged over 500 replications. Coverage is the fraction of replications in which the true parameter lies in the 95\% credible interval. ``Refined'' means ABC-MCMC refinement (without sequential training) is applied. Calibration error is the median absolute deviation between empirical and nominal coverage over a grid of credibility levels. 
\vspace{-0.8em}
\end{table}

\subsection{Population Ecology Model With Vague Prior} \label{subsec: lv}

\begin{figure}[!t]
    \centering
    \includegraphics[width=1.0\textwidth]{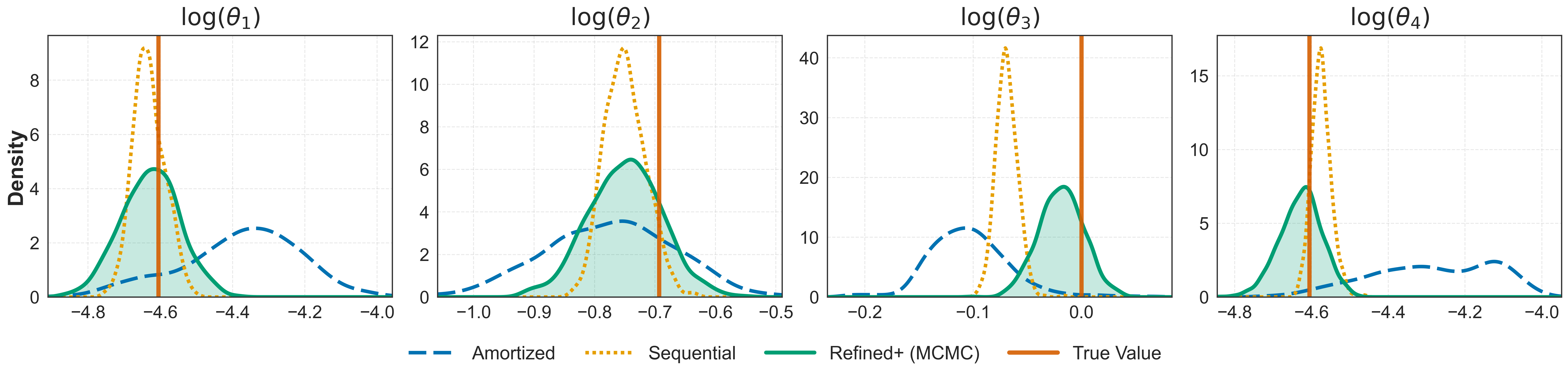}
    \caption{Posterior density plots across the stages of the \emph{Refine+} procedure (Algorithm~1) for the SMMD variant. ``Amortized,'' ``Sequential,'' and ``Refine+ (MCMC)'' denote the initial amortized posterior, the reweighted sequential posterior estimate, and the final estimate after ABC--MCMC refinement, respectively.}
    \label{fig:lv_refineplus}
\end{figure}

We follow the Lotka--Volterra benchmark of \citet{papamakarios2016fast}, a prey-predator model with $\theta\in\mathbb{R}^4$ controlling growth/decay rates and predation. Observations $\Xn$ are generated by the Gillespie algorithm, recording $d_X=2$ population states at 150 time points ($n=151$ including the initial state). We use a vague prior to ensure most simulated trajectories differ substantially from the observation. Prior SBI work addresses vague priors via sequential training \citep{papamakarios2016fast,papamakarios2019sequential}. Building on this idea, we adopt our \emph{Refine+} procedure, which adds a sequential training step on top of the ABC-MCMC local refinement. SNPE-C uses the standard hand-crafted $p=9$ summary vector from \citet{papamakarios2016fast}. We compare all methods under a matched simulation budget, defined as the total number of simulator calls across all stages (e.g., amortized training and refinement). Figure~\ref{fig:lv_refineplus} illustrates the contribution of each stage of \emph{Refine+} for SMMD. Sequential training produces a more concentrated posterior approximation than the initial amortized fit, while the subsequent ABC--MCMC refinement further reduces bias. Table~\ref{tab:lotka_volterra_model_performance_final} reports the quantitative results. Overall, the \emph{Refine+} methods achieve low bias and run substantially faster than SNPE-C, with SMMD attaining near-nominal coverage and the lowest calibration error. BayesFlow exhibits larger bias and calibration error than MMD and SMMD in this example. Our diagnostics suggest that this may be partly attributable to run-to-run variability in optimizing the KL-based NLL objective during amortized training. Because the posteriors are sharply concentrated relative to the vague prior, early misalignment may assign very low density to paired training parameters, leading to large NLL contributions and more variable stochastic updates. As a result, convergence can be sensitive to initialization, affecting the amortized fit and, in turn, the refined posterior.  DNN-ABC produces wider credible intervals and greater bias, indicating less precise inference under the same simulation budget.

\begin{table}[t]
\footnotesize
\centering
\caption{Results for the Lotka-Volterra model.}
\label{tab:lotka_volterra_model_performance_final}
\setlength{\tabcolsep}{4pt}
\renewcommand{\arraystretch}{0.95}
\begin{tabular}{@{}l S[table-format=2.2] *{2}{S[table-format=1.2]} *{2}{S[table-format=1.2]} *{4}{S[table-format=1.2]} *{4}{S[table-format=1.2]} c@{}}
\toprule
& \multicolumn{1}{c}{\textbf{Bias}} & \multicolumn{2}{c}{\textbf{CrI len}} & \multicolumn{2}{c}{\textbf{CrI len}} & \multicolumn{4}{c}{\textbf{Coverage}} & \multicolumn{4}{c}{\textbf{Cond Cal err}} & \multicolumn{1}{c}{\textbf{Time}} \\
& \multicolumn{1}{c}{($\times 10^{-1}$)} & \multicolumn{2}{c}{($\times 10^{-1}$)} & \multicolumn{2}{c}{($\times 10^{-2}$)} & \multicolumn{4}{c}{} & \multicolumn{4}{c}{} & \\
\cmidrule(lr){3-4} \cmidrule(lr){5-6} \cmidrule(lr){7-10} \cmidrule(lr){11-14}
& & {$\theta_1$} & {$\theta_3$} & {$\theta_2$} & {$\theta_4$} & {$\theta_1$} & {$\theta_2$} & {$\theta_3$} & {$\theta_4$} & {$\theta_1$} & {$\theta_2$} & {$\theta_3$} & {$\theta_4$} \\
\midrule
\shortstack[l]{\textbf{SMMD} \\ (Refine+)}    & {\bfseries 1.51} & 0.28 & 0.35 & 0.24 & 0.33 & 0.95 & 0.95 & 0.96 & 0.95 & 0.07 & 0.04 & 0.09 & 0.09 & 00:16:12 \\
\cmidrule{1-15}
\shortstack[l]{\textbf{MMD} \\ (Refine+)}     & 2.12 & 0.31 & 0.44 & 0.30 & 0.38 & 0.95 & 0.93 & 1.00 & 1.00 & 0.08 & 0.05 & 0.10 & 0.12 & 00:17:02 \\
\cmidrule{1-15}
\shortstack[l]{\textbf{BayesFlow} \\ (Refine+)} & 2.61 & 0.60 & 0.43 & 0.35 & 0.61 & 1.00 & 1.00 & 1.00 & 1.00 & 0.16 & 0.09 & 0.15 & 0.15 & 00:41:19 \\
\cmidrule{1-15}
\textbf{DNN-ABC}            & 12.02 & 2.76 & 5.92 & 4.14 & 2.91 & 1.00 & 1.00 & 1.00 & 1.00 & 0.26 & 0.23 & 0.23 & 0.17 & 00:05:02 \\
\cmidrule{1-15}
\textbf{SNPE-C}             & 3.18 & 0.39 & 0.51 & 0.71 & 0.53 & 0.95 & 0.90 & 1.00 & 0.95 & 0.13 & 0.13 & 0.20 & 0.14 & 01:00:01 \\
\bottomrule
\end{tabular}
\\[2pt]
\par\footnotesize\emph{Notes:}
 \emph{Cond cal err} denotes the median absolute deviation between empirical and nominal coverage over a grid of credibility levels. ``Cond'' indicates that datasets are generated at a fixed true parameter rather than over parameters drawn from the prior. \emph{CrI len} is the average length of the \(95\%\) credible intervals, and \emph{Time} is the average runtime. All metrics are averaged over \(100\) replications. Other notation and metric definitions follow Table~\ref{tab:sir}.
 \vspace{-0.5em}
\end{table}

\subsection{Real Data Application With A Misspecified Simulator}
\label{sec:radio}
\begin{figure}[ht]
  \centering
  \begin{minipage}{0.48\textwidth}
    \centering
    \includegraphics[width=\linewidth]{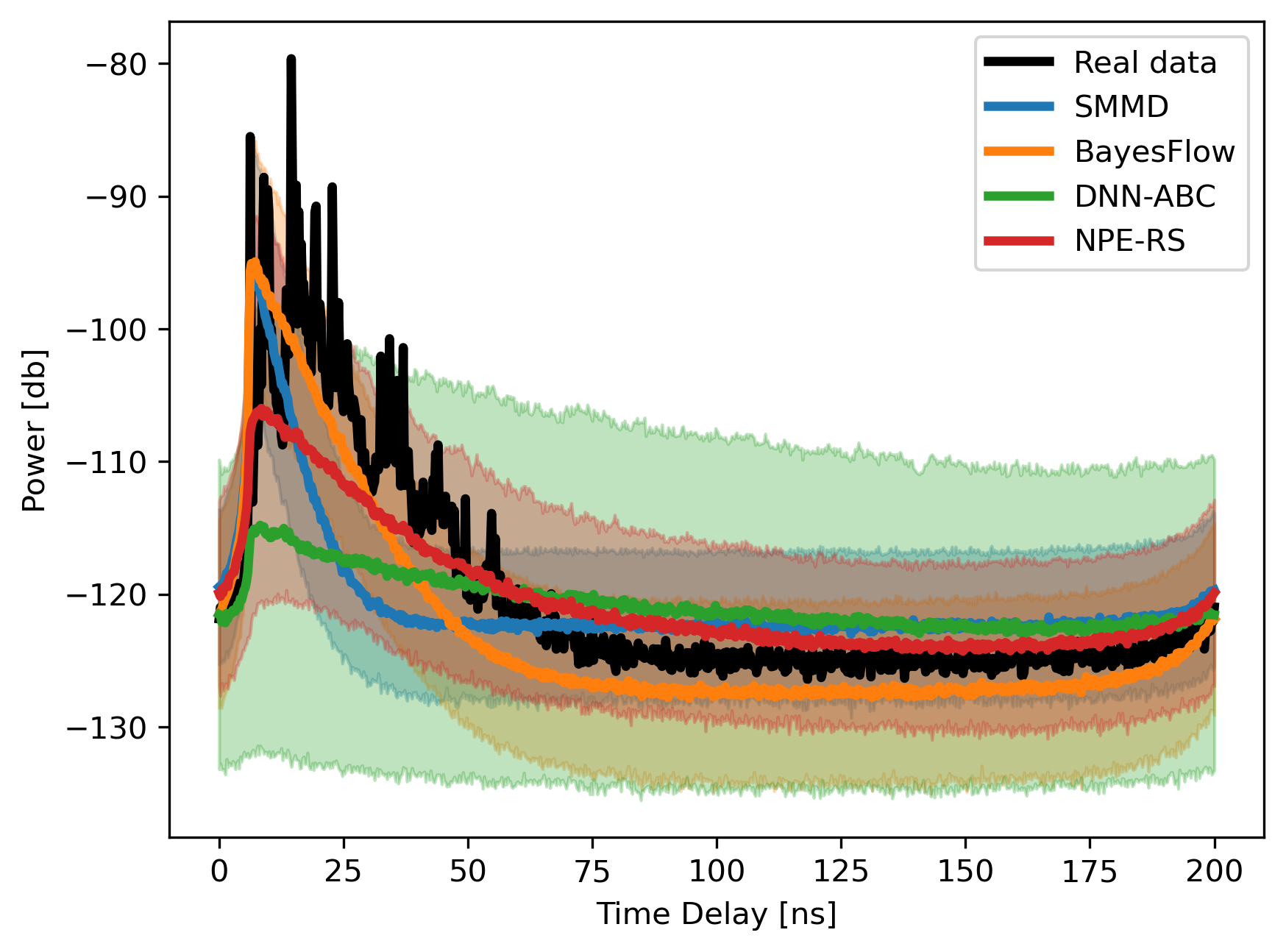}
    \subcaption{Amortized}
  \end{minipage}
  \hfill
  \begin{minipage}{0.48\textwidth}
    \centering
    \includegraphics[width=\linewidth]{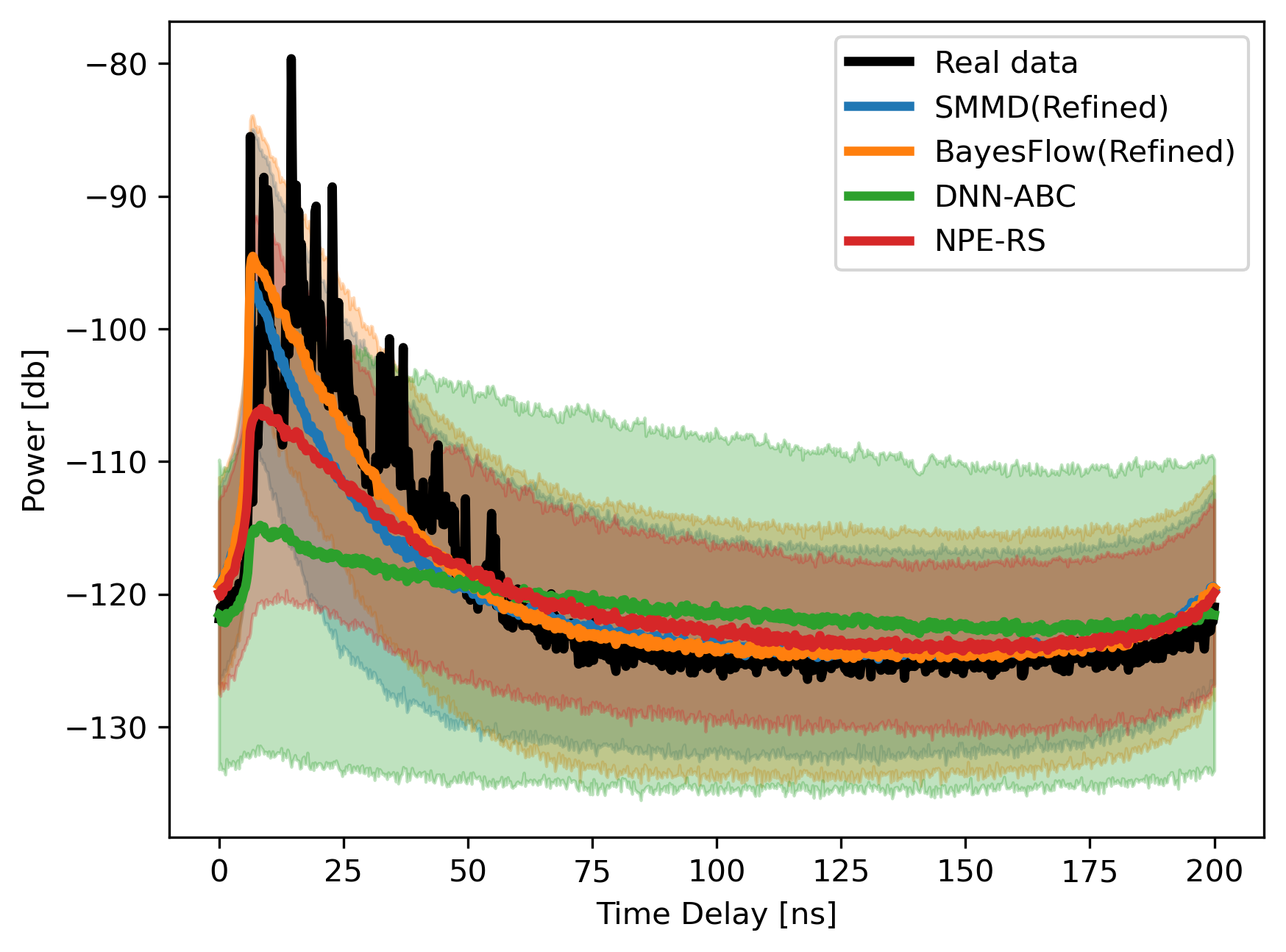}
    \subcaption{Refined}
  \end{minipage}
  \caption{Predictive performance for radio propagation benchmark. Refined means ABC-MCMC refinement is applied. For each $\Xn\in \mb R^{n\times d_X}$ drawn from the posterior predictive distribution, we first average its $n=100$ power-delay profiles to obtain a mean trace over the $d_X$ dimensions. Solid curves show predictive means and shaded regions show $\pm 1$ standard deviation. Each curve is plotted at all 801 delay points, covering approximately 0--200 ns.}
  \label{fig:radio}
 
\end{figure}

We evaluate our method on a real-world radio propagation dataset from~\cite{gustafson2016modeling}, using the Turin stochastic channel model as the simulator, following the setup in~\citet{huang2023learning}. The simulator has an intractable likelihood and generates high-dimensional complex-valued time series with four parameters.  We perform posterior inference on real measurements $\Xn_* \in \mathbb{R}^{n\times d_X}$ ($d_X=801, n=100$), and the simulator is known to be misspecified. We further increase the inference challenge by replacing the informative prior in~\citet{huang2023learning} with an uninformative one. We compare SMMD and BayesFlow with DNN-ABC and NPE-RS, a robust NPE variant from~\citet{huang2023learning} that adds penalization to encourage the learned summary statistics of the observed data to remain within the simulator-supported summary space, thereby improving NPE under model misspecification. Figure~\ref{fig:radio} shows
the resulting posterior predictive distributions with and without ABC-MCMC refinement.  Refinement markedly improves the predictive performance, suggesting that under model misspecification,  summary networks generalize better to ``unseen'' observations than inference networks. Moreover, despite lacking explicit robustness penalties, our refined methods fit the observed data better than NPE-RS. Quantitatively, we assess predictive accuracy using the 
  root mean squared error (RMSE) between  the posterior-predictive mean series and the observed mean series, averaged over the $20$ replications.  The RMSEs are \(8.94\) for DNN-ABC and \(6.48\) for NPE-RS. Before refinement, the RMSEs are \(5.08\) for SMMD and \(5.02\) for BayesFlow; after refinement, SMMD achieves the lowest RMSE at $4.12$, compared to $4.61$ for BayesFlow.

\vspace{-0.5em}
\section{Discussion}
\label{sec: Discussion}

In this work, we propose an SBI framework that jointly learns low-dimensional informative summaries and an amortized posterior estimator, and  we develop local refinement methods to mitigate the amortization gap. Our theory shows that operating through low-dimensional summaries can mitigate the curse of dimensionality caused by high-dimensional observations. Experiments validate the effectiveness of our framework in handling high-dimensional observations and demonstrate consistent gains from refinement. The results further suggest that the quality of the learned summary is a key factor in overall SBI performance, and that useful summaries may be learned with  fewer simulations than  typically required for the inference network. Given  that most of the SBI literature focuses on inference networks, we hope our results will draw more attention to summary networks.

Several directions remain for future work. Theoretically, although our formal large-sample corollary is stated for i.i.d. data, the Bernstein--von Mises phenomenon has been established for certain classes of dependent processes under mixing or ergodicity conditions~\citep{de2008asymptotic}. This suggests that dependent-data models may also admit low-dimensional representations that become asymptotically sufficient. Establishing corresponding approximation and learning guarantees for  dependent observations is left for future work. Methodologically, the ABC posterior is not the only possible target for local refinement.  A natural alternative is Bayesian synthetic likelihood (BSL), which uses simulations to approximate the sampling distribution of the summary statistics by a multivariate Gaussian distribution \citep{wood2010statistical,price2018bayesian}.   This avoids choosing an ABC tolerance and may improve computational efficiency, but it requires the learned summary to be approximately Gaussian conditional on the parameter \citep{frazier2023bayesian}, a property not guaranteed by our current objective. Jointly learning informative summaries that also satisfy this requirement is an interesting direction for future work.


\subsection*{Funding}
Rong Tang was supported by the Hong Kong RGC Grant ECS 26304226 and the NSFC grant NSFC26SC48.

\begingroup
\setlength{\bibsep}{0pt}
\bibliography{references/bibliography}
\endgroup

\clearpage
\appendix
\appendixstart
\appendixcontents

\medskip
 \noindent\textbf{Notation}:  We follow the notation in the main text and introduce additional symbols for technical proofs.  Let $\|\cdot\|_{\rm F}$ and $\|\cdot\|_{\rm op}$ denote the Frobenius and operator norms, respectively, and $\mb N_{+}$ the set of positive integers. For two $a,b \in \mathbb{R}$, write $a\vee b=\max(a,b)$ and $a\wedge b=\min(a,b)$.
The symbols $\lesssim$ and $\gtrsim$ denote inequalities up to an $n$-independent constant.   For any sequence $\{a_n:\,n\geq 1\}$,  $\m O(a_n)$ denotes the order of $a_n$ up to a multiplicative constant independent of $n$. The symbols $C$, $c$, $C_0$, $c_0$, $C_1$, $c_1$, $C_2$, $c_2$,\ldots are generically used to denote positive constants whose values might change from one line to another, but are independent of everything else. Define the element-wise clip function ${\rm clip}(x; a, b) = \max(\min(x, b), a)$ and the softmax function ${\rm softmax}(w)=\frac{\exp(w)}{\sum_j\exp(w_j)}$. We also formalize the feedforward ReLU network class used in our proofs. 

\medskip
\noindent\textbf{Definition (Neural network class):}
The class $\Phi_{d_1}^{d_2}(L,U,R)$ with input dimension $d_1$, output dimension $d_2$, depth $L\in\mb N_+$, width $U=(u_0,u_1,\ldots,u_L)\in\mb N_+^{L+1}$, and sparsity bound $R\in\mb N_+$, where $u_0=d_1$ and $u_L=d_2$, is defined as
\begin{equation*}
\begin{aligned}
&\Phi_{d_1}^{d_2}(L,U,R)=\bigg\{(A^{(L)}{\rm ReLU}(\cdot)+b^{(L)})\circ\cdots\circ
(A^{(2)}{\rm ReLU}(A^{(1)}x+b^{(1)})+b^{(2)})\,:\\
&\qquad A^{(i)}\in\mb R^{u_i\times u_{i-1}},\quad b^{(i)}\in\mb R^{u_i},\quad 1\leq i\leq L,  \sum_{i=1}^{L}\big(\|A^{(i)}\|_0+\|b^{(i)}\|_0\big)\leq R\bigg\},
\end{aligned}
\end{equation*}
where ${\rm ReLU}(x)=\max(0,x)$ acts coordinatewise and $\|\cdot\|_0$ denotes the number of nonzero entries. We use $\ms E(U)=\sum_{i=1}^L u_i(u_{i-1}+1)$ to denote the total number of parameter positions; $R$ bounds the number of nonzero weights and biases.

For the network constructions used throughout the proofs, we adopt the
following convention. Each constructed network has at most $R$ nonzero
weights and biases in total. Hidden units whose incoming weights and
biases are all zero have identically zero output and can be removed.
Each remaining hidden unit accounts for at least one nonzero parameter,
so each hidden layer contains at most $R$ such units. By relabeling units
and padding with zero weights and biases, we can therefore place these
networks in a common architecture with each hidden width at most $R$,
without changing the functions they represent. We enlarge $R$, if necessary, so that
$R\geq\max\{d_1,d_2,L,2\}$; for the constructions below, this does not
change the stated orders of the sparsity bounds. Thus every layer,
including the input and output layers, has width at most $R$, and
\[
\ms E(U)
=\sum_{\ell=1}^{L}u_\ell(u_{\ell-1}+1)
\leq LR(R+1)
\leq 2R^3.
\]
Accordingly, in the neural network approximation results below, we
state bounds on the depth $L$ and sparsity $R$; the corresponding bound on the total number of parameter positions $\ms E(U)$ follows from
the inequality above.

\section{Assumptions and Formal Theoretical Statements}\label{appendix: theory}

\subsection{Exponential Family Models}
\label{appendix: exponential_family}

We impose the following regularity conditions on the prior and the exponential family model.

\begin{assumption}\label{assump:exponential}
 Suppose the model $\Pi_{\theta}\m P_{\theta}^{(n)}$ satisfies the following conditions:
\begin{enumerate}
    \item (Positive density and Lipschitz continuity of prior). The prior $\Pi_{\theta}$ is supported on $\Theta=[0,1]^d$ and admits a density function $\pi(\theta)$ with $\pi(\theta)\geq c>0$ for all $\theta\in \Theta$. Moreover, $\log\pi(\cdot)$ is $L$-Lipschitz continuous on $\Theta$, i.e., for all $\theta_1,\theta_2 \in \Theta$, \( |\log\pi(\theta_1) - \log\pi(\theta_2)| \leq L\norm{\theta_1 - \theta_2}.\)
\item (Regularity for exponential family model). It holds that $ \m P_{\theta}^{(n)}=\m P_{\theta}^{\otimes n}$, where there exist functions $\eta:\Theta\to \mb R^k$, $A: \Theta\to \mb R$, $t: \mb R^{d_X}\to \mb R^k$ and $h: \mb R^{d_X} \to \mb R$ so that for any $\theta\in \Theta$, $\m P_{\theta}$ has a density $p(x|\theta)=h(x)\exp(\eta(\theta)^Tt(x)-A(\theta))$ on $\mb R^{d_X}$, and
    \begin{enumerate}
 \item (Bounded sufficient-statistic dimension). It holds that $k\leq p$ for a fixed $p\in \mb N^+$.
         \item (Compact data space). The support $\m X$ of $h(x)$ is compact and contained in a ball $ B(0_{d_X},R_{\mathcal{X}})$ for a constant $R_{\mathcal{X}}$.
        \item (Boundedness). There exists a constant $C$ so that $\sup_{\theta\in \Theta}\norm{\eta(\theta)} \leq C$, $\sup_{\theta\in \Theta}|A(\theta)| \leq C$ and $\sup_{x\in \m X}\|t(x)\|\leq C$.
    \item (Lipschitzness). There exists a constant $L_1$ so that $t(\cdot)$ is $L_1$-Lipschitz continuous on $\m X$, and both $\eta(\cdot)$ and $A(\cdot)$ are $L_1$-Lipschitz continuous on $\Theta$.
\end{enumerate}
\end{enumerate}
\end{assumption}

Note that we only assume the existence of functions $\eta(\cdot)$, $t(\cdot)$, $A(\cdot)$, $h(\cdot)$; their precise forms, as well as the sufficient-statistic dimension $k$, are unknown. Thus, Assumption~\ref{assump:exponential} is essentially nonparametric, as it imposes only boundedness and Lipschitz regularity on these unknown functions without specifying any parametric form. Then we present the  formal version of Corollary~\ref{thm:Exponential family result}, which shows that, with suitably chosen approximation families, any target error $\varepsilon$ can be achieved using a training sample size $N$ that depends polynomially on $\varepsilon^{-1}$, $n$ and $h^{-1}$. 
\begin{corollary}[Exponential Family; formal version]\label{thm:expfam:formal}  
Let $\{(\theta_j,\Xn_j)\}_{j=1}^N$ be i.i.d. samples from $\Pi_{\theta}\m P_{\theta}^{(n)}$.  For any $\varepsilon>0$ and $0<h<1$, there exists a model class $\cS\times \cW$ (depending only on $n$, $\varepsilon$, $h$, $d$, $d_X$ and constants $c,L,p,R_{\m X},C,L_1$ in Assumption~\ref{assump:exponential}), such that for any $\Pi_{\theta}\m P_{\theta}^{(n)}$ satisfying Assumption~\ref{assump:exponential}, if the training size satisfies
\begin{align*}
    N\geq C \log^5 \big(1+\frac{n}{\min(\varepsilon,1) h}\big) \Big(1+h\sqrt{\log\big(1+\frac{1}{\varepsilon}\big)}\Big)^{\frac{d\ov d}{2}+d}\varepsilon^{-\frac{\ov d}{2}-1}h^{-d(\ov d+1)} \max (n^d,n^{d_X+1},h^{-2d}),
\end{align*}
where $\ov d=\max(d,d_X)$ and $C$ is a constant depending only on $d,d_X$ and constants $c,L,p,R_{\m X},C,L_1$ in Assumption~\ref{assump:exponential}, then with probability at least $1-\frac{1}{N}$, 
\begin{equation*}\begin{aligned}
    &\E_{X^{(n)}\sim \m P^{(n)}} [\mmd^2_h(\mu^*_{\theta|\Xn}(\cdot|\Xn), \mu^*_{\theta|\wh S(\Xn)}(\cdot|\wh S(\Xn)))]\\
&\leq \E_{X^{(n)}\sim \m P^{(n)}} [\mmd^2_h(\mu^*_{\theta|\Xn}(\cdot|\Xn), \mu_{\wh{W}}(\cdot|\wh S(X^{(n)})))] \leq \varepsilon.
\end{aligned}
\end{equation*}
\end{corollary}

\subsection{Large-Sample Regime}
\label{appendix: large_sample}
We impose the following regularity conditions on the prior and the likelihood model.
\begin{assumption}\label{assump:large_sample}
Suppose the model $\Pi_{\theta}\m P_{\theta}^n$ satisfies that $\m P_{\theta}^n=\m P_{\theta}^{\otimes n}$, where for any $\theta\in \mb R^d$, $\m P_{\theta}$ admits a density function $p(x|\theta)$ on $\m X\subseteq \mb R^{d_X}$. Let $l(x,\theta)=-\log p(x|\theta)$ denote the negative log-density. Assume
\begin{enumerate}
\item (Smooth negative log-density). There exist constants $c_1 > 0$ and $\alpha_0,\alpha_1\geq 0$ such that for all $\theta \in \mb R^d$, $x \in \m X$, and multi-indices $j \in \N^d_0$ with $|j| \leq 3$, $   \abs{\frac{\partial^{|j|} l(x,\theta)}{\partial \theta^j}} \leq c_1 (1+\|x\|)^{\alpha_0}\left(1+ \norm{\theta}  \right)^{\alpha_1}.$ Moreover, for any $x,x'\in \m X$ and $\theta\in \mb R^d$, $|l(x,\theta)-l(x',\theta)|+\|\nabla_{\theta} l(x,\theta)-\nabla_{\theta} l(x',\theta)\|\leq c_1(1+\|x\|+\|x'\|)^{\alpha_0}(1+\|\theta\|)^{\alpha_1} \|x-x'\|.$  
\item (Light tail and Lipschitz continuity of prior). The prior $\Pi_{\theta}$ admits a density function $\pi(\theta)$. Moreover, there exist constants $c_2 > 0$ and $\alpha_2 > \max(1, 2\alpha_1)$ such that for all $\theta \in \R^d$,  $ \frac{1}{c_2} \exp(-c_2 \norm{\theta}^{\alpha_2}) \leq \pi(\theta) \leq c_2 \exp\big( -\frac{1}{c_2} \norm{\theta}^{\alpha_2} \big).$ Furthermore, for all $\theta,\theta'\in \mb R^d$,   $|\log \pi(\theta)-\log \pi(\theta')|\leq c_2 (\|\theta\|+\|\theta'\|+1)^{\alpha_2-1} \|\theta-\theta'\|.$
\item (Tail control for the data distribution). There exist constants $c_3>0$ and $\alpha_3>0$, $\alpha_4\geq 0$ so that for any $\theta\in \mb R^d$, $    \mb{E}_{X\sim \m P_{\theta}}\big[\exp\big(\Big(\frac{\|X\|}{c_3(1+\|\theta\|)^{\alpha_4}}\big)^{\alpha_3}\big)\big]\leq 2.$
\item (KL divergence lower bound and information-matrix smoothness). There exists constant $c_4>0$ such that for all $\theta, \theta' \in \R^d$, ${\rm KL}(\m P_{\theta} \| \m P_{\theta'}) \geq c_4 \min(\norm{\theta - \theta'}^2, 1)$. Moreover, for any $\theta\in \mb R^d$, the Fisher information matrix $I(\theta)=\mb{E}_{X\sim \m P_{\theta}}[\nabla_{\theta} l(X,\theta)\nabla_{\theta} l(X,\theta)^T]=\mb{E}_{X\sim \m P_{\theta}}[{\rm Hess}_{\theta} l(X,\theta)]$ is well-defined. Moreover, there exist constants $c_5,\alpha_5 \geq 0$ such that for all $\theta, \theta' \in \R^d$, $    \norm{I(\theta) - I(\theta')}_{\rm F} \leq c_5 \left( \norm{\theta} + \norm{\theta'} + 1 \right)^{\alpha_5} \norm{\theta - \theta'}.$
\end{enumerate}
\end{assumption}
\begin{remark}
As in Example~1, here we assume only nonparametric regularity conditions on the underlying density function $p(x|\theta)$, while its exact form is unknown. Moreover, we do not assume compactness of the parameter or data spaces. Instead, we require a sufficiently light-tailed prior, with $\pi(\theta)$ exhibiting an exponential tail of the form $\exp(-c\|\theta\|^{\alpha_2})$ for a sufficiently large $\alpha_2$, and we control the tails of $X\sim \m P_{\theta}$ via an Orlicz-norm bound (see e.g.\ \cite{Vershynin_2018,dudley2014uniform}), allowing the scale to grow polynomially with $\|\theta\|$. Moreover, we do not require uniform boundedness of the $\theta$-derivatives of $l(x,\theta)$; these derivatives may grow polynomially in both $\|\theta\|$ and $\|x\|$, expanding the applicability of our analysis beyond compact settings.
\end{remark} 

\begin{corollary}[Large-Sample Regime; formal version] 
Let $n\geq n_b$ for a sufficiently large constant $n_b$, and let  $\{(\theta_i,\Xn_i)\}_{i=1}^N$ be  i.i.d. samples from $\Pi_{\theta}\m P_{\theta}^{(n)}$.  Let the bandwidth satisfy $h = \wt h/\sqrt{n}$ with $c_6 n^{-c_7} \leq \wt h \leq c_8$ for constants $c_6,c_7,c_8>0$. There exists a model class $\cS \times \cW$  depending only on $n$, $d$, $d_X$ and constants $(c_1,c_2,\cdots,c_5),(\alpha_0,\alpha_1,\cdots,\alpha_5)$ in Assumption~\ref{assump:large_sample} such that,  for any $\Pi_{\theta}\m P_{\theta}^{(n)}$ satisfying Assumption~\ref{assump:large_sample},  
if the training size satisfies
\begin{align*}
  N \geq C\,  \frac{n^{2+d_X+d/2}}{\wt h^d}(\log n)^{\delta+5}, \quad \delta=d_X\Big(1+\frac{\alpha_4}{\alpha_2}+\frac{6\alpha_0+1}{\alpha_3}\Big)+d\Big(\frac{1}{\alpha_2}+\frac{1}{2}+\frac{7\alpha_0}{2\alpha_3}\Big),
\end{align*}
then, with probability at least $1-\frac{1}{N}$, 
\begin{align*}
   & n^{-\frac{d}{2}}\E_{X^{(n)}\sim \m P^{(n)}} [\mmd^2_h(\mu^*_{\theta|\Xn}(\cdot|\Xn), \mu^*_{\theta|\wh S(\Xn)}(\cdot|\wh S(\Xn)))]\\
   & \leq n^{-\frac{d}{2}} \E_{X^{(n)}\sim \m P^{(n)}} [\mmd^2_{h}(\mu^*_{\theta|\Xn}(\cdot|\Xn), \mu_{\wh{W}}(\cdot|\wh S(X^{(n)})))] \leq C_1\frac{\log n}{n},
\end{align*}
where $C$, $C_1$ are constants depending only on $d,d_X$, $c_6,c_7,c_8$ and constants $(c_1,c_2,\cdots,c_5)$,\\
$(\alpha_0,\alpha_1,\cdots,\alpha_5)$  in Assumption~\ref{assump:large_sample}.
\end{corollary}




\section{Experimental Configurations and Additional Results} \label{appendix:Implementation Details}

Our central motivation is to show that the proposed framework can mitigate the curse of dimensionality from high-dimensional observations. In particular, by jointly learning  a low-dimensional informative summary statistic and an amortized posterior estimator, our approach allows SBI to scale to observations with large effective dimension while retaining accuracy and computational efficiency. To illustrate this, we increase the observation dimensionality in the standard SBI benchmarks of \cite{lueckmann2021benchmarking} and study three simulation examples: a toy example with a multimodal posterior, the SIR model, and the Lotka--Volterra model, plus a real-data radio propagation example.

\subsection{Local Refinement Algorithm}
\label{app: refinement algorithm}

For ease of reference, Algorithm~\ref{algo:refinement:appx} below restates the local refinement procedure presented in Algorithm~\ref{algo:refinement} of the main text. The experiment-specific configurations in the following sections use the notation introduced there, including the number of chains \(N_{\mathrm{chain}}\), the number of refinement iterations per chain \(L\), the quantile level \(\tau\), the sample size for bandwidth estimation \(N_{\mathrm{ref}}\), and the ABC bandwidth \(\epsilon\).
  
\begin{algorithm}[h]
\small
\caption{Local Refinement (\emph{Refine+} includes summary retraining)}\label{algo:refinement:appx}
\KwIn{Data $\Xn_*$, estimator $(\widehat S,\widehat W)$, simulator $\theta\mapsto\mathcal{P}_{\theta}^{(n)}$, prior density $\pi(\cdot)$, training size $N$; Params: $N_{\text{train}}$,  $N_{\text{ref}}$, $N_{\text{chain}}$, $\tau$, $L$, $k(\cdot)$, $f(\cdot|\cdot)$.}
\KwOut{Refined particles $\{\vartheta_i\}_{i=1}^{N_{\text{chain}}}$}
Let  $\wh \mu(\dd\theta)=\mu_{\widehat W}(\dd\theta|\widehat S(\Xn_*))$\;
\If{Refine+}{
    Draw new training set $\{(\theta'_j,\Xn_j{}')\}_{j=1}^{N_{\rm train}} \sim \wh \mu(\dd\theta) \mathcal{P}_{\theta}^{(n)}$;
fit a density estimator $q(\cdot)$ from $\{\theta'_j\}_{j=1}^{N_{\rm train}}$; merge with the original training set and retrain to obtain $(\widehat S_{\text{new}},\widehat W_{\text{new}})$; update
    $\widehat S = \widehat S_{\text{new}}$ and $\wh \mu(\dd\theta)\propto \mu_{\widehat W_{\text{new}}}(\dd\theta|\widehat S_{\text{new}}(\Xn_*)) \cdot \frac{\pi(\theta)}{N_{\rm train}q(\theta) +N\pi(\theta)}$;
} 

\BlankLine
\emph{Adaptive Threshold:} Draw $\{(\wt\theta_j,\wt X^{(n)}_j)\}_{j=1}^{N_{\text{ref}}} \sim \wh\mu(\dd\theta)\mathcal{P}_{\theta}^{(n)}$; set $\epsilon = \tau$-th quantile of $\{\|\widehat S(\wt X^{(n)}_j)-\widehat S(\Xn_*)\|\}_{j=1}^{N_{\text{ref}}}$;

\BlankLine
\emph{Initialize (Per Chain, Parallel):}
\For{$i=1$ \textbf{to} $N_{\text{chain}}$}{
    $\vartheta_i\sim \wh\mu$; $\ov{X}^{(n)}_i \sim \mathcal{P}_{\vartheta_i}^{(n)}$;
    $R_i = k\left(\frac{\|\widehat S(\ov{X}^{(n)}_i)-\widehat S(\Xn_*)\|}{\epsilon}\right)$;
    Repeat these initialization draws until $\pi(\vartheta_i)R_i>0$\;
}

\BlankLine
\emph{Parallel MCMC:}
\For{$i=1$ \textbf{to} $N_{\text{chain}}$ (parallel) and $l=1$ \textbf{to} $L$ {(Iter per chain)}}{
    $\vartheta_{i\Delta} \sim f(\cdot|\vartheta_i)$; $\Xn_{i\Delta} \sim \mathcal{P}_{\vartheta_{i\Delta}}^{(n)}$;
    ${R}_{i\Delta} = k\Big(\frac{\|\widehat S(\Xn_{i\Delta})-\widehat S(\Xn_*)\|}{\epsilon}\Big)$\; 

    Update $(\vartheta_i,R_i) =(\vartheta_{i\Delta}, {R}_{i\Delta})$ with probability
    $ \min\left(1,\frac{\pi(\vartheta_{i\Delta})f(\vartheta_i|\vartheta_{i\Delta})}{\pi(\vartheta_i)f(\vartheta_{i\Delta}|\vartheta_i)}\frac{{R}_{i\Delta}}{R_i}\right)$\;
  
}

\Return $\{\vartheta_i\}_{i=1}^{N_{\text{chain}}}$
\end{algorithm}

\subsection{Experimental Setup}
\label{subsec:exp_setup}

All experiments were conducted on a server with a single AMD EPYC 7452 CPU (32 cores) and a single NVIDIA GeForce RTX 3090 GPU (24 GB memory). The summary network and amortized posterior estimator for SMMD, MMD, BayesFlow and SNPE-C were jointly trained on the GPU; the same GPU was also used to train the summary network for DNN-ABC. In contrast, all MCMC-refinement  and rejection-sampling procedures were performed on the CPU. To stabilize regularized optimization and conform to standard training practices, we utilize the AdamW optimizer \citep{loshchilov2017decoupled} for all gradient-based updates.  We initialize network parameters following the LLaMA training protocol \citep{touvron2023llama}, with all weights independently sampled from $\mathcal{N}(0, 0.2^2)$.

We also specify hyper-parameters that were consistent across all experiments. For the MMD training loss, we use a mixture of
two Gaussian kernels: $\kappa(x,y)=\frac{1}{2}\sum_{i=1}^{2} \frac{1}{h_i^d}\exp(-\frac{\|x-y\|^2}{2h_i^2})$, with $(h_1^2,h_2^2) = (10/n, 15/n)$  to balance discriminative sensitivity and training stability. For the SMMD training loss, we use a single Gaussian kernel $\kappa(x,y)= \exp(-\frac{\|x-y\|^2}{2h^2})$ with $h^2=1/n$. Since the MMD and SMMD losses involve expectations over  $\mu_{W}(\cdot|S(\Xn))$, we draw $M_2=50$ samples  $\{z_m\}_{m=1}^{M_2}$ from  the base distribution $N(0,I_d)$ at each $(S,W)$.  Given the current mini-batch  set $\{(\wt\theta_i, \wt X^{(n)}_i)\}_{i=1}^{M_1}$ from the full training set, the stochastic gradient of the MMD is
\begin{equation*}\label{eq:mmd_est}
\begin{aligned}
      & \nabla_{(\theta_S,\theta_W)} \frac{1}{M_1} \sum_{i=1}^{M_1} \Big[ \frac{1}{M_2(M_2-1)} \sum_{m=1}^{M_2} \sum_{m'=1}^{M_2} \kappa\big(W(z_m,S(\wt X^{(n)}_i)), W(z_{m'},S(\wt X^{(n)}_i))\big)\\
   &\qquad\qquad\qquad- \frac{2}{M_2} \sum_{m=1}^{M_2} \kappa\big(\wt\theta_i, W(z_m,S(\wt X^{(n)}_i))\big) \Big].
\end{aligned}
\end{equation*}
where $\theta_{W}, \theta_S$ denote the parameters of $W$ and $S$ respectively. For SMMD, each update of $(S,W)$ additionally involves sampling $M_3=20$ random directions $\{\lambda_l\}_{l=1}^{M_3}$ from the unit sphere $\mathbb{S}^{d-1}$. The stochastic gradient of the SMMD training loss is
 \begin{equation*}\label{eq:smmd_est}
\begin{aligned}
       & \nabla_{(\theta_S,\theta_W)} \frac{1}{M_1M_3} \sum_{i=1}^{M_1} \sum_{l=1}^{M_3} \Big[ \frac{1}{M_2(M_2-1)} \sum_{m=1}^{M_2} \sum_{m'=1}^{M_2} \kappa\big(\lambda_l^T W(z_m,S(\wt X^{(n)}_i)), \lambda_l^T W(z_{m'},S(\wt X^{(n)}_i))\big) \\
       &\qquad\qquad\qquad- \frac{2}{M_2} \sum_{m=1}^{M_2} \kappa\big(\lambda_l^T \wt\theta_i, \lambda_l^TW(z_m,S(\wt X^{(n)}_i))\big) \Big].
\end{aligned}
\end{equation*}

Across all experiments, for MMD, SMMD, and BayesFlow, we use a rule of thumb to set the summary dimension to 
$p=2d$, where $d$ is the parameter dimension, ensuring the networks have sufficient capacity to learn informative low-dimensional representations. 

 The implementation of the local refinement procedure 
(Algorithm~\ref{algo:refinement:appx}) is task-specific, with details  provided separately for each experiment. We set the quantile level $\tau$,  which determines the ABC kernel bandwidth, to $0.01$ for the SIR, 
Lotka--Volterra, and toy benchmarks and to $0.005$ for the radio benchmark.  We recommend $\tau=0.01$ as a practical starting point. A smaller value generally produces a narrower bandwidth and reduces smoothing relative to  the posterior conditional on the learned summary, but it may also lower the 
acceptance rate and increase Monte Carlo error. Because no universally optimal selection criterion is available, we recommend conducting a small pilot study that assesses calibration, acceptance rates, and effective sample sizes, and adjusting $\tau$ if the posterior appears oversmoothed or the chains mix poorly.

For the MCMC sampling, we execute multiple independent random-walk Metropolis-Hastings chains with a Gaussian proposal distribution. Each chain starts from a draw of the amortized posterior. The standard deviation of the proposal in component $i$ is $c_{\text{proposal}}\cdot\text{std}_i(\mu_{\wh W}(\cdot|\wh S(\Xn_*)))$, where $\text{std}_i(\mu_{\wh W}(\cdot|\wh S(\Xn_*)))$ denotes the standard deviation of the $i$-th component of $\theta\sim \mu_{\wh W}(\cdot|\wh S(\Xn_*))$. The scaling coefficient $c_{\text{proposal}}$ is task-specific and is adjusted to stabilize the Metropolis-Hastings acceptance ratio at approximately $0.1$--$0.2$.

For simplicity, Algorithm~\ref{algo:refinement} in the main text describes  a random walk directly on $\theta$. In practice, we adapt the  parameterization to the support of each task. When parameters are strictly positive, as in the SIR benchmark 
with log-normal priors, we sample in log space and include the corresponding 
Jacobian in the Metropolis--Hastings acceptance probability. This prevents proposals from crossing the boundary at zero and can improve mixing. For  parameters with bounded support, as in the toy benchmark with a uniform  prior on $[-3,3]^d$, we use componentwise reflected proposals that map  out-of-domain values back into the parameter space, thereby avoiding unnecessary boundary rejections.

\subsection{Toy example with Multimodal Posterior}
\label{implement_detail:exponential_family}

We consider a five-parameter bivariate Gaussian model with a nonlinear transformation. For \(i=1,\dots,n\), we generate latent i.i.d. samples 
  \begin{equation*}
  X'_i = (x_{i,1},x_{i,2})^\top \sim N(\mu(\theta), \Sigma(\theta)),
  \end{equation*}
  where \(\theta=(\theta_1,\dots,\theta_5)\), \(\mu(\theta)=(\theta_1,\theta_2)\), and  
  \begin{equation*}
  \Sigma(\theta)=\begin{pmatrix}
  \theta_3^2 & \tanh(\theta_5)|\theta_3\theta_4|\\
  \tanh(\theta_5)|\theta_3\theta_4| & \theta_4^2
  \end{pmatrix}.
\end{equation*}
Each latent sample $X'_i$ is mapped to the unit sphere using the inverse stereographic projection
\begin{equation*}
  \varphi(x,y) = \big(\frac{2x}{1+x^2+y^2}, \frac{2y}{1+x^2+y^2}, \frac{-1+x^2+y^2}{1+x^2+y^2}\big).
\end{equation*}
This nonlinear transformation breaks sufficiency of standard moment-based summaries (e.g., sample means and variances). The data-generation distribution $\m P^{(n)}_\theta$ is defined as the distribution of $\Xn=(X_1,X_2,\cdots,X_n)$ with $X_i = \varphi(X'_i)$ given $\theta$. We place a flat uniform prior $\mathrm{Unif}([-3,3])^{\otimes 5}$ on $\theta$; the true parameter is $\theta_{\text{true}}=(1.0,1.0,-1.0,-0.9,0.6)$, from which the observed data $\Xn_*$ are generated. We consider sample sizes $n\in\{25,50,100\}$ and average all reported metrics over $10$ replications.

\textbf{Evaluation metrics.}
Following \citet{lueckmann2021benchmarking} and \citet{linhart2024diffusion}, we evaluate the approximate posteriors with two distance-based metrics between $m=2000$ posterior draws $\{x_i\}_{i=1}^{m}$ and the $m=2000$ reference draws $\{y_i\}_{i=1}^{m}$: the (squared) maximum mean discrepancy (MMD) with a Gaussian RBF kernel and the sliced Wasserstein distance (sW). The squared MMD is computed in its V-statistic form,
\begin{equation}
\mathrm{MMD}^2=\frac{1}{m^2}\sum_{i,j=1}^{m}k(x_i,x_j)+\frac{1}{m^2}\sum_{i,j=1}^{m}k(y_i,y_j)-\frac{2}{m^2}\sum_{i,j=1}^{m}k(x_i,y_j),
\end{equation}
with the RBF kernel $k(x,y)=\exp(-\|x-y\|^2/h^2)$. The bandwidth $h$ is selected using the median heuristic: it is the
median of the pairwise Euclidean distances between the reference draws. To account for the sign nonidentifiability of $\theta_3$ and $\theta_4$
when selecting the bandwidth, these two components are replaced by their absolute values before the pairwise distances are computed.
  This evaluation metric is distinct from the MMD objective used in training.  Following \citet{linhart2024diffusion}, we also report the sliced Wasserstein distance (sW) as a complementary metric, approximated using
$1000$ random projections. To facilitate comparisons across different
sample sizes, we report $\sqrt{n}\times\mathrm{sW}$. Within each identifiable
posterior mode, the local posterior scale contracts at the rate
$n^{-1/2}$ as the number of observations increases. The unscaled sW may
therefore decrease mechanically with $n$, even when the relative
approximation accuracy remains unchanged. Since
\[
\mathrm{sW}\bigl(\{\sqrt{n}\,x_i\},\{\sqrt{n}\,y_i\}\bigr)
=
\sqrt{n}\,
\mathrm{sW}\bigl(\{x_i\},\{y_i\}\bigr),
\]
 the rescaled metric compares posterior approximations on the common
root-$n$ local scale.

\begin{figure}[ht]
  \centering
  \includegraphics[width=0.92\textwidth]{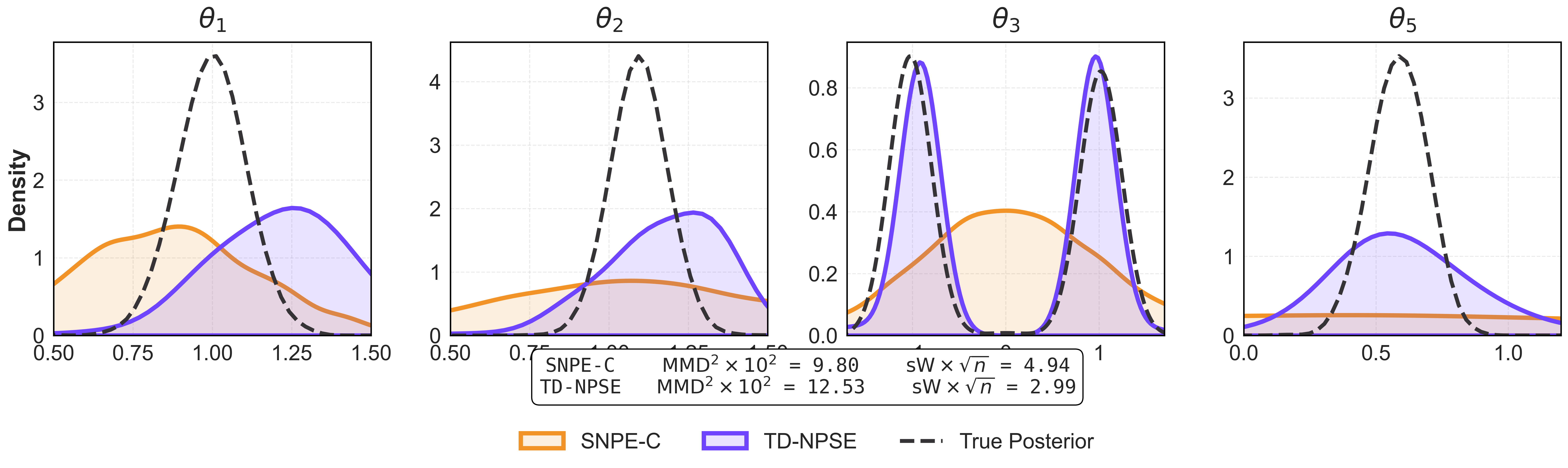}
  \caption{Marginal posteriors from SNPE-C and TD-NPSE compared with the MCMC reference posterior in the toy example at \(n=100\). Both approximations are overly diffuse and lie largely outside the Gaussian-kernel bandwidth, causing MMD to lose resolution. Consequently, MMD and sW rank the two methods oppositely, with sW favoring the more concentrated TD-NPSE posterior.}
  \label{fig:app_mmd_swd_illustration}
\end{figure}

\textbf{On the disagreement between the two evaluation metrics.} It is worth noting that MMD and sW measure discrepancy differently and may therefore rank approximate posteriors differently.  Figure~\ref{fig:app_mmd_swd_illustration} illustrates this at \(n=100\). SNPE-C is much more diffuse than the reference posterior, while TD-NPSE is visibly closer, although still overly dispersed. Nevertheless, MMD assigns a smaller error to SNPE-C than to TD-NPSE, whereas sW gives the more intuitive ranking. 
This reversal reflects the scale at which MMD compares distributions. With a median-heuristic bandwidth computed from the true posterior, MMD is sensitive to local shape differences when the estimated and true posteriors concentrate at similar scales and substantially overlap. It is therefore particularly informative for comparing better-fitting approximations from summary-based methods, and the improvements from refinement are often more visible in MMD than in sW. However, when an approximation is much more diffuse or has little overlap with the truth,  as may occur for SNPE-C, NLE, or TD-NPSE, most cross-distribution Gaussian-kernel similarities become nearly zero. MMD then becomes less responsive to further separation, which can produce a counterintuitive ranking among poorly fitted approximations. In contrast, sW measures how far probability mass must be transported and remains sensitive to large errors in location or spread. The two metrics therefore play complementary roles. MMD is sensitive to local
shape discrepancies when an approximation overlaps substantially with the
true posterior, whereas sW allows clearer comparisons among displaced or overly diffuse approximations.

\textbf{Training of the observation-wise baselines (TD-NPSE and NLE).}
TD-NPSE and NLE are observation-wise estimators. Each network is trained
on parameter--single-observation pairs and then reused, without retraining,
for datasets of any size \(n\). At inference time, the learned
single-observation quantities are combined across observations.

TD-NPSE learns a score network
\[
\widehat s(\theta_t,x,t)
\approx
\nabla_{\theta_t}\log p_t(\theta_t\mid x),
\]
where \(p_t(\theta_t\mid x)\) is the single-observation posterior after
diffusion to time \(t\). Let \(p_t^\lambda(\theta_t)\) and
\(p_t(\theta_t\mid X^{(n)})\) denote the diffused prior and diffused
tall-data posterior, respectively. Their scores satisfy
\begin{equation*}
\begin{aligned}
\nabla_{\theta_t}\log p_t(\theta_t\mid X^{(n)})
={}&(1-n)\nabla_{\theta_t}\log p_t^\lambda(\theta_t)
+\sum_{i=1}^{n}
\nabla_{\theta_t}\log p_t(\theta_t\mid X_i) \\
&+\nabla_{\theta_t}\log
\mathcal{L}_{\lambda}(\theta_t,X^{(n)}).
\end{aligned}
\end{equation*}
The diffused-prior score is available analytically for our uniform prior,
whereas the single-observation posterior scores are approximated by the
learned network. The remaining correction
\(\mathcal{L}_{\lambda}(\theta_t,X^{(n)})\), which depends on the backward
diffusion kernels, is generally intractable. We approximate it using the
GAUSS method of \citet{linhart2024diffusion} and generate posterior samples
using DDIM.

Training TD-NPSE is substantially more demanding than training the
dataset-level methods considered here. Each candidate is trained for
\(3{,}000\) epochs on an independently generated dataset
\[
\mathcal D_N^{(r)}
=
\{(\theta_i^{(r)},x_i^{(r)})\}_{i=1}^{N},
\qquad
(\theta_i^{(r)},x_i^{(r)})
\overset{\mathrm{i.i.d.}}{\sim}
\pi(\theta)p(x\mid\theta),
\qquad
N=50{,}000.
\]
At this training size, we observed substantial variation across fits, with
some candidates failing to represent all posterior modes. Increasing \(N\) may reduce sampling variability, but it does not eliminate optimization failures and would further increase an already substantial training cost.  Following the validation-based model-selection strategy of \citet{linhart2024diffusion}, we therefore train \(20\) candidates and retain the one with the lowest
held-out validation loss and a single TD-NPSE candidate
takes approximately \(2\)--\(3\) hours to train on our hardware. The   fitted networks are then reused for every sample size and evaluation replication.

NLE follows the same observation-wise training and selection protocol. It
learns a single-observation likelihood surrogate
\(\widehat p(x\mid\theta)\), which yields
\[
\widehat p(\theta\mid X^{(n)})
\propto
\pi(\theta)\prod_{i=1}^{n}\widehat p(X_i\mid\theta)
\]
under conditional independence. We draw posterior samples from this
approximation using slice sampling. NLE is trained for \(300\) epochs, by
which point its validation loss has effectively stabilized.

 This protocol follows the intended amortized use of the observation-wise methods. Their training objectives depend neither on the realized test dataset nor on the observation count \(n\), so the fitted networks can be reused across datasets and values of \(n\). For the other methods, we perform an independent end-to-end fit in each of the \(10\) replications for every value of \(n\), retaining all fits without selection.

\subsubsection{Architectural and Parameter Configuration}

\begin{itemize}
    \item \textbf{True Posterior (Reference) Sampling}:
        \begin{itemize}
            \item \textbf{Method}: We generate the reference posterior samples using random-walk
Metropolis--Hastings with $2000$ independent chains initialized from the
prior. Each chain is run for $1000$ burn-in iterations, after which its final state is retained, yielding $2000$ reference draws. Because the posterior has four symmetric modes, using many independently initialized chains improves coverage of all modes and reduces the risk of mode imbalance. The proposal scale is set to $0.55/\sqrt{n}$. This procedure is repeated independently for every replication and sample size.
        \end{itemize}
    \item \textbf{SMMD / MMD}:
        \begin{itemize}
            \item \textbf{Summary Network}: The Summary Network is a DeepSet-based summary network consisting of two stacked Equivariant Module blocks and one Invariant Module block, with each module containing two dense layers of 64 hidden units.
            \item \textbf{Generator}: The generator is an MLP with 3 hidden layers, each with 64 hidden units, and uses ReLU activations. We apply RMSNorm between linear layers to improve optimization stability and reduce sensitivity to simulator scaling. 
            \item \textbf{Hyper-parameters}: We run $2000$ independent chains, each with $119$ burn-in iterations,
and retain the final state of each chain (with $\mathrm{thin}=1$),
yielding $2000$ refined posterior samples. For parameter component $i$, the proposal scale is set to $0.2\,\widehat{\mathrm{sd}}_i$, where
$\widehat{\mathrm{sd}}_i$ is the corresponding amortized posterior
standard deviation. The ABC bandwidth is set to the $\tau=0.01$ quantile of the summary discrepancies computed from $20{,}000$ simulations. 
    \end{itemize}
    \item \textbf{BayesFlow}:
        \begin{itemize}
            \item \textbf{Summary Network}: DeepSets-like architecture. Same as SMMD/MMD.
            \item \textbf{Inference Network}: For the BayesFlow generator, we utilize an invertible neural network instantiated through the dedicated CouplingFlow module. The invertible network is structurally configured with 3 coupling layers and a fixed width of 64 hidden units per coupling layer.
            \item \textbf{Hyper-parameters}: Size of Training Datasets=19200, Epochs=200, Batch Size=256, Summary Dim=10.
            \item \textbf{Refine}: Same as SMMD/MMD.  

        \end{itemize}
    \item \textbf{DNN-ABC}:
        \begin{itemize}
            \item \textbf{Summary Network}:  We use the same architecture as in SMMD/MMD, but set the summary dimension to $d=5$.

            \item \textbf{Hyper-parameters}: Size of Training Datasets=19200, Epochs=200, Batch Size=256, Summary Dim=5.
            \item \textbf{Rejection Sampling}: Number of Proposals = 200000, Number of Accepted Samples = 2000. 
        \end{itemize}
    \item \textbf{SNPE-C}:
        \begin{itemize}
            \item \textbf{Method}: SNPE-C \citep{greenberg2019automatic} with $5$ rounds of sequential training, following the implementation of \citet{Deistler2025SimulationBasedIA}.
            \item \textbf{Hyper-parameters}: 5 Rounds, Simulations per round: 5000, Training Epochs per round: up to 100 (early stopping after 20 epochs without improvement), batch size 200.
            \item \textbf{Summary Stats}: Flattened raw data.
        \end{itemize}
    \item \textbf{NLE}:
        \begin{itemize}
            \item \textbf{Method}: A neural likelihood estimator (an autoregressive normalizing flow, MAF) trained on parameter--single-observation pairs, following \citet{papamakarios2019sequential}. At inference, the likelihood of a dataset $X^{(n)}=(x_1,\dots,x_n)$ is obtained by multiplying the per-observation surrogate likelihoods, exploiting the conditional independence of the i.i.d.\ observations, and posterior samples are drawn with a slice sampler (100 chains, 100 warm-up steps).
            \item \textbf{Hyper-parameters}: Trained on $50{,}000$ pairs; up to $500$ epochs, learning rate \(10^{-4}\), $80$ hidden features, $5$ transforms, $2$ blocks, batch size $256$, early stopping after $30$ epochs without improvement.
            \item \textbf{Training protocol}: $20$ candidate models were trained and the checkpoint with the lowest validation loss was retained; the retained checkpoint is used directly for inference on every dataset and every evaluation round without further retraining.
        \end{itemize}
    \item \textbf{TD-NPSE}:
        \begin{itemize}
            \item \textbf{Method}: A compositional score model trained by denoising score matching on single-observation contexts, following \citet{linhart2024diffusion}. At inference, the per-observation posterior scores are composed through a Gaussian approximation to the tall-posterior diffusion process, and posterior samples are drawn with $800$-step DDIM.
            \item \textbf{Hyper-parameters}: Trained on $50{,}000$ pairs; up to $5000$ epochs with early stopping (patience $500$, minimum $200$ epochs), learning rate $10^{-4}$, batch size $256$, an MLP with $3$ hidden layers of $256$ units, $3$ time-frequency features, $100$ diffusion steps, $800$ sampling steps.
            \item \textbf{Training protocol}: Same as NLE: $20$ candidate models were trained and the checkpoint with the lowest validation loss was retained.
        \end{itemize}
\end{itemize}

\subsubsection{Scaling with the observation count and training budget.}

To study how approximation error changes with the observation count \(n\) under different training budgets, we compare SMMD with NPE, implemented as SNPE-C without sequential training. We use the respective architectures and hyperparameters described above and train both methods for the same number of epochs. We vary
\(n\in\{10,75,150,225,300\}\) and
\(N=N_{\mathrm{train}}\in\{20{,}000,30{,}000,40{,}000\}\). For each \((n,N_{\mathrm{train}})\), we simulate a fresh reference table
\(\{(\theta_i,X_i^{(n)})\}_{i=1}^{N_{\mathrm{train}}}\), where
\(\theta_i\sim\Pi_\theta\) and
\(X_i^{(n)}\sim\mathbb P_{\theta_i}^{(n)}\), and use the same table to train both methods. We evaluate the fitted estimators on \(25\) independently simulated test datasets
\(X_*^{(n)}\sim\mathbb P_{\theta^*}^{(n)}\), using the fixed true parameter \(\theta^*\) from the main Gaussian experiment. For each test dataset, we compute the sliced-Wasserstein distance between the amortized approximation and the reference posterior \(p(\theta\mid X_*^{(n)})\), obtained using the same MCMC procedure as above. We report \(\sqrt{n}\times\mathrm{sW}\) to account for the \(n^{-1/2}\) posterior contraction, together with its mean and standard deviation over the \(25\) test datasets, in Figure~\ref{fig:toy_sample_size} of the main text.

\subsubsection{Principal Component Analysis of Learned Summary Statistics}
\label{appendix: pca_visualization}
\begin{figure}[h]
  \centering
  \includegraphics[width=1.\linewidth]{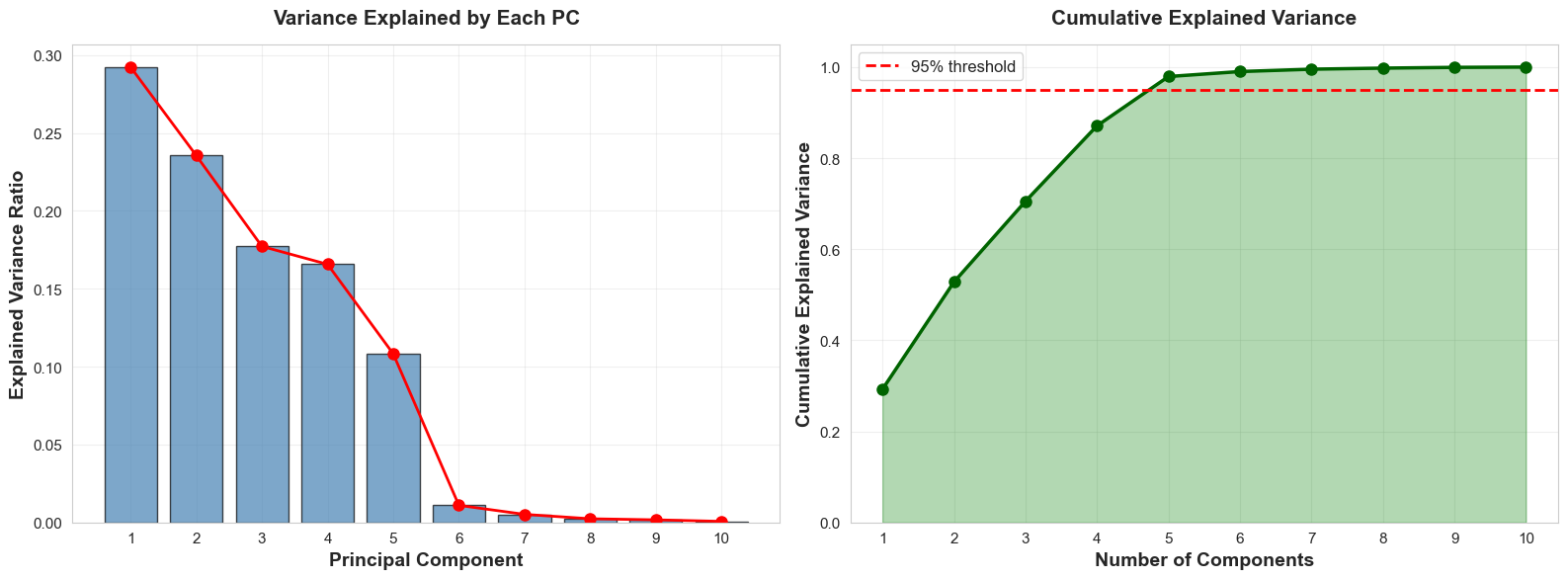}
  \caption{PCA results for SMMD approach}
    \label{figure:smmd_PCA}
\end{figure}

In the toy example, the true minimal sufficient statistic is $5$-dimensional and we set the summary dimension $p = 10$.      To empirically confirm that the summary network $S$ is  driven to learn representations that are both expressive and of low intrinsic dimension, we apply Principal Component Analysis (PCA) to the summary statistics learned from SMMD with $n = 50$. The results are shown in Figure~\ref{figure:smmd_PCA}.  
The PCA results show that the first $5$ principal components capture more than $95\%$ of the variance in the learned summary statistics, which is the dimension of the true minimal sufficient statistic.
\subsection{SIR Model}
\label{implement_detail:SIR}

We follow the SIR simulator in \citep{lueckmann2021benchmarking} with two adjustments to better match our high-dimensional setting: (i) we increase the observation dimension to 80; (ii) for each observation time we reduce the binomial trials from 1000 to 100. Following the simulation-based calibration protocol of Section~\ref{subsec: sir}, in each replication the true parameter is drawn from the prior below and the observation is generated from the corresponding simulator.

We generate observations according to the following data-generating process. The parameter is $\theta=(\beta,\gamma)\in\mathbb{R}^2$ with independent prior
\begin{equation*}
\beta \sim \mathrm{LogNormal}(\log(0.4),\,0.5), \qquad 
\gamma \sim \mathrm{LogNormal}(\log(1/8),\,0.2).
\end{equation*}
We fix the total population size in the dynamics to be $N_{\rm pop}=1{,}000{,}000$, simulate over a horizon $T=160$, and initialize $(S(0),I(0),R(0))=(N_{\rm pop}-1,1,0)$. Given $(\beta,\gamma)$, we generate the latent trajectory by solving the SIR ODE system
\begin{equation*}
\frac{dS}{dt}=-\beta\frac{SI}{N_{\rm pop}}, \qquad
\frac{dI}{dt}=\beta\frac{SI}{N_{\rm pop}}-\gamma I, \qquad
\frac{dR}{dt}=\gamma I.
\end{equation*}
We take observations at 80 time points and form $\mathbf{x}=(x_1,\ldots,x_{80})\in\mathbb{R}^{80}$, where at each time point we observe a noisy infection count via
\begin{equation*}
x_i \sim \mathrm{Binomial}\!\left(100,\,\frac{I}{N_{\rm pop}}\right), \qquad i=1,\ldots,80,
\end{equation*}
with $I$ denoting the latent infected population at that time.

 In each replication we draw a fresh true parameter $\theta^*$ from the prior above, simulate a corresponding dataset  $\Xn$, and perform posterior inference for that observation. The reported metrics are averaged over $500$ replications. The (prior-averaged) calibration error is computed as follows. For each credibility level $\alpha\in(0,1)$ on a predefined grid, we form the $100\alpha\%$ credible interval $\mathrm{CrI}_{j,\alpha}(\Xn)$ of the $j$-th marginal of the estimated posterior. The empirical coverage $\widehat p_{j,\alpha}$ is the proportion of replications in which this interval contains the corresponding prior-sampled true value $\theta^*_j$. The calibration error of component $j$ is the median of $|\widehat p_{j,\alpha}-\alpha|$ over the grid, and the calibration curves plot $\widehat p_{j,\alpha}-\alpha$ against $\alpha$, with positive values indicating overcoverage and negative values indicating undercoverage.

\subsubsection{Architectural and Parameter Configuration}

\begin{itemize}
    \item \textbf{SMMD / MMD}:
        \begin{itemize}
            \item \textbf{Summary Network}:  The SummaryNet is a 1D CNN-based time series summary network with a default hidden dimension of 64. It consists of two convolutional blocks with ReLU activations, kernel size 10, and max pooling for local feature extraction, followed by adaptive global average pooling for sequence aggregation.
            \item \textbf{Generator}:  The generator is an MLP with 4 hidden layers, each with 64 hidden units, and uses ReLU activations. We apply RMSNorm between linear layers to improve optimization stability and reduce sensitivity to simulator scaling. 
            \item \textbf{Hyperparameters}:
        Size of Training datasets =768,  Epochs=60, Batch Size = 256, Summary Dim=4, Slices ($L$)=20, Generator Samples ($M$)=50.
            \item  \textbf{Refine}:  $N_{chain}$ = 1000, $L$ = 50, $\tau = 0.01$, $c_{proposal}=0.3$, $N_{ref}=5000$. All notation is defined in Algorithm~\ref{algo:refinement:appx} of this appendix, and the proposal coefficient $c_{\text{proposal}}$ is defined in Section~\ref{subsec:exp_setup}.
        \end{itemize}
    \item \textbf{BayesFlow}:
        \begin{itemize}
            \item \textbf{Summary Network}: Same as SMMD/MMD.
            \item \textbf{Inference Network}: For the BayesFlow generator, we utilize an invertible neural network instantiated through the dedicated CouplingFlow module. The invertible network is structurally configured with 4 coupling layers and a fixed width of 64 hidden units per coupling layer.
            \item \textbf{Hyperparameters}:  Size of Training datasets =768, Epochs=100, Batch Size = 256, Summary Dim=4.
            \item  \textbf{Refine}: Same as SMMD/MMD
        \end{itemize}
    \item \textbf{DNN-ABC}:
        \begin{itemize}
            \item \textbf{Summary Network}: We use the same architecture as in SMMD/MMD, but set the summary dimension to $2$.
            \item \textbf{Hyperparameters}: Size of Training Datasets=768, Epochs=60, Batch Size=256, Summary Dim=2.
            \item \textbf{Rejection Sampling}: Number of Proposals = 50000, Number of Accepted Samples = 1000.
        \end{itemize}
\end{itemize}

\subsection{Lotka-Volterra Model With Vague Prior}
\label{implement_detail:lotkavolterra}

We consider the Lotka-Volterra model \citep{Wilkinson2011}, which describes predator-prey population dynamics, with our experimental setup referencing \citep{papamakarios2016fast} and simulations performed via Gillespie's algorithm \citep{gillespie1977exact}. Let $X$ (predators) and $Y$ (prey) follow four reactions: (i) predator birth (rate $\theta_1XY$), (ii) predator death (rate $\theta_2 X$), (iii) prey birth (rate $\theta_3Y$), (iv) prey predation (rate $\theta_4XY$).

Gillespie's algorithm simulates the model by: (i) drawing the waiting time to the next
reaction from an exponential distribution with total rate $\theta_1XY + \theta_2X + \theta_3Y + \theta_4XY$; (ii) selecting a reaction proportionally to its rate; (iii) updating populations and repeating.

Our simulations start with $X=50$, $Y=100$, run for 30 time units, and record $X,Y$ every 0.2 units (151 values each). To construct summaries for SNPE-C, we compute 9 normalized statistics from these time series (mean, log variance, lag-1/2 autocorrelation for each series, cross-correlation), with normalization based on 1000 pilot simulations.   This allows us to directly compare the performance of our learned summary statistics against handcrafted summaries based on domain expertise.

A key feature of this experiment is the use of a vague prior over 
${\theta}$ , uniform in the log domain: $p(\log \boldsymbol{\theta})\propto\prod_{i=1}^{4}\mathcal{U}(\log\theta_i\mid-5,2)$. Most prior draws produce trajectories incompatible with the observed oscillatory behavior, making a large fraction of simulations uninformative for inference. This setup is intentionally designed to test the robustness of our method under a highly uninformative prior. The true parameters are set to $\theta_1=0.01,\theta_2=0.5,\theta_3=1,\theta_4=0.01$ to generate observed data with stable oscillations.

To enable stable and well-scaled processing by the neural network, we employ a logarithmic transformation of the input parameters ${\theta}$ during training. The raw parameters span several orders of magnitude and include very small values,  which are difficult for a neural network to handle directly; taking 
$\log {\theta}$ rescales these inputs into a numerically favorable range, enabling reliable training and inference. The true parameter of this benchmark is fixed to the values above, so the calibration error reported in the main text is conditional on this parameter of interest. In each of the $100$ rounds, we draw a fresh observation from the simulator at the fixed true parameter,  conduct posterior inference using the \emph{Refine+} procedure, and evaluate the resulting posterior estimator. The (conditional) calibration error is computed as follows. For a grid of credibility levels $\alpha\in(0,1)$, we form the equal-tailed $100\alpha\%$ credible interval $\mathrm{CrI}_{j,\alpha}(\Xn)$ of the $j$-th marginal of the estimated posterior. The empirical coverage $\widehat p_{j,\alpha}$ is the proportion of replications in which the fixed true parameter $\theta^*_j$ falls inside this interval. The conditional calibration error of component $j$ is the median of $|\widehat p_{j,\alpha}-\alpha|$ over the grid.

\subsubsection{Architectural and Parameter Configuration}
In this experiment, we enforce a fixed total simulation budget across all methods to ensure a strictly fair comparison, where the budget denotes the total number of simulator calls across all stages (amortized training, refinement, and inference). We fix this total to $60,000$ for every method, including SMMD/MMD/BayesFlow+Refine, DNN-ABC, and SNPE-C, with hyperparameters selected so that each method exactly exhausts this budget. Under this constraint, our method achieves the best overall trade-off between accuracy and efficiency: it outperforms SNPE-C in posterior approximation quality while being substantially faster computationally.
\begin{itemize}
    \item \textbf{SMMD / MMD}:
        \begin{itemize}
            \item \textbf{Summary Network}: The SummaryNet is a 1D CNN-based time series summary network with a default hidden dimension of 64. It consists of two convolutional blocks with ReLU activations, kernel size 10, and max pooling for local feature extraction, followed by adaptive global average pooling for sequence aggregation.
            \item \textbf{Generator}: The generator is an MLP with 4 hidden layers, each with 64 hidden units, and uses ReLU activations. We apply RMSNorm between linear layers (i.e., normalization is placed inside the MLP blocks) to improve optimization stability and reduce sensitivity to simulator scaling. 
            \item \textbf{Hyper-parameters}:
            Size of Training datasets =12800,  Epochs=200, Batch Size = 128, Summary Dim=8, Slices ($L$)=20, Generator Samples ($M$)=50.
            \item  \textbf{\emph{Refine+}}: $N_{train}$ =12800,  $N_{chain}=1000$ , $L$ = 30, $\tau = 0.01$, $c_{proposal}=0.5$, $N_{ref}=4400$.
        \end{itemize}
    \item \textbf{BayesFlow}:
        \begin{itemize}
            \item \textbf{Summary Network}: Same as SMMD/MMD.
            \item \textbf{Inference Network}: For the BayesFlow generator, we utilize an invertible neural network instantiated through the dedicated CouplingFlow module. The invertible network is structurally configured with 4 coupling layers and a fixed width of 64 hidden units per coupling layer.
            \item \textbf{Hyper-parameters}:  Size of Training datasets =12800, Epochs=200, Batch Size = 128, Summary Dim=8.
            \item  \textbf{\emph{Refine+}}:  Same as SMMD/MMD.

        \end{itemize}
    \item \textbf{DNN-ABC}:
        \begin{itemize}
            \item \textbf{Summary Network}: We use the same architecture as in SMMD/MMD, but set the summary dimension to $d=4$.
            \item \textbf{Hyper-parameters}: Size of Training Datasets=25600, Epochs=400, Batch Size=128, Summary Dim=4.
            \item \textbf{Rejection Sampling}: Number of Proposals = 34400, Number of Accepted Samples = 1000.

        \end{itemize}
    \item \textbf{SNPE-C}:
        \begin{itemize}
        \item \textbf{Methods}: SNPE-C (NPE). We use the default SNPE-C settings in ~\cite{Deistler2025SimulationBasedIA}.
            \item \textbf{Hyper-parameters}: 10 Rounds, Simulations per round: 6000, Training Epochs per round: 200.
            \item \textbf{Summary Stats}: Following~\cite{papamakarios2016fast}, we employ the following nine summary statistics: (i) The mean of each time series; (ii) The log variance of each time series; (iii) The lag-1 autocorrelation of each time series; (iv) The lag-2 autocorrelation of each time series; (v) The cross-correlation at lag 0 between the two time series.
        \end{itemize}
\end{itemize}

\subsection{Radio Propagation Model}

\begin{figure}[ht]
  \centering
  \begin{minipage}{0.48\textwidth}
    \centering
    \includegraphics[width=\linewidth]{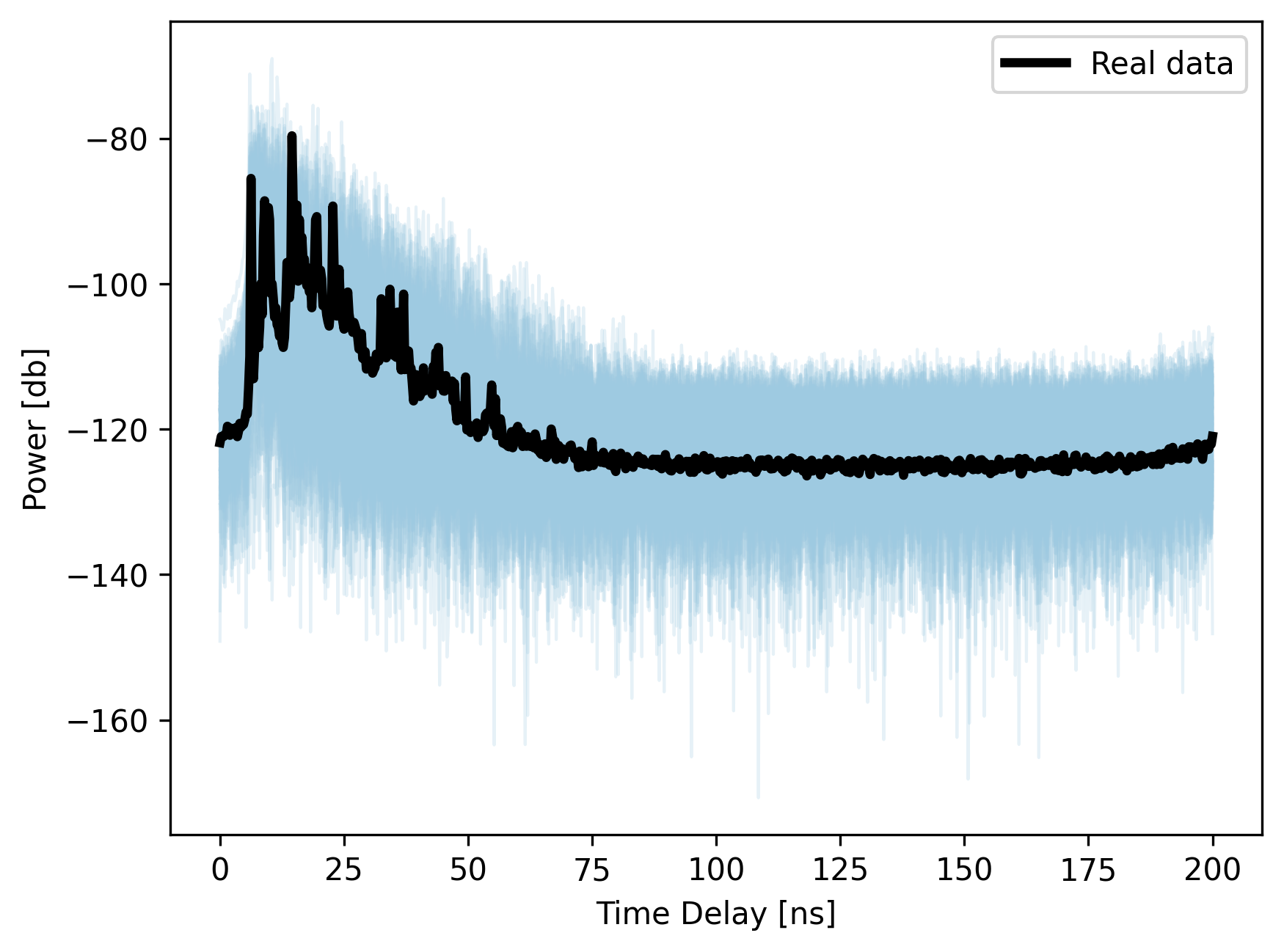}
    \subcaption{Informative prior}
  \end{minipage}
  \hfill
  \begin{minipage}{0.48\textwidth}
    \centering
    \includegraphics[width=\linewidth]{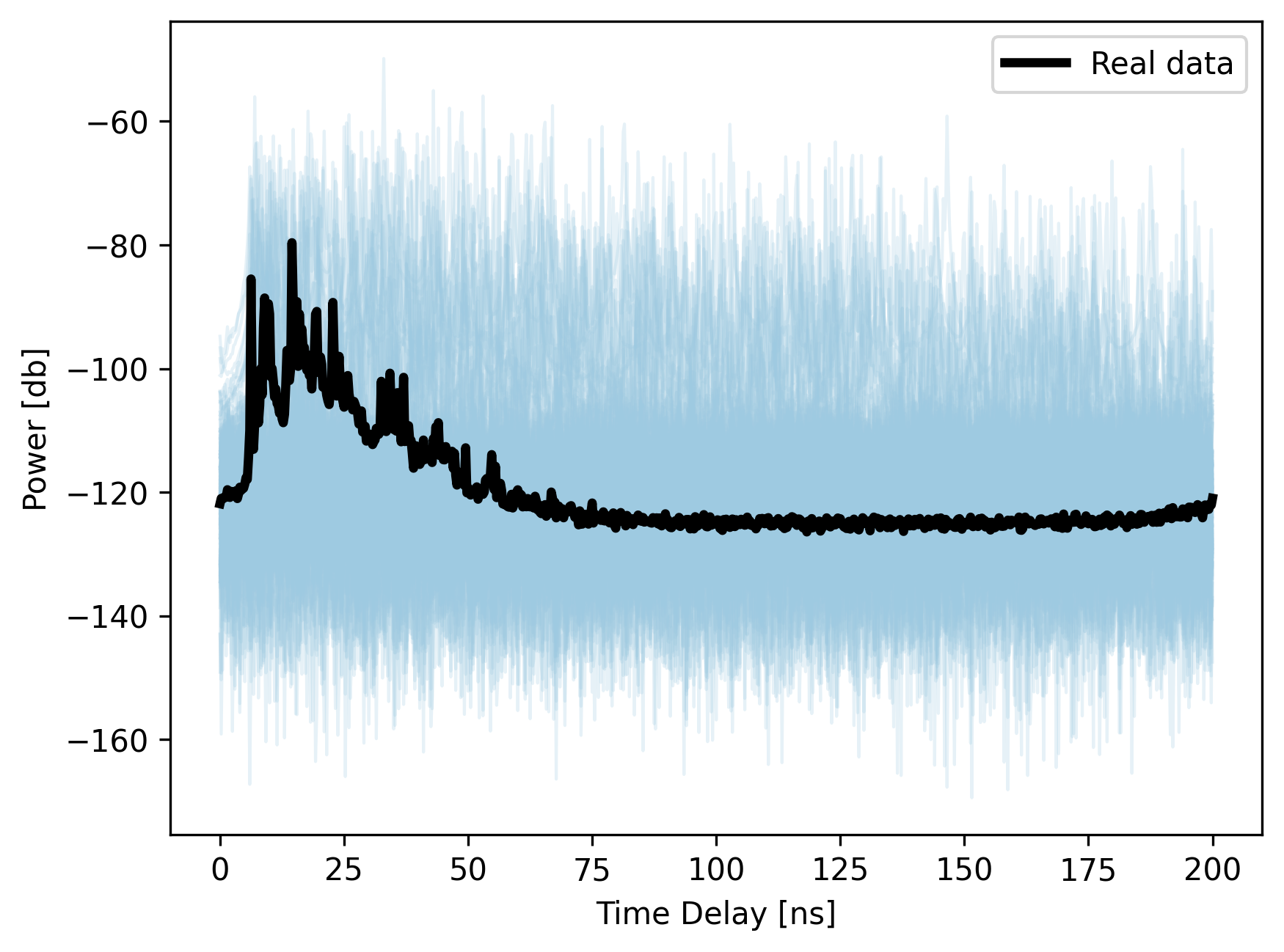}
    \subcaption{Uninformative prior}
  \end{minipage}
  \caption{Observations from different prior settings.}
  \label{fig:radio_prior}
\end{figure}

We employ the Turin stochastic channel model as our simulator, following the setup of \cite{huang2023learning} with one key adjustment to the prior specification: instead of the informative uniform prior used in \cite{huang2023learning}, we adopt a vaguer log prior to better evaluate the robustness of our inference method under weak prior information. Specifically, we generate model parameters $\boldsymbol{\theta}$ from the specified prior, then use these parameters to generate simulated observations, which are further compared with the true measured observations to illustrate the difference between the original informative prior and our proposed vague log prior (see Fig. \ref{fig:radio_prior} for a visual comparison).  The real measured observations originate from the measurement campaign of \citet{gustafson2016modeling}, and we obtained the data from the public repository accompanying \citet{huang2023learning}.

The Turin model generates a frequency-domain transfer function $H_k$ at the $k$-th frequency point as a sum of multipath components:
\[
H_k = \sum_l \alpha_l \exp(-j 2\pi \Delta f \, k \, \tau_l),
\]
where $\tau_l$ is the delay and $\alpha_l$ is the complex gain of the $l$-th path.
The arrival times of the paths $\{\tau_l\}$ are modeled as a one-dimensional homogeneous Poisson point process (PPP) on $\mathbb{R}_+$ with rate $\nu > 0$ and $\tau_0=6\times 10^{-9}$. Conditioned on the delays, the complex gains $\alpha_l$ are independent zero-mean complex Gaussian random variables. Their conditional variance is given by:
\[
\mathbb{E}[|\alpha_l|^2 \mid \tau_l] = \frac{G_0}{\nu} \exp(-\tau_l / T).
\]
The received signal at frequency $k$ is then:
\[
Y_k = H_k + W_k, \quad k = 0, 1, \dots, K-1,
\]
where $W_k$ is additive zero-mean circularly symmetric complex Gaussian noise with variance $\sigma_W^2$.  The model is then parameterized by $\boldsymbol{\theta} = [G_0, T, \nu, \sigma_W^2]^\top$, where the four parameters are defined as follows:
$G_0$: Initial power gain (related to the pathloss). $T$: Delay spread (controlling the exponential decay of power over delay). $\nu$: Path arrival rate of the PPP. $\sigma_W^2$: Noise variance.

Unlike the informative prior adopted in the NPE-RS paper \citep{huang2023learning}, we employ a less informative log-uniform prior, and both prior distributions are specified in Table~\ref{tab:priors}.

\begin{table}[ht]
    \centering
    \caption{Prior distributions for the parameters of the Turin model.}
    \label{tab:priors}
    \begin{tabular}{lll}
        \toprule
        Parameter & Informative Prior (\cite{huang2023learning}) & Uninformative  Prior (Ours) \\
        \midrule
        $G_0$ & $\mathcal{U}(10^{-9}, 10^{-8})$ & Log-Uniform ($10^{-10}, 10^{-7}$) \\
        $T$ & $\mathcal{U}(10^{-9}, 10^{-8})$ & Log-Uniform ($10^{-10}, 10^{-7}$) \\
        $\nu$ & $\mathcal{U}(10^7, 5 \times 10^9)$ & Log-Uniform ($10^7, 5 \times 10^9$) \\
        $\sigma_W^2$ & $\mathcal{U}(10^{-10}, 10^{-9})$ & Log-Uniform ($10^{-11}, 10^{-8}$) \\
        \bottomrule
    \end{tabular}
\end{table}

The true observation used in this study originates from a radio channel measurement campaign conducted in a small conference room with dimensions $3 \times 4 \times 3 \, \text{m}^3$ \citep{gustafson2016modeling}. A vector network analyzer was employed to measure the channel transfer function over a bandwidth of $B = 4$ GHz at $K = 801$ equidistant frequency points, resulting in a frequency separation of $\Delta f = B/(K-1)$. The observation consists of $n = 100$ independent realizations of complex-valued frequency-domain signals, denoted as $\ov{X}^{(n)} = (\ov X_1,\ov X_2,\cdots,\ov X_n)$, where each realization $\ov X_i \in \mathbb{C}^K$. Following Appendix~C of \citet{huang2023learning}, we apply the inverse
fast Fourier transform (IFFT) to each complex frequency response
$\overline X_i$ to obtain the delay-domain response
$y_i=\operatorname{IFFT}(\overline X_i)$. At delay $t_m$, its value is
\begin{equation*}
    (y_i)_m = \frac{1}{K}\sum_{k=0}^{K-1}(\overline X_i)_k
        \exp(j2\pi k\Delta f\,t_m),
    \qquad t_m = \frac{m}{K\Delta f},
    \qquad m=0,\ldots,K-1,
\end{equation*}
where $j^2=-1$. The delay samples are spaced by
$1/(K\Delta f)\approx 0.2497$~ns, ranging from $t_0=0$ to
$t_{800}\approx 199.75$~ns. The neural network takes as input the
power-delay profiles $X_i$ in dB, with entries
$(X_i)_m=10\log_{10}(|(y_i)_m|^2)$ and $d_X=K=801$ delay samples
per profile. Figure~\ref{fig:radio} shows these profiles over the
corresponding delay range of approximately $0$--$200$~ns.

The goal of the experiment is to infer the posterior distribution $p(\boldsymbol{\theta} \mid \Xn)$ given the measured data $\Xn$. The Turin model is known to be misspecified for this dataset, primarily because the measured channel exhibits non-i.i.d. characteristics and structural features (e.g., multiple peaks in the power-delay profile) that the simplified PPP-based model cannot fully capture. This experiment thus serves as a real-world testbed for evaluating the robustness of inference methods under model misspecification.

To quantify predictive accuracy, we evaluate each fitted posterior by the root mean squared error (RMSE) between the predicted and observed mean power-delay profiles. Concretely, given the fitted posterior, we draw $J = 1000$ posterior samples $\boldsymbol\theta^{(1)}, \ldots, \boldsymbol\theta^{(J)}$ and, for each draw, simulate one power-delay profile $X^{(j)} \in \mathbb{R}^K$ on the decibel scale via the Turin simulator. Their average, $    \bar X_{\mathrm{pred}} = \frac{1}{J}\sum_{j=1}^{J} X^{(j)}$, is the predicted mean series. On the observed side, we average the $n = 100$ measured realizations as $\bar X_{\mathrm{obs}} = \frac{1}{n}\sum_{i=1}^{n} X_i$, where $X_i$ denotes the $i$-th observed power-delay profile in dB. The RMSE between the two mean series is then computed pointwise over the $K = 801$ delay samples,
\begin{equation}
    \mathrm{RMSE} = \sqrt{\frac{1}{K}\sum_{k=1}^{K}\left(\bar X_{\mathrm{pred},k} - \bar X_{\mathrm{obs},k}\right)^2}.
\end{equation}
The reported RMSE is the average across $20$ independent retraining
rounds and measures in-sample posterior-predictive fit to the observed series.

\subsubsection{Architectural and Parameter Configuration}

In the radio propagation experiment, we adopt the summary network proposed in \citet{huang2023learning}. This network architecture consists of five one-dimensional convolutional layers, with the number of hidden channels configured as $[8, 16, 32, 64, 8]$ and a kernel size of 3. To guarantee permutation invariance across realizations, mean aggregation is adopted as the final operation to generate a low-dimensional summary statistic vector. Furthermore, a tanh activation function is utilized to constrain the output range of the summary statistics. The sole distinction among methods resides in the output dimensionality: the output dimension of SMMD, Bayesflow is $p=2d=8$, while that of DNN-ABC and NPE-RS is $p=4$.  Here NPE-RS (Neural Posterior Estimation with Robust Statistics) \citep{huang2023learning} is a method designed to handle model misspecification by learning robust summary statistics. It introduces a regularization term based on the Maximum Mean Discrepancy (MMD) into the standard NPE loss function. This regularizer penalizes summary statistics for which the observed data's statistic is far from the distribution of statistics that the model can simulate, thereby encouraging the learned statistics to be both informative about the parameters and replicable by the (potentially misspecified) model. For NPE-RS \citep{huang2023learning}, we modify the original configuration of  \cite{huang2023learning}  by reducing the summary dimension from $p=8$ to $p=4$, and replace the Masked Autoregressive Flow inference network with Neural Spline Flows.
 We do so because under our vague prior, the original configuration  of~\cite{huang2023learning}  trains unstably and preforms  worse than in the informative-prior setting considered in \citet{huang2023learning}. This revised configuration improves training stability and overall performance of NPE-RS.  Referring to the practice in \citet{huang2023learning}, we perform z-score standardization on the $\boldsymbol{\theta}$ of the training data.

\begin{itemize}
\item \textbf{SMMD}:
\begin{itemize}
\item \textbf{Generator}: Same as the Lotka-Volterra Model (The generator is an MLP with 4 hidden layers, each with 64 hidden units, and uses ReLU activations. We apply RMSNorm between linear layers to improve optimization stability and reduce sensitivity to simulator scaling).
\item \textbf{Hyperparameters}: Size of Training datasets = 19200, Epochs = 120, Batch Size = 128, Summary Dim = 8, Slices ($L$) = 20, Generator Samples ($M$) = 50.
\item \textbf{Refine}: $N_{chain}$ = 1000, $L$ = 50, $\tau = 0.005$, $c_{proposal}=1.0$, $N_{ref}=5000$. All notation is defined in Algorithm~\ref{algo:refinement:appx} of this appendix, and the proposal coefficient $c_{\text{proposal}}$ is defined in Section~\ref{subsec:exp_setup}.
\end{itemize}
\item \textbf{BayesFlow}:
\begin{itemize}
\item \textbf{Inference Network}: For the BayesFlow generator, we utilize an invertible neural network instantiated through the dedicated CouplingFlow module. The invertible network is structurally configured with 4 coupling layers and a fixed width of 64 hidden units per coupling layer.
\item \textbf{Hyperparameters}: Size of Training datasets = 19200, Epochs = 120, Batch Size = 128, Summary Dim = 8.
\item \textbf{Refine}: Same as SMMD/MMD.
\end{itemize}
\item \textbf{DNN-ABC}:
\begin{itemize}
\item \textbf{Hyperparameters}: Size of Training Datasets = 19200, Epochs = 120, Batch Size = 128, Summary Dim = 4.
\item \textbf{Rejection Sampling}: Number of Proposals = 50000, Number of Accepted Samples = 1000.
\end{itemize}
\item \textbf{NPE-RS}:
\begin{itemize}
\item \textbf{Distance Metric For The Penalty}: Maximum Mean Discrepancy (MMD).
\item \textbf{Regularization Weight}: 1.
\item \textbf{Hyperparameters}: Number of Simulations = 19200, Training Batch Size = 128, Epochs = 120.
\item \textbf{Inference Network}: A 5-layers conditional Neural Spline Flows with Hidden Features = 100.
\end{itemize}
\end{itemize}
\section{Proof of Main Results}

\subsection{Proof of Theorem~\ref{th1}}
Write $p(\theta|\Xn)$ for the (conditional) density function of the posterior $\mu^*_{\theta|\Xn}(\cdot|\Xn)$. Then by the tower rule, the conditional density function for  $\mu^*_{\theta|S(\Xn)}(\cdot|s)$ can be written as $p_{S}(\theta|s)= \mb{E}_{\Xn\sim \mu^*_{\Xn|S(\Xn)}(\cdot|s)}[p(\theta|\Xn)]$. Then it holds that
\begin{equation*}
\begin{aligned}
      &  \m R(S,W)=-\EXandth[\log W(\theta,S(\Xn))]\\
        &=-\mb{E}_{s\sim S_{\#}\m P^{(n)}}\mb{E}_{\theta\sim \mu^*_{\theta|S(\Xn)}(\cdot|s)}[\log W(\theta,s)]\\
        &=\mb{E}_{s\sim S_{\#}\m P^{(n)}}\Big[{\rm KL}(\mu^*_{\theta|S(\Xn)}(\cdot|s)\,||\,\mu_{W}(\cdot|s))\Big]-\mb{E}_{s\sim S_{\#}\m P^{(n)}}\mb{E}_{\theta\sim \mu^*_{\theta|S(\Xn)}(\cdot|s)}[\log (p_S(\theta|s))]\\
        &=\mb{E}_{\Xn\sim \m P^{(n)}}\Big[{\rm KL}\big(\mu^*_{\theta|S(\Xn)}(\cdot|S(\Xn))\,||\,\mu_{W}(\cdot|S(\Xn)) \big)\Big] 
        \\ & \quad -\mb{E}_{s\sim S_{\#}\m P^{(n)}}\mb{E}_{\theta\sim \mu^*_{\theta|S(\Xn)}(\cdot|s)}[\log (p_S(\theta|s))],
\end{aligned}
    \end{equation*}
  Since the last term $\mb{E}_{s\sim S_{\#}\m P^{(n)}}\mb{E}_{\theta\sim \mu^*_{\theta|S(\Xn)}(\cdot|s)}[\log (p_S(\theta|s))]$ is independent of $W$, it follows that for any $S\in \ms S$, 
\begin{equation*}
    \begin{aligned}
        {\arg\min}_{W\in \ms W} \m R(S,W)=      {\arg\min}_{W\in \ms W} \mb{E}_{\Xn\sim \m P^{(n)}}\Big[{\rm KL}(\mu^*_{\theta|S(\Xn)}(\cdot|S(\Xn))||\mu_{W}(\cdot|S(\Xn))) \Big].
    \end{aligned}
\end{equation*}
This proves Property~\ref{pro1}. Moreover, suppose that, for every $S\in\ms S$,
    \begin{equation*}
        {\min}_{W\in \ms W} \mb{E}_{X^{(n)}\sim \m P^{(n)}}[{\rm KL}\big(\mu^*_{\theta|S(\Xn)}(\cdot|S(\Xn))\,||\,\mu_{W}(\cdot|S(X^{(n)}))\big)]=0,
    \end{equation*}
    we have 
    \begin{equation*}
        \begin{aligned}
            &{\min}_{W\in \ms W} \m R(S,W)\\
            &={\min}_{W\in \ms W}  \mb{E}_{s\sim S_{\#}\m P^{(n)}}\Big[{\rm KL}(\mu^*_{\theta|S(\Xn)}(\cdot|s)||\mu_{W}(\cdot|s))\Big]-\mb{E}_{s\sim S_{\#}\m P^{(n)}}\mb{E}_{\theta\sim \mu^*_{\theta|S(\Xn)}(\cdot|s)}[\log (p_S(\theta|s))]\\
            &=-\mb{E}_{s\sim S_{\#}\m P^{(n)}}\mb{E}_{\theta\sim \mu^*_{\theta|S(\Xn)}(\cdot|s)}[\log (p_S(\theta|s))]\\
             &=-\mb{E}_{\theta\sim \Pi_{\theta}}\mb{E}_{\Xn\sim \m P_{\theta}^{(n)}}[\log (p_S(\theta|S(\Xn)))]\\
            &=-\mb{E}_{\Xn\sim\m P^{(n)}}\mb{E}_{\theta\sim \mu^*_{\theta|\Xn}(\cdot|\Xn)}[\log (p_S(\theta|S(\Xn)))]\\
             &=\mb{E}_{\Xn\sim\m P^{(n)}}[{\rm KL}(\mu^*_{\theta|\Xn}(\cdot|\Xn)||\mu^*_{\theta|S(\Xn)}(\cdot|S(\Xn)))]\\
             &\qquad\quad-\mb{E}_{\Xn\sim\m P^{(n)}}\mb{E}_{\theta\sim \mu^*_{\theta|\Xn}(\cdot|\Xn)}[\log (p(\theta|\Xn))].
        \end{aligned}
    \end{equation*}
     Since the last term $\mb{E}_{\Xn\sim\m P^{(n)}}\mb{E}_{\theta\sim \mu^*_{\theta|\Xn}(\cdot|\Xn)}[\log (p(\theta|\Xn))]$ is independent of $S$, it follows that, 
\begin{equation*}
    \begin{aligned}
        \arg\min_{S\in \mathcal{S}} \min_{W\in \mathcal{W}} R(S,W) 
        &= \arg\min_{S\in \mathcal{S}} 
           \mathbb{E}_{\Xn\sim P^{(n)}}\bigl[ \mathrm{KL}\bigl( \mu^*_{\theta|\Xn}(\cdot|\Xn) \big\| \mu^*_{\theta|S(\Xn)}(\cdot|S(\Xn))\bigr)\bigr].
    \end{aligned}
\end{equation*}
This proves Property~\ref{pro2}. Finally, note that for any $(S,W)\in \ms S \times \ms W$, it holds that 
\begin{equation*}
\begin{aligned}
        &\EX[{\rm KL}(\mu^*_{\theta|\Xn}(\cdot|\Xn)\,||\,\mu_{W}(\cdot|S(\Xn)) )]\\
        &=\EX[{\rm KL}(\mu^*_{\theta|\Xn}(\cdot|\Xn)\,||\,\mu^*_{\theta|S(\Xn)}(\cdot|S(\Xn)))]\\
        &\qquad+\EXandth[\log p_S(\theta|S(\Xn))-\log W(\theta,S(\Xn)) ]\\
        &=\EX[{\rm KL}(\mu^*_{\theta|\Xn}(\cdot|\Xn)\,||\,\mu^*_{\theta|S(\Xn)}(\cdot|S(\Xn)))]\\
        &\qquad+\mb{E}_{s\sim S_{\#}\m P^{(n)},\theta\sim \mu^*_{\theta|S(\Xn)}(\cdot|s)}[\log p_S(\theta|s)-\log W(\theta,s)]\\
        &=\EX[{\rm KL}(\mu^*_{\theta|\Xn}(\cdot|\Xn)\,||\,\mu^*_{\theta|S(\Xn)}(\cdot|S(\Xn)))]\\
        &\qquad\qquad+\EX[{\rm KL}(\mu^*_{\theta|S(\Xn)}(\cdot|S(\Xn))\,||\,\mu_{W}(\cdot|S(\Xn))) ].
\end{aligned}
\end{equation*}
 Hence it holds that
 \begin{equation*}
     \begin{aligned}
         \EX[{\rm KL}(\mu^*_{\theta|\Xn}(\cdot|\Xn)\,||\,\mu^*_{\theta|S(\Xn)}(\cdot|S(\Xn)))]\leq \EX[{\rm KL}(\mu^*_{\theta|\Xn}(\cdot|\Xn)\,||\,\mu_{W}(\cdot|S(\Xn)) )],
     \end{aligned}
 \end{equation*}
and this completes the proof.

\subsection{Proof of Theorem~\ref{th2} and Theorem~\ref{th3}} 
For any fixed $S$, it holds that 
\begin{equation*}
    \begin{aligned}
        &\m R(S,W)=\EXandth\bigg[\mb{E}_{\theta',\theta^{''}\sim \mu_{W}(\cdot|S(X^{(n)}))}[\kappa(\theta',\theta^{''})]-2\cdot \mb{E}_{\theta'\sim  \mu_{W}(\cdot|S(X^{(n)}))}[\kappa(\theta, \theta')]\bigg]\\
        &= \mb{E}_{s\sim S_{\#}\m P^{(n)}}
       \bigg[\mb{E}_{\theta',\theta^{''}\sim \mu_{W}(\cdot|s)}[\kappa(\theta',\theta^{''})]-2\cdot \mb{E}_{\theta\sim \mu^*_{\theta|S(\Xn)}(\cdot|s),\theta'\sim  \mu_{W}(\cdot|s)}[\kappa(\theta, \theta')] \\
       &\qquad +\mb{E}_{\theta',\theta^{''}\sim \mu^*_{\theta|S(\Xn)}(\cdot|s)}[\kappa(\theta',\theta^{''})]\bigg] - \mb{E}_{s\sim S_{\#}\m P^{(n)}}\mb{E}_{\theta',\theta^{''}\sim \mu^*_{\theta|S(\Xn)}(\cdot|s)}[\kappa(\theta',\theta^{''})]\\
        &=\mb{E}_{\Xn\sim \m P^{(n)}}\Big[{\rm MMD}^2_{\kappa}\big(\mu^*_{\theta|S(\Xn)}(\cdot|S(\Xn)),\mu_{W}(\cdot|S(\Xn))  \big)\Big] \\
       &\qquad-\mb{E}_{s\sim S_{\#}\m P^{(n)}}\mb{E}_{\theta',\theta^{''}\sim \mu^*_{\theta|S(\Xn)}(\cdot|s)}[\kappa(\theta',\theta^{''})].
    \end{aligned}
\end{equation*}
Since the last term $\mb{E}_{s\sim S_{\#}\m P^{(n)}}\mb{E}_{\theta',\theta^{''}\sim \mu^*_{\theta|S(\Xn)}(\cdot|s)}[\kappa(\theta',\theta^{''})]$ is independent of $W$, it follows that for any $S\in \ms S$, 
\begin{equation*}
    \begin{aligned}
        {\arg\min}_{W\in \ms W} \m R(S,W)=      {\arg\min}_{W\in \ms W} \mb{E}_{\Xn\sim \m P^{(n)}}\Big[{\rm MMD}^2_{\kappa}(\mu^*_{\theta|S(\Xn)}(\cdot|S(\Xn)),\mu_{W}(\cdot|S(\Xn))) \Big].
    \end{aligned}
\end{equation*}
This proves Property~\ref{pro1}. Moreover, if
    \begin{equation*}
        {\min}_{W\in \ms W} \mb{E}_{X^{(n)}\sim \m P^{(n)}}[{\rm MMD}^2_{\kappa}(\mu^*_{\theta|S(\Xn)}(\cdot|S(\Xn)),\mu_{W}(\cdot|S(X^{(n)})))]=0,
    \end{equation*}
    we have 
    \begin{equation*}
        \begin{aligned}
            &{\min}_{W\in \ms W} \m R(S,W)\\
            &={\min}_{W\in \ms W}  \mb{E}_{\Xn\sim \m P^{(n)}}[{\rm MMD}^2_{\kappa}(\mu^*_{\theta|S(\Xn)}(\cdot|S(\Xn)),\mu_{W}(\cdot|S(\Xn))  )] \\
           &\qquad-\mb{E}_{s\sim S_{\#}\m P^{(n)}}\mb{E}_{\theta',\theta^{''}\sim \mu^*_{\theta|S(\Xn)}(\cdot|s)}[\kappa(\theta',\theta^{''})]\\
                &=-\mb{E}_{s\sim S_{\#}\m P^{(n)}}\mb{E}_{\theta',\theta^{''}\sim \mu^*_{\theta|S(\Xn)}(\cdot|s)}[\kappa(\theta',\theta^{''})].
        \end{aligned}
    \end{equation*}
    Note that 
    \begin{equation}\label{eqn:equivmmd}
        \begin{aligned}
           & \mb{E}_{\Xn\sim \m P^{(n)}}\mb{E}_{\theta'\sim \mu^*_{\theta|\Xn}(\cdot|\Xn),\theta^{''}\sim \mu^*_{\theta|S(\Xn)}(\cdot|S(\Xn))}[\kappa(\theta',\theta^{''})]\\
            &= \mb{E}_{\Xn\sim \m P^{(n)},\theta'\sim \mu^*_{\theta|\Xn}(\cdot|\Xn)} \mb{E}_{\theta^{''}\sim \mu^*_{\theta|S(\Xn)}(\cdot|S(\Xn))}[\kappa(\theta',\theta^{''})]\\
            &= \mb{E}_{s\sim S_{\#}\m P^{(n)},\theta'\sim \mu^*_{\theta|S(\Xn)}(\cdot|s)} \mb{E}_{\theta^{''}\sim \mu^*_{\theta|S(\Xn)}(\cdot|s)}[\kappa(\theta',\theta^{''})]\\
            &=\mb{E}_{s\sim S_{\#}\m P^{(n)}}\mb{E}_{\theta',\theta^{''}\sim \mu^*_{\theta|S(\Xn)}(\cdot|s)}[\kappa(\theta',\theta^{''})]
        \end{aligned}
    \end{equation}
    Therefore, we have 
    \begin{equation*}
        \begin{aligned}
              &{\min}_{W\in \ms W} \m R(S,W)=-\mb{E}_{s\sim S_{\#}\m P^{(n)}}\mb{E}_{\theta',\theta^{''}\sim \mu^*_{\theta|S(\Xn)}(\cdot|s)}[\kappa(\theta',\theta^{''})]\\
              &=\mb{E}_{\Xn\sim \m P^{(n)}}\mb{E}_{\theta',\theta^{''}\sim \mu^*_{\theta|S(\Xn)}(\cdot|S(\Xn))}[\kappa(\theta',\theta^{''})]\\
              &\qquad-2\cdot \mb{E}_{\Xn\sim \m P^{(n)}}\mb{E}_{\theta'\sim \mu^*_{\theta|\Xn}(\cdot|\Xn),\theta^{''}\sim \mu^*_{\theta|S(\Xn)}(\cdot|S(\Xn))}[\kappa(\theta',\theta^{''})]\\
              &=\mb{E}_{\Xn\sim \m P^{(n)}}\big[\mmd^2_{\kappa}(\mu^*_{\theta|\Xn}(\cdot|\Xn),\mu^*_{\theta|S(\Xn)}(\cdot|S(\Xn)))\big] \\
           &\qquad-\mb{E}_{\Xn\sim \m P^{(n)}}\mb{E}_{\theta',\theta^{''}\sim \mu^*_{\theta|\Xn}(\cdot|\Xn)}[\kappa(\theta',\theta^{''})]
        \end{aligned}
    \end{equation*}
     Since the last term $\mb{E}_{\Xn\sim \m P^{(n)}}\mb{E}_{\theta',\theta^{''}\sim \mu^*_{\theta|\Xn}(\cdot|\Xn)}[\kappa(\theta',\theta^{''})]$ is independent of $S$, it follows that, 
\begin{equation*}
    \begin{aligned}
        \arg\min_{S\in \ms S} \min_{W\in \ms W}\m R(S,W) 
        &= \arg\min_{S\in \ms S} \mb{E}_{\Xn\sim \m P^{(n)}}\bigl[ \mmd^2_{\kappa}\bigl( \mu^*_{\theta|\Xn}(\cdot|\Xn), \mu^*_{\theta|S(\Xn)}(\cdot|S(\Xn))
        \bigr)\bigr].
    \end{aligned}
\end{equation*}
This proves Property~\ref{pro2}. Finally, for any $(S,W)\in \ms S \times \ms W$,  it holds that
\begin{equation*}
    \begin{aligned}
 & \EX\mb{E}_{\theta'\sim \mu^*_{\theta|\Xn}(\cdot|\Xn),\theta^{''}\sim \mu_{W}(\cdot|S(\Xn))}\big[ \kappa(\theta',\theta^{''})\big]\\
     &= \mb{E}_{\Xn\sim \m P^{(n)},\theta'\sim \mu^*_{\theta|\Xn}(\cdot|\Xn)} \mb{E}_{\theta^{''}\sim \mu_{W}(\cdot|S(\Xn))}[\kappa(\theta',\theta^{''})]\\
            &= \mb{E}_{s\sim S_{\#}\m P^{(n)},\theta'\sim \mu^*_{\theta|S(\Xn)}(\cdot|s)} \mb{E}_{\theta^{''}\sim \mu_{W}(\cdot|s)}[\kappa(\theta',\theta^{''})]\\
            &=  \mb{E}_{\Xn\sim \m P^{(n)}} \mb{E}_{\theta'\sim \mu^*_{\theta|S(\Xn)}(\cdot|S(\Xn)),\theta^{''}\sim \mu_{W}(\cdot|S(\Xn))}[\kappa(\theta',\theta^{''})].
    \end{aligned}
\end{equation*}
Therefore,
 \begin{equation*}
\begin{aligned}
        &\EX\Big[{\rm MMD}^2_{\kappa}(\mu^*_{\theta|\Xn}(\cdot|\Xn),\mu_{W}(\cdot|S(\Xn)) )\Big]\\
        &=\EX\Big[\mb{E}_{\theta',\theta^{''}\sim \mu_{W}(\cdot|S(\Xn))}\big[ \kappa(\theta',\theta^{''})\big]-2\cdot\mb{E}_{\theta'\sim \mu^*_{\theta|\Xn}(\cdot|\Xn),\theta^{''}\sim \mu_{W}(\cdot|S(\Xn))}\big[ \kappa(\theta',\theta^{''})\big]\\
        &\qquad+\mb{E}_{\theta',\theta^{''}\sim \mu^*_{\theta|\Xn}(\cdot|\Xn)}\big[ \kappa(\theta',\theta^{''})\big] 
        \Big]\\
        &=  \EX\Big[{\rm MMD}^2_{\kappa}(\mu^*_{\theta|S(\Xn)}(\cdot|S(\Xn)),\mu_{W}(\cdot|S(\Xn)) )\Big]\\
        &+\EX\Big[\mb{E}_{\theta',\theta^{''}\sim \mu^*_{\theta|\Xn}(\cdot|\Xn)}\big[ \kappa(\theta',\theta^{''})\big]-\mb{E}_{\theta',\theta^{''}\sim \mu^*_{\theta|S(\Xn)}(\cdot|S(\Xn))}\big[ \kappa(\theta',\theta^{''})\big]\Big]. 
\end{aligned}
\end{equation*}
Finally, by equation~\eqref{eqn:equivmmd}, 
\begin{equation*}
\begin{aligned}
        &\EX\Big[\mb{E}_{\theta',\theta^{''}\sim \mu^*_{\theta|\Xn}(\cdot|\Xn)}\big[ \kappa(\theta',\theta^{''})\big]-\mb{E}_{\theta',\theta^{''}\sim \mu^*_{\theta|S(\Xn)}(\cdot|S(\Xn))}\big[ \kappa(\theta',\theta^{''})\big]\Big]\\
        &=\EX\big[\mmd^2_{\kappa}(\mu^*_{\theta|\Xn}(\cdot|\Xn),\mu^*_{\theta|S(\Xn)}(\cdot|S(\Xn)))\big].
\end{aligned}
\end{equation*}
 Therefore, 
 \begin{align*}
     &\EX\Big[{\rm MMD}^2_{\kappa}(\mu^*_{\theta|\Xn}(\cdot|\Xn),\mu_{W}(\cdot|S(\Xn)) )\Big]\\
     &\geq \EX\big[\mmd^2_{\kappa}(\mu^*_{\theta|\Xn}(\cdot|\Xn),\mu^*_{\theta|S(\Xn)}(\cdot|S(\Xn)))\big], 
 \end{align*}
 which completes the proof of Theorem~\ref{th2}. The proof of Theorem~\ref{th3} follows identically after replacing $\kappa(\theta',\theta^{''})$ by $\mb{E}_{\lambda\sim {\rm Unif}(\mb S_1^{d-1})} \big[\kappa(\lambda^T\theta',\lambda^T\theta^{''})\big]$.

\subsection{Proof of Theorem \ref{thm:Generalization Error}}
 
 Consider the  loss function
\begin{align*}
\ell(S, W, \theta, X^{(n)}) =
\mathbb{E}_{\theta', \theta^{''} \sim \muws}
[\kappa_h(\theta', \theta^{''})] 
- 2 \cdot \mathbb{E}_{\theta' \sim \muws}[\kappa_h(\theta', {\theta})].
\end{align*} 
The estimated posterior $\hmuws$ is obtained by empirical risk minimization:
\begin{align*}
(\hS,\hW) \in {\arg\min}_{S \in \cS, W \in \cW} \frac{1}{N}\sum_{j=1}^N \ell(S,W,{\theta}_j,\Xn_j).
\end{align*}
Consider the population risk minimizer $(S^*,W^*)$ so that 
\begin{equation*}
    (S^*,W^*) \in {\arg\min}_{S \in \cS, W \in \cW} \m R(S,W),\quad \m R(S,W)=\mb{E}_{ \theta\sim \Pi_{\theta},X^{(n)}\sim \m P_{\theta}^{(n)}}[\ell(S,W,{\theta},\Xn)].
\end{equation*}
Notice that for any $(S,W)\in \ms S\times \ms W$,  it holds that
\begin{equation*}
\begin{aligned}
        & \EX\Big[\mmd^2_{h}\left(\mu^*_{\theta|\Xn}(\cdot|\Xn),\mu_{W}(\cdot|S(X^{(n)}))\right)\Big] \\
        &=\EX \Big[\mathbb{E}_{\theta', \theta^{''} \sim \muws}
[\kappa_h(\theta', \theta^{''})] 
\\
&\quad- 2 \cdot \mathbb{E}_{\theta\sim\mu^*_{\theta|\Xn}(\cdot|\Xn),\theta' \sim \muws}[\kappa_h(\theta', {\theta})]+\mathbb{E}_{\theta', \theta^{''} \sim  \mu^*_{\theta|\Xn}(\cdot|\Xn)}
[\kappa_h(\theta', \theta^{''})]
\Big]\\
&=\EX \Big[\mathbb{E}_{\theta', \theta^{''} \sim \muws}
[\kappa_h(\theta', \theta^{''})]\Big]  -2 \cdot\EXandth\Big[ \mathbb{E}_{\theta' \sim \muws}[\kappa_h(\theta', {\theta})]\Big]\\
&\quad +\EX \Big[\mathbb{E}_{\theta', \theta^{''} \sim  \mu^*_{\theta|\Xn}(\cdot|\Xn)}
[\kappa_h(\theta', \theta^{''})]\Big]\\
&=\m R(S,W)+ \EX \Big[\mathbb{E}_{\theta', \theta^{''} \sim \mu^*_{\theta|\Xn}(\cdot|\Xn)}
[\kappa_h(\theta', \theta^{''})]\Big].
\end{aligned}
\end{equation*}
The last term $\EX \big[\mathbb{E}_{\theta', \theta^{''} \sim  \mu^*_{\theta|\Xn}(\cdot|\Xn)}
[\kappa_h(\theta', \theta^{''})]\big]$ is independent of $(S,W)$. Therefore, using  $\frac{1}{N}\sum_{j=1}^N \ell(\wh S,\wh W,{\theta}_j,\Xn_j)\leq \frac{1}{N}\sum_{j=1}^N \ell(S^*,W^*,{\theta}_j,\Xn_j)$, we have
\begin{equation*}
\begin{aligned}
  &  
  \EX\Big[\mmd^2_h\left( \pos,\hmuws \right)- \mmd^2_h\left(\pos,\smuws \right) \Big]
  \\
  &=\m R(\wh S,\wh W)-\m R(S^*,W^*)\\
  & \leq \m R(\wh S,\wh W)-\m R(S^*,W^*)-
  \Big(\frac{1}{N} \sum_{i=1}^N \ellhati -
    \frac{1}{N} \sum_{i=1}^N \ellstari\Big)
  \\
  & =\EXandth\left[\ellhat - \ellstar \right] \\
    &\quad - \frac{1}{N}\sum_{i=1}^N
  \left( \ellhati - \ellstari
  \right).
\end{aligned}
\end{equation*}
Since $0\leq\kappa_h(x,y)\leq\frac{1}{(2\pi h^2)^{\frac{d}{2}}}$ for any $x,y\in \mb R^d$, we have for any  $S \in \cS,W \in \cW$, $\theta\in \mb R^d$ and $\Xn\in \m X^n$ that
\begin{align*}
  &\left|
    \ellws - \ellstar 
  \right|
  \\
  \leq &
  \left|
\mathbb{E}_{\theta', \theta^{''} \sim \muws}
[\kappa_h(\theta', \theta^{''})] - \mathbb{E}_{\theta', \theta^{''} \sim \smuws}[\kappa_h(\theta', \theta^{''})]\right|\\
&\qquad+2 \cdot \left|\mathbb{E}_{\theta' \sim \muws}[\kappa_h(\theta', {\theta})]-\mathbb{E}_{\theta' \sim\smuws}[\kappa_h(\theta', {\theta})]
  \right|\leq 
  \frac{3}{(2\pi)^{d/2}h^d}.
\end{align*}
Then note that $\min_{W\in \ms W, S\in \ms S} \E_{X^{(n)}\sim \m P^{(n)}}[\mmd^2_h(\mu^*_{\theta|\Xn}(\cdot|\Xn), \mu_{W}(\cdot|S(X^{(n)})))]\leq \varepsilon.$ If $\varepsilon\geq c h^{-d}$,   it holds that
\begin{equation*}
\begin{aligned}
       & \EX\Big[\mmd^2_h\left( \pos,\hmuws \right)\Big]\leq \frac{4}{(2\pi)^{\frac{d}{2}}h^d}\leq \frac{4}{(2\pi)^{\frac{d}{2}}c}\varepsilon.  
\end{aligned}
\end{equation*}
 Similarly, if $\frac{T_2\log(T_1N(2+h^d))}{N}\geq c$, then 
\begin{equation*}
    \begin{aligned}
        &\EX\bigl[ \mmd^2_h\left( \pos,\hmuws \right)\bigr]\\
        &\leq \frac{4}{(2\pi)^{\frac{d}{2}}h^d}  
        \leq \frac{4}{c(2\pi)^{\frac{d}{2}}} \cdot \frac{T_2\log\bigl(T_1N(2+h^d)\bigr)}{Nh^d}.
    \end{aligned}
\end{equation*}
Hence, we assume  $\varepsilon\leq c h^{-d}$ and $\frac{T_2\log(T_1N(2+h^d))}{N}\leq c$ for a sufficiently small absolute constant $c$. For a given $(S,W)\in \ms S \times \ms W$, denote the star-shifted and rescaled function as 
\begin{align*}
  \sgws(\theta,\Xn) = 
  \frac{(2\pi)^{d/2}h^d}{3}
  \left(
    \ellws - \ellstar
  \right),
\end{align*}
and the corresponding $1$-bounded function class 
$\cG^* = \{ \sgws : W \in \ms{W}, S \in \ms{S}\}$.
With the new notation, the inequalities can be further extended to:
\begin{equation*}
\begin{aligned}
  &   \EX\Big[\mmd^2_h\left( \pos,\hmuws \right) \\
  & \qquad \qquad \quad- 
  \mmd^2_h\left(\pos,\smuws \right) \Big]\\
  & \leq \frac{3}{(2\pi)^{d/2}h^d}  
  \left|\frac{1}{N}\sum_{i=1}^N  \sgwsh(\ttheta_i,\Xn_i)- \EXandth[\sgwsh(\theta,\Xn)]
  \right|.
\end{aligned}
\end{equation*}
Consider the star-shaped hull of $\mathcal{G}^*$, denoted by 
$\ov{\cG}^* = \{ a\sgws: a \in[0,1], \sgws \in \mathcal{G}^*\}$. Define the localized Rademacher complexity 
\begin{equation*}
    \ov {\ms R}_N(\delta, \ov{\cG}^*)=\mb{E}\left[\sup_{g\in \ov{\cG}^*\atop \mb{E}[(g(\theta,X^{(n)}))^2] \leq\delta^2 }\Big|\frac{1}{N}\sum_{j=1}^N \epsilon_j g(\theta_j, \Xn_j)\Big|\right].
\end{equation*}
Here $(\theta,X^{(n)})$ is distributed according to $\Pi_{\theta}\m P_{\theta}^{(n)}$,  $\{(\theta_j, \Xn_j)\}_{j=1}^N$
are $N$ i.i.d. copies of $(\theta,X^{(n)})$, and $\{\epsilon_j\}_{j=1}^N$ are i.i.d. Rademacher
variables taking values in $\{-1,1\}$ equiprobably, independent of everything else. For brevity, we omit subscripts on the expectation operator $\mb{E}$. Let $\delta_N$ be a solution to the inequality
  $
  \overline{\ms{R}}_N(\delta; \ov{\cG}^* ) \leq \delta^2
  $ in terms of $\delta$, which will be computed later.  Using the uniform law for Lipschitz cost functions (Theorem 14.20 of~\cite{wainwright2019high}), there exist universal constants $c_1, c_2$ so that it holds with probability at least $1 - c_1 e^{-c_2 N \delta_N^2}$ that for any $(S,W)\in \ms S\times \ms W$,
 \begin{align*}
  &\left|\frac{1}{N}\sum_{i=1}^N  \sgws(\ttheta_i,\Xn_i)- \EXandth[\sgws(\theta,\Xn)]
  \right|  \\ 
  & \leq
10\delta_N 
\left(\delta_N + \sqrt{  \EXandth \left[\sgws(\theta,\Xn)^2\right]
  }  
\right).
\end{align*}
Then since $(\wh S,\wh W) \in \ms S\times \ms W$, we have
\begin{equation}\label{eqn:main1}
    \begin{aligned}
      &  \EX\Big[\mmd^2_h(\pos,\hmuws) \Big]  \\ & \qquad \qquad -\EX\Big[\mmd^2_h(\pos,\smuws)\Big] \\
  &\leq    \frac{3}{(2\pi)^{d/2}h^d}  
  \left|\frac{1}{N}\sum_{i=1}^N  \sgwsh(\ttheta_i,\Xn_i)- \EXandth[\sgwsh(\theta,\Xn)]
  \right|\\
  &\leq \frac{30}{(2\pi)^{d/2}h^d}  \delta_N 
\left(\delta_N + \sqrt{  \EXandth \left[\sgwsh(\theta,\Xn)^2\right]
  }  \right).
    \end{aligned}
\end{equation}
Furthermore, since $\kappa_h(\theta,\theta')$ is the density function of $N(\theta',h^2I_d)$ evaluated at $\theta$, using the density of Gaussian convolution, we have 
\begin{equation*}
    \kappa_h(\theta,\theta')=\int \kappa_{h/\sqrt{2}}(\theta,y)\kappa_{h/\sqrt{2}}(\theta',y)\,\dd y.
\end{equation*}
We can then obtain 
\begin{small}
\begin{align*}
    & \EXandth \left[\sgwsh(\theta,\Xn)^2\right] \\
    & =\EXandth
    \left[\left(\frac{(2\pi)^{d/2}h^d}{3}
      \left(   \ell(\wh S,\wh W,\theta, \Xn) - \ell(S^*,W^*,\theta,\Xn) \right) \right)^2\right]
    \\
    & =
    \left(\frac{(2\pi)^{d/2}h^d}{3}\right)^2
   \EXandth
    \bigg[
    \\
    & \quad \Big(\int\left( \mathbb{E}_{\theta_1 \sim \hmuws}   [\kappa_{h/\sqrt{2}}(\theta_1,y)]   -  \mathbb{E}_{\theta_1 \sim \smuws} [\kappa_{h/\sqrt{2}}(\theta_1,y)]  \right)\\
        &\qquad \qquad\cdot
    \left( \mathbb{E}_{\theta_1 \sim \hmuws}   [\kappa_{h/\sqrt{2}}(\theta_1,y)]   + \mathbb{E}_{\theta_1 \sim \smuws} [\kappa_{h/\sqrt{2}}(\theta_1,y)]  \right) \,\dd y
    \\ & \quad - 2 \int
    \kappa_{h/\sqrt{2}}(\theta,y) \left( \mathbb{E}_{\theta_1 \sim \hmuws}
          [\kappa_{h/\sqrt{2}}(\theta_1,y)]  - 
          \mathbb{E}_{\theta_1 \sim \smuws}[\kappa_{h/\sqrt{2}}(\theta_1,y)]\right)
    \,\dd y\Big)^2\bigg] \\
    & \leq
    \left(\frac{(2\pi)^{d/2}h^d}{3}\right)^2
   \EXandth
    \Bigg[ \\ 
    & \quad \bigg(
    \sqrt{
      \int \left( \mathbb{E}_{\theta_1 \sim \hmuws}   [\kappa_{h/\sqrt{2}}(\theta_1,y)]   -  \mathbb{E}_{\theta_1 \sim \smuws} [\kappa_{h/\sqrt{2}}(\theta_1,y)]  \right)^2 \,\dd y
     } \\ & \quad \cdot
    \bigg\{
          \sqrt{
      \int \left( \mathbb{E}_{\theta_1 \sim \hmuws}   [\kappa_{h/\sqrt{2}}(\theta_1,y)]   + \mathbb{E}_{\theta_1 \sim \smuws} [\kappa_{h/\sqrt{2}}(\theta_1,y)]  \right)^2 \,\dd y
     }
     \\
     & \qquad
      +2
    \sqrt{
        \int \kappa^2_{h/\sqrt{2}}(\theta,y) \dd y}
      \bigg\}
    \bigg)^2\Bigg]
    \\
    & \overset{(i)}{\leq}
     \left(\frac{(2\pi)^{d/2}h^d}{3}\right)^2 
     \cdot    
        \frac{16}{(\pi h^2)^{d/2}} 
   \EX
    \left[
     \mmd^2_h(\hmuws,\smuws) 
    \right]
    \\
    & \leq
\frac{32  (2\sqrt{\pi}h)^d}{9} \cdot
   \EX \Big[
       \mmd^2_h\big(\hmuws,\pos\big) \\
       & \qquad \qquad \qquad  \qquad \qquad \qquad +  \mmd^2_h\big(\pos,\smuws\big) 
    \Big]
    \\
    & \leq
\frac{32  (2\sqrt{\pi}h)^d}{9}  \cdot
    \left(
        \EX \Big[\mmd^2_h(\hmuws,\pos) \Big]
      + \varepsilon
    \right),
\end{align*}
\end{small}
where $(i)$ uses 
\begin{equation*}
    \begin{aligned}
      &   \int \left( \mathbb{E}_{\theta_1 \sim \hmuws}   [\kappa_{h/\sqrt{2}}(\theta_1,y)]   -  \mathbb{E}_{\theta_1 \sim \smuws} [\kappa_{h/\sqrt{2}}(\theta_1,y)]  \right)^2 \,\dd y\\
      &=  \mmd^2_h(\hmuws,\smuws),
    \end{aligned}
\end{equation*}
and $\sup_{x,y}|\kappa_{h/\sqrt{2}}(x,y)|\leq \frac{1}{(\pi h^2)^{\frac{d}{2}}}$.
Plugging this back into the main inequality of~\eqref{eqn:main1}, we get
\begin{align*}
   &  \EX\Big[\mmd^2_h(\pos,\hmuws) \Big]  -\varepsilon 
  \\
  & \leq
  \frac{30}{(2\pi)^{d/2}h^d}
  \delta_N \\
  & \quad
  \left(
  \delta_N + \sqrt{
\frac{32  (2\sqrt{\pi}h)^d}{9}
    \left(
       \EX \Big[\mmd^2_h(\hmuws,\pos) \Big]+ \varepsilon
    \right)
  }
  \right)
  \\
  & \leq
  \frac{30}{(2\pi)^{d/2}h^d} \delta_N^2 \\ 
  & \quad +
  \frac{40\sqrt{2}}{\pi^{d/4}h^{d/2}} \delta_N\cdot\Big(
  \sqrt{
   \EX \Big[\mmd^2_h( \pos,\hmuws) \Big]
     }+\sqrt{\varepsilon}\Big).
\end{align*}
Let $Z = \sqrt{\EX \Big[\mmd^2_h( \pos,\hmuws) \Big]}$. The inequality is of the form $Z^2 -BZ-A\leq 0$, where $A = \frac{30}{(2\pi)^{d/2}h^d}\delta_N^2+\frac{40\sqrt{2}}{\pi^{d/4}h^{d/2}} \delta_N\sqrt{\varepsilon}+\varepsilon$ and $B = \frac{40\sqrt{2}}{\pi^{d/4}h^{d/2}} \delta_N$. We have 
\begin{equation*}
    Z^2\leq (\frac{B+\sqrt{B^2+4A}}{2})^2\leq (B+\sqrt{A})^2\leq 2\big( \frac{40\sqrt{2}}{\pi^{d/4}h^{d/2}} \delta_N\big)^2+2\big(\frac{30}{(2\pi)^{d/2}h^d}\delta_N^2+\frac{40\sqrt{2}}{\pi^{d/4}h^{d/2}} \delta_N\sqrt{\varepsilon}+\varepsilon\big).
\end{equation*}
Thus, we can write the bound as: it holds with probability at least $1 - c_1 e^{-c_2 N \delta_N^2}$ that
\begin{equation}\label{eqn:MMDbound}
    \begin{aligned}
        \EX \Big[\mmd^2_h( \pos,\hmuws) \Big] \leq C_d(\varepsilon+\frac{\delta_N^2}{h^d}),
    \end{aligned}
 \end{equation}
where $C_d$ is a constant that depends only on $d$.

Next, we compute $\delta_N$, which is the solution to $\overline{\ms{R}}_N(\delta; \ov{\mathcal{G}}^*) \leq \delta^2$. For $g\in \ov{\cG}^*$, write $\|g\|=\sqrt{\EXandth[(g(\theta,X^{(n)}))^2]}$. 
 Let $$r_N(\delta)= \sup_{g, g' \in \ov{\cG}^*\atop \|g\|,\|g'\|\leq \delta} d_N^{{\cG}^*}(g, g')$$ be the local diameter of $\ov{\cG}^*$ for a given sample $\{(\theta_i, \Xn_i)\}_{i=1}^N$, where $$d_N^{{\cG}^*}(g, g') = \sqrt{\frac{1}{N}\sum_{i=1}^N (g(\theta_i,\Xn_i) - g'(\theta_i,\Xn_i))^2}.$$ Note that $\ov{\cG}^*$ is $1-$bounded, so $r_N(\delta)\leq 2$ for any $\delta$.
The local Rademacher complexity $\overline{\mathcal{R}}_N(\delta; \ov{\cG}^*)$ can be bounded using Dudley's entropy integral (see for example, Theorem 5.22 of~\cite{wainwright2019high}):
\begin{equation}\label{eqn:dud}
   \begin{aligned}
\overline{\mathcal{R}}_N(\delta; \ov{\cG}^*) 
& \leq  16(2\pi)^{d/2}h^d \varepsilon+
\EXandth \left[  \frac{32}{\sqrt{N}} \int_{ \min\{2(2\pi)^{d/2}h^d \varepsilon,r_N(\delta)\}}^{r_N(\delta)} \sqrt{\log N(\ov{\cG}^*, d_N^{{\cG}^*}, u)} \dd u \right].
\end{aligned} 
\end{equation}

Now we bound the covering number of $\ov{\cG}^*$. Since $\ov{\cG}^* = \{a \cdot g : a \in [0,1], g \in \cG^*\}$, for any $u > 0$, we can cover $[0,1]$ with $\lceil 4/u \rceil$ intervals of size at most $u/4$. Then using the fact that for any $a,a'\in [0,1]$ and $g,g'\in \cG^*$,
\begin{equation*}
    \begin{aligned}
      &d_N^{{\cG}^*}(ag, a'g') \leq \sqrt{\frac{2}{N}\sum_{i=1}^N \left((ag(\theta_i,\Xn_i) - ag'(\theta_i,\Xn_i))^2+(ag'(\theta_i,\Xn_i) - a'g'(\theta_i,\Xn_i))^2\right)}\\
      &\leq \sqrt{2}\sqrt{\frac{1}{N}\sum_{i=1}^N (g(\theta_i,\Xn_i) - g'(\theta_i,\Xn_i))^2}+\sqrt{2}|a-a'|.
    \end{aligned}
\end{equation*}
This gives:
\begin{align*}
\log N(\ov{\cG}^*, d_N^{\cG^*}, u) \leq \log N(\cG^*, d_N^{\cG^*}, u/4) + \log(\lceil 4/u \rceil).
\end{align*}
The metric entropy on $\cG^*$ is related to the metric entropy on $\ms S\times \ms W$ by $$d_N^{\cG^*}(\sgws,  g^*_{W',S'}) = \frac{(2\pi)^{d/2}h^d}{3} d_N((S,W), (S',W')).$$
Using the complexity assumption on $\ms S\times \ms W$, we can then get for any $2(2\pi)^{d/2}h^d\varepsilon \leq u\leq r_N(\delta)$, 
\begin{align*}
\log N(\cG^*, d_N^{\cG^*}, u/4)  \leq T_2 \log\left(1\vee \frac{4T_1 (2\pi)^{d/2}h^d}{3u}\right).
\end{align*}
Let $C_h =2\vee \frac{(2\pi)^{d/2}h^d}{3}$. Then for any $u\leq r_N(\delta)\leq 2$, we have $\frac{4T_1C_h}{u}\geq 2T_1C_h\geq 4$. Since $\lceil4/u\rceil\leq 8/u\leq4T_1C_h/u$ for $0<u\leq2$, the integral term in~\eqref{eqn:dud} is bounded by
\begin{align*}
&\sqrt{2T_2}\int_0^{r_N(\delta)}
 \sqrt{\log\left(\frac{4T_1C_h}{u}\right)}\,\dd u\\
&\leq \sqrt{2T_2}\,r_N(\delta)
 \sqrt{\log\left(\frac{T_1C_h}{r_N(\delta)}\right)}
 +4\sqrt{2\pi T_2}\,r_N(\delta),
\end{align*}
where the second term follows by substituting $u=4r_N(\delta)v$ and using $\int_0^{1/4}\sqrt{-\log v}\,\dd v\leq\sqrt{\pi}$.
Note that $\sqrt{-\frac{y}{2} \log y  + \frac{1}{2} y}$ is concave and increasing on $[0,1]$. 
Let $y =  (\frac{r_N(\delta)}{T_1C_h})^2$, then 
\begin{equation*}
    \begin{aligned}
      &  r_N(\delta) \sqrt{T_2 \log\left( \frac{T_1 C_h}{r_N(\delta)}\right)}=r_N(\delta)\sqrt{-\frac{T_2}{2}\log(y)}=\sqrt{T_2}T_1C_h\frac{r_N(\delta)}{T_1C_h}\sqrt{-\frac{1}{2}\log(y)}\\
      &=\sqrt{T_2}T_1C_h \sqrt{-\frac{y}{2}\log y}\leq \sqrt{T_2}T_1C_h \sqrt{-\frac{y}{2}\log y+\frac{1}{2}y},
    \end{aligned}
\end{equation*}
 So the first part of the integral is bounded by:
\begin{align*}
& \quad
\EXandth \left[ r_N(\delta) \sqrt{T_2 \log\left(\frac{T_1 C_h}{r_N(\delta)}\right)} \right] \\
& \leq
\sqrt{T_2}T_1 C_h \EXandth \left[\sqrt{ - \frac{1}{2} \left(\frac{r_N(\delta)}{T_1C_h}\right)^2  \log\left(\frac{r_N(\delta)}{T_1C_h}\right)^2 + \frac{1}{2} \left(\frac{r_N(\delta)}{T_1C_h}\right)^2 } \right] \\
& \leq
\sqrt{T_2}  \EXandth \left[\left( r_N(\delta) \right)^2 \right]^{1/2} \cdot
\sqrt{  -\frac{1}{2}   \log\EXandth \left[\left(\frac{r_N(\delta)}{T_1C_h}\right)^2\right] + \frac{1}{2} },
\end{align*}
By symmetrization and contraction (see~\cite{wainwright2019high}),
we have $\EXandth[r_N(\delta)^2] \leq 4\delta^2 + 32\overline{\ms{R}}_N(\delta; \ov{\cG}^*)$.  Hence using the fact that $\sqrt{-\frac{y}{2} \log y  + \frac{1}{2} y}$ is  monotonically increasing on $[0,1]$,  there exists a constant $C$ so that for any $\delta\leq 1$,
\begin{align*}
&\overline{\ms{R}}_N(\delta; \ov{\cG}^*)\leq C\, \Big(h^d\varepsilon+
\sqrt{\frac{T_2}{N}}\big(\delta^2+\overline{\ms{R}}_N(\delta; \ov{\cG}^*)\big)^{\frac{1}{2}}\sqrt{\log \frac{T_1C_h}{\delta}}\Big)
\end{align*}
Choose $\delta_{N}=C_1 \sqrt{h^d\varepsilon+\frac{T_2(\log N+\log T_1+\log C_h)}{N}}$ with $C_1\geq \sqrt{\frac{1}{\log 2}}$. Then $\delta_N\geq C_1\sqrt{\frac{\log 2}{N}}\geq \frac{1}{\sqrt{N}}\geq \frac{1}{N}$. Moreover, $\delta_N\leq 1$ if $\varepsilon\leq c h^{-d}$ and $\frac{T_2\log(T_1N(2+h^d))}{N}\leq c$ for a sufficiently small absolute constant $c$. If $\overline{\ms{R}}_N(\delta_N; \ov{\cG}^*)> \delta_{N}^2$,  then 
 \begin{equation*}
 \begin{aligned}
\overline{\ms{R}}_N(\delta_N; \ov{\cG}^*)\leq C\Big(h^d\varepsilon+\sqrt{\frac{2T_2}{N}   \overline{\ms{R}}_N(\delta_N; \ov{\cG}^*)\log (T_1 C_h N)}\Big),
  \end{aligned}
 \end{equation*}
 which means
 \begin{equation*}
 \begin{aligned}
\overline{\ms{R}}_N(\delta_N; \ov{\cG}^*)  \leq 4C^2 \frac{T_2}{N} \Big(\log N+\log T_1+\log C_h\Big) +2C h^d\varepsilon \leq \frac{4C^2+2C}{C_1^2}\delta_{N}^2. \\
 \end{aligned}
 \end{equation*}
So $\overline{\ms{R}}_N(\delta_N; \ov{\cG}^*)  \leq \delta_{N}^2$ holds if $C_1>\sqrt{4C^2+2C} \vee  \sqrt{\frac{1}{\log 2}} $. Moreover, when $C_1$ is large enough, we have $ c_1 e^{-c_2 N \delta_N^2}\leq N^{-1}$.  Therefore, plugging $\delta_N$ into~\eqref{eqn:MMDbound}, we find that it holds with probability at least $1-N^{-1}$ that
\begin{align*}
& \EX[\mmd^2_h(\pos,\hmuws)] \\
  & \leq C_d \varepsilon + \frac{C_d}{h^d}\Big(C_1^2 h^d\varepsilon+C_1^2 \frac{T_2(\log N+\log T_1+\log C_h)}{N}\Big)\\
  &\leq   
  C_2\varepsilon 
  + \frac{C_3}{h^d}  \frac{T_2(\log N+\log T_1+\log (2+h^d))}{N}.
\end{align*}

\subsection{Proof of Corollary~\ref{thm:expfam:formal}}\label{proof:thm:Exponential family result}
 
Since $|\kappa_h(x,y)|\leq \frac{1}{(2\pi h^2)^{\frac{d}{2}}}$ and ${\mmd}_{h}^2(\mu,\nu)\leq \frac{2}{(2\pi h^2)^{\frac{d}{2}}}$, the results for $\varepsilon\geq \frac{2}{(2\pi h^2)^{\frac{d}{2}}}$ trivially hold. So we assume $0<\varepsilon<\frac{2}{(2\pi h^2)^{\frac{d}{2}}}.$ Let $(\theta^*_1,\theta^*_2,\cdots,\theta^*_J)\subset \Theta$ be the largest $\varepsilon_{\theta}$-packing set of $\Theta$, where 
$$\varepsilon_{\theta}=C_{\theta}\Big(1+h\sqrt{\log(1+\frac{1}{\varepsilon})}\Big)^{-\frac{d}{2}-1}\sqrt{\varepsilon}h^d \min(n^{-1},h^{2}),$$
for a sufficiently small absolute constant $C_{\theta}$. Since $\Theta= [0,1]^d$, we have $J\lesssim \varepsilon_{\theta}^{-d}$. Choose $$\ms S=
\Phi_{nd_X}^{p}(L_1,U_1,R_1),$$ and
\begin{equation*}
    \ms W=\Big\{W(\cdot)={\rm clip}\Big(W'(\cdot);-c_1\log(\frac{n}{\min(\varepsilon,1) h}),c_2\Big)\,:\,W'(\cdot)\,\in  \Phi_{p+d}^{1}(L_2,\, U_2,R_2)\Big\},
\end{equation*}
where $c_1,c_2$ are sufficiently large absolute constants independent of $n,N,h$, and the neural network size and sparsity parameters $L_1,U_1,L_2,U_2,R_1,R_2$ will be chosen  later. Then given $s\in \mb R^p$ and $W\in \ms W$, we define $$\mu_{W}(\cdot|s)=\sum_{j=1}^J p_{W,j}(s)\delta_{\theta^*_j}(\cdot),\quad p_W(s)=(p_{W,1}(s),\cdots,p_{W,J}(s))={\rm softmax}(\{W(s,\theta_j^*)\}_{j=1}^J),$$
where $\delta_{\theta^*_j}(\cdot)$ denotes the Dirac measure at $ \theta^*_j$ and $p_{W,j}(s)$ denotes the $j$th component in the vector of  $p_{W}(s)$. We now apply Theorem~\ref{thm:Generalization Error}  to obtain the desired result. First, for  fixed neural network size parameters, we identify the quantities $T_1$, $T_2$ required by the assumption in Theorem~\ref{thm:Generalization Error}. Next, we control the approximation error to determine the appropriate network size. Finally, we combine these two parts to conclude the proof. Note that for $(S,W),(S',W')\in \ms S\times \ms W$, 
\begin{equation*}
    \begin{aligned}
        &\frac{1}{N}\sum_{i=1}^N (\ell(S,W,\theta_i,X^{(n)}_i)-\ell(S',W',\theta_i,X^{(n)}_i))^2\\
       & \lesssim \frac{1}{N}\sum_{i=1}^N\bigg(\sum_{j_1=1}^J \sum_{j_2=1}^J \Big(p_{W,j_1}(S(X_i^{(n)}))p_{W,j_2}(S(X_i^{(n)}))-p_{W',j_1}(S'(X_i^{(n)}))p_{W',j_2}(S'(X_i^{(n)})) \Big)\cdot \kappa_h(\theta_{j_1}^*,\theta_{j_2}^*)\bigg)^2\\
&\qquad+\frac{1}{N}\sum_{i=1}^N\bigg(\sum_{j=1}^J \Big(p_{W,j}(S(X_i^{(n)}))-p_{W',j}(S'(X_i^{(n)})) \Big)\cdot \kappa_h(\theta_i,\theta_{j}^*)\bigg)^2\\
&\lesssim \frac{1}{h^{2d}} \frac{1}{N}\sum_{i=1}^N  \Big(\sum_{j=1}^J \big|p_{W,j}(S(X_i^{(n)}))-p_{W',j}(S'(X_i^{(n)})) \big|\Big)^2.
    \end{aligned}
\end{equation*}
Then notice that 
\begin{equation*}
    \begin{aligned}
        &\sum_{j=1}^J \big|p_{W,j}(S(X_i^{(n)}))-p_{W',j}(S'(X_i^{(n)})) \big|=\sum_{j=1}^J \bigg|\frac{e^{W(S(\Xn_i),\theta_{j}^*)}}{\sum_{j_1=1}^Je^{W(S(\Xn_i),\theta_{j_1}^*)}}-\frac{e^{W'(S'(\Xn_i),\theta_{j}^*)}}{\sum_{j_1=1}^Je^{W'(S'(\Xn_i),\theta_{j_1}^*)}} \bigg|\\
        &\leq 2\exp(c_1\log(\frac{n}{\min(\varepsilon,1) h})+c_2) \sum_{j=1}^J \big|W(S(X_i^{(n)}),\theta_{j}^*)-W'(S'(X_i^{(n)}),\theta_{j}^*) \big|.
    \end{aligned}
\end{equation*}
So  there exist constants $C,C_1$ so that 
\begin{equation*}
    \begin{aligned}
           &\frac{1}{N}\sum_{i=1}^N (\ell(S,W,\theta_i,X^{(n)}_i)-\ell(S',W',\theta_i,X^{(n)}_i))^2\\
           &\leq CJ\exp\Big(C_1\log (\frac{n}{\min(\varepsilon,1) h})\Big) \sum_{j=1}^J \frac{1}{N}\sum_{i=1}^N \big|W(S(X_i^{(n)}),\theta_{j}^*)-W'(S'(X_i^{(n)}),\theta_{j}^*) \big|^2
    \end{aligned}
\end{equation*}
Then consider the class $\Omega=\{f(\Xn,\theta)=W(S(\Xn),\theta):\, W\in \ms W,S\in \ms S\}$.  For $f,f'\in \Omega$, define the empirical sup-metric $$d_{\infty|N}(f,f')=\sup_{i\in[N]\atop j\in [J]}|f(X_i^{(n)},\theta_j^*)-f'(X_i^{(n)},\theta_j^*)|.$$ Then it holds that 
$$\m N(\ms S\times \ms W,d_N,u)\leq \m N\Big(\Omega,d_{\infty|N},\frac{u}{J\sqrt{C\exp(C_1\log(\frac{n}{\min(\varepsilon,1) h}))}}\Big).$$  
 Moreover, note that the clip function can be realized exactly by a ReLU neural network. By concatenation of neural networks (see for example, Lemma F.1 of~\cite{oko2023diffusion}), there exist $L,U,R$ with $L\lesssim L_1+L_2$ and $R\lesssim L+R_1+R_2$ so that $\Omega\subset \Phi_{nd_X+d}^1(L,U,R)$.  
There are at most $(\ms E(U)+1)^R$ choices for the positions of at most $R$ nonzero parameters. For each choice, the pseudo-dimension is $O(RL\log R)$ by Theorem 7 of~\cite{bartlett2019nearly}. Taking the union of the corresponding covers adds at most $R\log(\ms E(U)+1)=O(R\log R)$ to the logarithm of the covering number. 

By the pseudo-dimension bound for ReLU networks
(Theorem 7 of~\cite{bartlett2019nearly})  and the covering-number bound in terms of pseudo-dimension  (Theorem 12.2 of~\cite{anthony2009neural}),  there exist constants  $C_2,C_3$ so that
\begin{equation*}
    \begin{aligned}
        \m N(\ms S\times \ms W,d_N,u) & \leq  \m N\Big(\Omega,d_{\infty|N},\frac{u}{J\sqrt{C\exp(C_1\log(\frac{n}{\min(\varepsilon,1) h}))}}\Big) \\ &\leq \Big(C_2 \frac{NJ\exp(\frac{1}{2}C_1\log(\frac{n}{\min(\varepsilon,1) h}))}{u}\Big)^{C_3 RL\log R}.
    \end{aligned}
\end{equation*}
Therefore, using Theorem~\ref{thm:Generalization Error},   when 
\begin{equation}\label{eqn:N}
    N\geq C_4 \frac{(RL\log R)(\log (\frac{nJ}{\min(\varepsilon,1) h})+\log(RL))}{\varepsilon h^d}
\end{equation}
with a sufficiently large constant $C_4$, it holds that
  \begin{align*}
   & \E_{X^{(n)}\sim \m P^{(n)}} \big[\mmd^2_h\big(\mu^*_{\theta|\Xn}(\cdot|\Xn), \mu^*_{\theta|\wh S(\Xn)}(\cdot|\wh S(X^{(n)}))\big)\big]\\&\leq
  \E_{X^{(n)}\sim \m P^{(n)}} [\mmd^2_h(\mu^*_{\theta|\Xn}(\cdot|\Xn), \mu_{\wh{W}}(\cdot|\wh S(X^{(n)})))]\\
  &\leq 
  C \min_{W\in \ms W, S\in \ms S} \E_{X^{(n)}\sim \m P^{(n)}}[\mmd^2_h(\mu^*_{\theta|\Xn}(\cdot|\Xn), \mu_{W}(\cdot|S(X^{(n)})))]   +\frac{\varepsilon}{2}.
  \end{align*}
Then we will find $R,L$ and $J$ so that 
\begin{align*} 
\min_{W\in \ms W, S\in \ms S} \E_{X^{(n)}\sim \m P^{(n)}}[\mmd^2_h(\mu^*_{\theta|\Xn}(\cdot|\Xn), \mu_{W}(\cdot|S(X^{(n)})))]\leq \frac{\varepsilon}{2C}.
\end{align*}

Consider a fixed model $\Pi_{\theta}\m P^{(n)}_{\theta}$ satisfying Assumption~\ref{assump:exponential}, where $\m P_{\theta}$ has a density function $p(x|\theta)=h(x)\exp(\eta(\theta)^Tt(x)-A(\theta))$. Define the unnormalized posterior density $f(\theta,\Xn) = \pi(\theta)\exp\left(\eta(\theta)^T \sum_{i=1}^n t(X_i) - n A(\theta)\right)$.  Then the posterior $\mu^*_{\theta|\Xn}(\cdot|\Xn)$ has a density 
\begin{equation*}
    p(\theta|\Xn)=\frac{\pi(\theta)\exp\left(\eta(\theta)^T \sum_{i=1}^n t(X_i) - n A(\theta)\right)}{\int_{\Theta} \pi(\theta)\exp\left(\eta(\theta)^T \sum_{i=1}^n t(X_i) - n A(\theta)\right)\,\dd\theta}=\frac{f(\theta,\Xn)}{\int_{\Theta}f(\theta,\Xn)\,\dd\theta}.
\end{equation*}
Define $\wt f(\theta,T_X)=\pi(\theta)\exp\left(\eta(\theta)^T T_X - n A(\theta)\right)$.   We first construct a discrete distribution to approximate the posterior. Consider  the function
$$
k_b(x)=\left\{\begin{array}{cc}
1 & |x|\leq 1 \\
0 & |x|>2 \\
2-|x| & 1<|x| \leq 2
\end{array}\right.
$$
Then $k_b(x)={\rm ReLU}(2-|x|)-{\rm ReLU}(1-|x|)$ with $|x|={\rm ReLU}(x)+{\rm ReLU}(-x)$. So $k_b(\cdot)$ can be exactly realized by  a ReLU neural network. For $j\in [J]$, let $$\rho_j(\theta) =\frac{ k_b(\|\theta - \theta^*_j\|^2/\epth^2)} {\sum_{j=1}^J k_b(\|\theta - \theta^*_j\|^2/\epth^2)}.$$  Then $\sum_{j=1}^J \rho_j(\theta)=1$ for every $\theta\in \Theta$ and each $\rho_j$ is zero if $\|\theta-\theta_j^*\|^2\geq 2\varepsilon_{\theta}^2$. Fix an arbitrary $\Xn\in \m X^{n}$ and $T_X=T(\Xn)=\sum_{i=1}^n t(X_i)$, define the discrete measure
\begin{align*}
\nu(\cdot|T_X) = \sum_{j=1}^J 
\frac{\int_{\Theta}\rho_j(\theta)\,\dd\theta \wt f(\theta^*_j,T_X)}{\sum_{j_1=1}^J \int_{\Theta}\rho_{j_1}(\theta)\,\dd\theta \wt f(\theta^*_{j_1},T_X)} \delta_{\theta_j^*}(\cdot).
\end{align*}
We then first control the MMD between $\nu(\cdot|T(\Xn))$ and $\pos$, and then construct $S^*$, $W^*$ so that $\mu_{W^*}(\cdot|S^*(\Xn))$ is close to  $\nu(\cdot|T(\Xn))$.

Using the identity $\kappa_{h}(\theta,\theta')=\int_{\mb R^d}\kappa_{h/\sqrt{2}}(\theta,y)\kappa_{h/\sqrt{2}}(\theta',y)\,\dd y$,
for any two distributions $\mu$ and $\nu$ on $\Theta$, the squared MMD can be expressed as the squared $L_2$ norm of the difference of the Gaussian kernel means:
\begin{align*}
&\mmd_h^2(\mu, \nu)=\mb{E}_{\theta,\theta'\sim \mu}[\kappa_h(\theta,\theta')]+\mb{E}_{\theta,\theta'\sim \nu}[\kappa_h(\theta,\theta')]-2\cdot\mb{E}_{\theta\sim \mu,\theta'\sim \nu}[\kappa_h(\theta,\theta')]\\
&=\int_{\mb R^{d}} \mb{E}_{\theta,\theta'\sim \mu}[\kappa_{h/\sqrt{2}}(\theta,y)\kappa_{h/\sqrt{2}}(\theta',y)]+\mb{E}_{\theta,\theta'\sim \nu}[\kappa_{h/\sqrt{2}}(\theta,y)\kappa_{h/\sqrt{2}}(\theta',y)]\\
&\qquad-2\cdot\mb{E}_{\theta\sim \mu,\theta'\sim \nu}[\kappa_{h/\sqrt{2}}(\theta,y)\kappa_{h/\sqrt{2}}(\theta',y)]\,\dd y\\
&=\int_{\mb R^{d}}  \Big(\mb{E}_{\theta\sim \mu}[\kappa_{h/\sqrt{2}}(\theta,y)]-\mb{E}_{\theta\sim \nu}[\kappa_{h/\sqrt{2}}(\theta,y)]\Big)^2\,\dd y.
\end{align*}
Suppose both the support of $\mu$ and the support of $\nu$ are contained in $[0,1]^d$. Then there exist constants $C_1,C_2$ so that
\begin{equation}\label{eqn:MMDkernel}
\begin{aligned}
&\mmd_h^2(\mu, \nu)=\int_{\mb R^{d}}  \Big(\mb{E}_{\theta\sim \mu}[\kappa_{h/\sqrt{2}}(\theta,y)]-\mb{E}_{\theta\sim \nu}[\kappa_{h/\sqrt{2}}(\theta,y)]\Big)^2\,\dd y\\
&\leq \int_{B(0,C_1+C_2h\sqrt{ \log(1+\frac{1}{\varepsilon})})}  \Big(\mb{E}_{\theta\sim \mu}[\kappa_{h/\sqrt{2}}(\theta,y)]-\mb{E}_{\theta\sim \nu}[\kappa_{h/\sqrt{2}}(\theta,y)]\Big)^2\,\dd y\\
&+ 2\cdot\mb{E}_{\theta,\theta'\sim \mu}\Big[ \int_{\mb R^d\setminus B(0,C_1+C_2h\sqrt{ \log(1+\frac{1}{\varepsilon})})}  \kappa_{h/\sqrt{2}}(\theta,y)\kappa_{h/\sqrt{2}}(\theta',y)\,\dd y\Big]\\
&+ 2\cdot\mb{E}_{\theta,\theta'\sim \nu}\Big[ \int_{\mb R^d\setminus B(0,C_1+C_2h\sqrt{ \log(1+\frac{1}{\varepsilon})})}  \kappa_{h/\sqrt{2}}(\theta,y)\kappa_{h/\sqrt{2}}(\theta',y)\,\dd y\Big]\\
&\leq \int_{B(0,C_1+C_2h\sqrt{ \log(1+\frac{1}{\varepsilon})})}  \Big(\mb{E}_{\theta\sim \mu}[\kappa_{h/\sqrt{2}}(\theta,y)]-\mb{E}_{\theta\sim \nu}[\kappa_{h/\sqrt{2}}(\theta,y)]\Big)^2\,\dd y+\frac{\varepsilon}{64C}\\
&\leq V_d\Big(C_1+C_2h\sqrt{ \log(1+\frac{1}{\varepsilon})}\Big)^d
\\ 
&\quad \cdot \underset{y\in B(0,C_1+C_2h\sqrt{ \log(1+\frac{1}{\varepsilon})})}{\sup} \Big(\mb{E}_{\theta\sim \mu}[\kappa_{h/\sqrt{2}}(\theta,y)]-\mb{E}_{\theta\sim \nu}[\kappa_{h/\sqrt{2}}(\theta,y)]\Big)^2 +\frac{\varepsilon}{64C},
\end{aligned}
\end{equation}
where $V_d$ is the volume of the unit ball in $\mb R^d$, and we have used the fact that 
\begin{equation*}
    \begin{aligned}
        &\int_{\mb R^d\setminus B(0,C_1+C_2h\sqrt{ \log(1+\frac{1}{\varepsilon})})}  \kappa_{h/\sqrt{2}}(\theta,y)\kappa_{h/\sqrt{2}}(\theta',y)\,\dd y\\
        &=\frac{1}{(2\pi)^{\frac{d}{2}}h^d} \exp(-\frac{\|\theta-\theta'\|^2}{2h^2})
        \\
        & \quad \cdot \int_{\mb R^d\setminus B(0,C_1+C_2h\sqrt{ \log(1+\frac{1}{\varepsilon})})}  \frac{1}{(2\pi)^{\frac{d}{2}}(\frac{h}{2})^d} \exp\Big(-\frac{\|y-\frac{1}{2}(\theta+\theta')\|^2}{2(h/2)^2}\Big)\,\dd y,
    \end{aligned}
\end{equation*}
and $h\sqrt{\log \frac{1}{h}}$ is upper bounded for $h<1$. So $\int_{\mb R^d\setminus B(0,C_1+C_2h\sqrt{ \log(1+\frac{1}{\varepsilon})})}  \kappa_{h/\sqrt{2}}(\theta,y)\kappa_{h/\sqrt{2}}(\theta',y)\,\dd y$ can be controlled via  the sub-exponential tail bound (see for example,~\cite{wainwright2019high}).
We then show that, by setting $\varepsilon_{\theta}=C_{\theta}\Big(1+h\sqrt{\log(1+\frac{1}{\varepsilon})}\Big)^{-\frac{d}{2}-1}\sqrt{\varepsilon}h^d \min(n^{-1},h^{2})$ for some small enough absolute constant $C_{\theta}$,  the squared MMD between the discrete approximation $\nu(\cdot|T(\Xn))$ 
and  the true posterior $\pos$  is within  $\frac{\varepsilon}{32C}$.   For any $y \in B(0,C_1+C_2h\sqrt{ \log(1+\frac{1}{\varepsilon})})$, we first decompose the error using the triangle inequality:
\begin{equation}\label{eqn:trierror}
    \begin{aligned}
          & \quad
  \left|
  \int_{\Theta} \kappa_{h/\sqrt{2}} (\theta,y)\post \dd\theta 
  -\int_{\Theta} \kappa_{h/\sqrt{2}} (\theta,y)\nu(\dd\theta|T(\Xn))
  \right|
  \\
  &\leq
  \left|
    \int_{\Theta} \sum_{j=1}^J \rho_j(\theta)\kappa_{h/\sqrt{2}} (\theta,y)\post \dd\theta
  -
  \int_{\Theta} \sum_{j=1}^J \rho_j(\theta)\kappa_{h/\sqrt{2}} (\theta_j^*,y)\post \dd\theta
  \right|
  \\
  &\quad+
  \left|
    \int_{\Theta} \sum_{j=1}^J \rho_j(\theta)\kappa_{h/\sqrt{2}} (\theta_j^*,y)\post \dd\theta
  -
  \int_{\Theta} \kappa_{h/\sqrt{2}} (\theta,y)\nu(\dd\theta|T(\Xn))  
  \right|.
  \end{aligned}
\end{equation}
 For the first term in~\eqref{eqn:trierror}, which captures the error from approximating the kernel function, we have
\begin{align*}
  &\quad
  \left|
    \int_{\Theta} \sum_{j=1}^J \rho_j(\theta)\kappa_{h/\sqrt{2}} (\theta,y)\post \dd\theta
  -
  \int_{\Theta} \sum_{j=1}^J \rho_j(\theta)\kappa_{h/\sqrt{2}} (\theta^*_j,y)\post \dd\theta
  \right|
  \\
  &\leq 
  \int_{\Theta}
  \sum_{j=1}^J 
  \rho_j(\theta)
  \post 
  \left|
    \kappa_{h/\sqrt{2}} (\theta,y) - \kappa_{h/\sqrt{2}} (\theta^*_j,y)
  \right|
  \dd\theta\\
  &\leq  \frac{1}{(2\pi)^{d/2} (h/\sqrt{2})^d} \int_{\Theta}
  \sum_{j=1}^J 
  \rho_j(\theta)
  \post 
    \exp\left(-\frac{\Vert \theta^*_j - y\Vert^2}{h^2}\right)
   \\ 
  & \quad \cdot \left| 
    \exp\left(\frac{\Vert \theta^*_j - y\Vert^2 - \Vert \theta - y\Vert^2}{h^2}\right)
    - 1
  \right|
  \dd\theta
\end{align*}
Since $\rho_j$ is supported on $B(\theta^*_j,\sqrt{2}\varepsilon_\theta)$, the difference in the exponent can be bounded as:
\begin{align*}
  &\left|
    \frac{1}{h^2}\Vert \theta^*_j - y\Vert^2-\frac{1}{h^2}\Vert \theta - y\Vert^2
  \right|
  \leq
  \frac{1}{h^2}
  \left|\Vert \theta_j^* - y\Vert - \Vert \theta - y\Vert \right|
  (\Vert \theta_j^* - y\Vert + \Vert \theta - y\Vert )
  \\
  & \leq
  \frac{2 (\sqrt{d}+C_1+C_2h\sqrt{ \log(1+\frac{1}{\varepsilon})})}{h^2}
  \Vert \theta^*_j - \theta\Vert \\
  & \leq
  \frac{C_3(1+h\sqrt{ \log(1+\frac{1}{\varepsilon})}) \varepsilon_\theta}{h^2}.
\end{align*}
Then by taking $C_{\theta}$ to be small enough,   we have
$\varepsilon_{\theta}=C_{\theta}\Big(1+h\sqrt{\log(1+\frac{1}{\varepsilon})}\Big)^{-\frac{d}{2}-1}\sqrt{\varepsilon}h^d \min(n^{-1},h^{2})\leq C_{\theta}\big(1+h\sqrt{\log(1+\frac{1}{\varepsilon})}\big)^{-1}\frac{\sqrt{2}h^{\frac{d}{2}}}{(2\pi)^{\frac{d}{4}}} h^2\leq \frac{h^2}{C_3(1+h\sqrt{ \log(1+\frac{1}{\varepsilon})})}$ (recall $\varepsilon\leq \frac{2}{(2\pi h^2)^{\frac{d}{2}}}$ and $h<1$). This ensures $\big|
    \frac{1}{h^2}\Vert \theta^*_j - y\Vert^2-\frac{1}{h^2}\Vert \theta - y\Vert^2
  \big|$ is bounded by 1. Using the inequality $|e^x - 1| \leq e|x|$ for $|x|\leq 1$, we get
\begin{align*}
  \left|
    \exp\left(\frac{\Vert \theta_j^* - y\Vert^2 - \Vert \theta - y\Vert^2}{h^2}\right)
    -
    1
  \right|
  \leq
  \frac{C_4(1+h\sqrt{ \log(1+\frac{1}{\varepsilon})}) \varepsilon_\theta}{h^2}.
\end{align*}
Hence the first term in~\eqref{eqn:trierror} can be upper bounded by
\begin{align*}
  & \Big|
    \int_{\Theta} \sum_{j=1}^J \rho_j(\theta)\kappa_{h/\sqrt{2}} (\theta,y)\post \dd\theta
  -
  \int_{\Theta}\sum_{j=1}^J \rho_j(\theta)\kappa_{h/\sqrt{2}} (\theta_j^*,y)\post \dd\theta
  \Big| \\
  & \leq
  C_5\frac{ (1+h\sqrt{ \log(1+\frac{1}{\varepsilon})})\varepsilon_\theta}{h^{d+2}}.
\end{align*}
For the second term in~\eqref{eqn:trierror},  we have 
\begin{align*}
&  \left|
    \int_{\Theta} \sum_{j=1}^J \rho_j(\theta)\kappa_{h/\sqrt{2}} (\theta_j^*,y)\post \dd\theta
  -
  \int_{\Theta} \kappa_{h/\sqrt{2}} (\theta,y)\nu(\dd\theta|T(\Xn))  
  \right|\\
  & =
  \left|
    \int_{\Theta} \sum_{j=1}^J \rho_j(\theta)\kappa_{h/\sqrt{2}} (\theta^*_j,y)\post \dd\theta
    -
    \sum_{j=1}^J \kappa_{h/\sqrt{2}} (\theta^*_j,y)\nu(\theta_j^*|T(\Xn)) 
  \right|
  \\
  & =
  \left|
    \sum_{j=1}^J \kappa_{h/\sqrt{2}} (\theta_j^*,y)
    \left(
      \int_{\Theta} \rho_j(\theta) \post \dd\theta - \nu(\theta^*_j|T(\Xn))
    \right)
  \right|
  \\
  & \leq
  \frac{1}{(2\pi)^{d/2}(h/\sqrt{2})^d}
  \sum_{j=1}^J
  \left|
    \int_{\Theta} \rho_j(\theta) \frac{f(\theta,\Xn)}{\int_{\Theta} f(\theta',\Xn)\dd\theta'} \dd\theta
    -
    \frac{\int_{\Theta} \rho_j(\theta)\dd\theta \cdot f(\theta^*_j,\Xn)}{\sum_{k=1}^J \int_{\Theta} \rho_k(\theta)\dd\theta \cdot f(\theta^*_k,\Xn)}
  \right|.
\end{align*}
Note that there exists a constant $C_6$ so that for any $\theta,\theta'\in \Theta$,
 \begin{equation*}
 \begin{aligned}
          |\log f(\theta,\Xn)-\log f(\theta',\Xn)|&\leq |\log \pi(\theta)-\log \pi(\theta')| \\
          & \quad +\big\|\sum_{i=1}^n t(X_i)\big\|\cdot\|\eta(\theta)-\eta(\theta')\|+n\cdot|A(\theta)- A(\theta')|\\
          &\leq C_6 n\cdot\|\theta-\theta'\|.
 \end{aligned}
 \end{equation*}
Furthermore, denoting $Z = \int_{\Theta} f(\theta,\Xn)\dd\theta$ and $Z_A = \sum_{k=1}^J \int_{\Theta} \rho_k(\theta)\dd\theta \cdot f(\theta^*_k,\Xn)$, we can bound the term inside the sum by
\begin{align*}
&  \left|
    \int_{\Theta} \rho_j(\theta) \frac{f(\theta,\Xn)}{\int_{\Theta} f(\theta',\Xn)\dd\theta'} \dd\theta
    -
    \frac{\int_{\Theta} \rho_j(\theta)\dd\theta \cdot f(\theta^*_j,\Xn)}{\sum_{k=1}^J \int_{\Theta} \rho_k(\theta)\dd\theta \cdot f(\theta^*_k,\Xn)}
  \right|\\
  & =
  \left|
    \int_{\Theta} \frac{\rho_j(\theta) f (\theta,\Xn)}{Z} \dd\theta
    -
    \frac{\int_{\Theta} \rho_j(\theta)\dd\theta \cdot f (\theta_j^*,\Xn)}{Z_A}
  \right|
  \\
  & =
  \left|
    \frac{1}{Z Z_A}
    \left(
      Z_A \int_{\Theta} \rho_j(\theta) f(\theta,\Xn) \dd\theta - Z \int _{\Theta}\rho_j(\theta)\dd\theta\cdot f (\theta_j^*,\Xn)
    \right)
  \right|
  \\
  & \leq
  \frac{1}{Z Z_A}
  \left(
    Z_A \int_{\Theta} \rho_j(\theta)\cdot |f(\theta,\Xn) - f (\theta^*_ j,\Xn)| \dd\theta
    +
    |Z_A - Z| \int \rho_j(\theta)\dd\theta \cdot f (\theta_j^*,\Xn)
  \right)\\
  &\leq \frac{1}{Z}\int_{\Theta}\rho_j(\theta)f(\theta,\Xn)\, \Big| \exp\big(\log (f(\theta_j^*,\Xn))-\log (f(\theta,\Xn))\big)-1\Big|\dd\theta \\
  &\qquad+\frac{\int_{\Theta} \rho_j(\theta)\dd\theta \cdot f (\theta_j^*,\Xn)}{ZZ_A} 
  \\
  & \qquad \qquad  \cdot \sum_{k=1}^J \int_{\Theta}\rho_k(\theta) f(\theta,\Xn)
  \Big| \exp\big(\log (f(\theta_k^*,\Xn))-\log (f(\theta,\Xn))\big)-1\Big|\dd\theta\\
  &\lesssim   n\cdot\varepsilon_{\theta}\cdot\bigg(\frac{\int_{\Theta}\rho_j(\theta)f(\theta,\Xn)\,\dd\theta}{Z}+\frac{\int_{\Theta} \rho_j(\theta)f (\theta_j^*,\Xn)\dd\theta }{Z_A} \bigg).
  \end{align*}
So we finally bound the second term in~\eqref{eqn:trierror} by 
\begin{equation*}
    \begin{aligned}
        &  \left|
    \int_{\Theta} \sum_{j=1}^J \rho_j(\theta)\kappa_{h/\sqrt{2}} (\theta_j^*,y)\post \dd\theta
  -
  \int_{\Theta} \kappa_{h/\sqrt{2}} (\theta,y)\nu(\dd\theta|T(\Xn))  
  \right|\\
  & \leq
  \frac{1}{(2\pi)^{d/2}(h/\sqrt{2})^d}
  \sum_{j=1}^J
  \left|
    \int_{\Theta} \rho_j(\theta) \frac{f(\theta,\Xn)}{\int_{\Theta} f(\theta',\Xn)\dd\theta'} \dd\theta
    -
    \frac{\int_{\Theta} \rho_j(\theta)\dd\theta \cdot f(\theta^*_j,\Xn)}{\sum_{k=1}^J \int_{\Theta} \rho_k(\theta)\dd\theta \cdot f(\theta^*_k,\Xn)}
  \right|\\
  &\lesssim \frac{n\varepsilon_{\theta}}{h^d}\sum_{j=1}^J \bigg(\frac{\int_{\Theta}\rho_j(\theta)f(\theta,\Xn)\,\dd\theta}{Z}+\frac{\Big(\int_{\Theta} \rho_j(\theta)f (\theta_j^*,\Xn)\dd\theta \Big)}{Z_A} \bigg)\\
  &\lesssim \frac{n\varepsilon_{\theta}}{h^d}.
    \end{aligned}
\end{equation*}
Combining the two error terms, we have:
\begin{align*}
  &\left|
  \int_{\Theta} \kappa_{h/\sqrt{2}} (\theta,y)\post \dd\theta 
  -\int_{\Theta} \kappa_{h/\sqrt{2}} (\theta,y)\nu(\dd\theta|T(\Xn))  
  \right| \\
 &  \lesssim  \frac{ \big(1+h\sqrt{ \log(1+\frac{1}{\varepsilon})}\big)\varepsilon_\theta}{h^{d+2}} +  \frac{n\varepsilon_\theta}{h^d}.
\end{align*}
By choosing $\varepsilon_{\theta}=C_{\theta}\Big(1+h\sqrt{\log(1+\frac{1}{\varepsilon})}\Big)^{-\frac{d}{2}-1}\sqrt{\varepsilon}h^d \min(n^{-1},h^{2})$ for a sufficiently small constant $C_\theta$, we can ensure that 
\begin{align*}
&\mmd_h^2(\pos, \nu(\cdot|T(\Xn)))\\
&\leq \frac{\varepsilon}{64C} + V_d\Big(C_1+C_2h\sqrt{ \log(1+\frac{1}{\varepsilon})}\Big)^d\\
&\quad\cdot\underset{y\in B(0,C_1+C_2h\sqrt{ \log(1+\frac{1}{\varepsilon})})}{\sup} \Big(\mb{E}_{\theta\sim\mu^*_{\theta|\Xn}(\cdot|\Xn)}[\kappa_{h/\sqrt{2}}(\theta,y)]-\mb{E}_{\theta\sim \nu(\cdot|T(\Xn))}[\kappa_{h/\sqrt{2}}(\theta,y)]\Big)^2 \\
&\leq \frac{\varepsilon}{32C}.
\end{align*}

Next, we demonstrate the existence of a summary network $S^*\in \ms S$ and an auxiliary network $W^*\in \ms W$,  such that the induced distribution $\smuws$  can approximate the distribution $\nu(\cdot|T(\Xn))$ with the desired accuracy.  Recall $ \wt f(\theta,T_X)=\pi(\theta)\exp(\eta(\theta)^T T_X-nA(\theta))$. 
We define the logit for each grid point $\theta_j^*$
\begin{align*}
v_j(T(\Xn)) = \log\left(\int_{\Theta} \rho_j(\theta)\dd\theta\right) + \log \wt f(\theta^*_j, T(\Xn)).
\end{align*}
and 
\begin{equation*}
  v'_j(T(\Xn))=v_j(T(\Xn))-\max_{j\in [J]} v_j(T(\Xn))
\end{equation*}
Note that the construction of $W\in\ms W$ involves a clip function, so we  truncate the  logit to consider $\forall j\in [J]$:
\begin{equation}\label{def:w}
    \wt v_j(T(\Xn)) = {\rm clip}\Big( v'_j(T(\Xn));-c_1\log(\frac{n}{\min(\varepsilon,1) h}),c_2\Big)=\max\big(v'_j(T(\Xn)),-c_1\log(\frac{n}{\min(\varepsilon,1) h})\big),
\end{equation}
 where the last equation uses $v'_j(T(\Xn))\leq 0$. Let $\widetilde\nu(\cdot|T(\Xn)) $ be the distribution from the truncated logits $\wt v_j$:
\begin{align*}
\wt\nu(\cdot|T(\Xn)) &= \sum_{j=1}^J\frac{\exp(\wt v_j(T(\Xn)))}{\sum_{k=1}^J \exp(\wt v_k(T(\Xn)))}\delta_{\theta_j^*}(\cdot).
\end{align*}
Moreover, note that $\nu(\cdot|T(\Xn)) $  can be written as 
\begin{align*}
\nu(\cdot|T(\Xn)) = \sum_{j=1}^J\frac{\exp(v_j(T(\Xn)))}{\sum_{k=1}^J \exp(v_k(T(\Xn)))}\delta_{\theta_j^*}(\cdot)=\sum_{j=1}^J\frac{\exp( v'_j(T(\Xn)))}{\sum_{k=1}^J \exp( v'_k(T(\Xn)))}\delta_{\theta_j^*}(\cdot).
\end{align*}
The rationale for comparing $\nu(\cdot|T(\Xn)) $ and $\widetilde\nu(\cdot|T(\Xn)) $  is that when a logit $v'_j$ is very small (i.e., very negative), its contribution to the softmax sum is negligible. By choosing an appropriate $c_1$, we can ensure that    $\wt\nu(\cdot|T(\Xn))$ is close to $\nu(\cdot|T(\Xn)) $. Specifically, we have 
\begin{align*}
& \left| \mathbb{E}_{\theta \sim \nu(\cdot|T(\Xn))}[\kappa_{h/\sqrt{2}}(y, \theta)] - \mathbb{E}_{\theta \sim \widetilde\nu(\cdot|T(\Xn))}[\kappa_{h/\sqrt{2}}(y, \theta)] \right| \\
& \leq \frac{1}{(2\pi)^{d/2}(h/\sqrt{2})^d} \sum_{j=1}^J |\nu(\theta_j^*|T(\Xn)) -\widetilde\nu(\theta^*_j|T(\Xn))|.
\end{align*}
Write  $v'_j=v_j'(T(\Xn))$, $\wt v_j=\wt v_j(T(\Xn))$,
$Z_{v'} = \sum_{j=1}^J \exp(v'_j)$ and $Z_{\wt v} = \sum_{j=1}^J \exp(\wt v_j)$. Then we can write
\begin{equation}\label{eqn:TVbound}
\begin{aligned}
   & \sum_{j=1}^J |\nu(\theta_j^*|T(\Xn)) -\widetilde\nu(\theta^*_j|T(\Xn))|\\
&=\sum_{j=1}^J \left| \frac{\exp(v'_j)}{Z_{v'}} - \frac{\exp(\wt v_j)}{Z_{\wt v}} \right| \leq \sum_{j=1}^J \left| \frac{\exp(v'_j)}{Z_{v'}} - \frac{\exp(v'_j)}{Z_{\wt v}} \right| + \sum_{j=1}^J \left| \frac{\exp(v'_j)}{Z_{\wt v}} - \frac{\exp(\wt v_j)}{Z_{\wt v}} \right| \\
&= \frac{|Z_{\wt v} - Z_{v'}|}{Z_{v'} Z_{\wt v}} \sum_{j=1}^J \exp(v'_j) + \frac{1}{Z_{\wt v}} \sum_{j=1}^J |\exp(v'_j) - \exp(\wt v_j)| \\
&= \frac{|Z_{\wt v} - Z_{v'}|}{Z_{\wt v}} + \frac{1}{Z_{\wt v}} \sum_{j=1}^J |\exp(v'_j) - \exp(\wt v_j)|.
\end{aligned} 
\end{equation}
Since $|Z_{\wt v} - Z_{v'}| = |\sum_{k=1}^J (\exp(\wt v_k) - \exp(v'_k))| \leq \sum_{k=1}^J |\exp(\wt v_k) - \exp(v'_k)|$, we have:
\begin{align*}
 \sum_{j=1}^J |\nu(\theta_j^*|T(\Xn)) -\widetilde\nu(\theta^*_j|T(\Xn))| \leq \frac{2}{Z_{\wt v}} \sum_{j=1}^J |\exp(v'_j) - \exp(\wt v_j)|.
\end{align*}
The difference $|\exp(v'_j) - \exp(\wt v_j)|$ is non-zero only if truncation occurs, i.e., $v'_j < -C_{\text{trunc}}=-c_1\log(\frac{n}{\min(\varepsilon,1) h})$. In this case, $\wt v_j = -C_{\text{trunc}}$ and $\exp(v'_j) < \exp(-C_{\text{trunc}})$.
\begin{align*}
\sum_{j=1}^J |\exp(v_j') - \exp(\wt v_j)| &= \sum_{j: v'_j < -C_{\text{trunc}}} (\exp(-C_{\text{trunc}}) - \exp(v'_j)) \\
&\leq \sum_{j: v'_j < -C_{\text{trunc}}} \exp(-C_{\text{trunc}}) 
\end{align*}
Plugging this back, and using that $Z_{\wt v}\geq \exp(0)=1$, the total variation distance  is bounded by:
\begin{align*}
 \sum_{j=1}^J |\nu(\theta_j^*|T(\Xn)) -\widetilde\nu(\theta^*_j|T(\Xn))|  \leq \frac{2}{Z_{\wt v}} \sum_{j: v'_j < -C_{\text{trunc}}} \exp(-C_{\text{trunc}}) \leq  2 J (\frac{\min(\varepsilon,1) h}{n})^{c_1},
\end{align*}
and, using the RKHS norm and $\kappa_h(\theta,\theta)=(2\pi h^2)^{-d/2}$, the MMD is bounded by
\begin{equation*}
\mmd^2_h(\nu(\cdot|T(\Xn)),\wt\nu(\cdot|T(\Xn)))
\leq \frac{1}{(2\pi h^2)^{d/2}}
\Big(2J\big(\tfrac{\min(\varepsilon,1)h}{n}\big)^{c_1}\Big)^2.
\end{equation*}
Therefore, when $c_1$ is large enough, we have 
\begin{equation*}
\begin{aligned}
         &\mmd^2_h(\pos,\wt\nu(\cdot|T(\Xn))) \\
         &\leq 2\mmd^2_h(\pos,\nu(\cdot|T(\Xn)))+2\mmd^2_h(\nu(\cdot|T(\Xn)),\wt\nu(\cdot|T(\Xn))) \leq \frac{\varepsilon}{8C}.
\end{aligned}
\end{equation*}
This justifies that our neural network only needs to approximate the truncated logits $\wt v_j$. The networks $S$ and $W$ will be constructed to approximate $T(\Xn)$ and the mapping from the summary to these logits respectively. The result is stated in the following lemma.

\begin{lemma}\label{lemma: approximate summary statistics}
 There exist $L_1,L_2,U_1,U_2,R_1,R_2$ with $L_1=\m O(\log^2 (\frac{n}{\min(\varepsilon,1) h}))$, $L_2=\m O(\log (\frac{n}{\min(\varepsilon,1) h}))$ and,  
 \begin{equation*}
    \begin{aligned}
        &R_1=\m O\left(n^{d_X+1} h^{-dd_X}\varepsilon^{-d_X/2}\cdot\log \big(\frac{n}{\min(\varepsilon,1) h}\big)\cdot\Big(1+h\sqrt{\log(1+\frac{1}{\varepsilon})}\Big)^{\frac{d_Xd}{2}}\right),\quad \ms E(U_1)=\m O(R_1^3),\\
        &R_2=\m O\left(\Big(1+h\sqrt{\log(1+\frac{1}{\varepsilon})}\Big)^{\frac{d^2}{2}+d}\varepsilon^{-\frac{d}{2}} h^{-d^2} \max(n^{d},h^{-2d})\log \big(\frac{n}{\min(\varepsilon,1) h}\big)\right),\quad \ms E(U_2)=\m O(R_2^3),
    \end{aligned}
\end{equation*}
  such that for any $\Pi_{\theta}\m P_{\theta}^{(n)}$ satisfying Assumption~\ref{assump:exponential} and corresponding $\wt\nu(\cdot|T(\Xn))$, there exist a summary map  $S^*\in \ms S=\Phi_{nd_X}^p(L_1,U_1,R_1)$ and an auxiliary function 
 \begin{equation*}
   W^*\in \ms W=\Big\{W(\cdot)={\rm clip}\Big(W'(\cdot);-c_1\log(\frac{n}{\min(\varepsilon,1) h}),c_2\Big)\,:\,W'(\cdot)\,\in  \Phi_{p+d}^{1}(L_2, U_2,R_2)\Big\},
\end{equation*}
 so that  for any $\Xn\in \m X^{n}$,
 \begin{equation*}
    \mmd^2_h(\wt\nu(\cdot|T(\Xn)),\mu_{W^*}(\cdot|S^*(\Xn)))\leq\frac{\varepsilon}{8C}.
\end{equation*}
\end{lemma}
The squared triangle inequality gives
\[
\begin{aligned}
&\mmd_h^2(\pos,\mu_{W^*}(\cdot|S^*(\Xn)))\\
&\leq 2\mmd_h^2(\pos,\wt\nu(\cdot|T(\Xn)))\\
&\quad+2\mmd_h^2(\wt\nu(\cdot|T(\Xn)),\mu_{W^*}(\cdot|S^*(\Xn)))
\leq \frac{\varepsilon}{2C}.
\end{aligned}
\]
\noindent Then, plugging $L\lesssim L_1+L_2$ and $R\lesssim L+R_1+R_2$ into~\eqref{eqn:N} and writing $\ov d=\max(d,d_X)$, we find that there exists a constant $C$ so that when 
\begin{equation*}
    N\geq C\,  \log^5 (\frac{n}{\min(\varepsilon,1) h})  \Big(1+h\sqrt{\log(1+\frac{1}{\varepsilon})}\Big)^{\frac{d\ov d}{2}+d}\varepsilon^{-\frac{\ov d}{2}-1}h^{-d(\ov d+1)} \max (n^d,n^{d_X+1},h^{-2d}),
\end{equation*}
  it holds with probability at least $1-\frac{1}{N}$ that, 
\begin{align*}
   & \E_{X^{(n)}\sim \m P^{(n)}} [\mmd^2_h(\mu^*_{\theta|\Xn}(\cdot|\Xn), \mu^*_{\theta|\wh S(\Xn)}(\cdot|\wh S(\Xn)))]\\
   &\leq
  \E_{X^{(n)}\sim \m P^{(n)}} [\mmd^2_h(\mu^*_{\theta|\Xn}(\cdot|\Xn), \mu_{\wh{W}}(\cdot|\wh S(X^{(n)})))] \leq \varepsilon.
\end{align*} 
\subsection{Proof of Corollary \ref{corollary: large sample setting}}\label{sec: proof of large sample setting results}

We first outline the construction for $\ms S$ and $\ms W$. For a vector $x\in \mb R^{d^2}$, we denote by ${\rm mat}_d(x)$ the $d\times d$ matrix formed by stacking the entries of $x$ column-wise. Then we choose
\begin{equation*}
    \ms S=\big\{S(\cdot)={\rm clip}(S'(\cdot);-c_1(\log n)^{\alpha_2},c_1(\log n)^{\alpha_2})
    \,:\, S'(\cdot)\in \Phi_{nd_X}^d(L_1,U_1,R_1)\big\},
\end{equation*} 
\begin{equation*}
\begin{aligned}
     \ms W=\Big\{W(\cdot)={\rm mat}_d\big({\rm clip}(W'(\cdot);-c_2(\log n)^{\alpha_2}, c_2(\log n)^{\alpha_2})\big)
    \,:\, W'(\cdot)\in \Phi_{d}^{d^2}(L_2,U_2,R_2)\Big\},
\end{aligned}
\end{equation*}
where $c_1$, $c_2$ are sufficiently large constants, and the neural network size parameters will be chosen later. For a given $s\in \mb R^d$ and $W\in \ms W$, we define 
\begin{equation*}
    \mu_{W}(\cdot|s)= N(s,n^{-1}W(s)W(s)^T).
\end{equation*}
We now apply Theorem~\ref{thm:Generalization Error}  to obtain the desired result. First, for  fixed neural network size parameters, we identify the quantities $T_1$, $T_2$ required by the assumption in Theorem~\ref{thm:Generalization Error}. Next, we control the approximation error to determine the appropriate network size. Finally, we combine these two parts to conclude the proof.  Since $ \mu_{W}(\cdot|s)= N(s,n^{-1}W(s)W(s)^T)$, the loss function has a closed-form expression as
\begin{equation*}
    \begin{aligned}
       & \ell(S,W,\theta,\Xn)=\mathbb{E}_{\theta'
,\theta^{''} \sim N(s,n^{-1}W(s)W(s)^T)|_{s=S(\Xn)} }[\kappa_{h}(\theta',\theta^{''})] \\
       & -2\cdot\mathbb{E}_{\theta' \sim  N(s,n^{-1}W(s)W(s)^T)|_{s=S(\Xn)} }[\kappa_{h}(\theta,\theta')]\\
        &=(\frac{n}{2\pi})^{d/2} \Big(  \big| \wt h^2I_d+2W(S(\Xn))W(S(\Xn))^T\big|\Big)^{-1/2}\\
        &\qquad-2(\frac{n}{2\pi})^{d/2} \Big(  \big| \wt h^2I_d+W(S(\Xn))W(S(\Xn))^T\big|\Big)^{-1/2}\\
        &\qquad\qquad\qquad\cdot\exp\Big(-\frac{n}{2}(\theta-S(\Xn))^T \big( \wt h^2I_d+W(S(\Xn))W(S(\Xn))^T\big)^{-1}(\theta-S(\Xn))\Big).
    \end{aligned}
\end{equation*}
Therefore, for $S,S'\in \ms S$, $W,W'\in \ms W$, we have 
\begin{equation*}
\begin{aligned}
    &\frac{1}{N}\sum_{i=1}^N \Big(\ell(S,W,\theta_i,\Xn_i)-\ell(S',W',\theta_i,\Xn_i)\Big)^2\\
    &\lesssim \frac{n^{d}}{N}\sum_{i=1}^N \bigg( \Big(  \big| \wt h^2I_d+2W(S(\Xn_i))W(S(\Xn_i))^T\big|\Big)^{-1/2}\\
    &\qquad\qquad-\Big(   \big|\wt h^2I_d+2W'(S'(\Xn_i))W'(S'(\Xn_i))^T\big|\Big)^{-1/2}\bigg)^2\\
    &+ \frac{n^{d}}{N}\sum_{i=1}^N \bigg( \Big(  \big| \wt h^2I_d+ W(S(\Xn_i))W(S(\Xn_i))^T\big|\Big)^{-1/2}\\
    &\qquad\qquad-\Big(   \big|\wt h^2I_d+W'(S'(\Xn_i))W'(S'(\Xn_i))^T\big|\Big)^{-1/2}\bigg)^2\\
    &+\frac{n^{d}}{N}\sum_{i=1}^N  \Big(   \big|  \wt h^2I_d+W'(S'(\Xn_i))W'(S'(\Xn_i))^T\big|\Big)^{-1}\\
    &\quad\cdot\bigg( \exp\Big(-\frac{n}{2}(\theta_i-S(\Xn_i))^T \big( \wt h^2I_d+W(S(\Xn_i))W(S(\Xn_i))^T\big)^{-1}(\theta_i-S(\Xn_i))\Big)\\
    &\qquad\qquad\qquad-\exp\Big(-\frac{n}{2}(\theta_i-S'(\Xn_i))^T \big( \wt h^2I_d+W'(S'(\Xn_i))W'(S'(\Xn_i))^T\big)^{-1}(\theta_i-S'(\Xn_i))\Big)\bigg)^2\\
    &\lesssim n^{d+1}   \wt h^{-6d}(\log n)^{2/\alpha_2}\frac{1}{N}\sum_{i=1}^N \big(\|W(S(\Xn_i))-W'(S'(\Xn_i))\|_{\rm F}^2+\|S(\Xn_i)-S'(\Xn_i)\|^2\big).
\end{aligned}
\end{equation*}
So, as in the proof of Corollary~\ref{thm:expfam:formal}, including the union over nonzero-parameter positions, by the pseudo-dimension bound for ReLU networks
(Theorem 7 of~\cite{bartlett2019nearly})  and the covering-number bound in terms of pseudo-dimension  (Theorem 12.2 of~\cite{anthony2009neural}),   and using $  \wt h\gtrsim n^{-c_7}$, there exists a constant $C $ so that
\begin{equation*}
    \begin{aligned}
        &\log \m N(\ms S\times \ms W,d_N,u)\leq   C \, (RL\log R) \cdot\log\big(\frac{Nn}{u}\big),\\
        &\text{ where } R=R_1+R_2\text{ and } L=L_1+L_2.
        \end{aligned}
\end{equation*}
So using Theorem~\ref{thm:Generalization Error}, when 
\begin{equation}\label{eqn:N2}
    N\geq \frac{n\cdot (RL\log R)\cdot(\log (n)+\log (RL))}{\wt h^d},
\end{equation}
it holds with probability at least $1-\frac{1}{N}$ that 
\begin{equation*}
\begin{aligned}
    &
    n^{-\frac{d}{2}} \mb{E}_{\Xn\sim \m P^{(n)}}\big[\mmd^2_h(p(\cdot|\Xn),\mu_{\wh W}(\cdot|\wh S(\Xn)))\big]\\
    &\lesssim \min_{S\in \ms S,W\in \ms W}   n^{-\frac{d}{2}} \mb{E}_{\Xn\sim \m P^{(n)}}\big[\mmd^2_h(p(\cdot|\Xn),\mu_{W}(\cdot|S(\Xn)))\big]+\frac{1}{n}.
\end{aligned}
\end{equation*}
Then we will control the approximation error  and obtain the required $R$ and $L$. The proof leverages the Bernstein-von Mises theorem, which, 
under suitable regularity conditions, states that the posterior distribution  $\mu^*_{\theta|\Xn}(\cdot|\Xn)$ is well-approximated by a Gaussian distribution. Specifically, 
the distribution of the scaled parameter $\sqrt{n}(\theta -\wh{\theta}(\Xn))$, where $\wh{\theta}(\Xn)$ is the maximum likelihood estimator (MLE), converges in distribution to a Gaussian $N(0, I(\wh{\theta})^{-1})$, with the covariance 
being the inverse of the Fisher information matrix evaluated at the MLE.
Based on this, our strategy is to construct the summary network $S^*(\Xn)$ to approximate 
the MLE $\wh{\theta}(\Xn)$, and the network $W^*$ to approximate the components 
of the square root $I(\wh{\theta})^{-1/2}$ of the covariance matrix, thereby identifying the target  Gaussian distribution. The construction of these networks follows a similar approach  to the exponential family case, utilizing grid-based approximations and kernel local  averaging. We first state the following lemma about the positive definiteness of the Fisher information matrix.
\begin{lemma}\label{lemmaFisher}
Under Assumption~\ref{assump:large_sample}, there exists a positive constant $c_I$ so that for any $\theta^* \in \mb R^d$,   $I(\theta^*)\succeq c_II_d$.
 \end{lemma}
\noindent  Consider a fixed model $\Pi_{\theta}\m P_{\theta}^{(n)}$ satisfying Assumption~\ref{assump:large_sample}, where the negative log-density function of $\m P_{\theta}$ is denoted by $l(x,\theta)$.
We bound the MMD by first constraining the parameter space to a main region 
$$\wt{\Theta}=B(0_d,r_0(\log n)^{1/\alpha_2})$$
with a sufficiently large $r_0$. Then we can decompose the MMD into two parts, a main part on the main region and a negligible tail term:
\begin{align*}
& \left(\frac{1}{\sqrt{n}}\right)^d
\EX\left[
\mmd^2_{h}
\left(
    \smuws,
    \pos
\right)\right]
\\
& = 
\left(\frac{1}{\sqrt{n}}\right)^d
\mathbb{E}_{\theta^* \sim \Pi_{\theta}}
\bE_{\Xn\sim \m P_{\theta^*}^{(n)}}
\left[
\mmd^2_{h}
\left(
    \smuws,
    \pos
\right)
\cdot
\mathbf{1}(\theta^* \in \wt{\Theta})
\right] \\
& \quad +
\left(\frac{1}{\sqrt{n}}\right)^d
\mathbb{E}_{\theta^* \sim \Pi_{\theta}}
\bE_{\Xn\sim \m P_{\theta^*}^{(n)}}
\left[
\mmd^2_{h}
\left(
    \smuws,
    \pos
\right)
\cdot
\mathbf{1}(\theta^* \notin \wt{\Theta})
\right]
\\
& \leq
\left(\frac{1}{\sqrt{n}}\right)^d
\mathbb{E}_{\theta^* \sim \Pi_{\theta}}
\bE_{\Xn\sim \m P_{\theta^*}^{(n)}}
\left[
\mmd^2_{h}
\left(
    \smuws,
    \pos
\right)
\cdot
\mathbf{1}(\theta^* \in \wt{\Theta})
\right] \\
& \quad +
\frac{4}{(2\pi)^{\frac{d}{2}} \wt h^{d}}
\mathbb P_{\theta^* \sim \Pi_{\theta}}(\theta^* \notin \wt{\Theta}),
\end{align*}
Then we bound the tail probability using the tail condition on $\pi(\theta)$. When $r_0$ is sufficiently large, we have
\begin{align*}
\frac{4}{(2\pi)^{\frac{d}{2}} \wt h^{d}} \mathbb P_{\theta^* \sim \Pi_{\theta}}(\theta^* \notin \wt{\Theta})
& \leq \frac{4}{(2\pi)^{\frac{d}{2}} \wt h^{d}}
\int_{\theta \notin \wt{\Theta}}
c_2 \exp \left(-\frac{1}{c_2} \Vert \theta \Vert^{\alpha_2} \right) d\theta
\\
& \leq
\frac{8 c_2^2 r_0^{d-\alpha_2}}{\alpha_2 \Gamma(d/2) \wt h^{d}} 
(\log n)^{\frac{d}{\alpha_2}-1} n^{-\frac{r_0^{\alpha_2}}{c_2}}\\
&\leq n^{-1}.
\end{align*}
Then we define the region $$\wt \Theta'=\{\theta\in \mb R^d,\|\theta\|\leq  (r_0+r_1)(\log n)^{1/\alpha_2}\}$$ that contains the main region $\wt \Theta$,  where $r_1$ is chosen so that $\frac{r_1}{r_0}$ is sufficiently large. Define the mapping $\wh\theta: \m X^{n}\to \wt\Theta'$ so that for each $\Xn \in \m X^{n}$,
\begin{equation*}
    \wh\theta(\Xn)\in {\arg\min}_{\theta\in \wt\Theta'} \sum_{i=1}^n l(X_i,\theta).
\end{equation*}
Consider the following subsets of $\wt\Theta\times \mathcal{X}^n$:
\begin{equation*}
    \begin{aligned}
    A_0&=\Bigg\{(\theta^*,\Xn)\in \wt \Theta\times \m X^{n}\,:\, \forall i\in [n],\, \|X_i\|\leq c\, (1+\|\theta^*\|)^{\alpha_4}(\log n)^{\frac{1}{\alpha_3}}\Big\},\\
                A_1 
        &= \bigg\{ (\theta^*,\Xn)\in \wt \Theta\times \m X^{n}\,:  \, \forall \theta, \theta' \in \wt\Theta',\, \\
        & \qquad \qquad \qquad
        \left| 
            \frac{1}{n}\sum_{i=1}^n
             \left(   l(X_i,\theta) -  l(X_i,\theta') \right) 
            - 
            \bE_{X\sim \m P_{\theta^*}}
            \left[
                 l(X,\theta)  -   l(X,\theta')
            \right]
             \right|  \\ 
        & \qquad \qquad \qquad
        \leq c\,(\log n)^{\frac{1}{2}+\frac{\alpha_1}{\alpha_2}}\cdot(1+\|\theta^*\|)^{\alpha_0\alpha_4}\frac{1}{\sqrt{n}}  \Big(\frac{1}{n}+ \Vert \theta -\theta'\Vert \Big)\bigg\}, 
        \\
        A_2 
        &= \bigg\{  
            (\theta^*, \Xn)\in \wt \Theta\times \m X^{n}\,: \,  \forall \theta,\theta' \in  B(\theta^*,1), \,  \\ 
        & \qquad \qquad \qquad
            \Big| 
            \frac{1}{n}\sum_{i=1}^n 
            \left(
             l(X_i, \theta) -  l(X_i, \theta') 
                - \nabla_{\theta} l(X_i, \theta')^T (\theta - \theta') 
            \right) 
            \\
            &  \qquad \qquad \qquad -
            \bE_{X\sim \m P_{\theta^*}} 
            \left[
                l(X, \theta) -  l(X, \theta') - \nabla_{\theta} l(X,\theta')^T(\theta - \theta')
            \right]\Big| \\
            & \qquad \qquad \qquad \qquad 
            \leq c\, (\log n)^{\frac{1}{2}}\cdot(1+\|\theta^*\|)^{\alpha_0\alpha_4+\alpha_1}\frac{1}{\sqrt{n}}  \Big(\frac{1}{n}+ \Vert \theta -\theta'\Vert^2 \Big)\bigg\}, \\
         A_3&=\bigg\{(\theta^*,\Xn)\in \wt \Theta\times \m X^{n}\,:\,\forall j\in \mb N_0^d \text{ with }|j|=2,\\
         &\qquad\qquad\qquad\,\Big|\frac{1}{n}\sum_{i=1}^n \frac{\partial^{|j|}  l(X_i,\theta^*)}{\partial \theta^j}-\mb{E}_{X\sim \m P_{\theta^*}}\big[\frac{\partial^{|j|}  l(X,\theta^*)}{\partial \theta^j}\big]\Big|  \\ 
        & \qquad \qquad \qquad \qquad \leq c
         \,\sqrt{\frac{\log n}{n}}(1+\|\theta^*\|)^{\alpha_1+\alpha_0\alpha_4} \bigg\},\\
A_4&=\bigg\{(\theta^*,\Xn)\in \wt \Theta\times \m X^{n}\,:\,\Big\|\frac{1}{n}\sum_{i=1}^n \nabla_{\theta}l(X_i,\theta^*)\nabla_{\theta}l(X_i,\theta^*)^T-I(\theta^*)\Big\|_{\rm F}  \\ 
        & \qquad \qquad \qquad  \leq c\,\sqrt{\frac{\log n}{n}}\cdot (1+\|\theta^*\|)^{2\alpha_1+2\alpha_0\alpha_4}\bigg\},\\
            \end{aligned}
\end{equation*}
\begin{equation*}
    \begin{aligned}
A_5&=\bigg\{(\theta^*,\Xn)\in \wt \Theta\times \m X^{n}\,:\,\Big\|I(\theta^*)^{-1}\frac{1}{n}\sum_{i=1}^n \nabla_{\theta}l(X_i,\theta^*)\Big\|_{\rm F}\leq c\,\sqrt{\frac{\log n}{n}}\bigg\},\\
A_6 & = \bigg\{ (\theta^*, \Xn)\in \wt \Theta\times \m X^{n}\,: 
        \int_{\theta \in \wt\Theta'^c} 
        \frac{\pi(\theta)}{\pi(\theta^*)} 
        \exp{\left[ - \sum_{i=1}^n l(X_i,\theta) + \sum_{i=1}^n l(X_i, \theta^*) \right]} \dd\theta \\
        & \qquad \qquad \qquad + 
        \int_{\theta \in \wt\Theta'^c} \left[ \frac{\pi(\theta)}{\pi(\theta^*)} 
        \exp{\left( -\sum_{i=1}^n l(X_i,\theta) + \sum_{i=1}^n l(X_i,\theta^*) \right)} \right]^2 \dd\theta  \leq \frac{c}{n^{\frac{d}{2}+1}} \bigg\},\\
       \wt{A} & = A_0\cap A_1 \cap A_2 \cap A_3\cap A_4\cap A_5\cap A_6.
    \end{aligned}
\end{equation*}

\begin{lemma}
    \label{A123 with high probability}
  There exist constants $c,r_0,r_1$ so that, for any $\Pi_{\theta}\m P_{\theta}^{(n)}$ satisfying Assumption~\ref{assump:large_sample} and any $\theta^*\in\widetilde{\Theta}=B(0_d,r_0(\log n)^{1/\alpha_2})$,
\begin{align*}
  \mathbb{E}_{\Xn\sim \m P_{\theta^*}^{(n)}}
\left[\mathbf{1}\left((\theta^*, \Xn)\in \wt A\right)\right]
   \geq 1- \frac{\wt h^d}{n}.
\end{align*}
\end{lemma}
 \noindent So we have 
\begin{align*}
& n^{-\frac{d}{2}}\mathbb{E}_{\theta^* \sim \Pi_{\theta}}
\bE_{\Xn\sim \m P_{\theta^*}^{(n)}}
\left[
\mmd^2_{h}
\left( 
    \smuws,
    \pos
\right)
\cdot
\mathbf{1}(\theta^* \in \wt{\Theta})
\right]\\
&\leq n^{-\frac{d}{2}}\mathbb{E}_{\theta^* \sim \Pi_{\theta}}
\bE_{\Xn\sim \m P_{\theta^*}^{(n)}}
\left[
\mmd^2_{h}
\left( 
    \smuws,
    \pos
\right)
\cdot
\mathbf{1}((\theta^*,\Xn) \in \wt A) 
\right]\\
& \quad + n^{-\frac{d}{2}}\mb{E}_{\theta^*\sim \Pi_{\theta}}\mb{E}_{\Xn\sim \m P^{(n)}_{\theta^*}}
\Big[
\mmd^2_{h}
\left( 
    \smuws,
    \pos
\right)
\\ 
& \qquad \qquad \qquad \qquad \qquad \qquad \cdot
\mathbf{1}(\theta^* \in \wt{\Theta})\mathbf{1}((\theta^*,\Xn) \notin \wt A)
\Big]\\
&\leq  n^{-\frac{d}{2}}\mathbb{E}_{\theta^* \sim \Pi_{\theta}}
\bE_{\Xn\sim \m P_{\theta^*}^{(n)}}
\left[
\mmd^2_{h}
\left( 
    \smuws,
    \pos
\right)
\cdot
\mathbf{1}((\theta^*,\Xn) \in \wt A) 
\right]  \\ 
& \quad 
+\frac{4}{(2\pi)^{\frac{d}{2}}n}.
 \end{align*}
We then use the 
triangle inequality for MMD to split the target MMD into two terms:
\begin{small}
\begin{align*}
&  
 n^{-\frac{d}{2}}
\mathbb{E}_{\theta^* \sim \Pi_{\theta}}
\bE_{\Xn\sim \m P_{\theta^*}^{(n)}}
\left[
\mmd^2_{h}
\left( 
    \smuws,
    \pos
\right)
\cdot
\mathbf{1}((\theta^*,\Xn) \in \wt A) 
\right]
\\
&=
\mathbb{E}_{\theta^* \sim \Pi_{\theta}}
\bE_{\Xn\sim \m P_{\theta^*}^{(n)}}
\Big[
\mmd^2_{\wt h}
\Big( 
      \big[\sqrt{n} (\cdot - \wh{\theta}(\Xn))\big]_{\#} \mu_{W^*} \left( \cdot| S^*(\Xn) \right),  \\ 
        & \qquad \qquad \qquad \qquad \qquad \qquad
        \big[\sqrt{n} (\cdot - \wh{\theta}(\Xn))\big]_{\#}\pos
\Big)
\mathbf{1}((\theta^*,\Xn) \in \wt A) 
\Big]\\
&\lesssim
\mathbb{E}_{\theta^* \sim \Pi_{\theta}}
\bE_{\Xn\sim \m P_{\theta^*}^{(n)}}
\Big[
\mmd^2_{\wt h}
\left(
     \big[\sqrt{n} (\cdot - \wh{\theta}(\Xn))\big]_{\#}\pos
    ,
    N(0, I^{-1}(\wh{\theta}(\Xn)))
\right)
 \\ 
& \qquad \qquad \qquad \qquad \qquad 
\cdot
\mathbf{1}((\theta^*,\Xn) \in \wt A) 
\Big]
\\
&   +
\mathbb{E}_{\theta^* \sim \Pi_{\theta}}
\bE_{\Xn\sim \m P_{\theta^*}^{(n)}}
\Big[
\mmd^2_{\wt h}
\left(
    N(0, I^{-1}(\wh{\theta}(\Xn)))
    ,
   \big[\sqrt{n} (\cdot - \wh{\theta}(\Xn))\big]_{\#}\mu_{W^*} \left(  \cdot| S^*(\Xn) \right)
\right)
 \\ 
 & \qquad \qquad \qquad \qquad \qquad 
\cdot
\mathbf{1}((\theta^*,\Xn) \in \wt A) 
\Big]
.
\end{align*}
\end{small}
This split reduces the problem to bounding two simpler approximations 
(Gaussian fitting and neural network fitting). To analyze the two MMD terms above, we introduce 
the following lemma to quantify Gaussian approximation error.


\begin{lemma}\label{lemma: lst gaussian approximation}
There exists a constant $C$ such that for any $\Pi_{\theta}\m P_{\theta}^{(n)}$ satisfying Assumption~\ref{assump:large_sample},
    \begin{align*}
  & \mathbb{E}_{\theta^* \sim \Pi_{\theta}}
\bE_{\Xn\sim \m P_{\theta^*}^{(n)}}
\Big[
\mmd^2_{\wt h}
\left(
     \big[\sqrt{n} (\cdot - \wh{\theta}(\Xn))\big]_{\#}\pos
    ,
    N(0, I^{-1}(\wh{\theta}(\Xn)))
\right)\\
&\qquad\qquad\qquad\qquad\quad
\cdot
\mathbf{1}((\theta^*,\Xn) \in \wt A) 
\Big] \leq
    C\,\frac{\log n}{n}.
    \end{align*}
\end{lemma}

\noindent Then we introduce 
the following lemma to quantify the neural network approximation error.
\begin{lemma}\label{lemma:approximateNN}
  Define $c_x=d_X(1+\frac{\alpha_4}{\alpha_2}+\frac{6\alpha_0+1}{\alpha_3})$ and $c_{\theta}=d(\frac{1}{\alpha_2}+\frac{1}{2}+\frac{7\alpha_0}{2\alpha_3})$. There exist $L_1,L_2,U_1,U_2,R_1,R_2$ with $L_1,L_2=\m O(\log^2 n)$, $R_1=\m O(n^{1+d_X+d/2}(\log n)^{c_x+c_{\theta}+1})$, $R_2=\m O(n^{d/2}(\log n)^{c_{\theta}+1})$ and $\ms E(U_1)=\m O(R_1^3)$, $
\ms E(U_2)=\m O(R_2^3)$, such that for any $\Pi_{\theta}\m P_{\theta}^{(n)}$ satisfying Assumption~\ref{assump:large_sample},
there exist  $S^*\in \Phi_{nd_X}^{d}(L_1,U_1,R_1)$  and $W^*\in \Phi_{d}^{d^2}(L_2,U_2,R_2)$  so that 
  \begin{equation*}
    \begin{aligned}
     &   \mathbb{E}_{\theta^* \sim \Pi_{\theta}}
\bE_{\Xn\sim \m P_{\theta^*}^{(n)}}
\Big[
\mmd^2_{\wt h}
\left(
    N(0, I^{-1}(\wh{\theta}(\Xn)))
    ,
      \big[\sqrt{n} (\cdot - \wh{\theta}(\Xn))\big]_{\#} \mu_{W^*} \left(\cdot | S^*(\Xn) \right)
\right)\\
&\qquad\qquad\qquad\qquad\quad
\cdot
\mathbf{1}((\theta^*,\Xn) \in \wt A) 
\Big]\leq C\,\frac{\log n}{n}
    \end{aligned}
\end{equation*}
for a constant $C$.
\end{lemma}
Then  by substituting $R=R_1+R_2=\m O(n^{1+d_X+d/2}(\log n)^{c_x+c_{\theta}+1})$ and $L=L_1+L_2=\m O(\log^2 n)$ into~\eqref{eqn:N2}, we obtain that whenever
\begin{equation*}
    N\gtrsim \frac{n^{2+d_X+d/2}}{\wt h^d}(\log n)^{c_x+c_{\theta}+5},
\end{equation*}
it holds that
\begin{equation*}
    \begin{aligned}
        &   n^{-\frac{d}{2}} \mb{E}_{\Xn\sim \m P^{(n)}}\big[\mmd^2_h(p(\cdot|\Xn),\mu_{\wh W}(\cdot|\wh S(\Xn)))\big] 
 \lesssim \frac{\log n}{n}.
    \end{aligned}
\end{equation*}

\section{Proof of Technical Lemmas}

\subsection{Proof of Lemma \ref{lemma: approximate summary statistics}}\label{proof: lemma approximate summary statistics}

Consider a fixed model $\Pi_{\theta}\m P^{(n)}_{\theta}$ satisfying Assumption~\ref{assump:exponential}, where $\m P_{\theta}$ has a density function $p(x|\theta)=h(x)\exp(\eta(\theta)^Tt(x)-A(\theta))$ and $t(x)=(t_1(x),t_2(x),\cdots, t_k(x))$ has $k\leq p$ components.  Define the map  $\Pi_{1:k}: \mb R^p\to \mb R^k$, where $\Pi_{1:k}(s)=(s_1,\cdots, s_k)$ for $s=(s_1,\cdots, s_p)$. Moreover,  for any $S^*\in \ms S$, we denote $S^*_{1:k}(\Xn)=\Pi_{1:k}(S^*(\Xn))$.
We will proceed by first showing that there exists $S^*\in \ms S$ so that $\mmd^2_h\big(\wt\nu(\cdot|T(\Xn)),\wt\nu(\cdot|S^*_{1:k}(\Xn))\big)\leq\frac{\varepsilon}{32C},$ then showing that there exists $W^*\in \ms W$ so that   $\mmd^2_h(\wt\nu(\cdot|S^*_{1:k}(\Xn)), \mu_{W^*}(\cdot|S^*(\Xn)))\leq \frac{\varepsilon}{32C}$.
For a fixed $S^*$, consider the two distributions
  \begin{align*}
\wt\nu(\cdot|T(\Xn)) &= \sum_{j=1}^J\frac{\exp(\wt v_j(T(\Xn)))}{\sum_{j=1}^J \exp(\wt v_j(T(\Xn)))}\delta_{\theta_j^*}(\cdot).
\end{align*}
 and 
   \begin{align*}
\wt\nu(\cdot|S^*_{1:k}(\Xn)) &= \sum_{j=1}^J\frac{\exp(\wt v_j(S^*_{1:k}(\Xn)))}{\sum_{j=1}^J \exp(\wt v_j(S^*_{1:k}(\Xn)))}\delta_{\theta_j^*}(\cdot),
\end{align*}
where recall that 
\begin{equation*} 
\begin{aligned}
       & \wt v_j(s) =\max\big(v_j(s)-\max_{j\in [J]} v_j(s),-c_1\log(\frac{n}{\min(\varepsilon,1) h})\big),\quad \forall j\in [J],\\
       &\text{where } v_j(s) = \log\left(\int_{\Theta} \rho_j(\theta)\dd\theta\right) + \log \Big(\pi(\theta_j^*)\exp\big(\eta(\theta_j^*)^Ts-nA(\theta_j^*)\big)
       \Big).
\end{aligned}
\end{equation*}
For $j\in [J]$, we write $\wt\nu(\theta^*_j|T(\Xn)) = \frac{\exp(\wt v_j(T(\Xn)))}{\sum_{j=1}^J \exp(\wt v_j(T(\Xn)))}$ and $\wt\nu(\theta^*_j|S^*_{1:k}(\Xn)) = \frac{\exp(\wt v_j(S^*_{1:k}(\Xn)))}{\sum_{j=1}^J \exp(\wt v_j(S^*_{1:k}(\Xn)))}$. For any $\theta' \in \mb R^d$,  we have
\begin{align*}
& \left| \sum_{j=1}^J \kappa_{h/\sqrt{2}}(\theta^*_j, \theta') \left( \wt\nu(\theta^*_j|T(\Xn)) -\wt\nu(\theta^*_j|S^*_{1:k}(\Xn)) \right) \right| \\
& \leq \sup_{j\in [J]} \kappa_{h/\sqrt{2}}(\theta^*_j, \theta') \sum_{j=1}^J |\wt\nu(\theta^*_j|T(\Xn)) - \wt\nu(\theta^*_j|S^*_{1:k}(\Xn))|.
\end{align*}
Similarly to the analysis for equation~\eqref{eqn:TVbound}, we can further bound
\begin{equation*}
    \sum_{j=1}^J |\wt\nu(\theta^*_j|T(\Xn)) - \wt\nu(\theta^*_j|S^*_{1:k}(\Xn))|\leq \frac{2\sum_{j=1}^J \big|\exp(\wt v_j(T(\Xn)))-\exp(\wt v_j(S^*_{1:k}(\Xn)))\big|}{\sum_{j=1}^J \exp(\wt v_j(T(\Xn)))}.
\end{equation*}
Using the inequality $|\exp(a) - \exp(b)| \leq \exp(\max(a,b))|a-b|$ and first assuming $|\wt v_j(T(\Xn))-\wt v_j(S^*_{1:k}(\Xn))| \leq 1$ for any $j\in [J]$, we can bound the sum:
\begin{align*}
&\sum_{j=1}^J |\exp(\wt v_j(T(\Xn))) - \exp(\wt v_j(S^*_{1:k}(\Xn)))| \\
&\leq \sum_{j=1}^J \exp(\max(\wt v_j(T(\Xn)), \wt v_j(S^*_{1:k}(\Xn)))) |\wt v_j(T(\Xn)) - \wt v_j(S^*_{1:k}(\Xn))| \\
&\leq \sum_{j=1}^J \exp(\wt v_j(T(\Xn))) \exp\big(|\wt v_j(S^*_{1:k}(\Xn)) - \wt v_j(T(\Xn))|\big)\cdot |\wt v_j(T(\Xn))- \wt v_j(S^*_{1:k}(\Xn))| \\
&\leq e \cdot \max_{j \in [J]} |\wt v_j(T(\Xn)) - \wt v_j(S^*_{1:k}(\Xn))| \sum_{j=1}^J \exp(\wt v_j(T(\Xn))).  
\end{align*}
This gives a bound on the total variation distance:
\begin{align*}
   \sum_{j=1}^J |\wt\nu(\theta^*_j|T(\Xn)) - \wt\nu(\theta^*_j|S^*_{1:k}(\Xn))| \leq 2e \cdot   \max_{j \in [J]} |\wt v_j(T(\Xn)) -\wt v_j(S^*_{1:k}(\Xn))|.
\end{align*}
Furthermore, the difference in logits $\wt v_j$ is bounded by:
\begin{align*}
&|\wt v_j(T(\Xn)) - \wt v_j(S^*_{1:k}(\Xn))|  \\ & \leq |v_j(T(\Xn)) - v_j(S^*_{1:k}(\Xn))| + |\max_{j\in [J]} v_j(T(\Xn)) -\max_{j\in [J]} v_j(S^*_{1:k}(\Xn))|.
\end{align*}
Let $j_1^* = \arg\max_{j\in [J]} v_j(T(\Xn))$ and $j^*_2 = \arg\max_{j\in [J]} v_j(S^*_{1:k}(\Xn))$. We have:
\begin{align*}
\max_{j\in [J]} v_j(T(\Xn)) -\max_{j\in [J]} v_j(S^*_{1:k}(\Xn)) &= v_{j^*_1}(T(\Xn)) - v_{j_2^*}(S^*_{1:k}(\Xn)) \\
&\leq v_{j_1^*}(T(\Xn)) - v_{j_1^*}(S^*_{1:k}(\Xn)).
\end{align*}
By symmetry, $\max_{j\in [J]} v_j(S^*_{1:k}(\Xn)) - \max_{j\in [J]} v_j(T(\Xn)) \leq v_{j^*_2}(S^*_{1:k}(\Xn)) - v_{j^*_2}(T(\Xn))$. This implies:
\begin{align*}
|\max_{j\in [J]} v_j(T(\Xn)) - \max_{j\in [J]} v_j(S^*_{1:k}(\Xn))| \leq \max_{j\in [J]} |v_j(T(\Xn)) - v_j(S^*_{1:k}(\Xn))|.
\end{align*}
Thus, the logit difference is bounded by:
\begin{align*}
\max_{j\in[J]} |\wt v_j(T(\Xn)) - \wt v_j(S^*_{1:k}(\Xn))| &\leq 2 \max_{j \in [J]}|v_j(T(\Xn)) - v_j(S^*_{1:k}(\Xn))| \\
&= 2 \max_{j\in[J]} \left| \eta(\theta^*_j)^T\Big( \sum_{i=1}^n t(X_i) - S^*_{1:k}(\Xn)\Big) \right| \\
&\leq 2 \max_{j\in[J]} \|\eta(\theta^*_j)\|_2 \cdot   \Big\|\sum_{i=1}^n t(X_i) - S^*_{1:k}(\Xn)\Big\|_2 \\
&\leq 2 C_\eta   \Big\|\sum_{i=1}^n t(X_i) - S^*_{1:k}(\Xn)\Big\|_2.
\end{align*}
Combining these results,  if $\max_{j\in [J]}|\wt v_j(T(\Xn))-\wt v_j(S^*_{1:k}(\Xn))| \leq 1$, we have for any $\theta' \in \mb R^d$,   
\begin{align*}
& \left| \sum_{j=1}^J k_{h/\sqrt{2}}(\theta^*_j, \theta') \left( \wt\nu(\theta^*_j|T(\Xn)) -\wt\nu(\theta^*_j|S^*_{1:k}(\Xn)) \right) \right|  \\
&\leq \frac{1}{(2\pi)^{d/2}(h/\sqrt{2})^d} \cdot 2e \cdot 2 C_\eta    \Big\|\sum_{i=1}^n t(X_i) - S^*_{1:k}(\Xn)\Big\|_2.
\end{align*}
Then let $S^*_{1:k}(\Xn)=\sum_{i=1}^n s^*_{1:k}(X_i)$, where $s^*_{1:k}=\Pi_{1:k}\circ s^*(\cdot)$ for a mapping $s^*:\mb R^{d_X}\to \mb R^p$.
Using~\eqref{eqn:MMDkernel}, if $\max_{j\in [J]}|\wt v_j(T(\Xn))-\wt v_j(S^*_{1:k}(\Xn))| \leq 2C_\eta n\cdot \sup_{X\in \m X} \| t(X) -s^*_{1:k}(X)\|\leq 1$,   there exists a constant $C_1$ so that
\begin{equation*}
    \begin{aligned}
 &\mmd^2_h(\wt\nu(\cdot|T(\Xn)),\wt\nu(\cdot|S^*_{1:k}(\Xn))) \\
 &\leq  C_1\,  \Big(1+h\sqrt{ \log(1+\frac{1}{\varepsilon})}\Big)^d\cdot  h^{-2d}\cdot n^2 \sup_{X\in \m X} \| t(X) -s^*_{1:k}(X)\|^2_2+\frac{\varepsilon}{64C}
    \end{aligned}
\end{equation*}
Let $$\varepsilon_T =\Big(\frac{1}{8\sqrt{C_1C}}\Big(1+h\sqrt{ \log(1+\frac{1}{\varepsilon})}\Big)^{-d/2}\cdot \frac{h^d \sqrt{\varepsilon}}{n}\Big)\wedge \Big(\frac{1}{2nC_{\eta}}\Big).$$
In the remainder of this section, we construct the summary network $S^*(\Xn)=(S^*_{1:k}(\Xn),0_{p-k})$ with $S^*_{1:k}(\Xn)=\sum_{i=1}^n s^*_{1:k}(X_i)$ 
such that: 
\begin{align*}
 \sup_{X \in \mathcal{X}} \|t(X) - s^*_{1:k}(X)\|\leq\varepsilon_T.
\end{align*}
If this is the case, then we have $\mmd^2_h(\wt\nu(\cdot|T(\Xn)),\wt\nu(\cdot|S^*_{1:k}(\Xn)))\leq \frac{\varepsilon}{64C}+\frac{\varepsilon}{64C}=\frac{\varepsilon}{32C}$.
Let $U_{\mathcal{X}} = \{X^*_1, \ldots, X^*_M\}$ be the largest $\epx$-packing set of $\cX$.
We first consider the following approximation to $t(\cdot)$:
\begin{align*}
  t_A(x) = \sum_{m=1}^M t(X^*_m) 
  \frac{k_b \left( \| x - X_m^* \|^2 / \epx^2 \right) }{ \sum_{m=1}^M k_b\left( \| x - X^*_m \|^2 / \epx^2 \right) },
\end{align*}
where 
$$
k_b(x)=\left\{\begin{array}{cc}
1 & |x|\leq 1 \\
0 & |x|>2 \\
2-|x| & 1<|x| \leq 2.
\end{array}\right.
$$
Since $t(\cdot)$ is $C_t$-Lipschitz on $\cX$, for all $X \in \cX$, 
\begin{align*}
& \quad
\left\|
  t_{A}(X) - t(X)
\right\| \\
& \leq
\sum_{m=1}^M 
\left\| t(X^*_m) - t(X) \right\|
\frac{k_b\left( \| X - X^*_m \|^2 / \epx^2 \right) }{ \sum_{m=1}^M k_b\left( \| X - X^*_m \|^2 / \epx^2 \right) }
\\
& \leq
\sqrt{2}C_t\epx.
\end{align*}
Setting $
\epx=\frac{\varepsilon_{T}}{2\sqrt{2}C_t}
$ ensures
\begin{align*}
 \sup_{X \in \mathcal{X}} \|t(X) -t_A(X)\|\leq\frac{\varepsilon_T}{2}.
\end{align*}
Note that ReLU neural networks can approximate multiplication $\phi_{\text{mult}}(x,y) \approx xy$ and reciprocal $\phi_{\text{rec}}(x) \approx 1/x$.  By Lemma F.6 and Lemma F.7 of~\cite{oko2023diffusion},  for any positive constants $c_1,C_1$,
\begin{enumerate}
    \item There exist $L_3,U_3,R_3$ with $L_3=\m O(\log \varepsilon_T^{-1})$ and $R_3=\m O(\log \varepsilon_T^{-1}) $, and a network $\phi_{\text{mult}}\in \Phi_2^1(L_3,U_3,R_3)$ such that for any $x_1,x_2\in [-C_1\,\varepsilon_T^{-1} , C_1\,\varepsilon_T^{-1}]$,  $|\phi_{\text{mult}}(x_1,x_2)-x_1x_2|\leq c_1\varepsilon_T^{1+d_X}$.
       \item There exist $L_4,U_4,R_4$ with  $L_4=\m O(\log^2 \varepsilon_T^{-1})$ and $R_4=\m O(\log^5 \varepsilon_T^{-1})$, and a network $\phi_{\text{rec}}\in \Phi_1^1(L_4,U_4,R_4)$  such that for any $x\in [C_1^{-1}, C_1]$,  $|\phi_{\text{rec}}(x)-\frac{1}{x}|\leq c_1\varepsilon_T$.
\end{enumerate}
We will take $c_1$ to be a small enough absolute constant and $C_1$ to be a large enough absolute constant, to be chosen later.  Let $t_i(\cdot)$ represent the $i$th component of $t(\cdot)=(t_1(\cdot),t_2(\cdot),\cdots, t_k(\cdot))$. For $i\in [k]$, we consider
\begin{align*}
s^*_i(x)
&=\phi_{\text{mult}} 
  \Bigg[ 
  \sum_{m=1}^M k_b
\left( \sum_{l=1}^{d_X}\phi_{\text{mult}} \left(
  \frac{x_l - X^*_{m,l}}{\epx},
  \frac{x_l - X^*_{m,l}}{\epx}
  \right)
\right) t_i(X_m^*),\\
&\qquad\qquad
  \phi_{\text{rec}} 
  \left( 
   \sum_{m=1}^M k_b
\left( \sum_{l=1}^{d_X}\phi_{\text{mult}} \left(
  \frac{x_l - X^*_{m,l}}{\epx},
  \frac{x_l - X^*_{m,l}}{\epx}
  \right)
\right)
  \right)
  \Bigg],
\end{align*}
where  recall that  $k_b(x)={\rm ReLU}(2-|x|)-{\rm ReLU}(1-|x|)$ with $|x|={\rm ReLU}(x)+{\rm ReLU}(-x)$ can be realized exactly through the ReLU neural network. Then define 
\begin{equation*}
    s^*(x)=(s_1^*(x),s_2^*(x),\cdots, s_k^*(x),0_{p-k}).
\end{equation*}
To bound 
$ \sup_{X \in \mathcal{X}} \|t_A(X) -s^*_{1:k}(X)\|$, we decompose the error into three parts:
\begin{small}
\begin{equation}\label{eqn:threedecom}
\begin{aligned}
& \quad
\left\|
  s^*_{1:k}(x) - t_{A}(x)
\right\| \\
& \leq
\left\|
  \sum_{m=1}^M
  t(X^*_m)\left(
  \frac{k_b
\left( \sum_{l=1}^{d_X}\phi_{\text{mult}} \left(
  \frac{x_l - X^*_{m,l}}{\epx},
  \frac{x_l - X^*_{m,l}}{\epx}
  \right)\right)}{ \sum_{m=1}^M k_b
\left( \sum_{l=1}^{d_X}\phi_{\text{mult}} \left(
  \frac{x_l - X^*_{m,l}}{\epx},
  \frac{x_l - X^*_{m,l}}{\epx}
  \right)\right) }
  -
  \frac{k_b \left( \| x - X^*_m \|^2/ \epx^2 \right) }{ \sum_{m=1}^M k_b\left( \| x - X^*_m \|^2 / \epx^2 \right) }
\right)
  \right\|
\\
& \quad +
\Bigg\|
\sum_{m=1}^M
t(X^*_m)
k_b
\left( \sum_{l=1}^{d_X}\phi_{\text{mult}} \left(
  \frac{x_l - X^*_{m,l}}{\epx},
  \frac{x_l - X^*_{m,l}}{\epx}
  \right)\right)
  \cdot  \\ 
        &  \qquad
  \phi_{\text{rec}}
  \left( 
    \sum_{m=1}^M k_b
\left( \sum_{l=1}^{d_X}\phi_{\text{mult}} \left(
  \frac{x_l - X^*_{m,l}}{\epx},
  \frac{x_l - X^*_{m,l}}{\epx}
  \right)\right)
  \right)
\\
& \qquad
-
\sum_{m=1}^M
  t(X^*_m)
  \frac{k_b
\left( \sum_{l=1}^{d_X}\phi_{\text{mult}} \left(
  \frac{x_l - X^*_{m,l}}{\epx},
  \frac{x_l - X^*_{m,l}}{\epx}
  \right)\right) }{ \sum_{m=1}^Mk_b
\left( \sum_{l=1}^{d_X}\phi_{\text{mult}} \left(
  \frac{x_l - X^*_{m,l}}{\epx},
  \frac{x_l - X^*_{m,l}}{\epx}
  \right)\right) }
\Bigg\|
\\
& \quad +
\Bigg\|
  \phi_{\text{mult}} 
  \Big[ 
  \sum_{m=1}^M
  t(X_m^*)
 k_b
\left( \sum_{l=1}^{d_X}\phi_{\text{mult}} \left(
  \frac{x_l - X^*_{m,l}}{\epx},
  \frac{x_l - X^*_{m,l}}{\epx}
  \right)\right)
  ,  \\ 
& \qquad \qquad \qquad \qquad \qquad \qquad
  \phi_{\text{rec}} 
  \left( 
  \sum_{m=1}^M k_b
\left( \sum_{l=1}^{d_X}\phi_{\text{mult}} \left(
  \frac{x_l - X^*_{m,l}}{\epx},
  \frac{x_l - X^*_{m,l}}{\epx}
  \right)\right)
  \right)
  \Big]
    \\
    & \qquad -
    \sum_{m=1}^M
 t(X^*_m)
  k_b
\left( \sum_{l=1}^{d_X}\phi_{\text{mult}} \left(
  \frac{x_l - X^*_{m,l}}{\epx},
  \frac{x_l - X^*_{m,l}}{\epx}
  \right)\right)  \\ 
& \qquad \qquad \qquad \qquad 
  \cdot
  \phi_{\text{rec}}
  \left( 
    \sum_{m=1}^M k_b
\left( \sum_{l=1}^{d_X}\phi_{\text{mult}} \left(
  \frac{x_l - X^*_{m,l}}{\epx},
  \frac{x_l - X^*_{m,l}}{\epx}
  \right)\right) 
  \right)
\Bigg\|.
\end{aligned}
\end{equation}
\end{small}
The three error terms correspond to the errors from approximating the kernel, reciprocal, and multiplication functions using ReLU neural networks, respectively.  Recalling that $\varepsilon_X=\frac{\varepsilon_T}{2\sqrt{2}C_t}$, we have 
\begin{align*}
  & \left\|
  \sum_{m=1}^M
  t(X^*_m)\left(
  \frac{k_b
\left( \sum_{l=1}^{d_X}\phi_{\text{mult}} \left(
  \frac{x_l - X^*_{m,l}}{\epx},
  \frac{x_l - X^*_{m,l}}{\epx}
  \right)\right)}{ \sum_{m=1}^Mk_b
\left( \sum_{l=1}^{d_X}\phi_{\text{mult}} \left(
  \frac{x_l - X^*_{m,l}}{\epx},
  \frac{x_l - X^*_{m,l}}{\epx}
  \right)\right) }
  -
  \frac{k_b\left( \| x - X^*_m \|^2/ \epx^2 \right) }{ \sum_{m=1}^M k_b\left( \| x - X^*_m \|^2 / \epx^2 \right) }
\right)
  \right\|
    \\
    & \leq 
      \sum_{m=1}^M  
      \|t(X_m^*) \| \cdot
      k_b
\left( \sum_{l=1}^{d_X}\phi_{\text{mult}} \left(
  \frac{x_l - X^*_{m,l}}{\epx},
  \frac{x_l - X^*_{m,l}}{\epx}
  \right)\right) \\
  &\qquad\cdot 
  \frac{
          \sum_{m=1}^M
          \left|
k_b
\left( \sum_{l=1}^{d_X}\phi_{\text{mult}} \left(
  \frac{x_l - X^*_{m,l}}{\epx},
  \frac{x_l - X^*_{m,l}}{\epx}
  \right)\right) - k_b\left( \| x - X_m^* \|^2 / \epx^2 \right) 
          \right|}
        { \sum_{m=1}^M k_b
\left( \sum_{l=1}^{d_X}\phi_{\text{mult}} \left(
  \frac{x_l - X^*_{m,l}}{\epx},
  \frac{x_l - X^*_{m,l}}{\epx}
  \right)\right)
        \sum_{m=1}^M k_b\left( \| x - X_m^* \|^2 / \epx^2 \right) } 
    \\
    & +
    \sum_{m=1}^M \left\| t(X^*_m)\right\|\cdot  \frac{\left| k_b
\left( \sum_{l=1}^{d_X}\phi_{\text{mult}} \left(
  \frac{x_l - X^*_{m,l}}{\epx},
  \frac{x_l - X^*_{m,l}}{\epx}
  \right)\right) - k_b\left( \| x - X^*_m \|^2 / \epx^2 \right)  \right|}
      { \sum_{m=1}^M k_b\left( \| x - X_m^* \|^2 / \epx^2 \right) },
\end{align*}
For all $x\in \m X$, there exists  $X^*_{m^*} \in U_{\mathcal{X}}$ such that $\Vert x - X^*_{m^*} \Vert \leq \epx$, which means $k_b\left( \Vert x - X^*_{m^*} \Vert^2 / \epx^2 \right) = 1$ and thus $ \sum_{m=1}^M k_b\left( \| x - X_m^* \|^2 / \epx^2 \right)\geq 1$.
Therefore,  using $\sup_{X\in \m X} \|t(X)\|\leq C_T$ and  the compactness of $\m X$, there exists a constant $c$ so that the error can be upper bounded as follows:
\begin{align*}
& \left\|
  \sum_{m=1}^M
  t(X^*_m)\left(
  \frac{k_b
\left( \sum_{l=1}^{d_X}\phi_{\text{mult}} \left(
  \frac{x_l - X^*_{m,l}}{\epx},
  \frac{x_l - X^*_{m,l}}{\epx}
  \right)\right)}{ \sum_{m=1}^Mk_b
\left( \sum_{l=1}^{d_X}\phi_{\text{mult}} \left(
  \frac{x_l - X^*_{m,l}}{\epx},
  \frac{x_l - X^*_{m,l}}{\epx}
  \right)\right) }
  -
  \frac{k_b\left( \| x - X^*_m \|^2/ \epx^2 \right) }{ \sum_{m=1}^M k_b\left( \| x - X^*_m \|^2 / \epx^2 \right) }
\right)
  \right\|
  \\
& \leq 
    2 \sup_{X\in \m X} \|t(X)\|\cdot \sum_{m=1}^M
          \left|
k_b
\left( \sum_{l=1}^{d_X}\phi_{\text{mult}} \left(
  \frac{x_l - X^*_{m,l}}{\epx},
  \frac{x_l - X^*_{m,l}}{\epx}
  \right)\right) - k_b\left( \| x - X_m^* \|^2 / \epx^2 \right) 
          \right|\\
          &\leq 2C_T \sum_{m=1}^M
          \left|
k_b
\left( \sum_{l=1}^{d_X}\phi_{\text{mult}} \left(
  \frac{x_l - X^*_{m,l}}{\epx},
  \frac{x_l - X^*_{m,l}}{\epx}
  \right)\right) - k_b\left( \| x - X_m^* \|^2 / \epx^2 \right) 
          \right|\\
          &\leq 2C_T M\cdot\sup_{x\in \m X,\,m\in[M]} \sum_{l=1}^{d_X}\left|\phi_{\text{mult}} \left(
  \frac{x_l - X^*_{m,l}}{\epx},
  \frac{x_l - X^*_{m,l}}{\epx}
  \right) -  \frac{(x_l - X^*_{m,l})^2}{\epx^2}\right|\\
      &\leq  2C_T Md_X\cdot\sup_{x\in [-c\varepsilon_X^{-1},c\varepsilon_X^{-1}]} \sum_{l=1}^{d_X}\left|\phi_{\text{mult}} \left(x,x
  \right) -  x^2\right|.
    \end{align*}
    So using $M\lesssim \varepsilon_X^{-d_X}\lesssim \varepsilon_T^{-d_X}$, if $|\phi_{\text{mult}}(x_1,x_2)-x_1x_2|\leq c_1\varepsilon_T^{1+d_X}$ holds for $x_1,x_2\in [-C_1\,\varepsilon_T^{-1}, C_1\,\varepsilon_T^{-1}]$ with a small enough $c_1$ and large enough $C_1$, we can get 
\begin{equation*}
    \left\|
  \sum_{m=1}^M
  t(X^*_m)\left(
  \frac{k_b
\left( \sum_{l=1}^{d_X}\phi_{\text{mult}} \left(
  \frac{x_l - X^*_{m,l}}{\epx},
  \frac{x_l - X^*_{m,l}}{\epx}
  \right)\right)}{ \sum_{m=1}^Mk_b
\left( \sum_{l=1}^{d_X}\phi_{\text{mult}} \left(
  \frac{x_l - X^*_{m,l}}{\epx},
  \frac{x_l - X^*_{m,l}}{\epx}
  \right)\right) }
  -
  \frac{k_b\left( \| x - X^*_m \|^2/ \epx^2 \right) }{ \sum_{m=1}^M k_b\left( \| x - X^*_m \|^2 / \epx^2 \right) }
\right)
  \right\|\leq \frac{\varepsilon_T}{6}.
\end{equation*}
Then, we upper bound  the second part in~\eqref{eqn:threedecom}. The set $\{X_1^*,X_2^*,\cdots,X_M^*\}$ is the largest $\varepsilon_X$ packing set of $\m X$. For any $x\in \m X$, the number of $m\in [M]$ so that $\|x-X_m^*\|\leq c$ is upper bounded by $(1+2c/\varepsilon_X)^{d_X}$. Let $C_X=(1+2\sqrt{2})^{d_X}$. Therefore, 
\begin{equation*}
   1\leq \sum_{m=1}^M k_b(\|x-X_m^*\|^2/\varepsilon_X^2)\leq C_X. 
\end{equation*}
Furthermore, 
\begin{equation*}
     \frac{1}{2}\leq 
\sum_{m=1}^M k_b
\left( \sum_{l=1}^{d_X}\phi_{\text{mult}} \left(
  \frac{x_l - X^*_{m,l}}{\epx},
  \frac{x_l - X^*_{m,l}}{\epx}
  \right)\right) \leq C_X+\frac{1}{2}.
\end{equation*}
Hence,  if $|\phi_{\text{rec}}(x)-\frac{1}{x}|\leq c_1\varepsilon_T$ holds for  $x\in [C_1^{-1}, C_1]$ with a small enough $c_1$ and large enough $C_1$, the error can be upper bounded as follows:
\begin{align*}
& \Bigg\|
\sum_{m=1}^M
t(X^*_m)
k_b
\left( \sum_{l=1}^{d_X}\phi_{\text{mult}} \left(
  \frac{x_l - X^*_{m,l}}{\epx},
  \frac{x_l - X^*_{m,l}}{\epx}
  \right)\right)
  \cdot
  \phi_{\text{rec}}
  \left( 
    \sum_{m=1}^M k_b
\left( \sum_{l=1}^{d_X}\phi_{\text{mult}} \left(
  \frac{x_l - X^*_{m,l}}{\epx},
  \frac{x_l - X^*_{m,l}}{\epx}
  \right)\right)
  \right)
\\
& \qquad\qquad
-
\sum_{m=1}^M
  t(X^*_m)
  \frac{k_b
\left( \sum_{l=1}^{d_X}\phi_{\text{mult}} \left(
  \frac{x_l - X^*_{m,l}}{\epx},
  \frac{x_l - X^*_{m,l}}{\epx}
  \right)\right) }{ \sum_{m=1}^Mk_b
\left( \sum_{l=1}^{d_X}\phi_{\text{mult}} \left(
  \frac{x_l - X^*_{m,l}}{\epx},
  \frac{x_l - X^*_{m,l}}{\epx}
  \right)\right) }
\Bigg\|
  \\
  & 
  \leq C_T  (C_X+\frac{1}{2})\cdot\Bigg| \phi_{\text{rec}}
  \left( 
    \sum_{m=1}^M k_b
\left( \sum_{l=1}^{d_X}\phi_{\text{mult}} \left(
  \frac{x_l - X^*_{m,l}}{\epx},
  \frac{x_l - X^*_{m,l}}{\epx}
  \right)\right)
  \right)\\
  &\qquad\qquad\qquad\qquad\qquad-\frac{1}{{ \sum_{m=1}^Mk_b
\left( \sum_{l=1}^{d_X}\phi_{\text{mult}} \left(
  \frac{x_l - X^*_{m,l}}{\epx},
  \frac{x_l - X^*_{m,l}}{\epx}
  \right)\right) }}\Bigg|\\
  &    \leq C_T  (C_X+\frac{1}{2})\cdot\sup_{x\in [1/2,C_X+1/2]} \big|\phi_{\text{rec}}(x)-\frac{1}{x}\big| \leq \frac{\varepsilon_T}{6}.
\end{align*}
Finally, for the third part in~\eqref{eqn:threedecom}, note that 
\begin{equation*}
 \frac{2}{2C_X+1}\leq \frac{1}{
\sum_{m=1}^M k_b
\left( \sum_{l=1}^{d_X}\phi_{\text{mult}} \left(
  \frac{x_l - X^*_{m,l}}{\epx},
  \frac{x_l - X^*_{m,l}}{\epx}
  \right)\right) }\leq 2.
\end{equation*}
So when  $c_1$ is small enough, 
\begin{equation*}
 \frac{1}{2C_X+1}\leq \phi_{\text{rec}}
\left(\sum_{m=1}^M k_b
\left( \sum_{l=1}^{d_X}\phi_{\text{mult}} \left(
  \frac{x_l - X^*_{m,l}}{\epx},
  \frac{x_l - X^*_{m,l}}{\epx}
  \right)\right)\right) \leq 3.
\end{equation*}
Moreover, for any $i\in [k]$,
\begin{equation*}
    -C_T(C_X+\frac{1}{2})\leq  \sum_{m=1}^M
  t_i(X_m^*)
 k_b
\left( \sum_{l=1}^{d_X}\phi_{\text{mult}} \left(
  \frac{x_l - X^*_{m,l}}{\epx},
  \frac{x_l - X^*_{m,l}}{\epx}
  \right)\right)\leq C_T(C_X+\frac{1}{2}).
\end{equation*}
We have 
\begin{small}
\begin{equation*}
   \begin{aligned}
       &\Bigg\|
  \phi_{\text{mult}} 
  \left[ 
  \sum_{m=1}^M
  t(X_m^*)
 k_b
\left( \sum_{l=1}^{d_X}\phi_{\text{mult}} \left(
  \frac{x_l - X^*_{m,l}}{\epx},
  \frac{x_l - X^*_{m,l}}{\epx}
  \right)\right)
  ,
  \phi_{\text{rec}} 
  \left( 
  \sum_{m=1}^M k_b
\left( \sum_{l=1}^{d_X}\phi_{\text{mult}} \left(
  \frac{x_l - X^*_{m,l}}{\epx},
  \frac{x_l - X^*_{m,l}}{\epx}
  \right)\right)
  \right)
  \right]
    \\
    & \qquad -
    \sum_{m=1}^M
 t(X^*_m)
  k_b
\left( \sum_{l=1}^{d_X}\phi_{\text{mult}} \left(
  \frac{x_l - X^*_{m,l}}{\epx},
  \frac{x_l - X^*_{m,l}}{\epx}
  \right)\right) 
  \cdot
  \phi_{\text{rec}}
  \left( 
    \sum_{m=1}^M k_b
\left( \sum_{l=1}^{d_X}\phi_{\text{mult}} \left(
  \frac{x_l - X^*_{m,l}}{\epx},
  \frac{x_l - X^*_{m,l}}{\epx}
  \right)\right) 
  \right)
\Bigg\|\\
&\leq \sqrt{p}\cdot \sup_{x,y\in [-C_1\varepsilon_T^{-1},C_1\varepsilon_T^{-1}]} \big|\phi_{\rm mult}(x,y)-xy\big|\\
&\leq \frac{\varepsilon_T}{6}.
   \end{aligned}
\end{equation*}
\end{small}
So by combining all pieces, it holds that $ \sup_{X \in \mathcal{X}} \|t(X) -s^*_{1:k}(X)\|\leq \varepsilon_T$, and $$\mmd^2_h(\wt\nu(\cdot|T(\Xn)),\wt\nu(\cdot|S^*_{1:k}(\Xn)))\leq\frac{\varepsilon}{32C}.$$ Finally, by Lemmas F.1 and F.3 of~\cite{oko2023diffusion} about the Concatenation and Parallelization of neural networks,  there exist $L_1,U_1,R_1$ with $L_1=\m O(\log^2 (\frac{n}{\min(\varepsilon,1) h}))$ and $R_1=\m O\big(n^{d_X+1} h^{-dd_X}\varepsilon^{-d_X/2}\cdot\log (\frac{n}{\min(\varepsilon,1) h})\cdot(1+h\sqrt{\log(1+\frac{1}{\varepsilon})}\,)^{d_Xd/2}\big)$ so that for any $\Pi_{\theta}\m P_{\theta}^{(n)}$ satisfying Assumption~\ref{assump:exponential}, the corresponding $S^*(\Xn)=(\sum_{i=1}^n s^*_{1:k}(X_i),0_{p-k})\in \Phi_{nd_X}^p(L_1,U_1,R_1)$. This completes the first part of the proof.

\medskip

\noindent Then, we will show that there exists $W^*\in \ms W$ so that 
\begin{equation*}
    \mmd^2_h(\wt \nu(\cdot|S^*_{1:k}(\Xn)), \mu_{W^*}(\cdot|S^*(\Xn)))\leq \frac{\varepsilon}{32C},
\end{equation*}
where we recall that
   \begin{align*}
\mu_{W^*}(\cdot|S^*(\Xn)) &= \sum_{j=1}^J\frac{\exp(W^*(S^*(\Xn),\theta_j^*))}{\sum_{k=1}^J \exp(W^*(S^*(\Xn),\theta_k^*))}\delta_{\theta_j^*}(\cdot),
\end{align*}
Similar to the approach for bounding $\mmd^2_h(\wt\nu(\cdot|T(\Xn)),\wt\nu(\cdot|S^*_{1:k}(\Xn)))$,  we can show that when $|W^*(S^*(\Xn),\theta_j^*)-\wt v_j(S^*_{1:k}(\Xn))|\leq 1$ holds for any $j\in [J]$, it holds for a constant $C_1$ that
\begin{equation*}
    \begin{aligned}
 &\mmd^2_h(\mu_{W^*}(\cdot|S^*(\Xn)),\wt\nu(\cdot|S^*_{1:k}(\Xn))) \\
 &\leq  C_1\,  \Big(1+h\sqrt{ \log(1+\frac{1}{\varepsilon})}\Big)^d\cdot  h^{-2d}\cdot \max_{j \in [J]} | W^*(S^*(\Xn),\theta_j^*)-\wt v_j(S^*_{1:k}(\Xn))|^2+\frac{\varepsilon}{64C}.
    \end{aligned}
\end{equation*}
Then for $j\in [J]$, denote the constant $a_j=\log\left(\varepsilon_\theta^{-d}\int_{\Theta} \rho_j(\theta)\dd\theta\right)+\log \pi(\theta_j^*)-nA(\theta_j^*)$ and constant vector $b_j=\eta(\theta_j^*)$; we define a map $W_A^*:\mb R^{p+d}\to \mb R$ as 
\begin{equation*}
\begin{aligned}
W_A^*(S,\theta)={\rm clip}\Bigg(&\sum_{j=1}^J
 \big(a_j+b_j^T\Pi_{1:k}(S)\big)
 k_b\Big(\frac{2\|\theta-\theta_j^*\|^2}{\varepsilon_\theta^2}\Big)\\
&-\max_{j\in[J]}(a_j+b_j^T\Pi_{1:k}(S));
-c_1\log\Big(\frac{n}{\min(\varepsilon,1)h}\Big),c_2\Bigg).
\end{aligned}
\end{equation*}
 The common factor $\varepsilon_\theta^{-d}$ does not change the centered logits. Packing and the cube geometry give $c\varepsilon_\theta^d\leq\int_\Theta\rho_j(\theta)\dd\theta\leq C\varepsilon_\theta^d$. Then since $\{\theta_j^*\}_{j=1}^J$ is an $\varepsilon_\theta$-packing set of $\Theta$, it holds that $W_A^*(S^*(\Xn),\theta^*_j)= \wt v_j(S^*_{1:k}(\Xn))$.  Moreover, notice that for any $\Xn\in \m X^{n}$, $\|T(\Xn)\|\leq n\cdot C_T$ and  $ \sup_{\Xn \in \mathcal{X}^{n}} \|T(\Xn) -S^*_{1:k}(\Xn)\|\leq n\varepsilon_T$, we have $\|S^*_{1:k}(\Xn)\|\leq 2nC_T$ and there exists a constant $C_1$ so that for any $j\in [J]$ and $\Xn \in  \m X^{n}$,  $|a_j+b_j^TS^*_{1:k}(\Xn)|\leq C_1n$. Set
\[
\delta=\min\left\{1,\frac{h^d\sqrt{\varepsilon}}{8\sqrt{C_1C}}
\Big(1+h\sqrt{\log(1+1/\varepsilon)}\Big)^{-d/2}\right\}.
\]
Here $C_1$ is enlarged, if necessary, to also dominate the constant in the preceding MMD bound.
By Lemma F.6 of~\cite{oko2023diffusion}, 
 there exists  $\ov\phi_{\text{mult}}\in \Phi_2^1(L_5,U_5,R_5)$ with sparsity bound $R_5$, $L_5=\m O(\log \varepsilon_\theta^{-1})$ and $R_5=\m O(\log \varepsilon_\theta^{-1}) $  so that for any $x_1,x_2\in [-C_1\,n , C_1\,n]$,  $|\ov\phi_{\text{mult}}(x_1,x_2)-x_1x_2|\leq c\delta\varepsilon_\theta^{d+2}/n$ with a sufficiently small $c$. Then consider 
 \begin{equation*}
     \begin{aligned}
          W^*(S,\theta)&=  {\rm clip}\Bigg(\sum_{j=1}^J \ov\phi_{\rm mult}\bigg(\big(a_j+b_j^T\Pi_{1:k}(S)\big), k_b
\Big( \frac{2}{\varepsilon_\theta^2}\sum_{l=1}^{d}\ov\phi_{\text{mult}} \left(
 \theta_l - \theta^*_{j,l},
 \theta_l - \theta^*_{j,l}
  \right)\Big) \bigg)\\
  &\qquad- \max_{j\in [J]}(a_j+b_j^T\Pi_{1:k}(S));-c_1\log(\frac{n}{\min(\varepsilon,1) h}),c_2\Bigg).
     \end{aligned}
 \end{equation*}
It holds that 
\begin{equation*}
    \begin{aligned}
        &|W^*(S^*(\Xn),\theta)-W_A^*(S^*(\Xn),\theta)|\leq \frac{cJ\delta\varepsilon_\theta^{d+2}}{n}+C_2cJ\delta\varepsilon_\theta^d\leq\delta,\qquad \theta\in\Theta.
    \end{aligned}
\end{equation*}
The preceding MMD bound now gives $\mmd_h^2(\mu_{W^*}(\cdot|S^*(\Xn)),\wt\nu(\cdot|S^*_{1:k}(\Xn)))\leq\varepsilon/(32C)$. Furthermore, note that the projection map $\Pi_{1:k}(s)=(s_1,s_2,\cdots,s_k)$ can be  realized exactly using ReLU networks with constant order depth and widths. Moreover, the maximum function can also be realized exactly using ReLU networks.   The basic building block is the pairwise maximum, $\max(Z_1, Z_2)$: 
\begin{align*} \max(Z_1, Z_2) = Z_1 + \text{ReLU}(Z_2 - Z_1). \end{align*} 
This can be realized by a simple ReLU neural network,
with a single hidden layer and constant number of parameters. To compute the maximum over $J$ elements, we can arrange these pairwise operations in a binary tree. 
For example, for $\max(Z_1, Z_2, Z_3, Z_4)$, first compute $\max(Z_1, Z_2)$ and $\max(Z_3, Z_4)$  in parallel, then take the maximum of those results.  This structure yields a network depth $\m O(\log J)$ that scales logarithmically with $J$.  By Lemmas F.1 and F.3 of~\cite{oko2023diffusion} about the Concatenation and Parallelization of neural networks, 
the number of nonzero weights and biases in the complete network that realizes $\max(Z_1, \ldots, Z_J)$ is  $\mathcal{O}(J\log J)$. So
 there exist $L_2,U_2,R_2$ with  $L_2=\m O(\log \frac{n}{\min(\varepsilon,1) h})$ and $$R_2=\m O (J\log \varepsilon_{\theta}^{-1})= \m O\left(\Big(1+h\sqrt{\log(1+\frac{1}{\varepsilon})}\Big)^{\frac{d^2}{2}+d}\varepsilon^{-\frac{d}{2}} h^{-d^2} \max(n^{d},h^{-2d})\log\big(\frac{n}{\min(\varepsilon,1) h}\Big)\right),$$ 
 so that for any $\Pi_{\theta}\m P_{\theta}^{(n)}$ satisfying Assumption~\ref{assump:exponential}, the corresponding $W^*(\cdot)$  satisfies
$$W^*\in \Big\{W(\cdot)={\rm clip}\Big(W'(\cdot);-c_1\log(\frac{n}{\min(\varepsilon,1) h}),c_2\Big)\,:\,W'(\cdot)\,\in  \Phi_{p+d}^{1}(L_2, U_2,R_2)\Big\}.$$    The squared triangle inequality combines the two bounds $\varepsilon/(32C)$ into $\varepsilon/(8C)$, proving the lemma.

\subsection{Proof of Lemma~\ref{lemmaFisher}}
 Note that there exist constants $c_I,C>0$ so that for any $\theta^*\in \mb R^d$ and $\theta\in B(\theta^*,1)$,
\begin{equation*}
\begin{aligned}
      c_I\,\min(1,\|\theta^*-\theta\|^2)  &\leq{\rm KL}(\m P_{\theta^*}, \m P_{\theta})=\mb{E}_{X\sim \m P_{\theta^*}}[l(X,\theta)-l(X,\theta^*)]\\
        &\leq \frac{1}{2}(\theta-\theta^*)^T \mb{E}_{X\sim \m P_{\theta^*}}[{\rm Hess}_{\theta} l(X,\theta^*)](\theta-\theta^*)   \\ 
        & \quad + C\,(1+\|\theta^*\|)^{\alpha_1+\alpha_0\alpha_4}\|\theta-\theta^*\|^3\\
        &=\frac{1}{2}(\theta-\theta^*)^T I(\theta^*)(\theta-\theta^*)+C\,(1+\|\theta^*\|)^{\alpha_1+\alpha_0\alpha_4}\|\theta-\theta^*\|^3. 
\end{aligned}
\end{equation*}
Letting $\|\theta-\theta^*\|\to 0$, we obtain $I(\theta^*)\succeq c_I I_d$.

\subsection{Proof of Lemma \ref{A123 with high probability}}
Using the assumption
\begin{equation*}
    \mb{E}_{X\sim \m P_{\theta^*}}\Big[\exp\Big(\Big(\frac{\|X\|}{c_3(1+\|\theta^*\|)^{\alpha_4}}\Big)^{\alpha_3}\Big)\Big]\leq 2,
\end{equation*}
and Markov's inequality, it holds that 
\begin{equation*}
\begin{aligned}
        & \qquad  \m P_{\theta^*}\Big( \Big(\frac{\|X\|}{c_3(1+\|\theta^*\|)^{\alpha_4}}\Big)^{\alpha_3}\geq t\Big)\leq 2\exp(-t) \\ & \Leftrightarrow   \m P_{\theta^*}\Big( \|X\|\geq c_3 (1+\|\theta^*\|)^{\alpha_4}t^{\frac{1}{\alpha_3}} \Big)\leq   2\exp(-t).
\end{aligned}
\end{equation*}
So using $\wt h\geq c_6 n^{-c_7}$, when $c$ is large enough,  
\begin{equation*}
    \begin{aligned}
        \m P_{\theta^*}^{(n)}(A_0)\geq 1-n\cdot    \m P_{\theta^*}\Big( \|X\|> c\, (1+\|\theta^*\|)^{\alpha_4}(\log n)^{\frac{1}{\alpha_3}}\Big)\geq 1-\frac{\wt h^d}{13n}.
    \end{aligned}
\end{equation*}
 Then we define  $$\ov l(X,\theta)=l(X,\theta)\cdot \mathbf{1}(\|X\|\leq  c\, (1+\|\theta^*\|)^{\alpha_4}(\log n)^{\frac{1}{\alpha_3}}).$$  When $c$ is large enough, for any $\theta^*\in \wt \Theta$, $\theta,\theta'\in \wt\Theta'$,
 \begin{equation*}
     \begin{aligned}
          &\Big| \bE_{X\sim \m P_{\theta^*}}
            \left[
                l(X,\theta)  -  l(X,\theta')
            \right]-\bE_{X\sim \m P_{\theta^*}}
            \left[
                \ov l(X,\theta)  -  \ov l(X,\theta')
            \right]\Big|\\
            &\leq\bE_{X\sim \m P_{\theta^*}}
            \left[
               \Big| l(X,\theta)  -  l(X,\theta')\Big|\cdot  \mathbf{1}(\|X\|>  c\, (1+\|\theta^*\|)^{\alpha_4}(\log n)^{\frac{1}{\alpha_3}})
            \right]\\
            &\leq \sqrt{\bE_{X\sim \m P_{\theta^*}}
            \left[
               \Big( l(X,\theta)  -  l(X,\theta')\Big)^2\right]}\sqrt{ \m P_{\theta^*}\Big( \|X\|> c\, (1+\|\theta^*\|)^{\alpha_4}(\log n)^{\frac{1}{\alpha_3}}\Big)}\\
               &\leq\frac{1}{n^2}.
     \end{aligned}
 \end{equation*}
Similarly, we can also show that for sufficiently large $c$, for any $\theta^*\in \wt \Theta$, $\theta,\theta'\in B(\theta^*,1)$, 
\begin{equation*}
    \begin{aligned}
          &\Big|\bE_{X\sim \m P_{\theta^*}} 
            \left[
               l(X, \theta) - l(X, \theta') - \nabla_{\theta}l(X,\theta')^T(\theta - \theta')
            \right]\\
            &\qquad-\bE_{X\sim \m P_{\theta^*}} 
            \left[
               \ov l(X, \theta) - \ov l(X, \theta') - \nabla_{\theta}\ov l(X,\theta')^T(\theta - \theta')
            \right]\Big|\leq \frac{1}{n^2},
    \end{aligned}
\end{equation*}
\begin{equation*}
    \begin{aligned}
   &\forall j\in \mb N_0^d \text{ with }|j|=2,\,\Big|\mb{E}_{X\sim \m P_{\theta^*}}\big[\frac{\partial^{|j|}l(X,\theta^*)}{\partial \theta^j}\big]-\mb{E}_{X\sim \m P_{\theta^*}}\big[\frac{\partial^{|j|}\ov l(X,\theta^*)}{\partial \theta^j}\big]\Big|\leq n^{-2},
    \end{aligned}
\end{equation*}
\begin{equation*}
    \begin{aligned}
 &\Big\|\mb{E}_{X\sim \m P_{\theta^*}}\big[ \nabla_{\theta}\ov l(X,\theta^*)\nabla_{\theta}\ov l(X,\theta^*)^T\big]-I(\theta^*)\Big\|_{\rm F}\\
  &=\Big\|\mb{E}_{X\sim \m P_{\theta^*}}\big[ \nabla_{\theta}\ov l(X,\theta^*)\nabla_{\theta}\ov l(X,\theta^*)^T\big]-\mb{E}_{X\sim \m P_{\theta^*}}\big[ \nabla_{\theta}l(X,\theta^*)\nabla_{\theta} l(X,\theta^*)^T\big]\Big\|_{\rm F}\leq n^{-2},
    \end{aligned}
\end{equation*}
and 
\begin{equation*}
    \begin{aligned}
         &\|I(\theta^*)^{-1}\mb{E}_{X\sim \m P_{\theta^*}}[\nabla_{\theta}\ov l(X,\theta^*)]\| \\
         &=\|I(\theta^*)^{-1}\mb{E}_{X\sim \m P_{\theta^*}}[\nabla_{\theta}\ov l(X,\theta^*)]-I(\theta^*)^{-1}\mb{E}_{X\sim \m P_{\theta^*}}[\nabla_{\theta}l(X,\theta^*)]\|\\
         &\leq \frac{1}{n^2}.
    \end{aligned}
\end{equation*}
Then we define
\begin{equation*}
    \begin{aligned}
           \ov    A_1 
        &= \bigg\{ (\theta^*,\Xn)\in \wt \Theta\times \m X^{n}:  \, \forall \theta, \theta' \in \wt\Theta',\,  \\ 
        & \qquad \qquad \qquad
        \left| 
            \frac{1}{n}\sum_{i=1}^n
             \left( \ov l(X_i,\theta) -\ov l(X_i,\theta') \right) 
            - 
            \bE_{X\sim \m P_{\theta^*}}
            \left[
                \ov l(X,\theta)  -  \ov l(X,\theta')
            \right]
             \right|  \\ 
        & \qquad \qquad \qquad
        \leq \frac{c}{2} (\log n)^{\frac{1}{2}+\frac{\alpha_1}{\alpha_2}}\cdot(1+\|\theta^*\|)^{\alpha_0\alpha_4}\frac{1}{\sqrt{n}}  \Big(\frac{1}{n}+ \Vert \theta -\theta'\Vert \Big) \bigg\}, 
            \end{aligned}
\end{equation*}
\begin{equation*}
    \begin{aligned}
          \ov A_2 
        &= \bigg\{  
            (\theta^*, \Xn)\in \wt \Theta\times \m X^{n}\,: \,  \forall \theta,\theta' \in B(\theta^*,1), \,  \\ 
            & \qquad \qquad \qquad
            \Big| 
            \frac{1}{n}\sum_{i=1}^n 
            \left(
              \ov l(X_i, \theta) - \ov l(X_i, \theta') 
                - \nabla_{\theta}\ov l(X_i, \theta')^T (\theta - \theta') 
            \right) 
            \\
            &  \qquad \qquad \qquad -
            \bE_{X\sim \m P_{\theta^*}} 
            \left[
               \ov l(X, \theta) -\ov l(X, \theta') - \nabla_{\theta}\ov l(X,\theta')^T(\theta - \theta')
            \right]\Big| 
             \\ 
        & \qquad \qquad \qquad \qquad
             \leq  \frac{c}{2} (\log n)^{\frac{1}{2}}\cdot(1+\|\theta^*\|)^{\alpha_0\alpha_4+\alpha_1}\frac{1}{\sqrt{n}}  \Big(\frac{1}{n}+ \Vert \theta -\theta'\Vert^2 \Big)\bigg\}, 
               \end{aligned}
\end{equation*}
\begin{equation*}
    \begin{aligned}
          \ov  A_3&=\bigg\{(\theta^*,\Xn)\in \wt \Theta\times \m X^{n}\,:\,\forall j\in \mb N_0^d \text{ with }|j|=2,\\
         &\qquad\qquad \,\Big|\frac{1}{n}\sum_{i=1}^n \frac{\partial^{|j|}\ov l(X_i,\theta^*)}{\partial \theta^j}-\mb{E}_{X\sim \m P_{\theta^*}}\big[\frac{\partial^{|j|}\ov l(X,\theta^*)}{\partial \theta^j}\big]\Big|
         \leq \frac{c}{2} 
         \,\sqrt{\frac{\log n}{n}}(1+\|\theta^*\|)^{\alpha_1+\alpha_0\alpha_4}\bigg\},\\
            \end{aligned}
\end{equation*}
\begin{equation*}
    \begin{aligned}
   \ov A_4&=\bigg\{(\theta^*,\Xn)\in \wt \Theta\times \m X^{n}\,:\,
    \\ 
    & \qquad \qquad \qquad
   \Big\|\frac{1}{n}\sum_{i=1}^n \nabla_{\theta} \ov l(X_i,\theta^*)\nabla_{\theta} \ov l(X_i,\theta^*)^T-\mb{E}_{X\sim \m P_{\theta^*}}\big[ \nabla_{\theta}l(X,\theta^*)\nabla_{\theta} l(X,\theta^*)^T\big]\Big\|_{\rm F}\\
&\qquad \qquad \qquad \leq \frac{c}{2}\,\sqrt{\frac{\log n}{n}}\cdot (1+\|\theta^*\|)^{2\alpha_1+2\alpha_0\alpha_4}\bigg\},\\
   \end{aligned}
\end{equation*}
\begin{equation*}
    \begin{aligned}
   \ov A_5&=\bigg\{(\theta^*,\Xn)\in \wt \Theta\times \m X^{n}\,:\,
    \\ 
        & \qquad \qquad  
   \Big\|I(\theta^*)^{-1}\frac{1}{n}\sum_{i=1}^n \nabla_{\theta} \ov l(X_i,\theta^*) -I(\theta^*)^{-1}\mb{E}_{X\sim \m P_{\theta^*}}\big[ \nabla_{\theta}\ov l(X,\theta^*)\big]\Big\|_{\rm F}  
        \leq \frac{c}{2}\,\sqrt{\frac{\log n}{n}}\bigg\},\\
   \end{aligned}
\end{equation*}
\begin{equation*}
    \begin{aligned}
       \ov  A_6 & = \bigg\{ (\theta^*, \Xn)\in \wt \Theta\times \m X^{n}\,:  \\ 
        & \qquad \qquad  
        \int_{\theta \in \wt\Theta'^c} 
        \frac{\pi(\theta)}{\pi(\theta^*)} 
        \exp{\left[ - \sum_{i=1}^n \ov l(X_i,\theta) + \sum_{i=1}^n \ov l(X_i, \theta^*) \right]} \dd\theta \\
        & \qquad \qquad   + 
        \int_{\theta \in \wt\Theta'^c} \left[ \frac{\pi(\theta)}{\pi(\theta^*)} 
        \exp{\left( -\sum_{i=1}^n \ov l(X_i,\theta) + \sum_{i=1}^n \ov l(X_i,\theta^*) \right)} \right]^2 \dd\theta    
        \leq \frac{c}{n^{\frac{d}{2}+1}} \bigg\},\\
    \end{aligned}
\end{equation*}
Then  it holds for $i=1,2,\cdots,6$ that $A_0\cap \ov A_i\subset A_0\cap  A_i$, therefore,
\begin{equation*}
    \m P_{\theta^*}^{(n)}(\wt A)\geq 1-7 \m P_{\theta^*}^{(n)}(A_0^c)-\sum_{i=1}^6 \m P_{\theta^*}^{(n)}(\ov A_i^c).
\end{equation*} 
So it only remains to show that for $i=1,2,\cdots,6$, $\m P_{\theta^*}^{(n)}(\ov A_i)\geq 1- \frac{\wt h^d}{13n}$. Note that by  Assumption~\ref{assump:large_sample}, we can get 
\begin{enumerate}
    \item For any $\theta,\theta'\in \wt\Theta'$, it holds that $\sqrt{\mb{E}_{X\sim \m P_{\theta^*}}\big[\big(\ov l(X,\theta)-\ov l(X,\theta')\big)^2\big]}\lesssim (\log n)^{\frac{\alpha_1}{\alpha_2}}\cdot(1+\|\theta^*\|)^{\alpha_0\alpha_4}\|\theta-\theta'\|$ and for any $X\in \m X$, 
    $|\ov l(X,\theta)-\ov l(X,\theta')|\lesssim (\log n)^{\frac{\alpha_1}{\alpha_2}+\frac{\alpha_0}{\alpha_3}}\cdot(1+\|\theta^*\|)^{\alpha_0\alpha_4}\|\theta-\theta'\|$.
    \item For any $\theta,\theta'\in B(\theta^*,1)$,  
    \begin{equation*}
        \sqrt{\mb{E}_{X\sim \m P_{\theta^*}} \Big[\big(\ov l(X, \theta) -  \ov l(X, \theta')  - \nabla_{\theta}\ov l(X, \theta')^T (\theta -\theta')\big)^2\Big]}\lesssim  (1+\|\theta^*\|)^{\alpha_1+\alpha_0\alpha_4}\cdot\|\theta-\theta'\|^2.
    \end{equation*}
    and for any $X\in \m X$,
    $|\ov l(X, \theta) -  \ov l(X, \theta')  - \nabla_{\theta}\ov l(X, \theta')^T (\theta -\theta')|\lesssim  (\log n)^{\frac{\alpha_0}{\alpha_3}}\cdot(1+\|\theta^*\|)^{\alpha_1+\alpha_0\alpha_4}\cdot\|\theta-\theta'\|^2$.
    \item For any $X\in \m X$, $\theta,\theta'\in \wt \Theta'$, $\big\|\nabla_{\theta} \ov l(X,\theta)-\nabla_{\theta} \ov l(X,\theta')\big\|\lesssim  (\log n)^{\frac{\alpha_1}{\alpha_2}+\frac{\alpha_0}{\alpha_3}}\cdot(1+\|\theta^*\|)^{\alpha_0\alpha_4}\cdot\|\theta-\theta'\|$.
    \item For any $j\in \mb N_0^d$ with $|j|=2$, it holds that 
    \begin{equation*}
        \mb{E}_{X\sim \m P_{\theta^*}}\Big[\Big(\frac{\partial^{|j|}\ov l(X,\theta^*)}{\partial \theta^j}\Big)^2\Big]\lesssim (1+\|\theta^*\|)^{2\alpha_1+2\alpha_0\alpha_4},
    \end{equation*}
    and for any $X\in \m X$, $\Big|\frac{\partial^{|j|}\ov l(X,\theta^*)}{\partial \theta^j}\Big|\lesssim (1+\|\theta^*\|)^{\alpha_1+\alpha_0\alpha_4}  (\log n)^{\frac{\alpha_0}{\alpha_3}}$.
    \item It holds that  $\mb{E}_{X\sim \m P_{\theta^*}}\Big[\|\nabla_{\theta}l(X,\theta^*)\nabla_{\theta}l(X,\theta^*)^T\|^2_{\rm F}\Big]\lesssim (1+\|\theta^*\|)^{4\alpha_1+4\alpha_0\alpha_4} $, and for any $X\in \m X$,  we have $\|\nabla_{\theta}\ov l(X,\theta^*)\nabla_{\theta}\ov l(X,\theta^*)^T\|_{\rm F}\lesssim  (1+\|\theta^*\|)^{2\alpha_1+2\alpha_0\alpha_4}  (\log n)^{\frac{2\alpha_0}{\alpha_3}}$.
\end{enumerate}
For any $(\theta,\theta')\in \wt \Theta'\times \wt \Theta'$, define 
\begin{equation*}
    f(X,\theta,\theta')=\frac{\ov l(X,\theta)-\ov l(X,\theta')}{\|\theta-\theta'\|+\frac{1}{n}}.
\end{equation*}
Then we have 
\begin{equation*}
\begin{aligned}
    &P_{\Xn \sim \m P_{\theta^*}^{(n)}}\bigg( \forall \theta,\theta'\in \wt \Theta'\,,\, \Big| 
            \frac{1}{n}\sum_{i=1}^n
             \left( \ov l(X_i,\theta) -\ov l(X_i,\theta') \right) 
            - 
            \bE_{X\sim \m P_{\theta^*}}
            \left[
                \ov l(X,\theta)  -  \ov l(X,\theta')
            \right]
             \Big|  \\ 
        & \qquad \qquad \qquad
        \leq  \frac{c}{2} (\log n)^{\frac{1}{2}+\frac{\alpha_1}{\alpha_2}}\cdot(1+\|\theta^*\|)^{\alpha_0\alpha_4}\frac{1}{\sqrt{n}}  \Big(\frac{1}{n}+ \Vert \theta -\theta'\Vert \Big) \bigg)\\
        &=  P_{\Xn \sim \m P_{\theta^*}^{(n)}}\bigg( \forall \theta,\theta'\in \wt \Theta'\,,\, \Big| 
            \frac{1}{n}\sum_{i=1}^n
            f(X_i,\theta,\theta')
            - 
            \bE_{X\sim \m P_{\theta^*}}
            \left[f(X,\theta,\theta')
                \right] \Big|   \\
                &\qquad \qquad \qquad
        \leq  \frac{c}{2} (\log n)^{\frac{1}{2}+\frac{\alpha_1}{\alpha_2}}\cdot(1+\|\theta^*\|)^{\alpha_0\alpha_4}\frac{1}{\sqrt{n}}  \bigg).
\end{aligned}
\end{equation*}
Let $\m N_{\theta}$ be the largest $\frac{1}{n^3}$-packing set of $\wt \Theta'\times \wt\Theta'$. It holds that $|\m N_{\theta}|\lesssim n^{6d}(\log n)^{\frac{2d}{\alpha_2}}$. Moreover, since $\sqrt{\mb{E}_{X\sim \m P_{\theta^*}}[    f(X,\theta,\theta')^2]}\lesssim  (\log n)^{\frac{\alpha_1}{\alpha_2}}(1+\|\theta^*\|)^{\alpha_0\alpha_4}$ and $\sup_{X\in \m X}    |f(X,\theta,\theta')|\lesssim (\log n)^{\frac{\alpha_1}{\alpha_2}+\frac{\alpha_0}{\alpha_3}}(1+\|\theta^*\|)^{\alpha_0\alpha_4}$. Using Bernstein's inequality (see for example, Proposition 2.14 of~\cite{wainwright2019high}),  for a sufficiently large $c$, it holds that for any $(\theta,\theta')\in \m N_{\theta}$, 
\begin{equation*}
\begin{aligned}
    &   P_{\Xn \sim \m P_{\theta^*}^{(n)}}\bigg( \Big| 
            \frac{1}{n}\sum_{i=1}^n
            f(X_i,\theta,\theta')
            - 
            \bE_{X\sim \m P_{\theta^*}}
            \left[f(X,\theta,\theta')
                \right] \Big|   
        \leq  \frac{c}{4} (\log n)^{\frac{1}{2}+\frac{\alpha_1}{\alpha_2}}\cdot(1+\|\theta^*\|)^{\alpha_0\alpha_4}\frac{1}{\sqrt{n}}  \bigg)\\
        &\geq 1- \frac{\wt h^d}{13n |\m N_{\theta}|}.
\end{aligned}
\end{equation*}
So by a union bound,  it holds that
\begin{equation*}
\begin{aligned}
        & P_{\Xn \sim \m P_{\theta^*}^{(n)}}\bigg( \forall (\theta,\theta')\in \m N_{\theta}\,,\, \Big| 
            \frac{1}{n}\sum_{i=1}^n
            f(X_i,\theta,\theta')
            - 
            \bE_{X\sim \m P_{\theta^*}}
            \left[f(X,\theta,\theta')
                \right] \Big|
                 \\ 
        & \qquad \qquad \qquad \qquad \qquad
        \leq  \frac{c}{4} (\log n)^{\frac{1}{2}+\frac{\alpha_1}{\alpha_2}}\cdot(1+\|\theta^*\|)^{\alpha_0\alpha_4}\frac{1}{\sqrt{n}}  \bigg)\\
        &\geq 1- |\m N_{\theta}|\cdot \frac{\wt h^d}{13n |\m N_{\theta}|}\geq 1-\frac{\wt h^d}{13n}.
\end{aligned}
\end{equation*}
Moreover, for any $(\ov\theta,\ov\theta')\in \wt\Theta'\times \wt \Theta'$, there exists $(\theta, \theta')\in \m N_{\theta}$, so that $\|(\theta,\theta')-(\ov\theta,\ov \theta')\|\leq \frac{1}{n^3}$ and
\begin{equation*}
\begin{aligned}
   & \bigg|  \Big| 
            \frac{1}{n}\sum_{i=1}^n
            f(X_i,\ov\theta,\ov\theta')
            - 
            \bE_{X\sim \m P_{\theta^*}}
            \left[
           f(X,\ov\theta,\ov\theta')
            \right]\Big|-\Big| 
             \frac{1}{n}\sum_{i=1}^n
            f(X_i, \theta, \theta')
            - 
            \bE_{X\sim \m P_{\theta^*}}
            \left[
           f(X, \theta, \theta')
            \right]
             \Big| \bigg|\\
             &\leq C\, n (\log n)^{\frac{\alpha_0}{\alpha_3}+\frac{\alpha_1+\alpha_0\alpha_4}{\alpha_2}}\|(\theta,\theta')-(\ov\theta,\ov \theta')\|\\
             &\leq \frac{c}{4} (\log n)^{\frac{1}{2}+\frac{\alpha_1}{\alpha_2}}\cdot(1+\|\theta^*\|)^{\alpha_0\alpha_4}\frac{1}{\sqrt{n}}.
                         \end{aligned}
\end{equation*}
Therefore,  it holds that 
\begin{equation*}
    \begin{aligned}
     & P_{\Xn \sim \m P_{\theta^*}^{(n)}}(\ov A_1)=     P_{\Xn \sim \m P_{\theta^*}^{(n)}}\bigg( \forall \theta,\theta'\in \wt \Theta'\,,\, \Big| 
            \frac{1}{n}\sum_{i=1}^n
            f(X_i,\theta,\theta')
            - 
            \bE_{X\sim \m P_{\theta^*}}
            \left[f(X,\theta,\theta')
                \right] \Big|   \\
                &\qquad \qquad \qquad
        \leq  \frac{c}{2} (\log n)^{\frac{1}{2}+\frac{\alpha_1}{\alpha_2}}\cdot(1+\|\theta^*\|)^{\alpha_0\alpha_4}\frac{1}{\sqrt{n}}  \bigg)\\
      &\geq  P_{\Xn \sim \m P_{\theta^*}^{(n)}}\bigg( \forall (\theta,\theta')\in \m N_{\theta}\,,\, \Big| 
            \frac{1}{n}\sum_{i=1}^n
            f(X_i,\theta,\theta')
            - 
            \bE_{X\sim \m P_{\theta^*}}
            \left[f(X,\theta,\theta')
                \right] \Big|   
                 \\ 
        & \qquad \qquad \qquad \qquad \qquad \qquad
        \leq  \frac{c}{4} (\log n)^{\frac{1}{2}+\frac{\alpha_1}{\alpha_2}}\cdot(1+\|\theta^*\|)^{\alpha_0\alpha_4}\frac{1}{\sqrt{n}}  \bigg) \\
        &\geq 1-\frac{\wt h^d}{13n}.
    \end{aligned}
\end{equation*}
Using a similar argument we can obtain the result for $\ov A_2$.  The cases $\ov A_3$ and $\ov A_4$ follow directly from Bernstein's inequality. So we omit the detailed derivation here to avoid redundancy.
For $\ov A_5$, note that $I(\theta^*)^{-1}\preceq c_I^{-1} I_d$ and 
\begin{equation*}
  \Big\| I(\theta^*)^{-1} \mb{E}_{\m P_{\theta^*}}\big[\nabla_{\theta}\ov l(X,\theta^*)\nabla_{\theta}\ov l(X,\theta^*)^T\big]   I(\theta^*)^{-1}-I(\theta^*)^{-1}\Big\|\lesssim \frac{1}{n^2}.
\end{equation*}
So there exists a constant $C$ so that 
\begin{equation*}
    \mb{E}_{\m P_{\theta^*}}\big[\|I(\theta^*)^{-1}\nabla_{\theta}\ov l(X,\theta^*)\|^2\big]\leq C.
\end{equation*}
Moreover, it holds for any $X\in \m X$ that  $ \|I(\theta^*)^{-1} \nabla_{\theta}\ov l(X,\theta^*)\|\lesssim (1+\|\theta^*\|)^{\alpha_1+\alpha_0\alpha_4}(\log n)^{\frac{\alpha_0}{\alpha_3}}$. Therefore, the result for $\overline A_5$ also follows by an application of Bernstein’s inequality.
We next prove the result for $\ov A_6$.
To bound the integrand in $\ov A_6$, we first use the prior’s exponential decay, as well as the definition of $\wt\Theta$ and $\wt{\Theta}'$, to obtain that for any $\theta\in \wt{\Theta}'^c$ and $\theta^*\in \wt \Theta$:
\begin{align*}
    \log \pi(\theta) &\leq 
    \log{c}_2 
    - \frac{1}{c_2} \Vert \theta \Vert^{\alpha_2}
    \leq 
    \log{c}_2 - 
    \frac{1}{2c_2} (r_0+r_1)^{\alpha_2} \log{n}-\frac{1}{2c_2}\|\theta\|^{\alpha_2}, \\
    \log \pi(\theta^*) &\geq - \log c_2 - c_2 \Vert \theta^* \Vert^{\alpha_2} 
    \geq - \log c_2 - c_2 r_0^{\alpha_2} \log{n}.
\end{align*}
Combining these two bounds, when $\frac{r_1}{r_0}$ is large enough, we can get the following inequality for the prior ratio:
\begin{align*}
    \log \frac{\pi(\theta)}{ \pi(\theta^*) } &\leq 2\log c_2 - \frac{1}{2c_2} (r_0 +r_1)^{\alpha_2} \log n + c_2 r_0^{\alpha_2} \log{n} - \frac{1}{2c_2} \Vert \theta \Vert^{\alpha_2} \\
    &\leq - \frac{1}{2c_2} \Vert \theta \Vert^{\alpha_2}.
\end{align*}
To handle the integral over $\wt\Theta'^c$, we decompose it into a union of disjoint annuli.
Let $r=r_0+r_1$. We have:
\begin{align*}
    \wt \Theta'^c = \bigcup_{k=1}^{\infty}  B(0,2^k r (\log{n})^{\frac{1}{\alpha_2}}) \backslash B(0,2^{k-1} r (\log{n})^{\frac{1}{\alpha_2}}),
\end{align*}
Denote 
$\mathcal{C}_k 
= 
B(0,2^k r (\log{n})^{\frac{1}{\alpha_2}}) 
\backslash B(0,2^{k-1} r (\log{n})^{\frac{1}{\alpha_2}})$
, 
$
c' = \frac{1}{2^{\alpha_2 + 1} c_2} 
$
and $ \bar{k} = 2^k r$.
For $ \theta \in \mathcal{C}_k$, we have:
\begin{align*}
    \frac{\pi(\theta)}{\pi(\theta^*)} 
    &\leq \exp [-\frac{1}{2c_2} (2^{k-1}r (\log{n})^{\frac{1}{\alpha_2}})^{\alpha_2}]  = \exp[-\frac{1}{2c_2} (2^{k-1}r)^{\alpha_2} \log{n}] = \exp(-c' \bar{k}^{\alpha_2} \log{n}).
\end{align*}
For $(\theta^*,\Xn) \notin \ov A_6$, using the union bound over $\mathcal{C}_k$,
and the fact $\text{Vol}\left(\mathcal{C}_k\right) \sup_{\theta \in \mathcal{C}_k} f(\theta) \geq \int_{\theta \in \mathcal{C}_k} f(\theta) \dd \theta$ for any non-negative function $f(\theta)$, we can get
\begin{align*}
        & \mathbb{P}_{\Xn\sim \m P_{\theta^*}^{(n)}} ((\theta^*,\Xn) \notin \ov A_6) \\
        & = \mathbb{P}_{\Xn\sim \m P_{\theta^*}^{(n)}}  \Bigg( \int_{\theta \in \wt\Theta'^c} \frac{\pi(\theta)}{\pi(\theta^*)} \exp{\left( - \sum_{i=1}^n \ov  l(X_i,\theta) + \sum_{i=1}^n \ov l(X_i, \theta^*) \right)} d\theta \nonumber \\
        & \qquad + \int_{\theta \in \wt\Theta'^c} \left[ \frac{\pi(\theta)}{\pi(\theta^*)} \exp \left( -\sum_{i=1}^n \ov l(X_i,\theta) + \sum_{i=1}^n \ov l(X_i,\theta^*) \right) \right]^2 d\theta \geq  \frac{c_4}{n^{\frac{d}{2}+1}}  \Bigg)   \\
        &\overset{(i)}{ \leq} \sum_{k=1}^{\infty} \mathbb{P}_{\Xn\sim \m P_{\theta^*}^{(n)}}  \Big( \int_{\theta \in \mathcal{C}_k} \frac{\pi(\theta)}{\pi(\theta^*)} \exp{\left( - \sum_{i=1}^n \ov l(X_i,\theta) + \sum_{i=1}^n \ov l(X_i, \theta^*) \right)} d\theta  \nonumber \\
        & \qquad + \int_{\theta \in \mathcal{C}_k} \left[ \frac{\pi(\theta)}{\pi(\theta^*)} \exp\left( -\sum_{i=1}^n \ov l(X_i,\theta) + \sum_{i=1}^n \ov l(X_i,\theta^*) \right) \right]^2 d\theta \geq  \frac{c_4}{2^k n^{\frac{d}{2}+1}}  \Big) \nonumber \\
        & \leq \sum_{k=1}^{\infty} \mathbb{P}_{\Xn\sim \m P_{\theta^*}^{(n)}}  \Bigg(  \text{Vol}(\mathcal{C}_k)\sup_{\theta \in \mathcal{C}_k}  \left[\frac{\pi(\theta)}{\pi(\theta^*)} \exp{\left( - \sum_{i=1}^n \ov l(X_i,\theta) + \sum_{i=1}^n \ov l(X_i, \theta^*) \right)} \right] \\
        & \qquad + \text{Vol}(\mathcal{C}_k) \sup_{\theta \in \mathcal{C}_k} \left[ \frac{\pi(\theta)}{\pi(\theta^*)} \exp\left( -\sum_{i=1}^n \ov l(X_i,\theta) + \sum_{i=1}^n \ov l(X_i,\theta^*) \right) \right]^2  \geq  \frac{c_4}{2^k n^{\frac{d}{2}+1}}  \Bigg),
\end{align*}
where  $(i)$ uses 
\begin{equation*}
    \begin{aligned}
        &\bigcap_{k=1}^{\infty } \bigg\{\int_{\theta \in \mathcal{C}_k} \frac{\pi(\theta)}{\pi(\theta^*)} \exp{\left( - \sum_{i=1}^n \ov l(X_i,\theta) + \sum_{i=1}^n \ov l(X_i, \theta^*) \right)} d\theta  \nonumber \\
        & \qquad + \int_{\theta \in \mathcal{C}_k} \left[ \frac{\pi(\theta)}{\pi(\theta^*)} \exp\left( -\sum_{i=1}^n \ov l(X_i,\theta) + \sum_{i=1}^n \ov l(X_i,\theta^*) \right) \right]^2 d\theta \leq  \frac{c_4}{2^k n^{\frac{d}{2}+1}} \bigg\}\\
        &\subseteq \bigg\{\sum_{k=1}^{\infty} \bigg(\int_{\theta \in \mathcal{C}_k} \frac{\pi(\theta)}{\pi(\theta^*)} \exp{\left( - \sum_{i=1}^n \ov l(X_i,\theta) + \sum_{i=1}^n \ov l(X_i, \theta^*) \right)} d\theta  \nonumber \\
        & \qquad + \int_{\theta \in \mathcal{C}_k} \left[ \frac{\pi(\theta)}{\pi(\theta^*)} \exp\left( -\sum_{i=1}^n \ov l(X_i,\theta) + \sum_{i=1}^n \ov l(X_i,\theta^*) \right) \right]^2 d\theta \bigg)\leq   \sum_{k=1}^{\infty}\frac{c_4}{2^k n^{\frac{d}{2}+1}}= \frac{c_4}{  n^{\frac{d}{2}+1}}\bigg\}.\\
    \end{aligned}
\end{equation*}
Then using  
   $ \frac{\pi(\theta)}{\pi(\theta^*)} 
    \leq  \exp(-c' \bar{k}^{\alpha_2} \log{n})$, we have for sufficiently large $r$,
    \begin{small}

 \begin{align*}
        & \mathbb{P}_{\Xn\sim \m P_{\theta^*}^{(n)}} ((\theta^*,\Xn) \notin \ov A_6) \\
         & \leq \sum_{k=1}^{\infty} \mathbb{P}_{\Xn\sim \m P_{\theta^*}^{(n)}} \Bigg( \text{Vol}(\mathcal{C}_k) \exp(-c' \bar{k}^{\alpha_2} \log{n}) \cdot \sup_{\theta \in \mathcal{C}_k}\exp{\left( - \sum_{i=1}^n \ov l(X_i,\theta) + \sum_{i=1}^n \ov  l(X_i, \theta^*) \right)}   \\
        & \qquad  + \text{Vol}(\mathcal{C}_k) \exp(-2c' \bar{k}^{\alpha_2} \log{n}) \cdot \sup_{\theta \in \mathcal{C}_k}  \exp\left( - 2 \sum_{i=1}^n \ov  l(X_i,\theta) + 2 \sum_{i=1}^n \ov l(X_i,\theta^*) \right)  \geq  \frac{c_4}{2^k n^{\frac{d}{2}+1}}  \Bigg) \nonumber \\
        & \leq \sum_{k=1}^{\infty} \mathbb{P}_{\Xn\sim \m P_{\theta^*}^{(n)}} \Bigg(  \text{Vol}(\mathcal{C}_k) \sup_{\theta \in \mathcal{C}_k}\exp{\left( - \sum_{i=1}^n \ov l(X_i,\theta) +\sum_{i=1}^n \ov l(X_i, \theta^*)  - c' \bar{k}^{\alpha_2} \log{n} \right)}
          \geq \frac{c_4}{2^{k+1} n^{\frac{d}{2}+1}} \Bigg)  \\ 
        & \leq \sum_{k=1}^{\infty} \mathbb{P}_{\Xn\sim \m P_{\theta^*}^{(n)}}  \Big( \sup_{\theta \in \mathcal{C}_k}   \exp \Big(-  \sum_{i=1}^{n} \ov l(X_i,\theta) +\sum_{i=1}^n \ov l(X_i, \theta^*) - c'\bar{k}^{\alpha_2}\log n \Big)  \\ 
        & \qquad \qquad \qquad \geq \frac{c_5}{2^{k(d+1)} r^dn^{\frac{d}{2}+1} (\log n)^{\frac{d}{\alpha_2}}} \Big) \nonumber \\
        & \leq \sum_{k=1}^{\infty} \mathbb{P}_{\Xn\sim \m P_{\theta^*}^{(n)}}  \Big( \sup_{\theta \in \mathcal{C}_k}  \Big[ -  \sum_{i=1}^{n} \ov l(X_i,\theta) + \sum_{i=1}^n \ov l(X_i, \theta^*) -c'\bar{k}^{\alpha_2}\log n   \Big
        ]  \geq \log \Big( \frac{c_5}{2^{k(d+1)} r^dn^{\frac{d}{2}+1} (\log n)^{\frac{d}{\alpha_2}}} \Big) \Big) \nonumber \\
        & = \sum_{k=1}^{\infty} \mathbb{P}_{\Xn\sim \m P_{\theta^*}^{(n)}}  \Big( \sup_{\theta \in \mathcal{C}_k}  \Big[ - \frac{1}{n} \sum_{i=1}^n \ov l(X_i, \theta) + \frac{1}{n} \sum_{i=1}^n\ov l(X_i,\theta^*) \Big]  \\
        &\qquad\qquad\qquad\qquad\qquad\geq  \frac{1}{n} \log \left( \frac{c_5}{2^{k(d+1)} r^dn^{\frac{d}{2}+1} (\log n)^{\frac{d}{\alpha_2}}} \right) + \frac{1}{n} c' 2^{k\alpha_2}r^{\alpha_2}\log n  \Big)   \\
        & \leq \sum_{k=1}^{\infty} \mathbb{P}_{\Xn\sim \m P_{\theta^*}^{(n)}}  \Big( \sup_{\theta \in \mathcal{C}_k}  \Big[ - \frac{1}{n} \sum_{i=1}^n \ov l(X_i, \theta) + \frac{1}{n} \sum_{i=1}^n \ov  l(X_i,\theta^*) \Big] \geq   \frac{c'}{2n} \bar{k}^{\alpha_2}\log n  \Big).
\end{align*}
   \end{small}
Note that there exists a constant $C$ so that for all $\theta \in \mathcal{C}_k$:
\begin{align*}
    \left| \ov l(X, \theta) - \ov l(X,\theta^*) \right| &\leq C_1\,(\bar{k} (\log n)^{\frac{1}{\alpha_2}})^{\alpha_1} (\log n)^{\frac{\alpha_0}{\alpha_3}}\cdot(1+\|\theta^*\|)^{\alpha_0\alpha_4}
    \Vert \theta - \theta^* \Vert\\
   & \leq C\, \,(\bar{k} (\log n)^{\frac{1}{\alpha_2}})^{\alpha_1} (\log n)^{\frac{\alpha_0}{\alpha_3}+\frac{\alpha_0\alpha_4}{\alpha_2}}
    \Vert \theta - \theta^* \Vert, \\
    |\ov l(X,\theta)| & \leq C\, \,(\bar{k}(\log n)^{\frac{1}{\alpha_2}})^{\alpha_1}(\log n)^{\frac{\alpha_0}{\alpha_3}+\frac{\alpha_0\alpha_4}{\alpha_2}}.
\end{align*}
To bound the supremum of the empirical loss difference over $\mathcal{C}_k$, 
we first define a function class for $\theta \in \mathcal{C}_k$.
\begin{align*}
    \mathcal{F}_k = \left\{ f(x) = \ov l(x,\theta) - \ov l(x,\theta^*) : \theta \in \mathcal{C}_k \right\}
\end{align*}
For any $f,f' \in \mathcal{F}_k$ with corresponding parameters $\theta, \theta' \in \mathcal{C}_k$,
the empirical distance between $f$ and $f'$ can be bounded as:
\begin{align*}
    d_n^{\mathcal{F}}(f, f')  &= 
    \sqrt{\frac{1}{n} \sum_{i=1}^n [f(X_i) - f'(X_i)]^2} 
    \leq 2\sup_{\theta\in \m C_k\atop X\in \m X}|\ov l(X,\theta)|  \\ 
    & \leq  2C\,(\bar{k}(\log n)^{\frac{1}{\alpha_2}})^{\alpha_1} (\log n)^{\frac{\alpha_0}{\alpha_3}+\frac{\alpha_0\alpha_4}{\alpha_2}},
\end{align*}
and
\begin{align*}
    d_n^{\mathcal{F}}(f, f') 
    \leq \sup_{X\in \m X} |\ov l(X,\theta)-\ov l(X,\theta')|\leq
    C
    \bar{k}^{\alpha_1} 
 (\log n)^{\frac{\alpha_0}{\alpha_3}+\frac{\alpha_0\alpha_4+\alpha_1}{\alpha_2}} \Vert \theta - \theta'\Vert.
\end{align*}
The metric entropy of $\mathcal{F}_k$ can be bounded by:
\begin{align*}
    \log N(\mathcal{F}_k, d_n^{\mathcal{F}}, \varepsilon) &\leq \log N(\mathcal{C}_k, \Vert \cdot \Vert_2 , \frac{\varepsilon}{ C\,(\bar{k} (\log n)^{\frac{1}{\alpha_2}})^{\alpha_1} (\log n)^{\frac{\alpha_0}{\alpha_3}+\frac{\alpha_0\alpha_4}{\alpha_2}}})\\
    & \leq  
d \log\left(
      \frac{3C(\bar{k} (\log n)^{\frac{1}{\alpha_2}})^{\alpha_1+1}(\log n)^{\frac{\alpha_0}{\alpha_3}+\frac{\alpha_0\alpha_4}{\alpha_2}}}{\varepsilon} 
      \right).
\end{align*}
So by Dudley's inequality, it holds that
\begin{align*}
&\mb{E}_{\Xn\sim \m P_{\theta^*}^{(n)}}\bigg[\sup_{\theta\in \m C_k}\Big(-\frac{1}{n}\sum_{i=1}^n \ov l(X_i,\theta)+\frac{1}{n}\sum_{i=1}^n \ov l(X_i,\theta^*)+\mb{E}_{X\sim \m P_{\theta^*}}[\ov l(X,\theta)]-\mb{E}_{X\sim \m P_{\theta^*}}[\ov l(X,\theta^*)]\Big)\bigg]\\
  &\lesssim\int_{0}^{\infty} \sqrt{\frac{\log N(\mathcal{F}_k, d_n^{\mathcal{F}}, \varepsilon)}{n}} d\varepsilon =  \int_{0}^{2C\,(\bar{k}(\log n)^{\frac{1}{\alpha_2}})^{\alpha_1}(\log n)^{\frac{\alpha_0}{\alpha_3}+\frac{\alpha_0\alpha_4}{\alpha_2}}} \sqrt{\frac{\log N(\mathcal{F}_k, d_n^{\mathcal{F}}, \varepsilon)}{n}} d\varepsilon\\
  &\lesssim \frac{1}{\sqrt{n}} (\sqrt{\log \bar{k}} + \sqrt{\log \log{n}})\cdot (\bar{k} (\log n)^{\frac{1}{\alpha_2}})^{\alpha_1}\cdot (\log n)^{\frac{\alpha_0}{\alpha_3}+\frac{\alpha_0\alpha_4}{\alpha_2}}.
\end{align*}
Using the concentration inequality for empirical processes (see for example, Theorem 3.27 of~\cite{wainwright2019high}),
there exist constants $c_1,c_2$ so that  for any $\delta>0$,
\begin{equation*}
\begin{aligned}
   & \mathbb{P}_{\Xn\sim \m P_{\theta^*}^{(n)}} \bigg(\sup_{\theta\in \m C_k}\Big(-\frac{1}{n}\sum_{i=1}^n \ov l(X_i,\theta)+\frac{1}{n}\sum_{i=1}^n \ov l(X_i,\theta^*)+\mb{E}_{X\sim \m P_{\theta^*}}[\ov l(X,\theta)]-\mb{E}_{X\sim \m P_{\theta^*}}[\ov l(X,\theta^*)]\Big)\\
   &\qquad\geq \big(\bar{k} (\log n)^{\frac{1}{\alpha_2}}\big)^{\alpha_1}\cdot (\log n)^{\frac{\alpha_0}{\alpha_3}+\frac{\alpha_0\alpha_4}{\alpha_2}}\cdot \Big(  \frac{c_1\delta}{n} +c_2\frac{1+\sqrt{\delta}}{\sqrt n}(1+\sqrt{\log \bar{k}} + \sqrt{\log \log{n}})\Big)\bigg) \\ &\leq \exp(-\delta).
\end{aligned}
\end{equation*}
Moreover, the KL lower bound gives
$\bE_{X\sim\m P_{\theta^*}}[l(X,\theta)-l(X,\theta^*)]\geq c_3$
for $\theta\in\mathcal C_k$. By increasing the truncation constant in $A_0$, the tail and growth assumptions imply, uniformly over these shells,
\[
\left|\bE_{X\sim\m P_{\theta^*}}[(l-\ov l)(X,\theta)-(l-\ov l)(X,\theta^*)]\right|
\leq C\bar k^{\alpha_1}(\log n)^C n^{-3}
\leq \frac{c'\bar k^{\alpha_2}\log n}{4n},
\]
where the last inequality uses $\alpha_2>\alpha_1$. Thus we bound the probability as:
\begin{align*}
 &\mathbb{P}_{\Xn\sim \m P_{\theta^*}^{(n)}} \Big( \sup_{\theta \in \mathcal{C}_k}  \left[ - \frac{1}{n} \sum_{i=1}^n \ov l(X_i, \theta) + \frac{1}{n} \sum_{i=1}^n \ov l(X_i, \theta^*) \right] \geq   \frac{c'\bar{k}^{\alpha_2} \log n}{2n}  \Big)\\
&\leq
 \mathbb{P}_{\Xn\sim \m P_{\theta^*}^{(n)}} \bigg(\sup_{\theta\in \m C_k}\Big(-\frac{1}{n}\sum_{i=1}^n \ov l(X_i,\theta)+\frac{1}{n}\sum_{i=1}^n \ov l(X_i,\theta^*)+\mb{E}_{X\sim \m P_{\theta^*}}[\ov l(X,\theta)]-\mb{E}_{X\sim \m P_{\theta^*}}[\ov l(X,\theta^*)]\Big)\\
   &\qquad\geq \frac{c'\bar{k}^{\alpha_2} \log n}{4n}+c_3  \bigg).
\end{align*}
Then choose $\delta = \bar{k}^{\frac{\alpha_2 - 2\alpha_1}{2}} \log n$, where recall $\bar k=2^k r$ and $\alpha_2>2\alpha_1$. We can verify that when $r$ and $n$ are large enough,
\begin{equation*}
    \big(\bar{k} (\log n)^{\frac{1}{\alpha_2}}\big)^{\alpha_1}\cdot (\log n)^{\frac{\alpha_0}{\alpha_3}+\frac{\alpha_0\alpha_4}{\alpha_2}}\cdot \Big(  \frac{c_1\delta}{n} +c_2\frac{1+\sqrt{\delta}}{\sqrt n}(1+\sqrt{\log \bar{k}} + \sqrt{\log \log{n}})\Big)\leq \frac{c'\bar{k}^{\alpha_2} \log n}{4n}+c_3.
\end{equation*}
Therefore,
\begin{equation*}
\begin{aligned}
        &\mathbb{P}_{\Xn\sim \m P_{\theta^*}^{(n)}} \Big( \sup_{\theta \in \mathcal{C}_k}  \left[ - \frac{1}{n} \sum_{i=1}^n\ov  l(X_i, \theta) + \frac{1}{n} \sum_{i=1}^n \ov l(X_i, \theta^*) \right] \geq   \frac{c'\bar{k}^{\alpha_2} \log n}{2n}  \Big)\leq \exp(- \bar{k}^{\frac{\alpha_2 - 2\alpha_1}{2}} \log n)\\
        &=\exp(- 2^{\frac{k(\alpha_2 - 2\alpha_1)}{2}} r^{\frac{\alpha_2-2\alpha_1}{2}} \log n)
\end{aligned}
\end{equation*}
Since $\alpha_2 > 2\alpha_1$ and $\wt h\geq c_6 n^{-c_7}$, summing over $k \geq 1$, we can get, when $r$ is large enough,
\begin{align*}
    &          \mathbb{P}_{\Xn\sim \m P_{\theta^*}^{(n)}} ((\theta^*,\Xn) \notin \ov A_6)  \leq \sum_{k=1}^{\infty} \exp(- 2^{\frac{k(\alpha_2 - 2\alpha_1)}{2}} r^{\frac{\alpha_2-2\alpha_1}{2}} \log n)\leq \frac{\wt h^d}{13n}.
\end{align*}

\subsection{Proof of Lemma \ref{lemma: lst gaussian approximation}}
Notice that for two distributions on $\mb R^d$ with densities $p(\theta)$ and $q(\theta)$, it holds that 
\begin{align*}
\mmd^2_{\wt h}(p, q) & =
\int_{\mb R^d} \big(\mb{E}_{\theta\sim p}\big[\kappa_{\wt h/\sqrt{2}}(\theta,y)\big]-\mb{E}_{\theta\sim q}\big[\kappa_{\wt h/\sqrt{2}}(\theta,y)\big]\big)^2\,\dd y\\
&=\int_{\mb R^d} \Big(\int_{\mb R^d} \kappa_{\wt h/\sqrt{2}}(\theta,y)(p(\theta)-q(\theta))\,\dd\theta\Big)^2\,\dd y\\
&=\int_{\mb R^d}\left[\mathbb{E}_{\theta \sim N\left(y, \frac{\wt h^2}{2} I_d\right)}(p(\theta)-q(\theta))\right]^2\, \dd y \\
&\leq\int_{\mb R^d} \mathbb{E}_{\theta \sim N\left(y, \frac{\wt h^2}{2} I_d\right)}\left[(p(\theta)-q(\theta))^2\right] \,\dd y \\
& = \int_{\mb R^d}(p(\theta)-q(\theta))^2 \cdot\left(\int_{\mb R^d} \kappa_{\wt h/\sqrt{2}}(\theta,y) \,\dd y\right) \,\dd \theta \\
& = \int_{\mb R^d}(p(\theta)-q(\theta))^2\,\dd\theta:=\|p-q\|^2_{L_2}.
\end{align*}
Therefore, we can bound
\begin{equation*}
    \begin{aligned}
          & \mathbb{E}_{\theta^* \sim \Pi_{\theta}}
\bE_{\Xn\sim \m P_{\theta^*}^{(n)}}
\Big[
\mmd^2_{\wt h}
\left(
\big[\sqrt{n}(\cdot -\wh{\theta}(\Xn))\big]_{\#} \pos
    ,
    N(0, I^{-1}(\wh{\theta}(\Xn)))
\right)  \\ 
& \qquad \qquad \qquad \qquad
\cdot \mathbf{1}((\theta^*,\Xn) \in \wt A) 
\Big] \\
&\leq  \mathbb{E}_{\theta^* \sim \Pi_{\theta}}
\bE_{\Xn\sim \m P_{\theta^*}^{(n)}}
\Big[  \big\|  \big[\sqrt{n}(\cdot -\wh{\theta}(\Xn))\big]_{\#} \pos- N(0, I^{-1}(\wh{\theta}(\Xn)))\big\|_{L_2}^2  \\ 
& \qquad \qquad \qquad \qquad
\cdot
\mathbf{1}((\theta^*,\Xn) \in \wt A) 
\Big].
    \end{aligned}
\end{equation*}
We first use the following lemma to bound the distance between $\wh\theta(\Xn)$ and $\theta^*$.
\begin{lemma}\label{lemmathetahat}
There exists a positive constant $C_{\theta}$ so that for any $(\theta^*,\Xn)\in \wt A$,  it holds that $\|\theta^*-\wh\theta(\Xn)\|\leq C_{\theta}\sqrt{\frac{\log n}{n}}$.
 \end{lemma}
Then fix $(\theta^*,\Xn)\in \wt A$. For simplicity, we write $\wh\theta=\wh\theta(\Xn)$. When $n$ is large enough, $\wh\theta$ is an interior point of $\wt\Theta'$ and $n^{-1}\sum_{i=1}^n \nabla_{\theta} l(X_i,\wh\theta)=0$.  Under the change of variables $z = \sqrt{n}(\theta - \wh{\theta})$, we have
\begin{align*}
        & \quad
        \Big\Vert
       \big[\sqrt{n}(\cdot -\wh{\theta}(\Xn))\big]_{\#} \pos
        -
        N(0, I^{-1}(\wh{\theta}))
        \Big\Vert^2_{L_2} 
        \\
        & = 
        \int
        \left|
            \frac{
                \pi(\wh{\theta}+z/\sqrt{n}) \exp{\left( -\sum_{i=1}^{n} l(X_i, \wh{\theta}+ z/\sqrt{n}) \right)}}{
                \int \pi(\wh{\theta}+z/\sqrt{n}) \exp{\left( -\sum_{i=1}^{n} l(X_i, \wh{\theta}+ z/\sqrt{n}) \right)} dz
                }
            -
            \frac{
                \exp{\left( -\frac{1}{2} z^T I(\wh{\theta}) z \right)}
            }{
                \int \exp{\left( -\frac{1}{2} z^T I(\wh{\theta}) z \right)} dz
            }
        \right|^2 dz 
        \\
        & =
        \int 
        \Big| 
            \frac{ 
                \frac{\pi(\wh\theta+z/\sqrt{n})}{\pi(\theta^*)} \exp{ \left( \sum_{i=1}^n l(X_i,\wh\theta)-\sum_{i=1}^nl(X_i,\wh\theta +z/\sqrt{n}) \right)} 
                }{ 
                    \int \frac{\pi(\wh\theta+z/\sqrt{n})}{\pi(\theta^*)} \exp{ \left(\sum_{i=1}^n l(X_i,\wh\theta) -\sum_{i=1}^nl(X_i,\wh\theta +z/\sqrt{n}) \right)}dz 
                }   
             \\ 
            & \qquad \qquad
            - \frac{ \exp{\left( -\frac{1}{2} z^T I(\wh\theta)z \right)} 
                }{\int  \exp{\left(-\frac{1}{2} z^T I(\wh\theta)z \right)} dz  
                } 
        \Big|^2 dz,
\end{align*}
To simplify the notation, denote
\begin{align*}
A &= \frac{\pi(\wh\theta+z/\sqrt{n})}{\pi(\theta^*)} \exp{ \left( \sum_{i=1}^n l(X_i,\wh\theta)-\sum_{i=1}^n l(X_i,\wh\theta +z/\sqrt{n}) \right)}, \\
B &= \exp{\left( -\frac{1}{2} z^T I(\wh\theta)z \right)},
\end{align*}
Then the squared $L_2$ distance can be rewritten as
\begin{align*}
    &\Big\Vert
      \big[\sqrt{n}(\cdot -\wh{\theta}(\Xn))\big]_{\#} \pos
    -
    N(0, I^{-1}(\wh{\theta}))
    \Big\Vert^2_{L_2} \\
    &= 
    \int \left( \frac{A}{\int A dz} - \frac{B}{\int B dz} \right)^2 dz 
    \\
    &= \int 
    \left( 
        \frac{(A-B)}{\int A dz} 
        + 
        \frac{B\int(B-A)dz}{(\int A dz) (\int B dz)} 
    \right)^2 dz \\
    & \leq  
    2 \left( 
        \frac{\int(A-B)^2dz}{(\int A dz)^2} 
        + 
        \frac{(\int B^2 dz) \left(\int (B-A) dz\right)^2}{(\int A dz)^2 (\int B dz)^2} 
    \right) \\
    & =  
    2 \left( 
        \frac{\int(A-B)^2dz}{(\int A dz)^2}  
        + 
        \frac{ \left(\int (B-A) dz\right)^2 }{( (4\pi)^{d/2} | I^{-1}(\wh\theta) |^{1/2}) (\int A dz)^2 } \right).
\end{align*}
We now bound the terms $\int (A-B)^2 dz$ and $\left( \int (B-A) dz \right)^2$. Consider the set $\m Z_{\wh\theta}=\{z=\sqrt{n}(\theta-\wh\theta)\,:\, \theta\in \wt\Theta'\}$. Then we can decompose 
\begin{equation*}
    \begin{aligned}
        &\int (A-B)^2 dz\\
        &=\int_{z\in \m Z_{\wh\theta}\atop \|z\|> n^{\frac{1}{7}}}(A-B)^2 dz+
        \int_{z\in \m Z_{\wh\theta}\atop \|z\|\leq n^{\frac{1}{7}}}(A-B)^2 dz+ \int_{z\in \m Z_{\wh\theta}^c}(A-B)^2 dz.
    \end{aligned}
\end{equation*}
and
\begin{equation*}
    \begin{aligned}
    &\Big(\int (A-B) dz\Big)^2\leq \Big(\int |A-B| dz\Big)^2\\
        &\leq3\Big(\int_{z\in \m Z_{\wh\theta}\atop \|z\|> n^{\frac{1}{7}}}|A-B| dz\Big)^2+
        3\Big(\int_{z\in \m Z_{\wh\theta}\atop \|z\|\leq n^{\frac{1}{7}}}|A-B| dz\Big)^2+ 3\Big(\int_{z\in \m Z_{\wh\theta}^c} |A-B| dz\Big)^2.
    \end{aligned}
\end{equation*}
Denote $R_n(\theta) = \frac{1}{n} \sum_{i=1}^n l(X_i, \theta)$ 
and $R(\theta) = \mb E_{X\sim \m P_{\theta^*}}[l(X,\theta)]$. Then it holds that 
\begin{equation*}
    R(\wh\theta)-R(\theta^*)\lesssim \|\wh\theta-\theta^*\|^2 (\log n)^{\frac{\alpha_1+\alpha_0\alpha_4}{\alpha_2}},
\end{equation*}
where we have used  
\begin{equation*}
    \begin{aligned}
        &\|I(\theta^*)\|_{\rm F}=\big\|\mb{E}_{X\sim \m P_{\theta^*}}[{\rm Hess}_{\theta} l(X,\theta^*)]\|_{\rm F}\\
        &\lesssim (1+\|\theta^*\|)^{\alpha_1+\alpha_0\alpha_4}\mb{E}_{X\sim \m P_{\theta^*}}\Big[\Big(1+\frac{\|X\|}{(1+\|\theta^*\|)^{\alpha_4}}\Big)^{\alpha_0}\Big]\\
        &\lesssim  (1+\|\theta^*\|)^{\alpha_1+\alpha_0\alpha_4}\lesssim (\log n)^{\frac{\alpha_1+\alpha_0\alpha_4}{\alpha_2}}.
    \end{aligned}
\end{equation*}
So there exists a positive constant $c$ so that it holds for any $z\in \m Z_{\wh\theta}$ with $\|z\|\geq n^{\frac{1}{7}}$ that
\begin{align*}
R(\wh\theta + \frac{z}{\sqrt{n}}) -
R(\wh\theta)={\rm KL}(\m P_{\theta^*},\m P_{\wh\theta + \frac{z}{\sqrt{n}}})+R(\theta^*)-R(\wh\theta)\geq c\min(\frac{\|z\|^2}{n},1).
\end{align*}
Therefore, when $(\theta^*,\Xn)\in \wt A$,  it holds for any  $z\in \m Z_{\wh\theta}$ with $\|z\|\geq n^{\frac{1}{7}}$ that
\begin{equation*}
    \begin{aligned}
        &R_n(\wh\theta + \frac{z}{\sqrt{n}})-R_n(
        \wh\theta)\\
        &\geq R(\wh\theta + \frac{z}{\sqrt{n}}) -
R(\wh\theta)-\Big|R_n(\wh\theta + \frac{z}{\sqrt{n}})-R_n(\wh\theta)-R(\wh\theta + \frac{z}{\sqrt{n}}) +
R(
\wh\theta)\Big| \\
&\geq \frac{c}{2}\min(\frac{\|z\|^2}{n},1).
\end{aligned}
\end{equation*}
Moreover,  using the Lipschitzness assumption for $\log \pi(\theta)$, we have, for $z\in \m Z_{\wh\theta}$  with $\|z\|\geq n^{\frac{1}{7}}$ ,
\begin{align*}
& \left|
    \log \pi(\wh\theta+\frac{z}{\sqrt{n}})  - \log \pi(\theta^*)
\right|
\leq c_2(\|\wh\theta+\frac{z}{\sqrt{n}}\|+ \Vert \theta^* \Vert + 1)^{\alpha_2 - 1} \cdot (\Vert \wh\theta - \theta^* \Vert + \Vert \frac{z}{\sqrt{n}} \Vert)\\
&\leq c_3 (\log n)^{\frac{\alpha_2-1}{\alpha_2}} \frac{\|z\|}{\sqrt{n}}.
\end{align*}
Hence  for $z\in \m Z_{\wh\theta}$ with $\|z\|\geq n^{\frac{1}{7}}$, it holds that
\begin{align*}
&A=\frac{\pi(\wh{\theta}+\frac{z}{\sqrt{n}})}{\pi(\theta^*)}
\exp{\left( -\sum_{i=1}^n l(X_i,\wh{\theta}+\frac{z}{\sqrt{n}}) + \sum_{i=1}^n l(X_i,\wh{\theta}) \right)}\\
&\leq 
\exp
\left( c_3 (\log n)^{\frac{\alpha_2-1}{\alpha_2}} \frac{\|z\|}{\sqrt{n}}- \frac{c}{2}\min(\|z\|^2,n)
     \right)
 \leq 
2\exp(-\frac{cn^{\frac{2}{7}}}{2}),
\end{align*}
and 
\begin{equation*}
    \int_{z\in \m Z_{\wh\theta}\atop \|z\|> n^{\frac{1}{7}}}B^2 dz
 =\int_{z\in \m Z_{\wh\theta}\atop \|z\|> n^{\frac{1}{7}}}\exp(-z^T I(\wh\theta) z) dz
 \leq \frac{1}{n^2}.
\end{equation*}
So
\begin{align*}
\int_{z\in \m Z_{\wh\theta}\atop \|z\|> n^{\frac{1}{7}}}(A-B)^2 dz 
& \leq
2 \left(
    \int_{z\in \m Z_{\wh\theta}\atop \|z\|> n^{\frac{1}{7}}}A^2 dz
    +
\int_{z\in \m Z_{\wh\theta}\atop \|z\|> n^{\frac{1}{7}}}B^2 dz
\right)
\lesssim \frac{1}{n^2}.
\end{align*}
Similarly, we can show
\begin{equation*}
    \Big(\int_{z\in \m Z_{\wh\theta}\atop \|z\|> n^{\frac{1}{7}}}|A-B| dz \Big)^2\lesssim \frac{1}{n^2}.
\end{equation*}
Then we consider the integral over $\|z\|\leq n^{\frac{1}{7}}$.
Let's first bound $|\log A-\log B|$.  For any $z \in \mathbb{R}^d$, we have:
\begin{align*}
&| \log A - \log B| =
        \left| 
            \log{\frac{\pi(\wh\theta+z/\sqrt{n})}{\pi(\theta^*)}} 
            +  \sum_{i=1}^n l(X_i, \wh\theta) -\sum_{i=1}^n l(X_i,\wh\theta +z/\sqrt{n}) 
            + \frac{1}{2} z^T I(\wh\theta) z 
        \right| \\
        & \leq
        \left| 
            \log{\frac{\pi(\wh\theta+z/\sqrt{n})}{\pi(\theta^*)}} \right| \\
            & \quad + n \bigg| R_n(\wh\theta +z/\sqrt{n}) 
            - R_n(\wh\theta)   
            - \frac{1}{n} \sum_{i=1}^n \nabla_{\theta} l(X_i,\wh\theta)^T \frac{z}{\sqrt{n}}   
            - \Big( R(\wh\theta +z/\sqrt{n}) 
            - R(\wh\theta) 
             \\ 
            & \qquad  \qquad
            - \bE_{X\sim\m P_{\theta^*}} [\nabla_{\theta} l(X,\wh\theta)^T \frac{z}{\sqrt{n}}] \Big)
            \bigg| \\
        & \quad + \Big|  n\cdot R(\wh\theta +z/\sqrt{n}) 
            - n\cdot R(\wh\theta) 
            - n\cdot\bE_{X\sim\m P_{\theta^*}} [\nabla_{\theta} l(X,\wh\theta)^T \frac{z}{\sqrt{n}}] 
            - \frac{1}{2} z^T I(\wh\theta) z \Big|,
\end{align*}
We will bound the three terms separately.
For the first term,  note that
\begin{align*}
\left|
    \log \pi(\wh\theta+\frac{z}{\sqrt{n}})  - \log \pi(\theta^*)
\right|
\leq c_2(\Vert \wh\theta \Vert + \Vert \frac{z}{\sqrt{n}} \Vert + \Vert \theta^* \Vert + 1)^{\alpha_2 - 1} \cdot (\Vert \wh\theta - \theta^* \Vert + \Vert \frac{z}{\sqrt{n}} \Vert).
\end{align*}
Also by  Lemma~\ref{lemmathetahat},
\begin{align*}
\Vert \wh\theta - \theta^* \Vert \leq C_{\theta} \sqrt{\frac{\log n}{n}}.
\end{align*}
Hence for $z\in \m Z_{\wh\theta}$ with $\|z\|\leq n^{\frac{1}{7}}$, we can upper bound the first term as:
\begin{align*}
\left|
    \log \pi(\wh\theta+\frac{z}{\sqrt{n}})  - \log \pi(\theta^*)
\right|
  \lesssim  (1+\|\theta^*\|)^{\alpha_2-1} \frac{\|z\|+\sqrt{\log n}}{\sqrt{n}}.
\end{align*}
 For the second term, by   $(\theta^*,\Xn)\in A_2$, we have:
\begin{align*}
& \quad
n\cdot\bigg| R_n(\wh\theta +z/\sqrt{n}) 
            - R_n(\wh\theta)   
            - \frac{1}{n} \sum_{i=1}^n \nabla_{\theta} l(X_i,\wh\theta)^T \frac{z}{\sqrt{n}}   
             \\ 
            & \qquad  \qquad
            - \Big( R(\wh\theta +z/\sqrt{n}) 
            - R(\wh\theta) 
            - \bE_{X\sim\m P_{\theta^*}} [\nabla_{\theta} l(X,\wh\theta)^T \frac{z}{\sqrt{n}}] \Big)
            \bigg| 
\\
& \lesssim (\log n)^{\frac{1}{2}}\cdot(1+\|\theta^*\|)^{\alpha_0\alpha_4+\alpha_1}\frac{1+\|z\|^2}{\sqrt{n}}  
\end{align*}
For the third term, by Taylor's theorem, 
\begin{align*}
& \quad
\Big|
    n\cdot R(\wh\theta +z/\sqrt{n}) 
            - n\cdot R(\wh\theta) 
            - n\cdot\bE_{X\sim\m P_{\theta^*}} [\nabla_{\theta} l(X,\wh\theta)^T \frac{z}{\sqrt{n}}] 
            - \frac{1}{2} z^T I(\wh\theta) z
\Big|
\\
& \leq \sup_{t\in [0,1]} 
\left|
    \frac{1}{2}z^T 
  \bE_{X\sim \m P_{\theta^*}}\Big[   {\rm Hess}_{\theta}\, l(X,\wh\theta+t\frac{z}{\sqrt{n}}) \Big] 
    z
    - \frac{1}{2} z^T I(\wh\theta) z
\right|  
\\
& \leq \sup_{t\in [0,1]} 
\left|
    \frac{1}{2}z^T 
     \bE_{X\sim \m P_{\theta^*}}\Big[   {\rm Hess}_{\theta}\, l(X,\wh\theta+t\frac{z}{\sqrt{n}}) \Big] 
    z -\frac{1}{2}z^T 
    \bE_{X\sim \m P_{\theta^*}}\Big[   {\rm Hess}_{\theta}\, l(X,\theta^*) \Big] 
    z\right|  \\ 
    & \qquad  + 
    \left|
    \frac{1}{2}z^T I(\theta^*)
    z
    - \frac{1}{2} z^T I(\wh\theta) z
\right|
\\
& \lesssim (1+\|\theta^*\|)^{(\alpha_1+\alpha_0\alpha_4)\vee \alpha_5} \frac{\|z\|+\sqrt{\log n}}{\sqrt{n}}\|z\|^2.
\end{align*}
Hence, combining the three terms together, we have for $z\in \m Z_{\wh\theta}$ with $\|z\|\leq n^{\frac{1}{7}}$, 
\begin{align*}
\left| \log A - \log B \right|&\lesssim  (1+\|\theta^*\|)^{(\alpha_1+\alpha_0\alpha_4)\vee (\alpha_2-1)\vee \alpha_5} \frac{\sqrt{\log n}(1+\|z\|)^3}{\sqrt{n}}=o(1).
\end{align*}
Therefore, we can  bound
\begin{align*}
     &  \int_{z\in \m Z_{\wh\theta}\atop \|z\|\leq n^{\frac{1}{7}}} (A-B)^2 dz\\
        &=
        \int_{z\in \m Z_{\wh\theta}\atop \|z\|\leq n^{\frac{1}{7}}} B^2 \left(\exp(\log A - \log B)-1\right)^2 \, dz
        \\
        & \lesssim
        \int_{z\in \m Z_{\wh\theta}\atop \|z\|\leq n^{\frac{1}{7}}} B^2 \left(\log A - \log B\right)^2 \, dz
        \\
        &\lesssim
     \int (1+\|\theta^*\|)^{(2\alpha_1+2\alpha_0\alpha_4)\vee (2\alpha_2-2)\vee (2\alpha_5)} (1+\|z\|)^6\frac{\log n}{n}\exp(-z^T I(\wh\theta) z)\,\dd z
\end{align*}
Since $ c_I I_d \preceq I(\theta^*) \preceq C(1+\|\theta^*\|)^{\alpha_1+\alpha_0\alpha_4} I_d$ and 
$ \Vert I(\wh\theta)-I(\theta^*) \Vert_{\rm F}=o(1)$, we can get
\begin{align*}
 \int_{z\in \m Z_{\wh\theta}\atop \|z\|\leq n^{\frac{1}{7}}} (A-B)^2 dz\lesssim \frac{\log n}{n}    (1+\|\theta^*\|)^{(2\alpha_1+2\alpha_0\alpha_4)\vee (2\alpha_2-2)\vee (2\alpha_5)}.
\end{align*}
 Similarly, we can bound
\begin{align*}
     \Big(\int_{z\in \m Z_{\wh\theta}\atop \|z\|\leq n^{\frac{1}{7}}} |A-B| dz\Big)^2 &=
        \Big(\int_{z\in \m Z_{\wh\theta}\atop \|z\|\leq n^{\frac{1}{7}}} B \left|\exp(\log A - \log B)-1\right| \, dz\Big)^2 \\ &
    \lesssim \frac{\log n}{n}    (1+\|\theta^*\|)^{(2\alpha_1+2\alpha_0\alpha_4)\vee (2\alpha_2-2)\vee (2\alpha_5)}.
\end{align*}
Finally, for the integral over $\m Z_{\wh\theta}^c$, using $(\theta^*,\Xn)\in A_6$, we have
\begin{align*}
\frac{c}{n^{\frac{d}{2}+1}} 
& \geq
\int_{\theta \notin \wt{\Theta}'}
\frac{\pi(\theta)}{\pi(\theta^*)}
\exp{\left( - \sum_{i=1}^n l(X_i,\theta) + \sum_{i=1}^n l(X_i, \theta^*) \right)} d\theta
\\
& \quad +
\int_{\theta \notin\wt{\Theta}'}
\left[ \frac{\pi(\theta)}{\pi(\theta^*)}
\exp{\left( -\sum_{i=1}^n l(X_i,\theta) + \sum_{i=1}^n l(X_i,\theta^*) \right)} \right]^2 d\theta
\\
& \geq
\int_{\theta \notin \wt{\Theta}'}
\frac{\pi(\theta)}{\pi(\theta^*)}
\exp{\left( - \sum_{i=1}^n l(X_i,\theta) + \sum_{i=1}^n l(X_i,\wh\theta) \right)} d\theta
\\
& \quad +
\int_{\theta \notin\wt{\Theta}'}
\left[ \frac{\pi(\theta)}{\pi(\theta^*)}
\exp{\left( -\sum_{i=1}^n l(X_i,\theta) + \sum_{i=1}^n l(X_i,\wh\theta) \right)} \right]^2 d\theta
\\
& =
\left(\frac{1}{\sqrt{n}}\right)^d
\int_{\m Z_{\wh\theta}^c}
 A^2 +A \,\dd z.
\end{align*}
Moreover, using $\theta^*\in \wt \Theta$ and for any $\theta\notin \wt\Theta'$, $\|\theta-\theta^*\|\geq r_1(\log n)^{\frac{1}{\alpha_2}}$, we can verify that 
$\int_{\m Z_{\wh\theta}^c} B^2 +B \,\dd z \lesssim \frac{1}{n}$. Finally, using 
$(\int B \,\dd z)^{-1}\lesssim |I^{-1}(\wh\theta)|^{-\frac{1}{2}}\lesssim (1+\|\theta^*\|)^{\frac{(\alpha_1+\alpha_0\alpha_4)d}{2}}$, we can obtain by combining all pieces that,  for any $(\theta^*,\Xn)\in \wt A$,
\begin{equation*}
    \begin{aligned}
       &  \Big\Vert
      \big[\sqrt{n}(\cdot -\wh{\theta}(\Xn))\big]_{\#} \pos
    -
    N(0, I^{-1}(\wh{\theta}(\Xn)))
    \Big\Vert^2_{L_2} \\
    &\lesssim  \frac{\log n}{n}   (1+\|\theta^*\|)^{\frac{3d(\alpha_1+\alpha_0\alpha_4)}{2}+\big((2\alpha_1+2\alpha_0\alpha_4)\vee (2\alpha_2-2)\vee (2\alpha_5)\big)}.
    \end{aligned}
\end{equation*}
Therefore we have
 \begin{equation*}
     \begin{aligned}
  & \mathbb{E}_{\theta^* \sim \Pi_{\theta}}
\bE_{\Xn\sim \m P_{\theta^*}^{(n)}}
\Big[
\mmd^2_{\wt h}
\left(
     \big[\sqrt{n}(\cdot -\wh{\theta}(\Xn))\big]_{\#} \pos
    ,
    N(0, I^{-1}(\wh{\theta}(\Xn)))
\right)  \\ 
& \qquad \qquad \qquad \qquad \cdot \mathbf{1}((\theta^*,\Xn) \in \wt A) 
\Big] \\
&\leq  \mathbb{E}_{\theta^* \sim \Pi_{\theta}}
\bE_{\Xn\sim \m P_{\theta^*}^{(n)}}
\Big[  \big\|   \big[\sqrt{n}(\cdot -\wh{\theta}(\Xn))\big]_{\#} \pos- N(0, I^{-1}(\wh{\theta}(\Xn)))\big\|_{L_2}^2  \\ 
& \qquad \qquad \qquad \qquad
\cdot
\mathbf{1}((\theta^*,\Xn) \in \wt A) 
\Big]\\
 &\lesssim\frac{\log n}{n}    \cdot\mathbb{E}_{\theta^* \sim \Pi_{\theta}}\Big[(1+\|\theta^*\|)^{\frac{3d(\alpha_1+\alpha_0\alpha_4)}{2}+\big((2\alpha_1+2\alpha_0\alpha_4)\vee (2\alpha_2-2)\vee (2\alpha_5)\big)}\Big]\\
 &\lesssim \frac{\log n}{n}    .
     \end{aligned}
 \end{equation*}

\subsection{Proof of Lemma~\ref{lemma:approximateNN}}

  Define $    \varepsilon_{\theta}=n^{-\frac{1}{2}}(\log n)^{-\frac{7\alpha_0}{2\alpha_3}-\frac{1}{2}}$ and  $  \varepsilon_{X}=n^{-1}(\log n)^{-\frac{6\alpha_0}{\alpha_3}-1}$.  Let  $\m X'=B(0_{d_X}, c_1(\log n)^{\frac{\alpha_4}{\alpha_2}+\frac{1}{\alpha_3}})\cap \m X$, where $c_1=c(1+r_0)^{\alpha_4}$ so that for any $(\theta^*,\Xn)\in \wt A$,  we have $\Xn\in \m X'{}^{n}$. Let $\{\theta_1^*,\theta_2^*,\cdots,\theta_J^*\}$ be the largest $\varepsilon_{\theta}$-packing set of $\wt\Theta^{'}$ and $\{X_1^*,X_2^*,\cdots, X_M^*\}$ be the largest $\varepsilon_X$ packing set of $\m X'$. Then  $J\lesssim (\log n)^{\frac{d}{\alpha_2}}\varepsilon_{\theta}^{-d}$ and $M\lesssim  (\log n)^{\frac{\alpha_4d_X}{\alpha_2}+\frac{d_X}{\alpha_3}}\varepsilon_{X}^{-d_X}$. Consider  a  function
$$
k_b(x)=\left\{\begin{array}{cc}
1 & |x|\leq 1 \\
0 & |x|>2 \\
2-|x| & 1<|x| \leq 2.
\end{array}\right.
$$
For $j\in [J]$, define
\begin{equation*}
    l_j(x)=\frac{\sum_{m=1}^M l(X_m^*,\theta_j^*)k_b(\frac{\|x-X_m^*\|^2}{\varepsilon_X^2})}{\sum_{m=1}^M k_b(\frac{\|x-X_m^*\|^2}{\varepsilon_X^2})},
\end{equation*}
\begin{equation*}
    g_j(x)=\frac{\sum_{m=1}^M \nabla_{\theta}l(X_m^*,\theta_j^*)k_b(\frac{\|x-X_m^*\|^2}{\varepsilon_X^2})}{\sum_{m=1}^M k_b(\frac{\|x-X_m^*\|^2}{\varepsilon_X^2})},
\end{equation*}
and 
\begin{equation*}
    H_j(x)=\frac{\sum_{m=1}^M \nabla_{\theta}l(X_m^*,\theta_j^*)\nabla_{\theta}l(X_m^*,\theta_j^*)^Tk_b(\frac{\|x-X_m^*\|^2}{\varepsilon_X^2})}{\sum_{m=1}^M k_b(\frac{\|x-X_m^*\|^2}{\varepsilon_X^2})}.
\end{equation*}
Then using the Lipschitz continuity of $x\to l(x,\theta)$ and  $x\to \nabla_{\theta}l(x,\theta)$, it holds that  $$\sup_{x\in \m X'}\frac{|l_j(x)-l(x,\theta_j^*)|}{(1+\|x\|)^{\alpha_0}}\lesssim (\|\theta_j^*\|+1)^{\alpha_1}\varepsilon_X,$$  $$\sup_{x\in \m X'}\frac{\|g_j(x)-\nabla_{\theta}l(x,\theta_j^*)\|}{(1+\|x\|)^{\alpha_0}}\lesssim  (\|\theta_j^*\|+1)^{\alpha_1}   \varepsilon_X,$$ and
\begin{equation*}
    \sup_{x\in \m X'}\frac{\|H_j(x)-\nabla_{\theta}l(x,\theta_j^*)\nabla_{\theta}l(x,\theta_j^*)^T\|_{\rm F}}{(1+\|x\|)^{2\alpha_0}}\lesssim (\|\theta_j^*\|+1)^{2\alpha_1}   \varepsilon_X.
\end{equation*}
Furthermore, similarly to the proof of Lemma~\ref{lemma: approximate summary statistics}, using Lemmas F.6 and F.7 in~\cite{oko2023diffusion} for the approximation of multiplication and reciprocal functions, there exist $L_3,L_4,L_5,U_3,U_4,U_5,R_3,R_4,R_5$ with 
  $L_3,L_4,L_5=O(\log^2 n)$ and $R_3,R_4,R_5=O(M\log n)$ so that for any negative log-density $l(\cdot,\cdot)$ satisfying Assumption~\ref{assump:large_sample}, and any $j\in [J]$,  there exist $\phi_{l_j}\in \Phi_{d_X}^{1}(L_3,U_3,R_3)$, $\phi_{g_j}\in \Phi_{d_X}^{d}(L_4,U_4,R_4)$, $\phi_{H_j}\in \Phi_{d_X}^{d^2}(L_5,U_5,R_5)$ so that 
\begin{equation*}
    \sup_{j\in[J]\atop x\in \m X'}|l_j(x)-\phi_{l_j}(x)|+\|g_j(x)-\phi_{g_j}(x)\|+\|H_j(x)-{\rm mat}_d(\phi_{H_j}(x))\|_{\rm F}\leq n^{-4}.
\end{equation*}
We choose ${\rm mat}_d(\phi_{H_j}(x))$ symmetric by averaging it with its transpose; this preserves the error bound and changes the network size by at most a constant factor.
 For any $j\in [J]$,  define
 \begin{equation*}
     \m E_{\theta^*_{j}}(\Xn)=\theta_j^*-\big(n^{-1}\sum_{i=1}^n {\rm mat}_d(\phi_{H_j}(X_i))+n^{-3}I_d\big)^{-1}(n^{-1}\sum_{i=1}^n \phi_{g_j}(X_i)).
 \end{equation*}
Here note that $H_j(x)$ is positive semi-definite for any $x\in \m X'$, and that $   \sup_{j\in[J], x\in \m X'}\|H_j(x)-{\rm mat}_d(\phi_{H_j}(x))\|_{\rm F}\leq n^{-4}$. Hence, for large enough $n$, $n^{-1}\sum_{i=1}^n {\rm mat}_d(\phi_{H_j}(X_i))+n^{-3}I_d\succeq \frac{I_d}{2n^3}$ and its inverse is well-defined.  Then define 
\begin{equation*}
    \begin{aligned}
        \m E(\Xn)=\frac{\sum_{j=1}^J   \m E_{\theta^*_{j}}(\Xn)\cdot k_b\Big(\frac{1}{n\varepsilon_X}\big(\sum_{i=1}^n\phi_{l_j}(X_i)-\min_{j_1\in [J]}\sum_{i=1}^n\phi_{l_{j_1}}(X_i)\big)\Big)}{\sum_{j=1}^J    k_b\Big(\frac{1}{n\varepsilon_X}\big(\sum_{i=1}^n\phi_{l_j}(X_i)-\min_{j_1\in [J]}\sum_{i=1}^n\phi_{l_{j_1}}(X_i)\big)\Big)}.
    \end{aligned}
\end{equation*}
Similarly to the proof of Lemma~\ref{lemma: approximate summary statistics}, the minimization function can be realized exactly using ReLU networks, where the depth and number of nonzero weights and biases for a balanced pairwise-minimum network that realizes $\min(Z_1,\cdots,Z_J)$ are $\m O(\log J)$ and $\m O(J \cdot \log J)$ respectively. Moreover,  by the Cayley--Hamilton theorem, for any invertible $A\in \mb R^{d\times d}$, let $S_k={\rm tr}(A^k)$ and denote by $B_{k}$ the $k$th complete exponential Bell polynomial.\footnote{$
B_{k}\left(x_{1}, \ldots, x_{k}\right)=\sum_{w=1}^{k} B_{k, w}\left(x_{1}, x_{2}, \ldots, x_{k-w+1}\right)$ with $B_{k, w}\left(x_{1}, x_{2}, \ldots, x_{k-w+1}\right) \\ =\sum_{\substack{j_{1}+\ldots+j_{k-w+1}=w \\ j_{1}+2 j_{2}+\ldots+(k-w+1)j_{k-w+1}=k}} \frac{k !}{j_{1} ! j_{2} ! \ldots j_{k-w+1}!}\left(\frac{x_{1}}{1 !}\right)^{j_{1}}\left(\frac{x_{2}}{2 !}\right)^{j_{2}} \cdots \left(\frac{x_{k-w+1}}{(k-w+1)!}\right)^{j_{k-w+1}}$.}
Then we can express the determinant and inverse of $A$ as
 \begin{equation*}
     \begin{aligned}
         & {\rm det}(A)=\frac{1}{d!}B_d(S_1,-1!S_2,\cdots, (-1)^{d-1}(d-1)!S_d)\\
         &A^{-1}=\frac{1}{{\rm det}(A)}\sum_{k=0}^{d-1} (-1)^{d+k-1} \frac{A^{d-k-1}}{k!} B_k(S_1,-1!S_2,\cdots, (-1)^{k-1}(k-1)!S_k).
     \end{aligned}
 \end{equation*}
 So using Lemmas F.6 and F.7 in~\cite{oko2023diffusion} for the approximation of multiplication and reciprocal functions,  and  by Lemmas F.1 and F.3 of~\cite{oko2023diffusion} about the Concatenation and Parallelization of neural networks, there exist $L_1,U_1,R_1$ with $L_1=\m O(\log^2 n)$ and $R_1=\m O(nJM\log n)$, so that for any negative log-density $l(\cdot,\cdot)$ satisfying Assumption~\ref{assump:large_sample}, there exists $S^*\in \Phi_{nd_X}^{d}(L_1,U_1,R_1)$ so that for any $\Xn\in \m X'{}^{n}$,  
 \begin{equation}\label{eqnS-E}
    \| S^*(\Xn)-\m E(\Xn)\|\leq n^{-1}.
 \end{equation}
 Then we will control the distance between $\m E(\Xn)$ and $\wh\theta(\Xn)$. Let's fix an $(\theta^*,\Xn)$ pair, where $(\theta^*,\Xn)\in \wt A$.  For ease of notation, we denote $\wh\theta=\wh\theta(\Xn)$. Using Lemma~\ref{lemmathetahat}, we have $\|\wh\theta-\theta^*\|\lesssim \sqrt{\frac{\log n}{n}}$.  Then using the facts that $\theta^*\in \wt \Theta$ and $\wt\Theta'$ contains $\wt\Theta$, when $n$ is large enough, we have that $\wh\theta$ is an interior point of $\wt \Theta'$ and 
 $n^{-1}\sum_{i=1}^n \nabla_{\theta} l(X_i,\wh\theta)=0_d$. Then consider the index set 
 \begin{equation*}
     \m J=\Big\{j\in [J]\,:\,  \frac{1}{n}\sum_{i=1}^n\phi_{l_j}(X_i)-\frac{1}{n}\min_{j\in [J]}\sum_{i=1}^n\phi_{l_j}(X_i)\leq 2\varepsilon_X\Big\}.
 \end{equation*}
Since $\{\theta_1^*,\theta_2^*,\cdots, \theta_J^*\}$ is the largest $\varepsilon_{\theta}$ packing set of $\wt\Theta^{'}$, there exists $j^*\in [J]$ so that $\|\theta_{j^*}^*-\wh\theta\|\leq \varepsilon_{\theta}$ and 
\begin{equation*}
    \begin{aligned}
       & n^{-1}\sum_{i=1}^n l(X_i,\theta_{j^*}^*)-n^{-1}\sum_{i=1}^n l(X_i,\wh\theta)\lesssim (1+\|\theta^*\|)^{\alpha_1+\alpha_0\alpha_4} (\log n)^{\frac{\alpha_0}{\alpha_3}}\cdot\|\theta_{j^*}^*-\wh\theta\|^2\\
       &\lesssim  (1+\|\theta^*\|)^{\alpha_1+\alpha_0\alpha_4} (\log n)^{\frac{\alpha_0}{\alpha_3}}\cdot\varepsilon_{\theta}^2.
    \end{aligned}
\end{equation*}
Write $F_n(\theta)=n^{-1}\sum_{i=1}^n l(X_i,\theta)$,
$A_n=(1+\|\theta^*\|)^{\alpha_1+\alpha_0\alpha_4}(\log n)^{\alpha_0/\alpha_3}$, and
$C_n=C(\log n)^{(\alpha_1+\alpha_0\alpha_4)/\alpha_2+\alpha_0/\alpha_3}$.
The interpolation weights defining $l_j$ are independent of $\theta_j^*$.
Consequently, the Lipschitz assumption on $x\mapsto\nabla_\theta l(x,\theta)$ bounds the difference of the interpolation errors at $\theta_j^*$ and $\theta_{j^*}^*$ by
$C_n\varepsilon_X\|\theta_j^*-\theta_{j^*}^*\|$ after averaging over the observations. Thus, for $j\in\m J$,
\begin{equation*}
\begin{aligned}
 F_n(\theta_j^*)-F_n(\wh\theta)
 &\leq 2\varepsilon_X+F_n(\theta_{j^*}^*)-F_n(\wh\theta)+2n^{-4}
       +C_n\varepsilon_X\|\theta_j^*-\theta_{j^*}^*\|\\
 &\lesssim \varepsilon_X+A_n\varepsilon_\theta^2
       +C_n\varepsilon_X\|\theta_j^*-\theta_{j^*}^*\|.
\end{aligned}
\end{equation*}
Let $r_j=\|\theta_j^*-\theta^*\|$ and
$a_n=n^{-1/2}(\log n)^{1/2+\alpha_1/\alpha_2}(1+\|\theta^*\|)^{\alpha_0\alpha_4}$.
The KL lower bound and the concentration bound on $\wt A$ give
\begin{equation*}
 c_4\min(r_j^2,1)\lesssim
 a_n(n^{-1}+r_j)+\varepsilon_X+A_n\varepsilon_\theta^2
 +C_n\varepsilon_X(r_j+\|\wh\theta-\theta^*\|+\varepsilon_\theta).
\end{equation*}
Using Lemma~\ref{lemmathetahat}, $C_n\varepsilon_X=o(n^{-1/2})$, and
$\varepsilon_X+A_n\varepsilon_\theta^2\lesssim a_n^2$, this yields
$\|\theta_j^*-\theta^*\|\lesssim a_n$.
Moreover, using Taylor's theorem,  it holds that 
\begin{equation*}
    \begin{aligned}
        &n^{-1}\sum_{i=1}^n l(X_i,\theta_{j}^*)-n^{-1}\sum_{i=1}^n l(X_i,\wh\theta)\\
        &\geq \frac{1}{2}\min_{t\in [0,1]} (\theta_{j}^*-\wh\theta)^T \Big(\frac{1}{n}\sum_{i=1}^n{\rm Hess}_{\theta}\, l(X_i,t\wh\theta+(1-t)\theta_j^*)\Big)(\theta_{j}^*-\wh\theta).
    \end{aligned}
\end{equation*}
Then note that for any $t\in (0,1)$, it holds that 
\begin{equation}\label{eqn:diffHess}
    \begin{aligned}
       & \big\|\frac{1}{n}\sum_{i=1}^n{\rm Hess}_{\theta}\, l(X_i,t\wh\theta+(1-t)\theta_j^*)- I(\theta^*)\big\|_{\rm F}\\
        &\leq \big\|\frac{1}{n}\sum_{i=1}^n{\rm Hess}_{\theta}\, l(X_i,t\wh\theta+(1-t)\theta_j^*)- \frac{1}{n}\sum_{i=1}^n{\rm Hess}_{\theta}\, l(X_i,\theta^*)\big\|_{\rm F}\\
        &+ \big\| \frac{1}{n}\sum_{i=1}^n {\rm Hess}_{\theta}\, l(X_i,\theta^*)-\mb{E}_{\m P_{\theta^*}}[{\rm Hess}_{\theta}\, l(X,\theta^*)]\big\|_{\rm F}\\
        &\lesssim  \frac{1}{\sqrt{n}}(\log n)^{\frac{1}{2}+\frac{2\alpha_1+2\alpha_0\alpha_4}{\alpha_2}+\frac{\alpha_0}{\alpha_3}}.
    \end{aligned}
\end{equation}
Combined with $I(\theta^*)\succeq c_I I_d$, this gives a fixed positive lower bound on the empirical Hessian. In the preceding likelihood comparison, absorb the term $C_n\varepsilon_X\|\theta_j^*-\wh\theta\|$ into the resulting quadratic lower bound. Since $C_n^2\varepsilon_X^2=o(\varepsilon_X)$ and $C_n\varepsilon_\theta=o(1)$, we obtain, for any $j\in\m J$, 
\begin{equation*}
\begin{aligned}
       \|\wh\theta-\theta_{j}^*\|\lesssim  \sqrt{n^{-1}\sum_{i=1}^n l(X_i,\theta_{j}^*)-n^{-1}\sum_{i=1}^n l(X_i,\wh\theta)}\lesssim \sqrt{\varepsilon_X}+ (1+\|\theta^*\|)^{\frac{\alpha_1+\alpha_0\alpha_4}{2}} (\log n)^{\frac{\alpha_0}{2\alpha_3}} \varepsilon_{\theta},
\end{aligned}
 \end{equation*}
and thus 
\begin{equation*}
    \begin{aligned}
        &\Big\|\frac{1}{n}\sum_{i=1}^n \nabla_{\theta} l(X_i,\theta_j^*)\Big\|= \Big\|\frac{1}{n}\sum_{i=1}^n \nabla_{\theta} l(X_i,\theta_j^*)-\frac{1}{n}\sum_{i=1}^n \nabla_{\theta} l(X_i,\wh\theta)\Big\|\\
        &\lesssim (1+\|\theta^*\|)^{ \alpha_1+\alpha_0\alpha_4 } (\log n)^{\frac{\alpha_0}{\alpha_3}} \sqrt{\varepsilon_X}+ (1+\|\theta^*\|)^{\frac{3(\alpha_1+\alpha_0\alpha_4)}{2}} (\log n)^{\frac{3\alpha_0}{2\alpha_3}} \varepsilon_{\theta} \\
        &\lesssim (1+\|\theta^*\|)^{\frac{3(\alpha_1+\alpha_0\alpha_4)}{2}} (\log n)^{-\frac{2\alpha_0}{\alpha_3}-\frac{1}{2}}\frac{1}{\sqrt{n}} .
    \end{aligned}
\end{equation*}
Then consider 
\begin{equation*}
    \wt\theta_j^*=\theta_j^*-\Big(\frac{1}{n}\sum_{i=1}^n  {\rm Hess}_{\theta}\, l(X_i,\theta_j^*)\Big)^{-1} \frac{1}{n}\sum_{i=1}^n \nabla_{\theta}l(X_i,\theta_j^*). 
\end{equation*}
 It holds for any $j\in \m J$ that 
 \begin{equation*}
     \begin{aligned}
         \Big\|\frac{1}{n}\sum_{i=1}^n \nabla_{\theta} l(X_i,\wt\theta_j^*)\Big\| & \lesssim(1+\|\theta^*\|)^{\alpha_1+\alpha_0\alpha_4} (\log n)^{\frac{\alpha_0}{\alpha_3}} \|\wt\theta_j^*-\theta_j^*\|^2 \\
         & \lesssim   (\log n)^{-\frac{3\alpha_0}{\alpha_3}-1}(1+\|\theta^*\|)^{4\alpha_1+4\alpha_0\alpha_4}\frac{1}{n}  .
     \end{aligned}
 \end{equation*}
So using~\eqref{eqn:diffHess} and  $I(\theta^*)\succeq c_I I_d$, we can get 
\begin{equation*}
    \|\wh\theta-   \wt\theta_j^*\|\lesssim  (\log n)^{-\frac{3\alpha_0}{\alpha_3}-1}\frac{ (1+\|\theta^*\|)^{4\alpha_1+4\alpha_0\alpha_4}}{n},\quad \forall\, j\in \m J.
\end{equation*}
Then we will bound $\|\wt\theta_j^*-\m E_{\theta_j^*}(\Xn)\|$ for $j\in \m J$.  Note that 
\begin{equation*}
    \begin{aligned}
    & \|\wt\theta_j^*- \m E_{\theta^*_{j}}(\Xn)\|\\
        &= \Big\|\Big(\frac{1}{n}\sum_{i=1}^n  {\rm Hess}_{\theta}\, l(X_i,\theta_j^*)\Big)^{-1} \Big(\frac{1}{n}\sum_{i=1}^n \nabla_{\theta}l(X_i,\theta_j^*)\Big)  \\ 
        & \qquad -\Big(n^{-1}\sum_{i=1}^n {\rm mat}_d(\phi_{H_j}(X_i))+n^{-3}I_d\Big)^{-1}\Big(n^{-1}\sum_{i=1}^n \phi_{g_j}(X_i)\Big)\Big\|\\
        &\leq \Big\|\Big(\frac{1}{n}\sum_{i=1}^n  {\rm Hess}_{\theta}\, l(X_i,\theta_j^*)\Big)^{-1} -\Big(n^{-1}\sum_{i=1}^n {\rm mat}_d(\phi_{H_j}(X_i))+n^{-3}I_d\Big)^{-1}\Big\|_{\rm F} \\ 
        & \qquad \cdot \big\|n^{-1}\sum_{i=1}^n \phi_{g_j}(X_i)\big\|\\
        &+\Big\|\Big(\frac{1}{n}\sum_{i=1}^n  {\rm Hess}_{\theta}\, l(X_i,\theta_j^*)\Big)^{-1} \Big\|_{\rm F}\cdot  \Big\| \frac{1}{n}\sum_{i=1}^n \nabla_{\theta}l(X_i,\theta_j^*)-n^{-1}\sum_{i=1}^n \phi_{g_j}(X_i)\Big\|\\
        &\lesssim \Big\|\Big(\frac{1}{n}\sum_{i=1}^n  {\rm Hess}_{\theta}\, l(X_i,\theta_j^*)\Big)^{-1} -\Big(n^{-1}\sum_{i=1}^n {\rm mat}_d(\phi_{H_j}(X_i))+n^{-3}I_d\Big)^{-1}\Big\|_{\rm F} 
         \\ 
        & \qquad \cdot (1+\|\theta^*\|)^{\frac{3(\alpha_1+\alpha_0\alpha_4)}{2}} (\log n)^{-\frac{2\alpha_0}{\alpha_3}-\frac{1}{2}}\frac{1}{\sqrt{n}} +(1+\|\theta^*\|)^{\alpha_1+\alpha_0\alpha_4}(\log n)^{\frac{\alpha_0}{\alpha_3}}\varepsilon_X.
         \end{aligned}
\end{equation*}
 Furthermore, it holds that 
\begin{equation*}
    \begin{aligned}
        &\big\|n^{-1}\sum_{i=1}^n {\rm mat}_d(\phi_{H_j}(X_i))-n^{-1}\sum_{i=1}^n {\rm Hess}_{\theta} l(X_i,\theta_j^*)\big\|_{\rm F}\\
        & \leq \|n^{-1}\sum_{i=1}^n {\rm mat}_d(\phi_{H_j}(X_i))- n^{-1}\sum_{i=1}^n \nabla_{\theta}l(X_i,\theta_j^*)\nabla_{\theta}l(X_i,\theta_j^*)^T \big\|_{\rm F}\\
        &+\|n^{-1}\sum_{i=1}^n \nabla_{\theta}l(X_i,\theta_j^*)\nabla_{\theta}l(X_i,\theta_j^*)^T- n^{-1}\sum_{i=1}^n \nabla_{\theta}l(X_i,\theta^*)\nabla_{\theta}l(X_i,\theta^*)^T \big\|_{\rm F}\\
         &+\| n^{-1}\sum_{i=1}^n \nabla_{\theta}l(X_i,\theta^*)\nabla_{\theta}l(X_i,\theta^*)^T -I(\theta^*)\big\|_{\rm F}+\|  I(\theta^*)-n^{-1}\sum_{i=1}^n {\rm Hess}_{\theta} l(X_i,\theta^*)\big\|_{\rm F}\\
         &+\|  n^{-1}\sum_{i=1}^n {\rm Hess}_{\theta} l(X_i,\theta^*)- n^{-1}\sum_{i=1}^n {\rm Hess}_{\theta} l(X_i,\theta^*_j)\big\|_{\rm F}\\
         &\lesssim (1+\|\theta^*\|)^{2\alpha_1+2\alpha_0\alpha_4}(\log n)^{\frac{2\alpha_0}{\alpha_3}} \varepsilon_X+ (1+\|\theta^*\|)^{2\alpha_1+3\alpha_0\alpha_4} (\log n)^{\frac{1}{2}+\frac{\alpha_1}{\alpha_2}+\frac{2\alpha_0}{\alpha_3}}\frac{1}{\sqrt{n}}. \\
    \end{aligned}
\end{equation*}
Combined with   $ \big\|\frac{1}{n}\sum_{i=1}^n{\rm Hess}_{\theta}\, l(X_i,\theta_j^*)- I(\theta^*)\big\|_{\rm F}=o(1)$, $I(\theta^*)\succeq c_I I_d$  for $c_I>0$ and $2\alpha_1<\alpha_2$,   it holds for any $j\in  \m J$ that
\begin{equation*}
\begin{aligned}
               &\|\wt\theta_j^*- \m E_{\theta^*_{j}}(\Xn)\|\lesssim  \frac{\sqrt{\log n}}{n} (1+\|\theta^*\|)^{\frac{7\alpha_1}{2}+\frac{9\alpha_0\alpha_4}{2}},
\end{aligned}
\end{equation*}
and 
\begin{align*}
    \|\wh\theta- \m E_{\theta^*_{j}}(\Xn)\| &\lesssim
     (\log n)^{-\frac{3\alpha_0}{\alpha_3}-1}\frac{ (1+\|\theta^*\|)^{4\alpha_1+4\alpha_0\alpha_4}}{n}+\frac{\sqrt{\log n}}{n} (1+\|\theta^*\|)^{\frac{7\alpha_1}{2}+\frac{9\alpha_0\alpha_4}{2}} \\
     & \lesssim \frac{\sqrt{\log n}}{n} (1+\|\theta^*\|)^{\frac{7\alpha_1}{2}+\frac{9\alpha_0\alpha_4}{2}}.
\end{align*}
Since $k_b(x)=0$ if $x\geq 2$, we have
\begin{equation*}
    \begin{aligned}
     & \| \wh\theta-\m E(\Xn)\|\\
      &=\bigg\|\wh\theta-\frac{\sum_{j\in \m J}   \m E_{\theta^*_{j}}(\Xn)\cdot k_b\Big(\frac{1}{n\varepsilon_X}\big(\sum_{i=1}^n\phi_{l_j}(X_i)-\min_{j_1\in [J]}\sum_{i=1}^n\phi_{l_{j_1}}(X_i)\big)\Big)}{\sum_{j\in \m J}   k_b\Big(\frac{1}{n\varepsilon_X}\big(\sum_{i=1}^n\phi_{l_j}(X_i)-\min_{j_1\in [J]}\sum_{i=1}^n\phi_{l_{j_1}}(X_i)\big)\Big)}\bigg\|\\
        &=\bigg\|\frac{\sum_{j\in \m J}   \big(\wh\theta-\m E_{\theta^*_{j}}(\Xn)\big)\cdot k_b\Big(\frac{1}{n\varepsilon_X}\big(\sum_{i=1}^n\phi_{l_j}(X_i)-\min_{j_1\in [J]}\sum_{i=1}^n\phi_{l_{j_1}}(X_i)\big)\Big)}{\sum_{j\in \m J}   k_b\Big(\frac{1}{n\varepsilon_X}\big(\sum_{i=1}^n\phi_{l_j}(X_i)-\min_{j_1\in [J]}\sum_{i=1}^n\phi_{l_{j_1}}(X_i)\big)\Big)}\bigg\|\\
        &\leq \sup_{j\in \m J}  \|\wh\theta- \m E_{\theta^*_{j}}(\Xn)\|\lesssim       \frac{\sqrt{\log n}}{n} (1+\|\theta^*\|)^{\frac{7\alpha_1}{2}+\frac{9\alpha_0\alpha_4}{2}}.
    \end{aligned}
\end{equation*}
Hence combined with~\eqref{eqnS-E}, for any $\Xn$ for which there exists
 $\theta^*\in \wt \Theta$ with $(\theta^*,\Xn)\in \wt A$, we have
\begin{equation*}
\begin{aligned}
    & \| S^*(\Xn)-\wh\theta(\Xn)\|
     \lesssim  \frac{\sqrt{\log n}}{n} (1+\|\theta^*\|)^{\frac{7\alpha_1}{2}+\frac{9\alpha_0\alpha_4}{2}}.
\end{aligned}
\end{equation*}
Furthermore,  consider 
\begin{equation*}
    \m I(\theta)=\frac{\sum_{j=1}^J I^{-\frac{1}{2}}(\theta_j^*) k_b\big(\frac{\|\theta-\theta_j^*\|^2}{2\varepsilon_{\theta}{}^2}\big)}{\sum_{j=1}^J k_b\big(\frac{\|\theta-\theta_j^*\|^2}{2\varepsilon_{\theta}{}^2}\big)}.
\end{equation*}
There exist $L_2,U_2,R_2$ with $L_2=\m O(\log^2 n)$ and $R_2=\m O(J\log n)$, and $W^*\in \Phi_{d}^{d^2}(L_2,U_2,R_2)$ so that for any $\theta\in \cup_{j=1}^{J} B(\theta_j^*,\sqrt{2}\varepsilon_{\theta})$,
\begin{equation*}
    \big\|  \m I(\theta)-{\rm mat}_d\big(W^*(\theta)\big)\big\|_{\rm F}\leq  n^{-1}.
\end{equation*}
Moreover, since for any $(\theta^*,\Xn)\in \wt A$, it holds that  $\wh\theta(\Xn)\in \wt \Theta'$ and $\|S^*(\Xn)-\wh\theta(\Xn)\|=o(\varepsilon_{\theta})$, we have $S^*(\Xn)\in \cup_{j=1}^{J} B(\theta_j^*,\sqrt{2}\varepsilon_{\theta})$, and
\begin{equation*}
\begin{aligned}
      &\big\| I^{-\frac{1}{2}}(\wh\theta(\Xn))-{\rm mat}_d\big(W^*(S^*(\Xn))\big)\big\|_{\rm F}\\
      &\leq \big\| I^{-\frac{1}{2}}(\wh\theta(\Xn))- \m I(S^*(\Xn))\big\|_{\rm F}  +\big\|\m I(S^*(\Xn))-{\rm mat}_d\big(W^*(S^*(\Xn))\big)\big\|_{\rm F}\\
      &\lesssim   \big\| I^{-\frac{1}{2}}(\wh\theta(\Xn))- \m I(S^*(\Xn))\big\|_{\rm F} +n^{-1}\\
      &\lesssim (1+\|\theta^*\|)^{\alpha_5}\cdot\varepsilon_{\theta},
\end{aligned}
\end{equation*}
where the last inequality uses 
\begin{equation*}
    \begin{aligned}
        &  \big\| I^{-\frac{1}{2}}(\wh\theta(\Xn))- \m I(S^*(\Xn))\big\|_{\rm F}\\
        & =\Big\|\frac{\sum_{j=1}^J  \big(I^{-\frac{1}{2}}(\wh\theta(\Xn))-I^{-\frac{1}{2}}(\theta_j^*)\big)\cdot k_b\big(\frac{\|S^*(\Xn)-\theta_j^*\|^2}{2\varepsilon_{\theta}{}^2}\big)}{\sum_{j=1}^J k_b\big(\frac{\|S^*(\Xn)-\theta_j^*\|^2}{2\varepsilon_{\theta}{}^2}\big)}\Big\|_{\rm F}\\
        &\leq \underset{j\in [J]\atop \|S^*(\Xn)-\theta_j^*\|\leq 2\varepsilon_{\theta}}{\sup}\big\|I^{-\frac{1}{2}}(\wh\theta(\Xn))-I^{-\frac{1}{2}}(\theta_j^*)\big\|_{\rm F}\\
          &\leq \underset{j\in [J]\atop \|\wh\theta(\Xn)-\theta_j^*\|\leq 3\varepsilon_{\theta}}{\sup}\big\|I^{-\frac{1}{2}}(\wh\theta(\Xn))-I^{-\frac{1}{2}}(\theta_j^*)\big\|_{\rm F}\\
          &\lesssim  (1+\|\theta^*\|)^{\alpha_5}\cdot\varepsilon_{\theta}.
    \end{aligned}
\end{equation*}
Therefore, using $\|\wh\theta(\Xn)-\theta^*\|\lesssim \sqrt{\frac{\log n}{n}}$ and $I(\theta^*)\succeq c_I I_d$ with $c_I>0$, we have
\begin{equation*}
\begin{aligned}
      &\big\| I^{-1}(\wh\theta(\Xn))-{\rm mat}_d\big(W^*(S^*(\Xn))\big) {\rm mat}_d\big(W^*(S^*(\Xn))\big)^T\big\|_{\rm F}\lesssim    (1+\|\theta^*\|)^{\alpha_5}\varepsilon_{\theta}.
\end{aligned}
        \end{equation*}
Then consider  a fixed $(\theta^*,\Xn)\in \wt A$. For ease of notation, we write  $S^* = S^*({\Xn})$, $\wh\theta=\wh\theta(\Xn)$ and $\Sigma^* ={\rm mat}_d(W^*(S^*({\Xn}))) {\rm mat}_d(W^*(S^*({\Xn})))^T$. By the triangle inequality for MMD, we have
\begin{equation*}
\begin{aligned}
    & \mmd^2_{\wt h}
    \left(
        N\left(0, I^{-1}(\wh{\theta})\right)
        ,
       \big[\sqrt{n}(\theta - \wh{\theta})]_{\#} \mu_{W^*}\left(\cdot| S^*(\Xn) \right)
    \right)\\
& = \mmd^2_{\wt h}
    \left(
        N\left(0, I^{-1}(\wh{\theta})\right)
        ,
        N\left(\sqrt{n}(S^* - \wh{\theta}), \Sigma^*\right)
    \right)\\
& \leq 2\mmd^2_{\wt h}
    \left(
        N\left(0, I^{-1}(\wh{\theta})\right)
        ,
        N\left(\sqrt{n}(S^* - \wh{\theta}), I^{-1}(\wh{\theta}) \right)
    \right)\\
&  \qquad+ 2\mmd^2_{\wt h}
    \Big(
        N\left(\sqrt{n}(S^* - \wh{\theta}), I^{-1}(\wh{\theta}) \right),
        N\left(\sqrt{n}(S^* - \wh{\theta}), \Sigma^*\right)
    \Big).
  \end{aligned}
\end{equation*}
For the first term, note that 
\begin{equation*}
\begin{aligned}
    &\mmd^2_{\wt h}
    \left(
        N\left(0, I^{-1}(\wh{\theta})\right)
        ,
        N\left(\sqrt{n}(S^* - \wh{\theta}), I^{-1}(\wh{\theta}) \right)\right)  \\
        &=\frac{2}{(2\pi)^{d/2} \left| \wt h^2 I_d + 2I^{-1}(\wh\theta)\right|^{1/2}}
-2\frac{\exp\left( 
    -\frac{n}{2}\left( 
        \left(S^*-\wh\theta\right) ^T 
        \left(\wt h^2 I_d + 2I^{-1}(\wh\theta) \right)^{-1} 
        \left(S^*-\wh\theta \right)
        \right)\right)}{(2\pi)^{d/2} \left| \wt h^2 I_d + 2I^{-1}(\wh\theta) \right|^{1/2}}.
\end{aligned}   
\end{equation*}
Since $\|\wh\theta-\theta^*\|\lesssim \sqrt{\frac{\log n}{n}}$ and $\|I(\theta^*)\|_{\rm F}=\big\|\mb{E}_{X\sim \m P_{\theta^*}} [{\rm Hess}_{\theta} l(X,\theta^*)] \big\|_{\rm F}\lesssim  (1+\|\theta^*\|)^{\alpha_1+\alpha_0\alpha_4}$,  we have $\|I(\wh\theta)\|_{\rm F}\lesssim (1+\|\theta^*\|)^{\alpha_1+\alpha_0\alpha_4}$. We can then get
   \begin{align*}
    & 
    \left|
        1 - \exp\left( 
        -\frac{n}{2}\left( 
            \left(S^*-\wh\theta\right) ^T 
            \left(\wt h^2 I_d + 2I^{-1}(\wh\theta) \right)^{-1} 
            \left(S^*-\wh\theta \right)
            \right)\right)
    \right| 
    \\
    & \leq
    e\cdot
    \Big\Vert \left( \wt h^2 I_d + 2I^{-1}(\wh\theta) \right)^{-1} \Big\Vert_{\rm op}
    \cdot n
    \Vert S^*({\Xn}) - \wh{\theta}(\Xn) \Vert^2\\
    &\lesssim (1+\|\theta^*\|)^{\alpha_1+\alpha_0\alpha_4} \cdot n\cdot  \Vert S^*({\Xn}) - \wh{\theta}(\Xn) \Vert^2.
    \end{align*}
Hence, noting that $ \frac{1}{\left|I^{-1}(\wh\theta)\right|}\leq  (1+\|\theta^*\|)^{d(\alpha_1+\alpha_0\alpha_4)}$, we can get 
\begin{equation*}
\begin{aligned}
        &\mmd^2_{\wt h}
    \left(
        N\left(0, I^{-1}(\wh{\theta})\right)
        ,
           N\left(\sqrt{n}(S^*-\wh\theta),I^{-1}(\wh\theta)\right)
    \right)\\
    &\lesssim   (1+\|\theta^*\|)^{(\alpha_1+\alpha_0\alpha_4)(1+d/2)}\cdot n \cdot \Vert S^*({\Xn}) - \wh{\theta}(\Xn) \Vert^2\\
    &\lesssim \frac{\log n}{n} (1+\|\theta^*\|)^{(8+d/2)(\alpha_1+\alpha_0\alpha_4)+2\alpha_0\alpha_4} .
\end{aligned}
\end{equation*}
For the second term, we have
\begin{equation*}
\begin{aligned}
     &\mmd^2_{\wt h}
    \Big(
        N\left(\sqrt{n}(S^* - \wh{\theta}), I^{-1}(\wh{\theta}) \right),
        N\left(\sqrt{n}(S^* - \wh{\theta}), \Sigma^*\right)
    \Big)\\
    &=\mmd^2_{\wt h}
    \Big(
        N\left(0_d, I^{-1}(\wh{\theta}) \right),
        N\left(0_d, \Sigma^*\right)
    \Big)\\
    &=\int_{\mb R^{d}}  \Big(\mb{E}_{\theta\sim   N\big(0_d, I^{-1}(\wh{\theta}) \big)}[\kappa_{\wt h/\sqrt{2}}(\theta,y)]-\mb{E}_{\theta\sim  N\big(0_d, \Sigma^*\big)}[\kappa_{\wt h/\sqrt{2}}(\theta,y)]\Big)^2\,\dd y\\
    &=\int_{\mb R^d}\bigg(\frac{1}{(2\pi)^{d/2}\big|\frac{\wt h^2}{2}I_d+I^{-1}(\wh\theta)\big|^{\frac{1}{2}}}\exp\Big(-\frac{1}{2}y^T\big(\frac{\wt h^2}{2}I_d+I^{-1}(\wh\theta)\big)^{-1}y\Big)
    \\
    &\qquad\qquad-
    \frac{1}{(2\pi)^{d/2}\big|\frac{\wt h^2}{2}I_d+\Sigma^*\big|^{\frac{1}{2}}}\exp\Big(-\frac{1}{2}y^T\big(\frac{\wt h^2}{2}I_d+\Sigma^*\big)^{-1}y\Big)
    \bigg)^2\,\dd y\\
    &\leq 2\, \Big(\frac{1}{(2\pi)^{d/2}\big|\frac{\wt h^2}{2}I_d+I^{-1}(\wh\theta)\big|^{\frac{1}{2}}}-  \frac{1}{(2\pi)^{d/2}\big|\frac{\wt h^2}{2}I_d+\Sigma^*\big|^{\frac{1}{2}}}\Big)^2 \int_{\mb R^d} \exp\Big(-y^T\big(\frac{\wt h^2}{2}I_d+\Sigma^*\big)^{-1}y\Big)\,\dd y\\
    &+ \int_{\mb R^d} \bigg(\exp\Big(-\frac{1}{2}y^T\big(\frac{\wt h^2}{2}I_d+I^{-1}(\wh\theta)\big)^{-1}y\Big)-\exp\Big(-\frac{1}{2}y^T\big(\frac{\wt h^2}{2}I_d+\Sigma^*\big)^{-1}y\Big)\bigg)^2\,\dd y \\
    & \quad \cdot \frac{2}{(2\pi)^d\big|\frac{\wt h^2}{2}I_d+I^{-1}(\wh\theta)\big|}.
    \end{aligned}
\end{equation*}
 Furthermore,  note that $I^{-1}(\theta^*)\preceq c_I^{-1} I_d$ and  $   \big\| I^{-1}(\theta^*)-\Sigma^*\big\|_{\rm F}\lesssim    (1+\|\theta^*\|)^{\alpha_5}\big(\varepsilon_{\theta}+\sqrt{\frac{\log n}{n}}\big)$,
 we have $\int_{\mb R^d} \exp\Big(-y^T\big(\frac{\wt h^2}{2}I_d+\Sigma^*\big)^{-1}y\Big)\,\dd y=\pi^{\frac{d}{2}}|\frac{\wt h^2}{2}I_d+\Sigma^*|^{\frac{1}{2}}=\m O(1)$ and
 \begin{equation*}
     \begin{aligned}
        & 2\, \Big(\frac{1}{(2\pi)^{d/2}\big|\frac{\wt h^2}{2}I_d+I^{-1}(\wh\theta)\big|^{\frac{1}{2}}}-  \frac{1}{(2\pi)^{d/2}\big|\frac{\wt h^2}{2}I_d+\Sigma^*\big|^{\frac{1}{2}}}\Big)^2 \int_{\mb R^d} \exp\Big(-y^T\big(\frac{\wt h^2}{2}I_d+\Sigma^*\big)^{-1}y\Big)\,\dd y\\
        &\lesssim \Big(\frac{1}{\big|\frac{\wt h^2}{2}I_d+I^{-1}(\wh\theta)\big|^{\frac{1}{2}}}-  \frac{1}{ \big|\frac{\wt h^2}{2}I_d+\Sigma^*\big|^{\frac{1}{2}}}\Big)^2 \\
        &\lesssim  (1+\|\theta^*\|)^{3d(\alpha_1+\alpha_0\alpha_4)}
        \|I^{-1}(\wh\theta)-\Sigma^*\|^2_{\rm F}\\
        &\lesssim  \varepsilon_{\theta}^2(1+\|\theta^*\|)^{2\alpha_5+3d(\alpha_1+\alpha_0\alpha_4)} .
     \end{aligned}
 \end{equation*}
Set $A=\frac{\wt h^2}{2}I_d+I^{-1}(\wh\theta)$ and $B=\frac{\wt h^2}{2}I_d+\Sigma^*$.
Their largest eigenvalues are bounded, their smallest eigenvalues are bounded below by an inverse power of $\log n$, and $\|A-B\|_{\rm F}$ is at most $\varepsilon_\theta$ times a power of $\log n$. Thus, for large $n$, the mean value theorem gives
\begin{equation*}
\begin{aligned}
 &\frac{1}{(2\pi)^d|A|}\int_{\mb R^d}
  \left(e^{-y^TA^{-1}y/2}-e^{-y^TB^{-1}y/2}\right)^2\,\dd y\\
 &\lesssim \frac{\|A^{-1}-B^{-1}\|_{\rm F}^2}{|A|}
  \int_{\mb R^d}e^{-y^TA^{-1}y/2}\|y\|^4\,\dd y\\
 &\lesssim |A|^{-1/2}\|A^{-1}-B^{-1}\|_{\rm F}^2\\
 &\lesssim (1+\|\theta^*\|)^{(\alpha_1+\alpha_0\alpha_4)(4+d/2)}
        \|I^{-1}(\wh\theta)-\Sigma^*\|_{\rm F}^2\\
 &\lesssim \varepsilon_\theta^2
 (1+\|\theta^*\|)^{(\alpha_1+\alpha_0\alpha_4)(4+d/2)+2\alpha_5}.
\end{aligned}
\end{equation*}
Combining all the pieces together, we can get for any $(\theta^*,\Xn)$ with $(\theta^*,\Xn)\in \wt{ A}$, it holds that  
\begin{equation*}
    \begin{aligned}
      & \mmd^2_{\wt h}
    \left(
        N\left(0, I^{-1}(\wh{\theta})\right)
        ,
         \big[\sqrt{n}(\theta - \wh{\theta})\big]_{\#} \mu_{W^*}\left(\cdot| S^*(\Xn) \right)
    \right)\\
    &\lesssim \frac{\log n}{n} (1+\|\theta^*\|)^{(8+d/2)(\alpha_1+\alpha_0\alpha_4)+2\alpha_0\alpha_4}+ \varepsilon_{\theta}^2(1+\|\theta^*\|)^{2\alpha_5+3d(\alpha_1+\alpha_0\alpha_4)} \\ & \quad+\varepsilon_{\theta}^2(1+\|\theta^*\|)^{(\alpha_1+\alpha_0\alpha_4)(4+d/2)+2\alpha_5}\\
    &\lesssim \frac{\log n}{n} (1+\|\theta^*\|)^{(8+3d)(\alpha_1+\alpha_0\alpha_4)+2\alpha_0\alpha_4+2\alpha_5}.
    \end{aligned}
\end{equation*}
Hence, 
\begin{equation*}
    \begin{aligned}
     &   \mathbb{E}_{\theta^* \sim \Pi_{\theta}}
\bE_{\Xn\sim \m P_{\theta^*}^{(n)}}
\Big[
\mmd^2_{\wt h}
\left(
    N(0, I^{-1}(\wh{\theta}(\Xn)))
    ,
     \big[\sqrt{n}(\theta - \wh{\theta})\big]_{\#} \mu_{W^*}(\cdot| S^*(\Xn))
\right)\\
&\qquad\qquad\qquad\qquad\quad
\cdot
\mathbf{1}((\theta^*,\Xn) \in \wt A) 
\Big]\\
&\lesssim \frac{\log n}{n}\mb{E}_{\theta^*\sim \Pi_{\theta}}\big[(1+\|\theta^*\|)^{(8+3d)(\alpha_1+\alpha_0\alpha_4)+2\alpha_0\alpha_4+2\alpha_5}\big]\\
&\lesssim \frac{\log n}{n}.
    \end{aligned}
\end{equation*}

\subsection{Proof of Lemma~\ref{lemmathetahat}}
Consider a fixed $(\theta^*,\Xn)\in \wt A$ and write $\wh\theta=\wh\theta(\Xn)$.  Note that
 \begin{equation*}
     {\rm KL}(\m P_{\theta^*}, \m P_{\wh \theta})=\mb{E}_{X\sim \m P_{\theta^*}} [l(X,\wh\theta)-l(X,\theta^*)]\geq c_3\min(\|\theta^*-\wh\theta\|^2,1).
 \end{equation*}
Using $\wh\theta\in \wt\Theta'$, we have
\begin{equation*}
\begin{aligned}
      &c_3\,\min(\|\theta^*-\wh\theta\|^2,1)\leq \mb{E}_{X\sim \m P_{\theta^*}} [l(X,\wh\theta)-l(X,\theta^*)]\\
      &\leq \mb{E}_{X\sim \m P_{\theta^*}} [l(X,\wh\theta)-l(X,\theta^*)]-n^{-1}\sum_{i=1}^n l(X_i,\wh\theta)+n^{-1}\sum_{i=1}^n l(X_i,\theta^*)\\
      &\leq  c\,  (\log n)^{\frac{1}{2}+\frac{\alpha_1}{\alpha_2}}\cdot(1+\|\theta^*\|)^{\alpha_0\alpha_4}\frac{1}{\sqrt{n}}  \Big(\frac{1}{n}+ \Vert \wh\theta -\theta^*\Vert \Big) ,
\end{aligned}
\end{equation*}
and therefore, $\|\theta^*-\wh\theta\|\lesssim \frac{1}{\sqrt{n}}(\log n)^{\frac{1}{2}+\frac{\alpha_1}{\alpha_2}}(1+\|\theta^*\|)^{\alpha_0\alpha_4}$. So when $n$ is sufficiently large, it holds that $\wh\theta$ is an interior point of $\wt\Theta'$ and  $\frac{1}{n}\sum_{i=1}^n \nabla_{\theta} l(X_i,\wh\theta)=0$. Next we remove the 
factor $(\log n)^{\frac{\alpha_1}{\alpha_2}}(1+\|\theta^*\|)^{\alpha_0\alpha_4}$ from the upper bound on $\|\theta^*-\wh\theta\|$. Define 
\begin{equation*}
    \wh\theta^\diamond= \theta^*-I(\theta^*)^{-1}\frac{1}{n}\sum_{i=1}^n \nabla_{\theta} l(X_i,\theta^*).
\end{equation*}
 Using $(\theta^*,\Xn)\in A_5$, we have $\|\wh\theta^\diamond-\theta^*\|\lesssim \sqrt{\frac{\log n}{n}}$. Moreover, it holds that 
 \begin{equation*}
 \begin{aligned}
       &\big\|\frac{1}{n}\sum_{i=1}^n \nabla_{\theta} l(X_i,\wh\theta^\diamond)\big\|\\
       &\leq\Big\|   \frac{1}{n}\sum_{i=1}^n \nabla_{\theta} l(X_i,\theta^*)+  \frac{1}{n}\sum_{i=1}^n {\rm Hess}_{\theta} l(X_i,\theta^*)(\wh \theta^\diamond-\theta^*)\Big\|+C\, (\log n)^{1+\frac{\alpha_0\alpha_4+\alpha_1}{\alpha_2}+\frac{\alpha_0}{\alpha_3}} n^{-1}\\
       &\lesssim (\log n)^{1+\frac{\alpha_0\alpha_4+\alpha_1}{\alpha_2}+\frac{\alpha_0}{\alpha_3}} n^{-1}.
 \end{aligned}
 \end{equation*}
Then note that $\frac{1}{n}\sum_{i=1}^n \nabla_{\theta} l(X_i,\wh\theta)=0$,  $\|\theta^*-\wh\theta\|\lesssim \frac{1}{\sqrt{n}}(\log n)^{\frac{1}{2}+\frac{\alpha_1}{\alpha_2}}(1+\|\theta^*\|)^{\alpha_0\alpha_4}$ and   $\|\wh\theta^\diamond-\theta^*\|\lesssim \sqrt{\frac{\log n}{n}}$, we have 
\begin{equation*}
    \begin{aligned}
        \big\| \frac{1}{n}\sum_{i=1}^n \nabla_{\theta} l(X_i,\wh\theta^\diamond)- \frac{1}{n}\sum_{i=1}^n {\rm Hess}_{\theta} l(X_i,\wh\theta^\diamond) (\wh\theta^\diamond-\wh\theta)\big\|\lesssim (\log n)^{\frac{\alpha_0\alpha_4+\alpha_1}{\alpha_2}+\frac{\alpha_0}{\alpha_3}}  \|\wh\theta^\diamond-\wh\theta\|^2.
    \end{aligned}
\end{equation*}
 Then using 
\begin{equation*}
    \begin{aligned}
      &  \Big\|\frac{1}{n}\sum_{i=1}^n {\rm Hess}_{\theta} l(X_i,\wh\theta^\diamond) -I(\theta^*)\Big\|_{\rm F}\\
        &\leq    \Big\|\frac{1}{n}\sum_{i=1}^n {\rm Hess}_{\theta} l(X_i,\wh\theta^\diamond) -\frac{1}{n}\sum_{i=1}^n {\rm Hess}_{\theta} l(X_i,\theta^*)\Big\|_{\rm F} + \Big\| \frac{1}{n}\sum_{i=1}^n {\rm Hess}_{\theta} l(X_i,\theta^*)-I(\theta^*)\Big\|_{\rm F}\\
        &\lesssim (\log n)^{\frac{1}{2}+\frac{\alpha_0\alpha_4+\alpha_1}{\alpha_2}+\frac{\alpha_0}{\alpha_3}} n^{-\frac{1}{2}}.
    \end{aligned}
\end{equation*}
Combined with $I(\theta^*)\succeq c_I I_d$, we have $\frac{1}{n}\sum_{i=1}^n {\rm Hess}_{\theta} l(X_i,\wh\theta^\diamond)\succeq \frac{c_I}{2} I_d$ and 
\begin{equation*}
    \begin{aligned}
       & \big\|\frac{1}{n}\sum_{i=1}^n \nabla_{\theta} l(X_i,\wh\theta^\diamond)\big\|\geq \frac{c_I}{2} \|\wh\theta^\diamond-\wh\theta\|-C
    \,(\log n)^{\frac{\alpha_0\alpha_4+\alpha_1}{\alpha_2}+\frac{\alpha_0}{\alpha_3}}  \|\wh\theta^\diamond-\wh\theta\|^2\geq \frac{c_I}{4} \|\wh\theta^\diamond-\wh\theta\|,
    \end{aligned}
\end{equation*}
where the last inequality uses $ \|\wh\theta^\diamond-\wh\theta\|\lesssim \|\wh\theta^\diamond-\theta^*\|+\|\wh\theta-\theta^*\|\lesssim  \frac{1}{\sqrt{n}}(\log n)^{\frac{1}{2}+\frac{\alpha_1}{\alpha_2}}(1+\|\theta^*\|)^{\alpha_0\alpha_4}$. Therefore,
\begin{equation*}
    \|\wh\theta^\diamond-\wh\theta\|\lesssim \big\|\frac{1}{n}\sum_{i=1}^n \nabla_{\theta} l(X_i,\wh\theta^\diamond)\big\| 
       \lesssim (\log n)^{1+\frac{\alpha_0\alpha_4+\alpha_1}{\alpha_2}+\frac{\alpha_0}{\alpha_3}} n^{-1}.
\end{equation*}
Combined with $\|\wh\theta^\diamond-\theta^*\|\lesssim \sqrt{\frac{\log n}{n}}$, we can get $\|\wh\theta-\theta^*\|\lesssim \sqrt{\frac{\log n}{n}}.$

\end{document}